\documentclass[unnumsec,webpdf,contemporary,large]{oup-authoring-template}%

\graphicspath{{Fig/}}

\usepackage[a4paper]{geometry}
\usepackage{amssymb}
\usepackage{graphicx}

\usepackage{algorithm}
\usepackage{algpseudocode}

\usepackage{subfig}

\theoremstyle{thmstyleone}%
\theoremstyle{thmstyletwo}%

\theoremstyle{thmstylethree}%
\newtheorem{definition}{Definition}

\def\argmin{\mathop{\rm argmin}}

\def \ie {{\it i. e.}}
\def \real {\mathbb{R}}
\def \ltwo {\mathbb{L}^2}
\newcommand{\norm}[1]{\left\lVert#1\right\rVert}

\begin{document}

\journaltitle{}
\DOI{DOI HERE}
\copyrightyear{}
\pubyear{}
\access{Advance Access Publication Date: Day Month Year}
\appnotes{Paper}

\firstpage{1}


\title[Peak-Persistence Diagrams]{Peak-Persistence Diagrams for Estimating Shapes and Functions from Noisy Data}

\author[1]{Woo Min Kim}
\author[2]{Sutanoy Dasgupta}
\author[1,$\ast$]{Anuj Srivastava}

\authormark{Kim et al.}

\address[1]{\orgdiv{Dept of Statistics}, \orgname{Florida State University}, \orgaddress{\street{Tallahassee}, \postcode{32306}, \state{FL}, \country{USA}}}
\address[2]{\orgdiv{Dept of Statistics}, \orgname{Texas A\&M University}, \orgaddress{\street{College Station}, \postcode{77843}, \state{TX}, \country{USA}}}

\corresp[$\ast$]{Corresponding author. \href{email}{anuj@stat.fsu.edu}}

\received{Date}{0}{Year}
\revised{Date}{0}{Year}
\accepted{Date}{0}{Year}


%



\abstract{Estimating signals underlying noisy data is a significant problem in statistics and engineering. Numerous estimators are available in the literature, depending on the observation model and estimation criterion. This paper introduces a framework that estimates the shape of the unknown signal and the signal itself. The approach utilizes a peak-persistence diagram (PPD), a novel tool that explores the dominant peaks in the potential solutions and estimates the function's shape, which includes the number of internal peaks and valleys. It then imposes this shape constraint on the search space and estimates the signal from partially-aligned data. This approach balances two previous solutions: averaging without alignment and averaging with complete elastic alignment. From a statistical viewpoint, it achieves an optimal estimator under a model with both additive noise and phase or warping noise. We also present a computationally-efficient procedure for implementing this solution and demonstrate its effectiveness on several simulated and real examples. Notably, this geometric approach outperforms the current state-of-the-art in the field.}

\maketitle

\section{Introduction} \label{sec:Introduction}
Analyzing noisy data to estimate underlying signals or some relevant properties is a fundamental problem in statistics and engineering. One models the observed data as an actual underlying signal corrupted by noise, and the goal is to use data to estimate that signal. The choice of an estimator depends on the observation model, the noise distribution, the data structure, and the optimization criterion. The classical approach to function estimation is to identify a function space (typically a Hilbert space with an orthonormal basis), impose an objective function, and optimize it over the function space. Our approach is slightly different. We will focus on the shapes of functions and introduce a novel tool, the peak-persistence diagram (PPD), to search for the estimate. This PPD enables us to estimate the shape of an unknown function first, followed by a shape-constrained estimation of the function itself.   

\begin{figure*}[ht]
 \centering
 \includegraphics[height=2.54in]{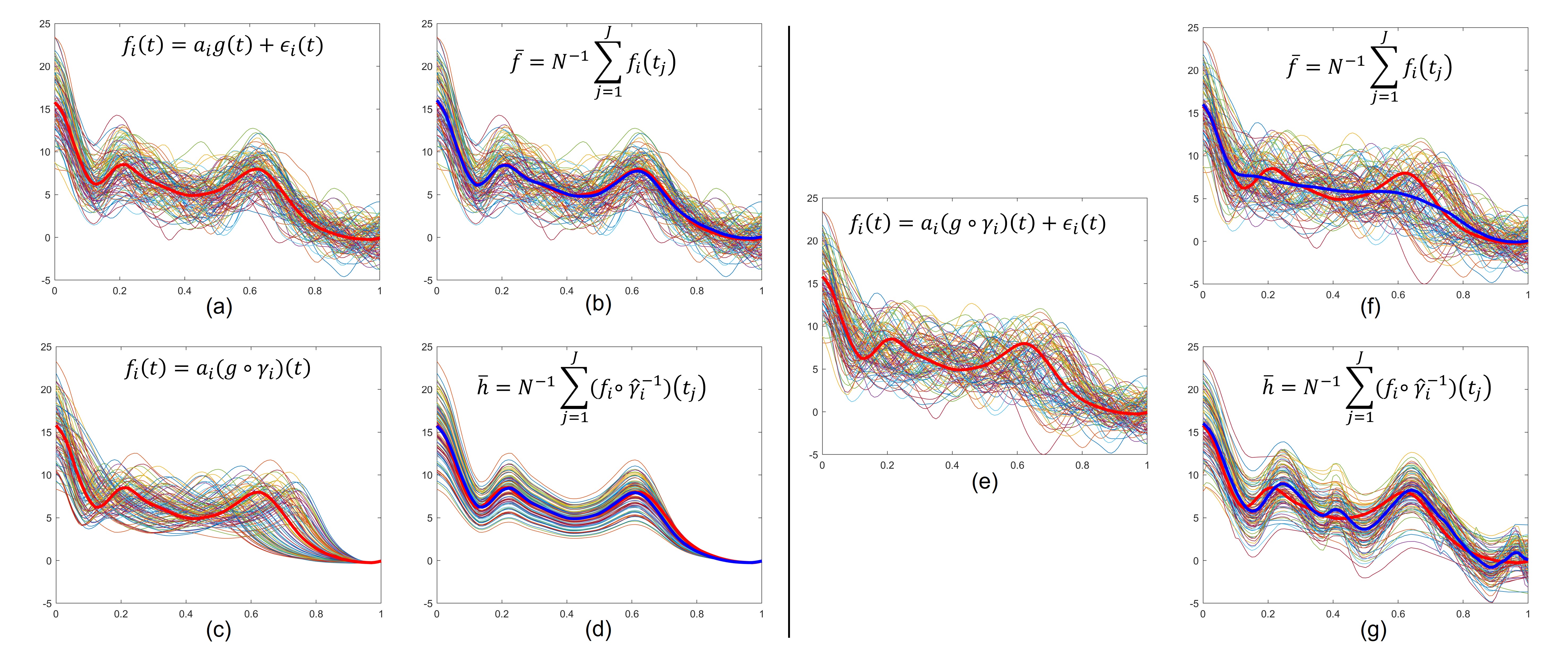}
 \caption{Bold red lines denote $g$, blue lines in (b),(f) denote $\bar f$, and blue lines in (d),(g) denote $\bar h$. Functions in (a) have additive noise only, and those in (c) have phase noise only. Functions in (e) contain both additive and phase noises. The estimations in (b) and (d) perform well, whereas those in (f) and (g) do not.}
 \label{fig:intro}
\end{figure*}

We start from a traditional perspective on the function estimation problem and later make a case for a geometric (shape-based) approach. A simple, traditional model for function estimation is the {\bf additive noise model}:
\begin{equation} \label{eq:additive-model}
f_i(t) = a_i g(t) + \epsilon_i(t)\ , i=1,2,\dots, n\ ,
\end{equation}
where $g \in {\mathcal F}$ is the unknown signal, $\epsilon_i \in {\mathcal F}$ is the zero mean, independent noise and $a_i \in \real$ is an independent random variable with mean one. Here ${\mathcal F}$ denotes a functional space of interest; most commonly, ${\mathcal F}$ is the Hilbert space of square-integrable functions on an interval. In a simple case, with dense time samples, an estimate of $g$ is given by the {\it cross-sectional mean} $\bar{f} = \frac{1}{n} \sum_{i=1}^n f_i$, because  $\frac{1}{n} \sum_{i=1}^n \epsilon_i \rightarrow 0$ and $\frac{1}{n} \sum_{i=1}^n a_i \rightarrow 1$. The top left part of Fig.~\ref{fig:intro} shows a pictorial illustration of this estimation. Plot (a) shows the data $\{f_i\}$ and the true function $g$ (in red), and (b) shows their cross-section mean superimposed in blue. In case of sparse data or other challenges, one can utilize a smooth basis for $g$ or add a roughness penalty (or a prior) to the standard least-squares objective function (\citet{ramsay1997functional,fan1996local,CHOUDHURI2005,green1994}). 

In recent years, there has been a growing recognition of a different kind of noise or variability in functional data, the so-called {\it phase} or {\it compositional} noise. The observation model for this noise is given by the {\bf phase noise model}:
\begin{equation} \label{eq:phase-model}
f_i(t) =  g(\gamma_i(t)) \ , i=1,2,\dots, n\ ,
\end{equation}
where $\gamma_i$s are random time-warping functions of the domain. (A precise mathematical definition of time-warping functions is presented later.) To have identifiable $g$, one assumes that the phase noise has identity mean, \ie, $\frac{1}{n}\sum_{i=1}^n \gamma_i(t) \rightarrow t$.  One can verify that, in this case, the cross-sectional mean $\bar{f}$ does not directly estimate $g$. This is because $\frac{1}{n} \sum_{i=1}^n g(\gamma_i(t))$ does not converge to $g(t)$, in general, despite the fact that $(\frac{1}{n} \sum_{i=1}^n \gamma_i(t)) \mapsto t$. The quantity $\lim_{n \rightarrow \infty} \frac{1}{n} \sum_{i=1}^n g(\gamma_i(t))$ can be approximated by $\int g(t + x) \pi(x)~dx$, where $\pi$ is a normal density, and thus, the result is a Gaussian blurring of $g$. The estimation solution comes from solving for the $\gamma_i$s explicitly and then performing an alignment of given data according to $\{\tilde{f}_i = f_i \circ \hat{\gamma}_i^{-1}\}$. This alignment is also called {\it phase-amplitude separation} in the literature (\citet{ramsayli1998,srivastava2016functional,srivastava2011registration,marron2015functional}). One can then estimate $g$ using cross-sectional averaging of the {\it aligned functions} $\frac{1}{n} \sum_{i=1}^n \tilde{f}_i$ (~\citet{kurtek-wu-NIPS}). We will call this average quantity the {\it fully-elastic} mean. The bottom left part of Fig.~\ref{fig:intro} shows a pictorial illustration of this estimation. Plot (c) shows the data $\{f_i\}$ and the true function $g$ (in red), and (d) shows the aligned functions $\{\tilde{f}_i\}$ and their mean superimposed in blue. While there are several approaches in the literature for functional alignment (see, e.g., \cite{kneip-ramsay:2008,muller-JASA:2004,marron2015functional}), a particularly efficient method is based on the nonparametric Fisher-Rao metric~(\citet{srivastava2016functional,srivastava2011registration}). It uses a square-root representation of functions and a dynamic programming (DP) algorithm to estimate $\{\hat{\gamma}_i\}$ and is remarkably successful in aligning peaks and valleys of $f_i$s. The example shown in (d) uses this method. 

\subsection{Problem Specification}
This paper focuses on a more general problem where {\it both} additive and phase noise are present in the data. The observation model is now given by the {\bf additive and phase noise model}:
\begin{equation} \label{eq: model specification}
f_i(t)=a_i(g\circ\gamma_i )(t)+\epsilon_i (t), \ i=1,2,\dots, n\ .
\end{equation}
The components of the model remain the same as earlier. Next, we discuss the current state-of-the-art in handling this model. 

\subsection{Past Methods}
There has been some research on function estimation under the model specified in Eqn.~\ref{eq: model specification}. 
A seemingly natural idea for estimating $g$ is the least square solution: 
\begin{equation} \label{eq:penalized-align-L2}
\argmin_{g \in {\mathcal F}} \left( \sum_{i=1}^n \min_{a_i \in \real, \gamma_i \in \Gamma} \left(\| f_i - a_i (g \circ \gamma_i) \|^2 + \kappa {\mathcal R}(\gamma_i) \right) \right) \ ,
\end{equation}
where ${\mathcal R}$ denotes a roughness penalty on $\gamma_i$s. 
However, this optimization has several problems. Firstly, for $\kappa = 0$, the optimization over $\gamma_i$s is degenerate; this phenomenon is called the {\it pinching effect} (\cite{ramsay1997functional,srivastava2016functional}). While adding a penalty term (setting $\kappa > 0$) may avoid degeneracy, it does not entirely solve the problem. It limits the search space for a solution and can have unintended consequences. Also, determining the optimal value for the penalty parameter $\kappa$ in Eqn.~\ref{eq:penalized-align-L2} is known to be a challenging task. Different values of $\kappa$ may lead to widely different solutions (examples are provided later in Section \ref{sec: l2 alignment}).

Some past papers (\cite{kneip-ramsay:2008,ramsayli1998,muller-biometrika:2008,VI-registration}) have used Eqn.~\ref{eq:penalized-align-L2} but with a focus on aligning functional data. \cite{marron2015functional} provide a survey of papers on functional alignment. Some others have studied a mixed-effects model by adding a random effect term to Eqn.~\ref{eq: model specification}~(\cite{RAKET20141,Claeskens-etal:EJS:2021}).
A similar problem has been investigated in the registration of images~(\cite{trouve-JRSSB:2007,SIMPSON20122438}), albeit with focus on preserving visual features. Almost all these papers use the $\ltwo$ objective function and must balance the two terms in Eqn.\ref{eq:penalized-align-L2}.
Also, some papers have pursued a Bayesian approach for the alignment of given data $\{f_i\}$ by imposing a prior on $\gamma_i$s~(\cite{kurtek-EJS:2017}). Some papers advocate smoothing the given functions and then computing averages (\citet{cheng2013, Lu2017, telesca2008}). However, these papers do not explicitly estimate the signal $g$. One can also impose a prior on $g$ to formulate a Bayesian solution, as in \cite{matuk2021,horton2020}. This approach requires specific prior information about $g$ to be effective. 

We are going to pursue a geometric approach and discuss related literature. 
Some papers have studied the estimation of the so-called Procrustes means of "shapes" of Euclidean curves \citep{elasticProcrustes2022}, but they do not provide an estimator of $g$. 
There also exists literature on shape-constrained density estimation but often restricted to a narrow shape class. This large body of work started with \cite{Grenander1956-nt} and followed by several (\cite{wang-berger:2016,Cheng1999-up,Wegman1970-ux,doss-wellner:2016}). This effort is restricted to unimodal or log-concave shape classes rather than general shapes. Please refer to the special issue~(\cite{samworth-bodhi}) for a recent overview of that field. We note that~\cite{dasgupta2021} generalized the problem to general shape classes but still restricted the estimation to densities. The current paper looks at general functions under arbitrary shape constraints. 

Lastly, we mention sparse literature on estimating the number of modes of probability density underlying given data. There are some papers (see e.g.~\cite{minnotte-annals:1997}) that study estimation of multiple modes in density functions but most of the past literature is focused on the narrow case of unimodal functions. Our problem differs from density estimation in that we are dealing with functional data. 

\subsection{Our Approach}
Our approach adopts a geometric perspective that focuses on function shapes. It is motivated by the fact that Eqns.~\ref{eq:additive-model},~\ref{eq:phase-model} are special cases of Eqn.~\ref{eq: model specification}. We aim to strike a balance between the non-elastic and fully elastic, as neither of these solutions is satisfactory. The cross-sectional mean over-smooths the data, while the fully-elastic mean aligns even the noise artifacts, resulting in spurious peaks and valleys in the estimate. The right part of Fig.~\ref{fig:intro} shows an example:  For the data in (e), plot (f) shows the non-elastic mean, and plot (g) shows the elastic mean. We argue that a good solution lies between these two extremes and is adaptive. To achieve this, we propose using partial alignments that control the elasticity of the functions during alignment via a parameter $\lambda \in \real_+$. The resulting partially-aligned functions are denoted $\tilde{f}_{\lambda,i}$, and their mean is denoted $\hat{g}_{\lambda} = \sum_{i=1}^n \tilde{f}_{\lambda,i}$. An important consideration is choosing $\lambda$ adaptive to the given data. Additionally, we must address the degeneracy from the pinching effect. To tackle these issues, we apply shape analysis techniques. Our approach varies the elasticity parameter to estimate the shape of $g$ and $g$ as follows: 
\begin{enumerate}
        \item {\bf Peak-Persistence Diagram} (PPD): The objective of this stage is to estimate the shape of $g$ by identifying its geometric features which remain invariant to phase noise. A novel tool called a Peak Persistence Diagram (PPD) has been developed for this purpose. When we change the parameter $\lambda$, the shape of the average function $\hat{g}_{\lambda}$ changes. Typically, the geometric features, such as peaks and valleys, are smoothed out as $\lambda$ increases. Therefore, a good choice of $\lambda$ can be identified by studying the persistence of the internal peaks of $\hat{g}_{\lambda}$ versus $\lambda$ and selecting the most persistent peaks. A graphic display of this persistence is called a Peak Persistence Diagram or PPD.  A set of heuristics is used to define criteria for selecting the most persistent peaks in a PPD. This process yields three quantities: an optimal $\lambda^*$, the number $m$ of persistent internal peaks, and ${ \tilde{f}_{\lambda^, i} }$, which is the partially aligned data for the optimal weight $\lambda^*$. 
    \item {\bf Shape-Constrained Function Estimation}: 
   The preceding step generates a partially-aligned mean $\hat{g}_{\lambda^*}$. This is an element of the correct shape class, but may not be the optimal estimate of $g$ in a precise sense. Therefore, in the second step, we aim to find the optimal element in the chosen class. We constrain the estimate of $g$ to have precisely $m$ internal peaks and utilize a geometric approach to perform a shape-constrained estimation of $g$. The estimation process involves minimizing the squared error over the appropriate constraint space ${\mathcal F}_m$, using the data ${ \tilde{f}_{\hat{\lambda}, i} }$. Here, ${\mathcal F}_m$ refers to the set of all elements of ${\mathcal F}$ that have the correct shape, \ie, $m$ internal peaks. This approach differs from previous shape-constrained estimators since it provides a penalized-MLE in ${\mathcal F}_m$, rather than an arbitrary element of ${\mathcal F}_m$.
\end{enumerate}
In this paper we have focused on only the {\it internal} peaks and valleys of $g$. The potential peaks at the boundaries are also relevant but they can be detected separately using simple tests and are ignored here. 

\section{Penalized \texorpdfstring{$\mathbb{L}^2$}{L2} Estimator}\label{sec: l2 alignment}

\begin{figure*}[h]
    \centering
    \hspace*{-0.125in}
    \subfloat[]{\includegraphics[width = 1.3in]{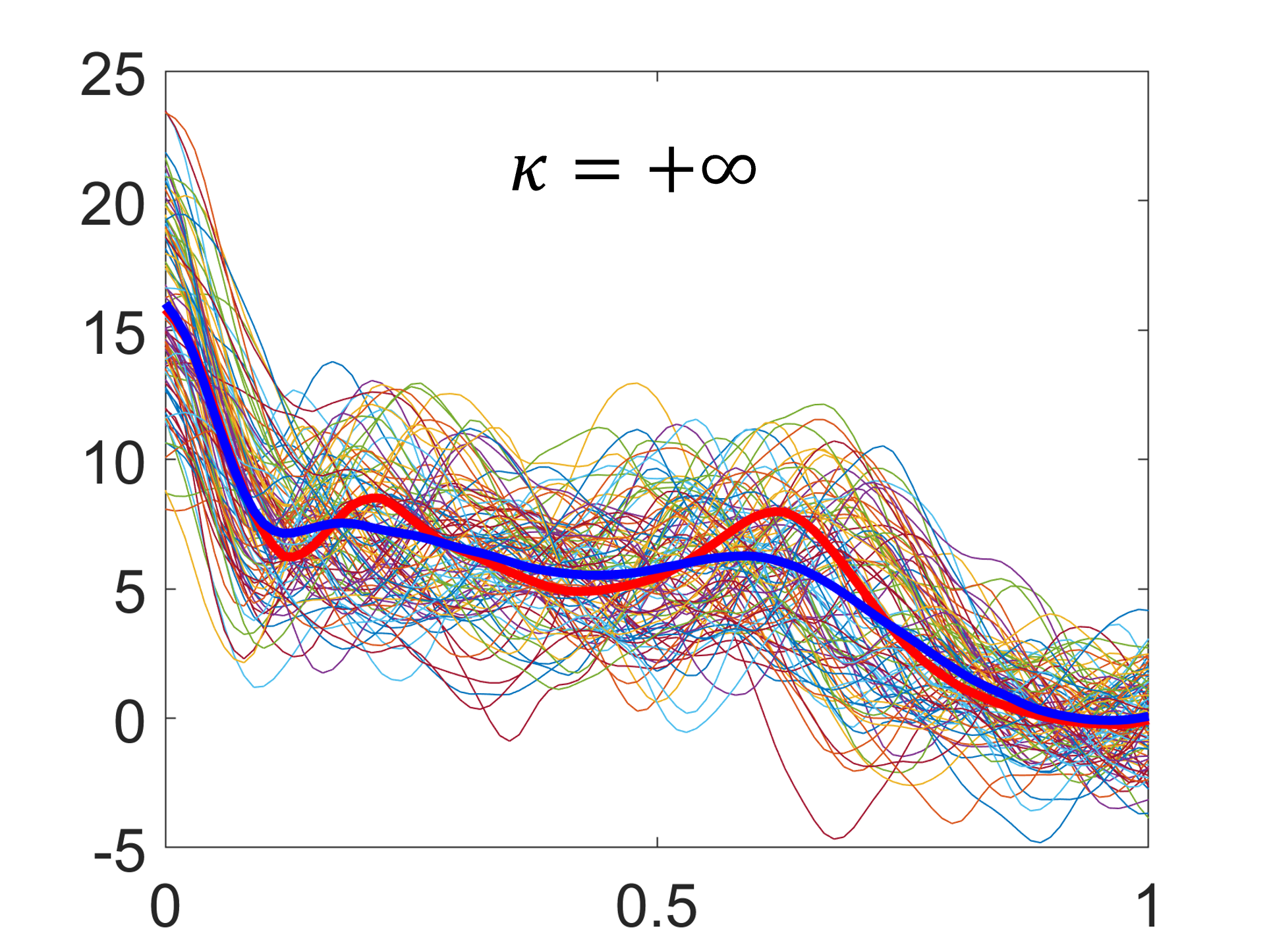}}
    \hspace*{-0.125in}
    \subfloat[]{\includegraphics[width = 1.3in]{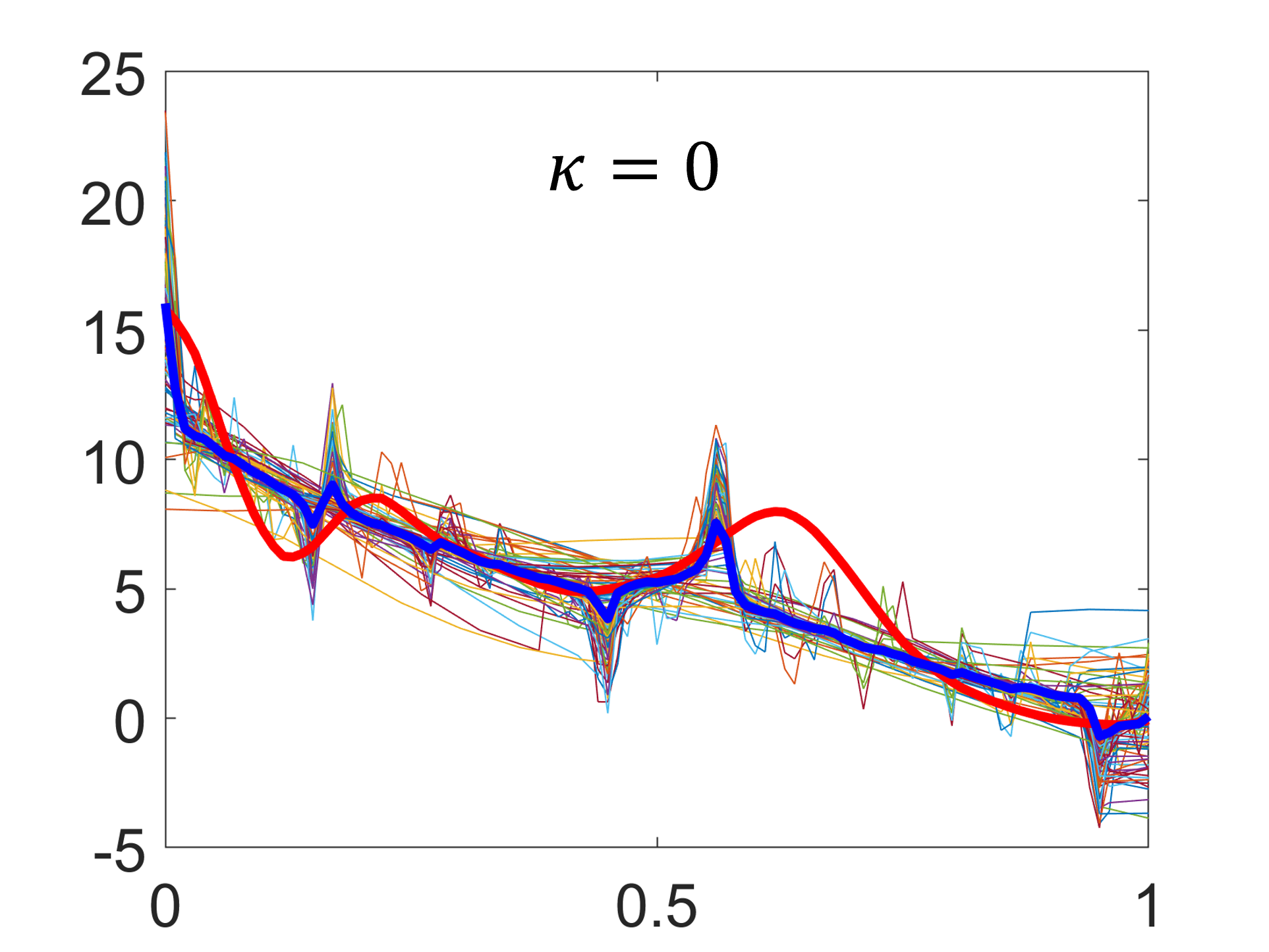}}
    \hspace*{-0.125in}
    \subfloat[]{\includegraphics[width = 1.3in]{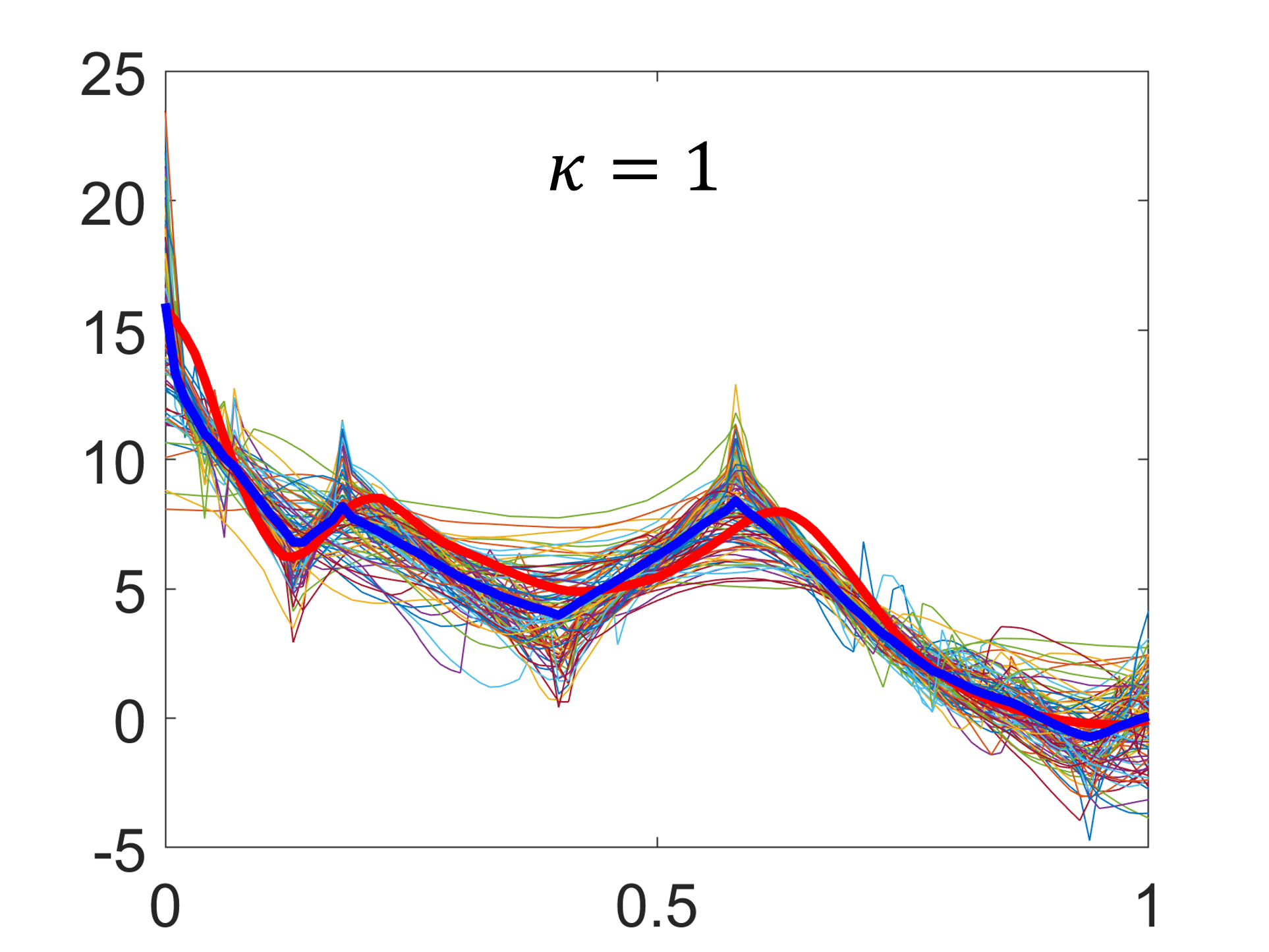}}
    \hspace*{-0.125in}
    \subfloat[]{\includegraphics[width = 1.3in]{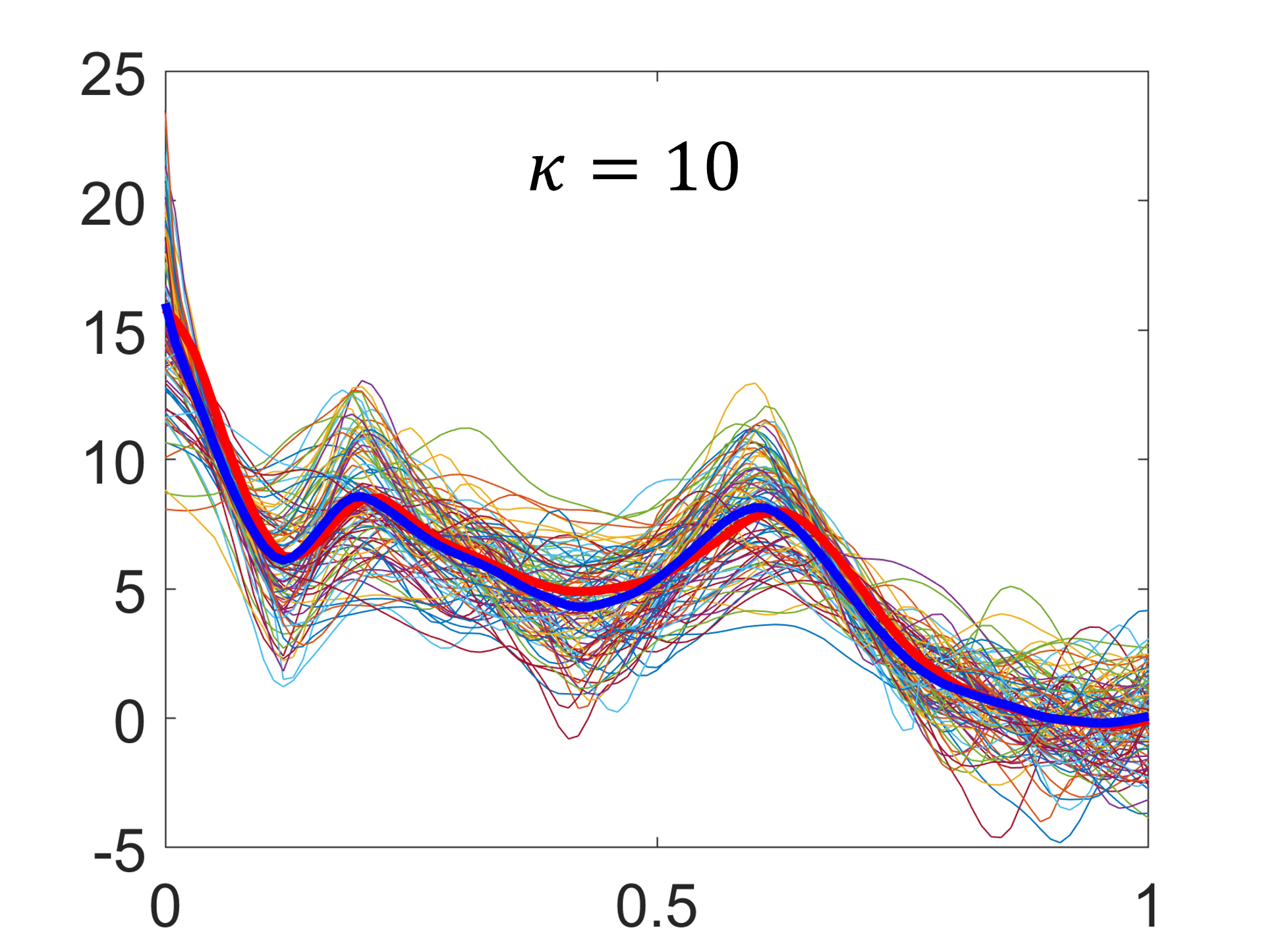}} 
    \hspace*{-0.125in}
    \subfloat[]{\includegraphics[width = 1.3in]{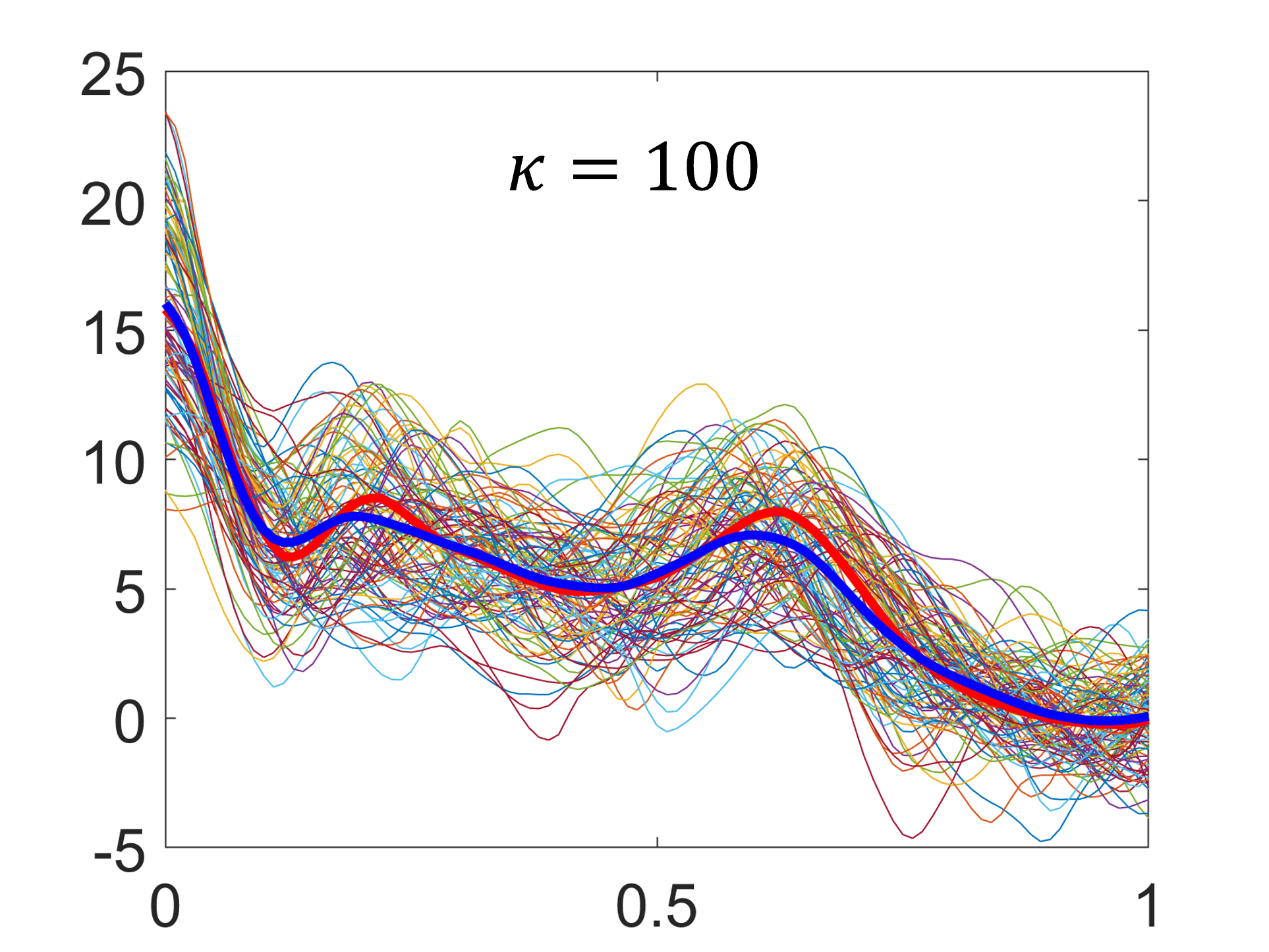}}
    \caption{Alignments of given functions and estimates of $g$ according to Eqn.~\ref{eq:penalized-align-L2}. Plot (a) show the original functions and their mean (blue). Plots (b)-(e) show the true $g$ (red), the partially-aligned set, and the estimates  (blue curves) for different $\kappa$.}
    \label{fig:partial-alignment-L2}
\end{figure*}

Before we lay out the proposed framework, we further elaborate on the issues facing a solution based on Eqn.~\ref{eq:penalized-align-L2}. There are two main issues: (1) The problem is degenerate for $\kappa = 0$, giving rise to the {\it pinching effect}, and (2) How should one set the value of $\kappa$ to avoid over smoothing and pinching? Although there are several ideas in the literature on choosing the smoothing parameter, 
pinching makes this selection complicated. One typically needs higher values of $\kappa$ to avoid pinching but that can result in oversmoothing. We illustrate this with an example using the first order roughness penalty $R(\gamma_i) = \int (1 - \sqrt{\dot{\gamma}_i(t)})^2dt$. (The detailed algorithm is presented in \cite{srivastava2016functional}.) Fig.~\ref{fig:partial-alignment-L2} (a) displays the given functions $\{ f_i\}$ and their cross-sectional mean $\bar{f}$ (depicted in blue). The red curves represent the ground truth, $g$, and the blue curves in plots (b)-(e) denote the cross-section mean of aligned curves, denoted by $\hat g_{L^2}$. At $\kappa = 0 \sim 1$, one can see the pinching effects in the estimate, but as $\kappa$ increases, the estimate gets smoother. This example shows that a carefully chosen $\kappa$ is needed to reach a good estimate of $g$. The challenge is finding an automated technique for optimal $\kappa$ while avoiding pinching.

Our shape-based framework differs from Eqn.~\ref{eq:penalized-align-L2} in that the formulation is developed using the Fisher-Rao distance rather than the $\ltwo$ distance, and it avoids the pinching problem altogether. However, the issue of choosing the smoothing parameter remains, and we tackle it using PPDs. For the remainder of this paper, 
we will assume that the functional data is defined on a compact interval $I$ and ${\mathcal F}$ is the set of all absolutely-continuous, real-valued functions on $I$. We will consider the model stated in Eqn.~\ref{eq: model specification}, 
with following independent components: (1) Scaling noise: $a_i\in \mathbb{R}^+ \sim N(1,\sigma_a^2)$;
(2) Additive noise: $\epsilon_i\in {\mathcal F}$, a random function with mean $E[\epsilon_i(t)] = 0$ for all $t \in I$;
(3) Phase noise: $\gamma_i\in \Gamma$ (where $\Gamma$ is defined below),  with the mean $E[\gamma_i(t)] = \gamma_{id}(t) = t$.
Given a set of independent observations $\{f_i\}$, our goal is to estimate the number of internal peaks in $g$ and the function $g$ itself.

\section{Step 1: Shape Estimation Using PPDs} \label{sec: step1}

The goal here is to estimate the shape of $g$, \ie, estimate the number of internal peaks and valleys in $g$, from the given $\{f_i\}$ and form an initial estimate of $g$. 
\\

\noindent{\bf Background Material}: We summarize a mechanism for partial elastic alignments of functions under the Fisher-Rao metric and refer the reader to ~\cite{srivastava2016functional} for details. For alignment purposes, a function $f \in {\mathcal F}$ is represented by its square-root velocity function (SRVF): 
$q(t) = \mbox{sign}(\dot{f}(t)) \sqrt{|\dot{f}(t)|}$.
Let $\Gamma$ denote the group of all boundary-preserving diffeomorphisms of $I$. Any element $\gamma \in \Gamma$ is a smooth function with a smooth inverse and preserves the boundaries of $I$. The set $\Gamma$ forms a group under composition, \ie, for any $\gamma_1, \gamma_2 \in \Gamma$, we have $\gamma_1 \circ \gamma_2 \in \Gamma$. This group has an identity element $\gamma_{id}(t) = t$. Let a function $f \in {\mathcal F}$ be time-warped by any $\gamma \in \Gamma$, resulting in $f \circ \gamma$. The SRVF of the warped function is given by $(q \circ \gamma) \sqrt{\dot{\gamma}}$, and we will denote it by $q \star \gamma$ for brevity. 

With this setup, we can solve for elastic alignment of any two functions $f_1, f_2 \in {\mathcal F}$ as follows. Let $q_1, q_2$ denote the SRVFs of $f_1, f_2$, respectively. Then, the optimal time warping to align $f_2$ to $f_1$ is given by: 
$\gamma^* = \argmin_{\gamma \in \Gamma} \| q_1 - (q_2 \star \gamma)\|^2$,
where $\| \cdot \|$ denotes the $\ltwo$ norm. However, if we want to penalize the level of elasticity in their alignment, we can do so using: 
$\argmin_{\gamma \in \Gamma} \left( \| q_1 - (q_2 \star \gamma)\|^2 + \lambda {\mathcal R}(\gamma)  \right)$.
The term ${\mathcal R}(\gamma)$ denotes a penalty on the roughness of $\gamma$ and forces it to be close to the identity element $\gamma_{id}$. (In this paper, we have chosen the first-order penalty ${\mathcal R} = \| 1 - \sqrt{\dot{\gamma}}\|^2$, but one can use other penalties instead.) The constant $\lambda > 0$ controls the amount of elasticity in the alignment: $\lambda = 0$ is fully elastic and $\lambda = \infty$ is non elastic. To align multiple functions, say $f_1, f_2, \dots, f_n$, let $q_1, q_2, \dots, q_n$ denotes their respective SRVFs. Then, a joint alignment of these functions is performed using the following optimization: 
\begin{equation} \label{eq:penalized-align}
\min_{q \in \ltwo} \left[ \sum_{i=1}^n \left( \min_{\gamma_i\in \Gamma} \left( \| q - (q_i \star \gamma_i)\|^2 + \lambda {\mathcal R}(\gamma_i) \right) \right) \right] \ .
\end{equation}
We can rearrange this equation using the facts that $\|q_1 - q_2 \star \gamma\| = \| q_1 \star \gamma^{-1} - q_2\|$, for all $q_1, q_2 \in \ltwo$ and $\gamma \in \Gamma$, and that $\Gamma$ is a group. This results in: 
$$
\min_{q \in \ltwo} \left[ \sum_{i=1}^n \left( \min_{\gamma_i\in \Gamma} \left( \| q_i - (q \star \gamma_i)\|^2 + \lambda {\mathcal R}(\gamma_i) \right) \right) \right] \ ,
$$
which can be directly compared to Eqn.~\ref{eq:penalized-align-L2}. (Note that we ignore the optimization over $a_i$s here as they average out in any mean-based estimation.) The main difference between the two solutions lies in the use of SRVFs and the invariance properties of the elastic metric. Fundamentally, the difference comes from the fact that $\|q\| = \| q \star \gamma \|$, for all $q \in \ltwo$ and $\gamma \in \Gamma$, while $\|f\| \neq \| f \circ \gamma \|$ in general.

\begin{figure*}[ht]
    \centering
    \hspace*{-0.125in}
    \subfloat[]{\includegraphics[width = 1.3in]{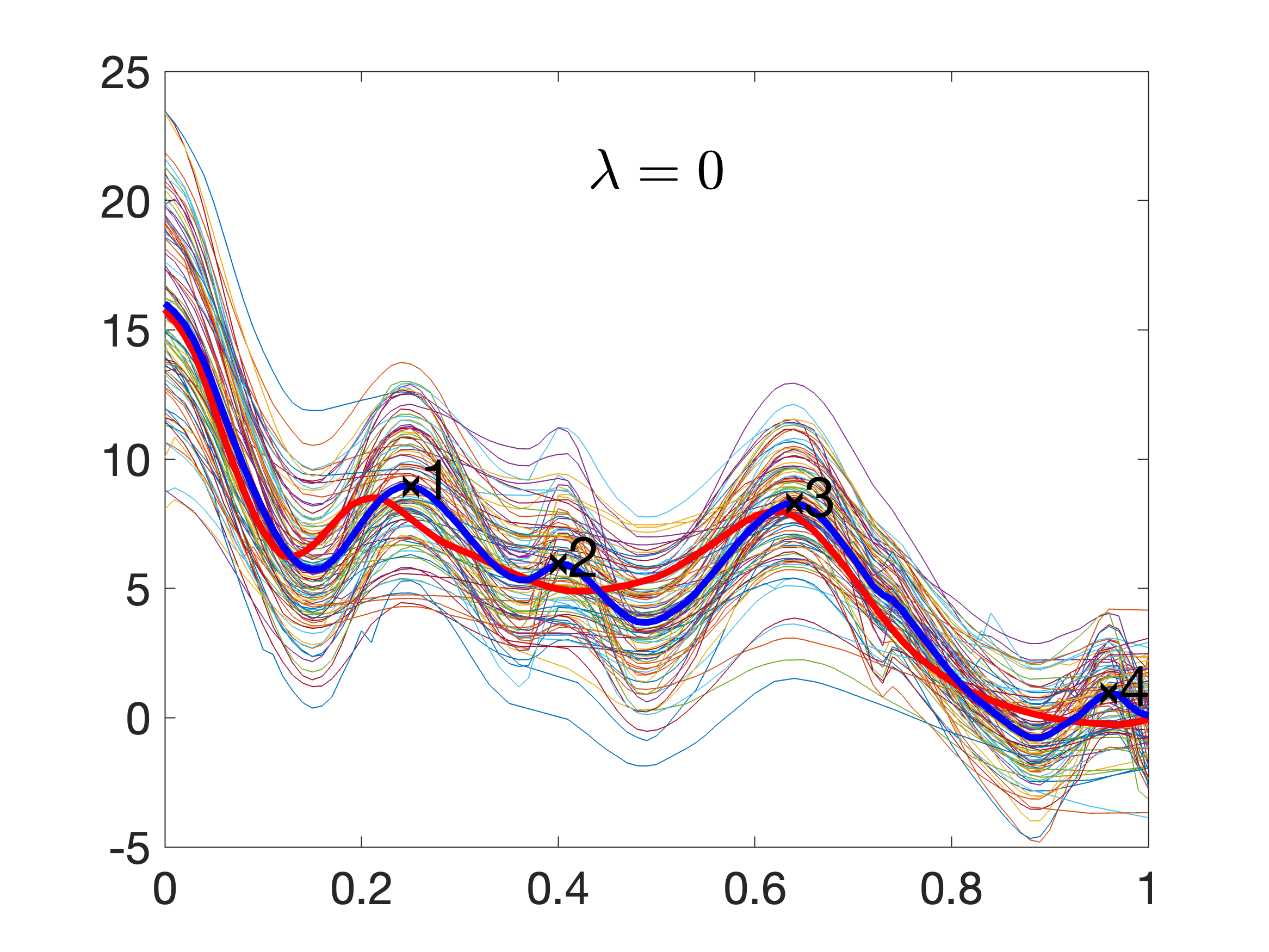}}
    \hspace*{-0.125in}
    \subfloat[]{\includegraphics[width = 1.3in]{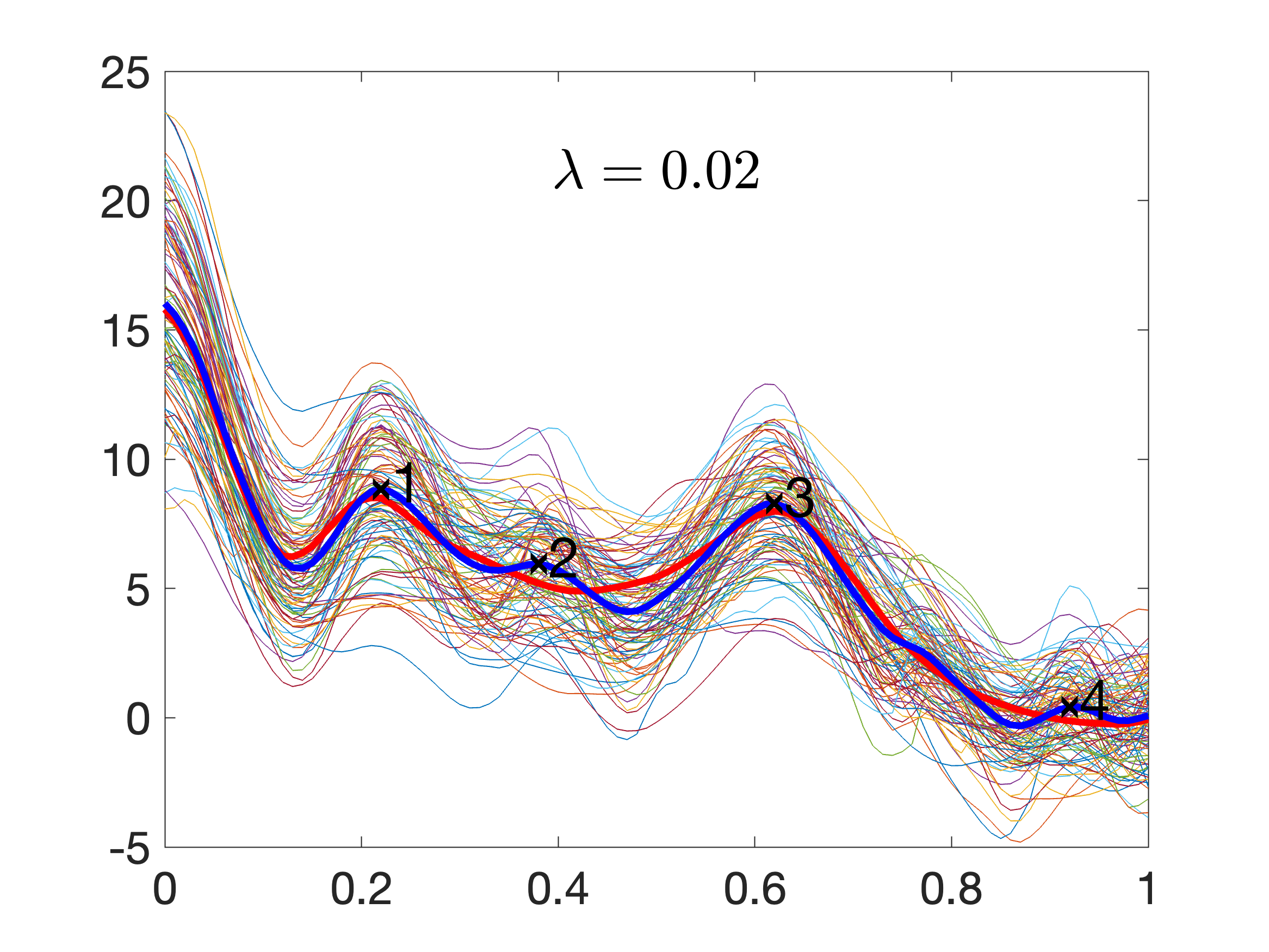}}
    \hspace*{-0.125in}
    \subfloat[]{\includegraphics[width = 1.3in]{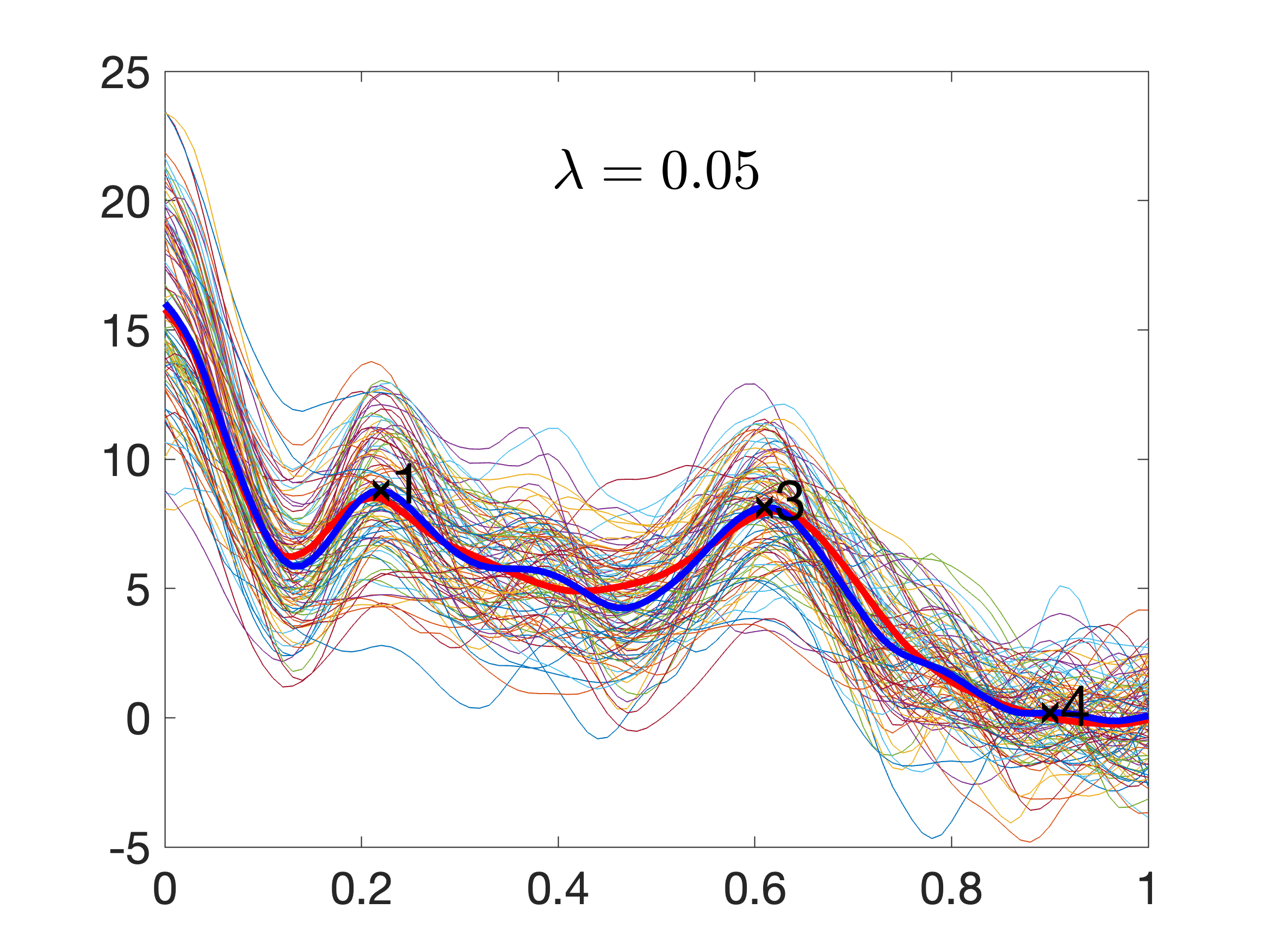}}
    \hspace*{-0.125in}
    \subfloat[]{\includegraphics[width = 1.3in]{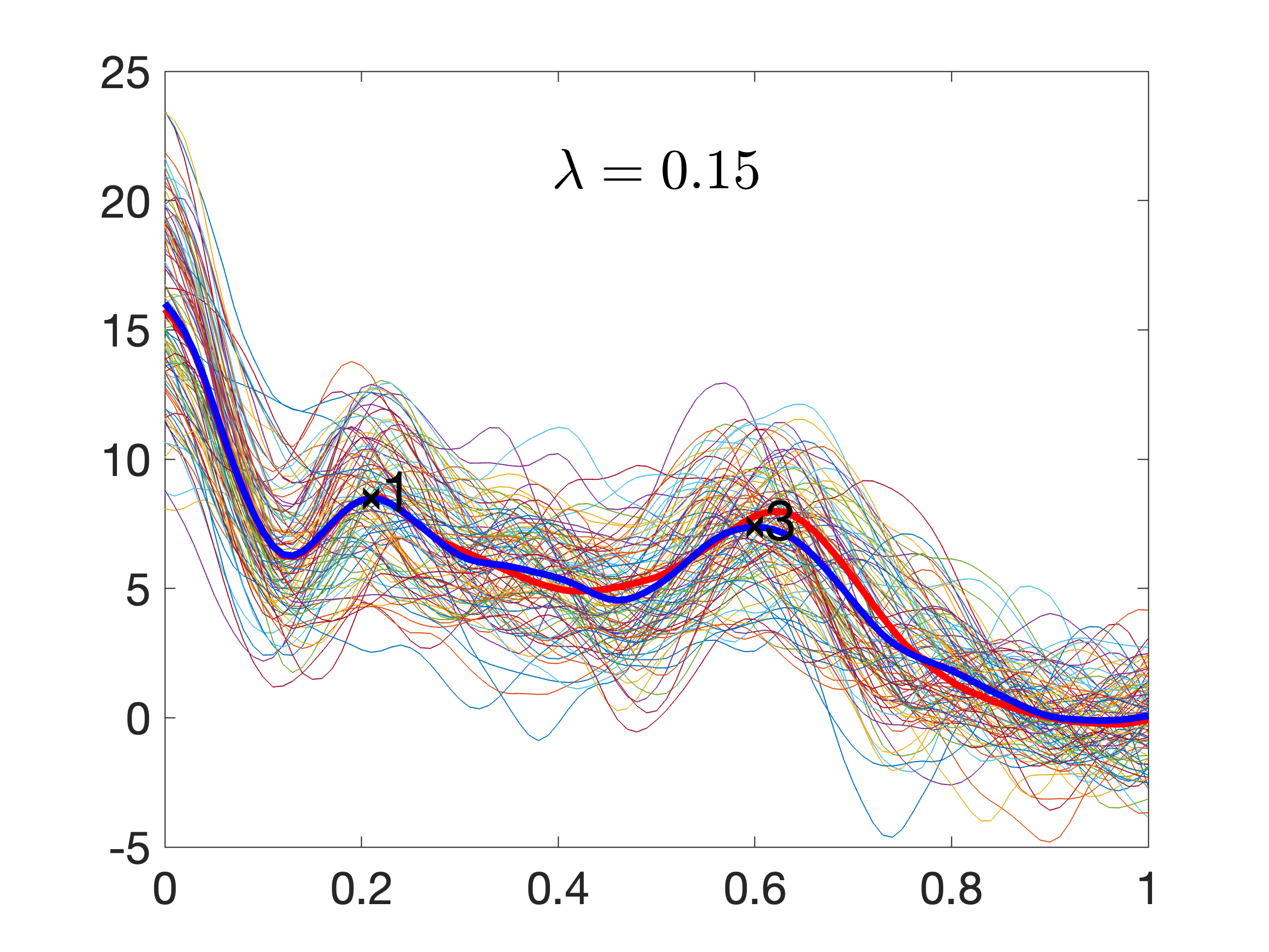}} 
    \hspace*{-0.125in}
    \subfloat[]{\includegraphics[width = 1.3in]{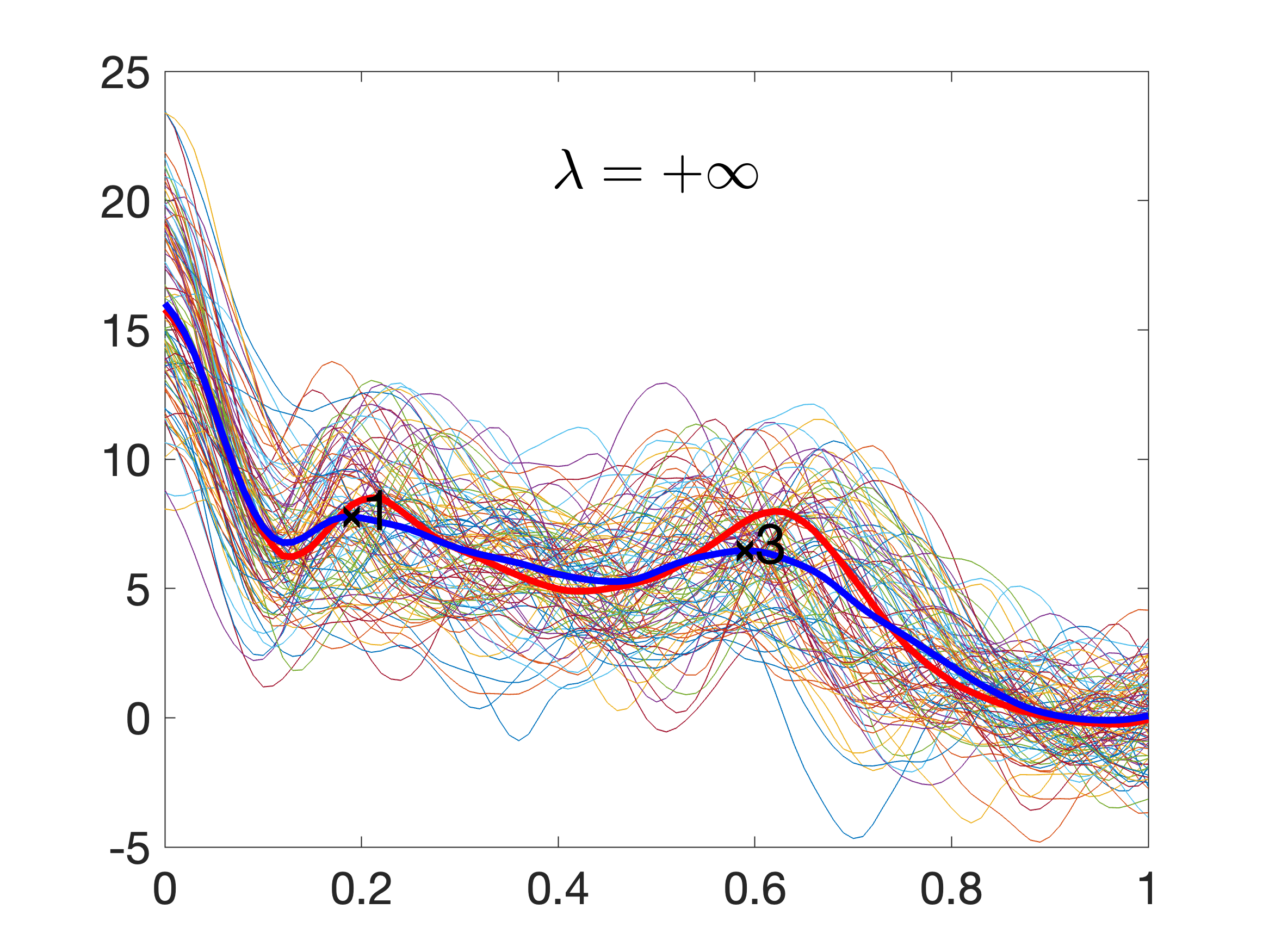}}
    \caption{Partial elastic alignments for different values of $\lambda$ according to Eqn.~\ref{eq:penalized-align}. Each plot shows the true $g$ (red), the partially aligned set $\{\tilde{f}_{\lambda,i}\}$, and their cross-sectional mean $\hat{g}_{\lambda}$ (blue curve).}
    \label{fig:partial-alignment}
\end{figure*}

The SRVF-based optimization (Eqn.~\ref{eq:penalized-align}) does not provide an estimate of $g$ directly but results in several quantities of interest: (1) If $\hat{\gamma}_{\lambda,i}$ denotes the optimal time warping inside the summation, for each $i$, then $\tilde{f}_{\lambda,i} = f_i \circ \hat{\gamma}_{\lambda,i}$ are the resulting partially-aligned functions; (2) Let $\hat{g}_{\lambda}$ be the cross-sectional mean of these $\tilde{f}_{\lambda,i}$. 
Algorithm~1 summarizes the main steps in this partial elastic alignment. 
\begin{algorithm}[ht] 
\begin{algorithmic}[1]
    \Require Data $\{f_1,f_2,\dots,f_n\}$
    \For{$i = 1,\dots,n$}
    \State $q_i \leftarrow \mbox{sign}(\dot{f}_i) \sqrt{|\dot f_i|}$
    \EndFor
    \State $\bar q \leftarrow \argmin_{q \in \{q_1,\dots,q_n\}}\norm{q- \frac{1}{n}\sum_{i=1}^n q_i}$
    \While{$\epsilon > tol$}
        \For{$i = 1,2,\dots,n$}
            \State $\hat\gamma_{\lambda,i} \leftarrow \arg\inf_{\gamma\in\Gamma} \left( \norm{\bar q - (q_i\star \gamma)}^2 + \lambda \mathcal R \right)$
            \State $f_i^*\leftarrow f_i \circ \hat\gamma_{\lambda,i}$,\ \ \mbox{and}\ \ 
            $q_i^* \leftarrow  \mbox{sign}(\dot{f}_i^*) \sqrt{|\dot{f}_i^*|}$
        \EndFor
        \State $\bar q^* \leftarrow \frac{1}{n}\sum_{i=1}^n q_i^*$,\ \ \ 
        $\epsilon \leftarrow \norm{\bar q -\bar q^*}^2$
        \If{$\epsilon > tol$}
            \State $\bar{q} \leftarrow \bar{q}^*$
        \EndIf
    \EndWhile
    \State $\bar{\gamma}^{-1} \leftarrow  (\frac{1}{n} \sum_{i=1}^n \hat\gamma_{\lambda,i})^{-1}$,\
    $\bar{q}^* \leftarrow (\bar{q}^* \star \bar{\gamma}^{-1})$
    \For{$i = 1,2,\dots,n$}
        \State $\hat{\gamma}_{\lambda,i} \leftarrow \arg\inf_{\gamma_{i}\in\Gamma} \left( \norm{\bar q^*-(q_i\star\gamma_i)}^2+ \lambda \mathcal R \right)$
        \State $\tilde{f}_{\lambda, i} \leftarrow f_i \circ \hat{\gamma}_{\lambda,i}$
    \EndFor
    \State \Return$\hat{g}_{\lambda} = \frac{1}{n} \sum_{i=1}^n \tilde{f}_{\lambda, i}$
\end{algorithmic}
 \caption{Function-Alignment Algorithm}\label{alg:Function-Alignment-Algorithm}
\end{algorithm}
Fig. \ref{fig:partial-alignment} shows an illustration of the output of Algorithm~1. It offers several results, each showing partially-aligned functions and their cross-sectional mean for a different value of $\lambda$. The original data is shown in the bottom right panel. On one extreme, $\lambda = 0$ results in a perfect alignment of peaks and valleys. Conversely, $\lambda = \infty$ provides no alignment at all. 


Given this context, the subsequent task involves discovering an automated method to select the best $\lambda$ from the available data. This is achieved by utilizing the peak persistence diagram, which is introduced next.

\subsection{Peak Persistence Diagrams} \label{sec: Peak Persistence Diagram}

As $\lambda$ changes from $0$ to $\infty$, some peaks and valleys in $\hat{g}_{\lambda}$ start to diminish and even disappear altogether. Sometimes new peaks can also be generated. We formalize this behavior using a peak persistence diagram. 

\begin{definition}
The peak persistence diagram (PPD) of a set of functions $\{ f_i, i=1,2,\dots, n\}$ is a visual presentation of the {\bf significant, internal peaks} of their partial elastic mean $\hat{g}_{\lambda}$ plotted versus $\lambda$. Essentially, a PPD serves to identify the existence, magnitudes, and positions of significant internal peaks of $\hat{g}_{\lambda}$.
\end{definition}
A PPD results in several displays. Plotting only the peak indicators across the range of $\lambda$, we obtain a PPD bar chart. If we are interested in peak locations as well, we can use a PPD surface. An important issue to address is what qualifies as a {\it significant} peak, which we discuss in the following subsection. The idea of tracking significant peaks has been previously utilized in density estimation~(\cite{SiZer}), although in the context of bandwidth selection for kernel-based methods.
A PPD is akin to persistence homology in topological data analysis~(\citet{zomorodian_2005}), where one traces the presence of topological features at different data resolutions. In contrast, a PPD tracks geometric features (peaks) for scalar functions. 

\begin{figure*}[ht]
    \centering
    \hspace*{-0.2in}
    \subfloat[]{\includegraphics[height = 1.4in]{fig/fns-by-lambda-inf.png}}
    \hspace*{-0.4in}
    \subfloat[]{\includegraphics[height = 1.4in]{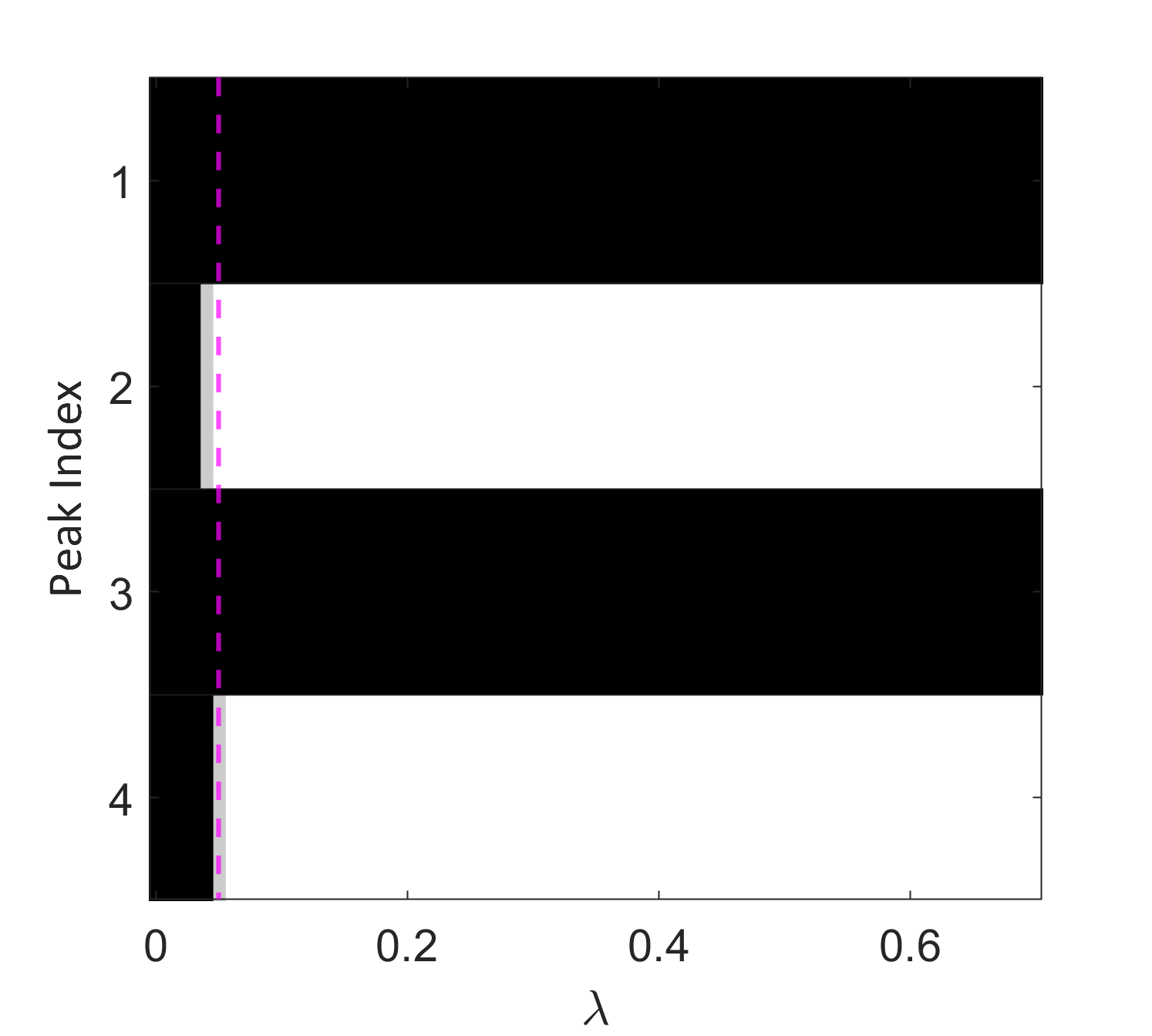}}
    \hspace*{-0.4in}
    \subfloat[]{\includegraphics[height = 1.4in]{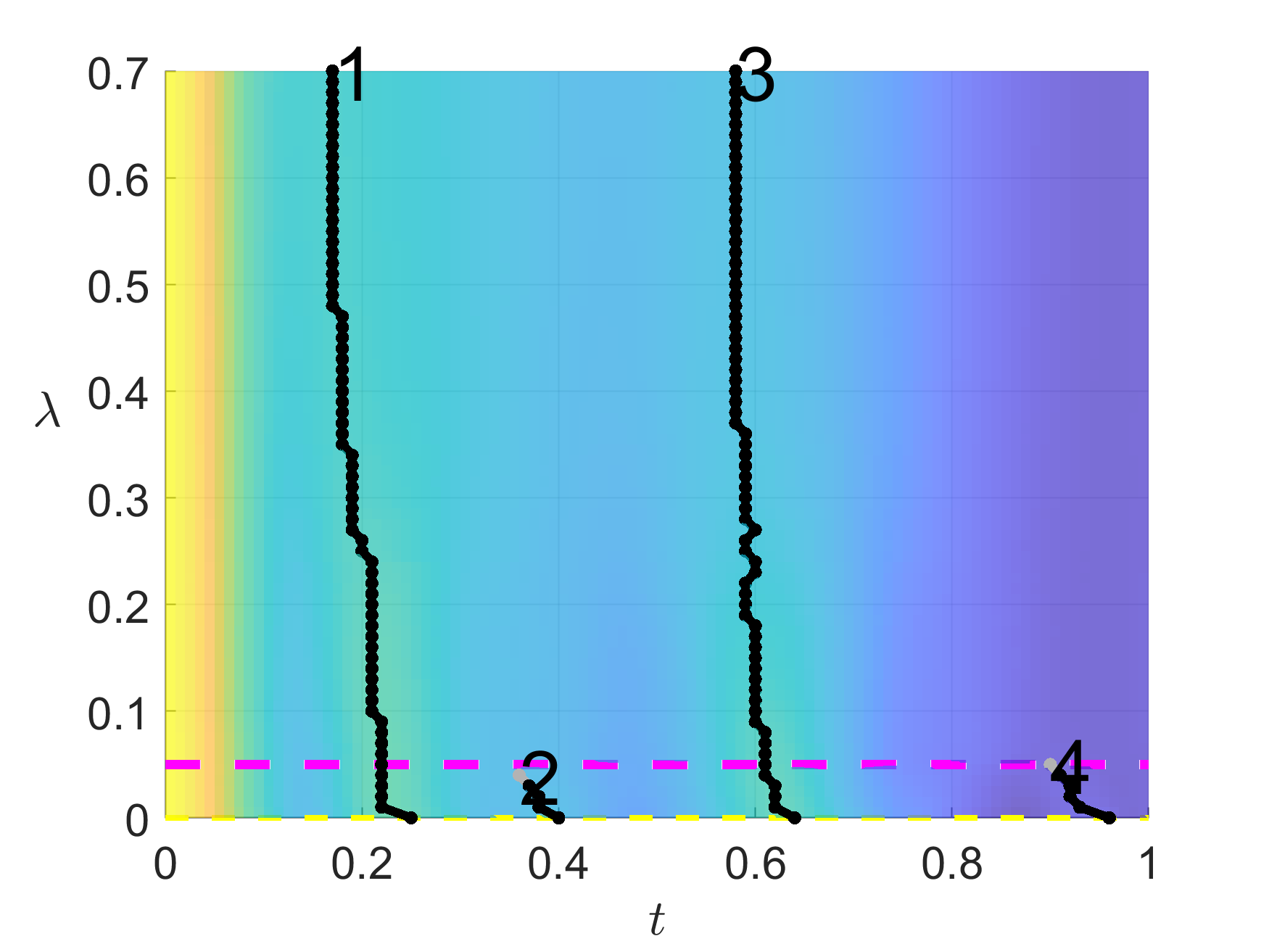}}     \hspace*{-0.4in}
    \subfloat[]{\includegraphics[height = 1.4in]{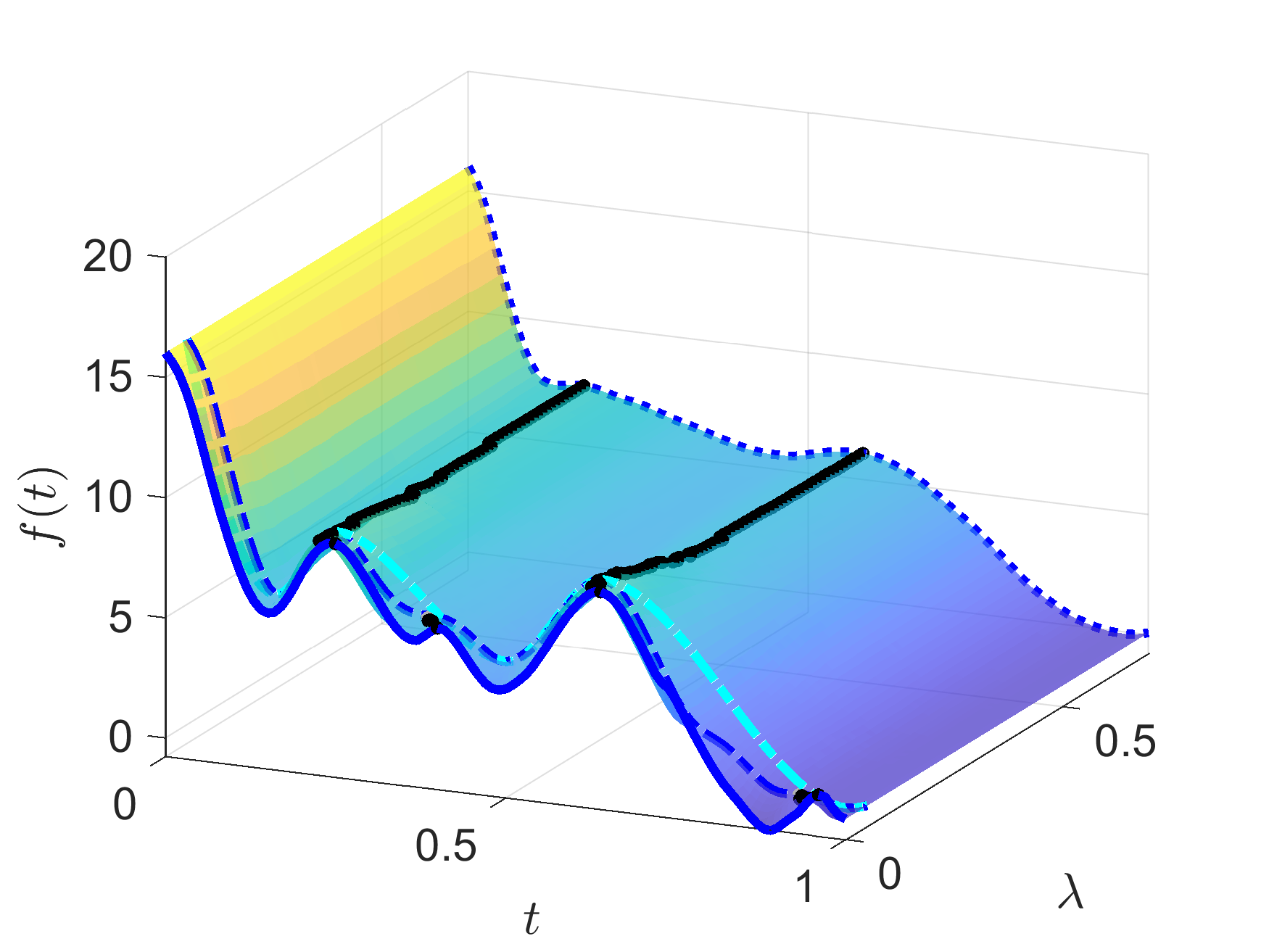}}     \hspace*{-0.4in}
    \caption{(a) The original functional data, (b) the PPD barchart, and (c), (d) PPD surface generated by interpolating $\hat{g}_{\lambda}$'s for different $\lambda$ and the movements of the peaks. 
    }
    \label{fig: PPD}
\end{figure*}

Figure \ref{fig: PPD} serves as an illustration of a PPD for the data shown in \ref{fig: PPD}(a). In panel (b), the PPD bar chart displays the presence of detected peaks for different values of $\lambda$. The $x$ axis represents $\lambda$, while the $y$ axis indicates peak labels. Panels (c) and (d) show a 3D surface plot, where the values of $\hat{g}_{\lambda}(t)$ are represented by colors ranging from blue to yellow, based on their height. The $y$ axis represents $\lambda$, the $x$ axis is $t$, and the labeled black lines indicate the positions of the peaks as $\lambda$ varies. At $\lambda = 0$, $\hat{g}_{\lambda}$ exhibits four internal peaks, which change as $\lambda$ increases. Notably, peaks labeled 2 and 4 disappear when $\lambda \sim 0.05$, while those labeled 1 and 3 persist.

A PPD is a useful tool for studying geometrical or shape features of the unknown function across a range of $\lambda$. Peaks that persist over a longer range are deemed significant, while those that are small or do not persist for long  are attributed to noise or alignment artifacts. However, determining what is significant and what is not can be a challenging issue. In the next section, we will discuss this and other related issues.

\subsection{Peak Significance and Persistence} \label{sec: criteria}

Choosing whether a peak in a PPD is actual or an artifact is a complicated process, and one has to make some {\it ad hoc} choices depending on the final goal. 
In this paper, we focus on larger, more global structures and accept the loss of some smaller, finer features. \\

\noindent {\bf Significance of a peak}: 
When is a peak considered significant? Small peaks occur in $\hat{g}_{\lambda}$, either due to noise or computational errors, that are not in the original $g$. This can happen, for example, in parts of the domain where $g$ is constant and an alignment of noise results in spurious peaks. For a peak at $t_0$, we define the {\it strength} of this peak to be $\frac{-\hat{g}''_{\lambda}(t_{0})}{\| \hat{g}''_{\lambda}\|}$. This quantity measures the curvature at point $t_0$, normalized appropriately. 
If this quantity is less than a predetermined value, say $\tau$, then the peak at $t_0$ is considered {\it insignificant} and is discarded. Otherwise, it is significant and kept in PPD. This paper uses a conservative value of $\tau = 0.03$, determined through extensive experimentation and has worked well across datasets.
\\

\noindent {\bf Persistence of a peak}: \\
For the peak labeled $k$, 
we define its {\it persistence} to be $p_k = |\{ \lambda \geq 0 \mid \mbox{peak $k$ in $\hat{g}_{\lambda}$ is significant}\}|$, where $| \cdot |$ denotes the length of the interval.  The next issue is deciding the minimum value for a peak for it to be {\it persistent}. For this, we select the most persistent peak, say $k_0$, and define persistence of other peaks relative to $p_{k_0}$. If the relative persistence of $k^{th}$ peak $p_k/p_{k_0}$ is larger than a threshold, say $\theta$, then that peak is considered {\it persistent}. We used experiments to find that $\theta \approx 0.28$ performs best on simulated and real data. The number of internal significant peaks is then given by: 
$m = | \{k \geq 0 \mid p_k/p_{k_0} > \theta \} |$. This number $m$ determines the estimated shape of $g$ and forms a constraint in estimating $g$. 

In summary, two hyper-parameters $\tau$ and $\theta$ are needed to determine significant and persistent peaks. The results are found to be relatively stable with respect to the choices we have made in this paper. 

\subsection{Selection of $\lambda$ and Initial Estimate of $g$} \label{sec: prarameter selection}
Once we have selected significant and persistent internal peaks, the next step is determining an optimal value of $\lambda$, say $\lambda^*$, to help in subsequent process. 
We select the smallest value of $\lambda$ that results in $m$ significant peaks in $\hat{g}_{\lambda}$. This choice is motivated by the observations that extrema of $\hat{g}_{\lambda}$ typically diminish as $\lambda$ increases.
For the example shown in Fig.~\ref{fig: PPD}, we select $m=2$ and $\lambda^*=0.05$. 
For this $\lambda^*$, we also compute the cross-sectional mean $\hat{g}_{\lambda^*} = \frac{1}{n} \sum_{i=1}^n \tilde{f}_{i, \lambda^*}$.


\subsection{Shape Estimator Properties}
It would be useful to investigate statistical properties of the estimator $m$ of the number of internal peaks or modes in $g$. Several papers in the past have studied the problem of mode estimation in the context of density estimation, especially when using nonparametric kernel estimators, see {\it e.g.}, \cite{minnotte-annals:1997,SiZer} and references therein. Our estimator is based on the geometric properties of the cross-sectional mean $\hat{g}_{\lambda^*}$ of partially-aligned functions. Given the non-Euclidean nature of the Riemannian elastic metric, and the geometric abstraction of peak persistence, we have not pursued any theoretical investigations of this estimator. Instead, we provide extensive experimental validation of this approach using both simulated and real datasets. 

\section{Experimental Results: Shape Estimation}

\begin{figure*}[htbp]
    \centering
    \subfloat[]{\includegraphics[height = 1.3in]{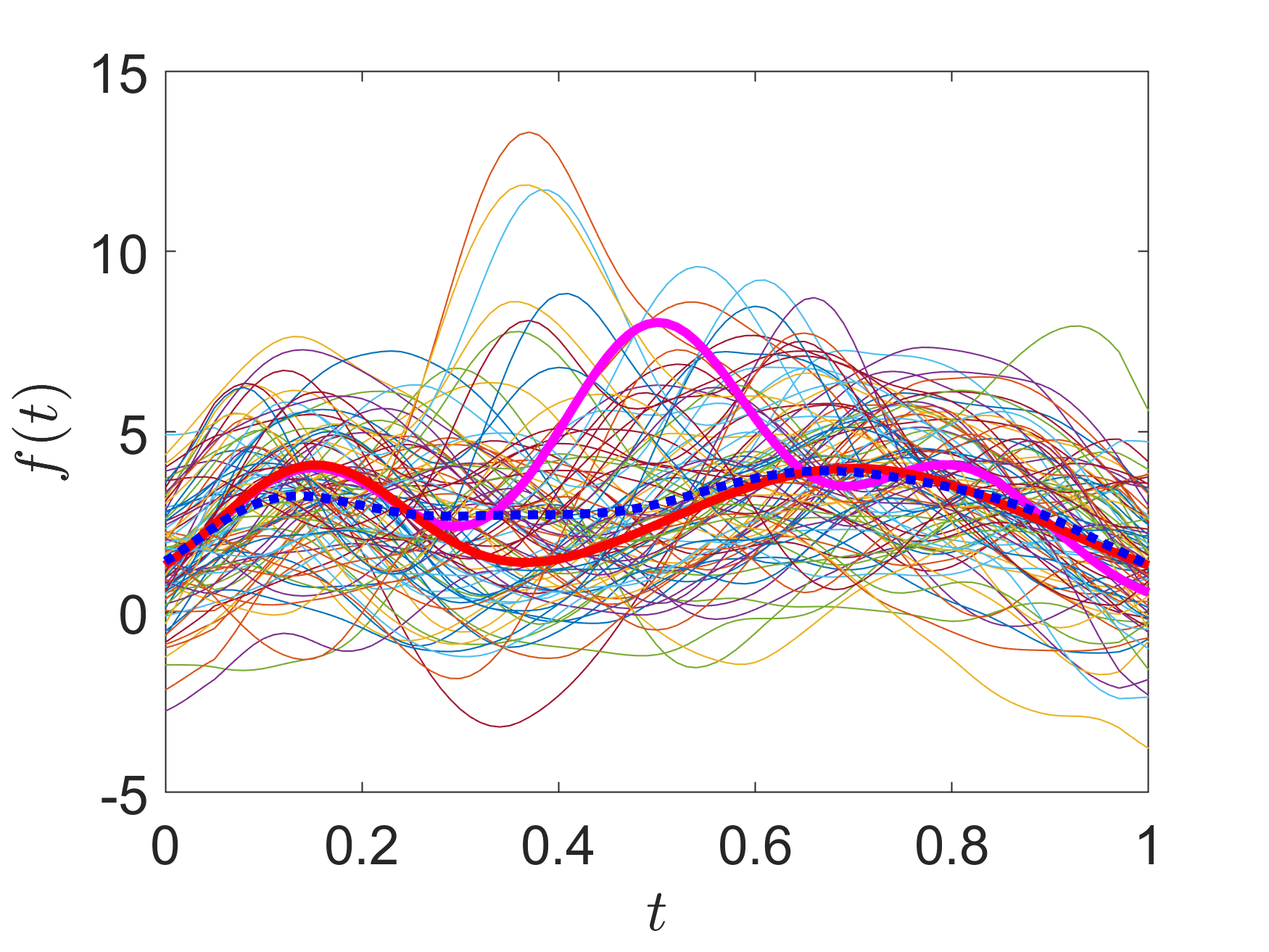}}
    \hspace{-0.2in}
    \subfloat[]{\includegraphics[height = 1.3in]{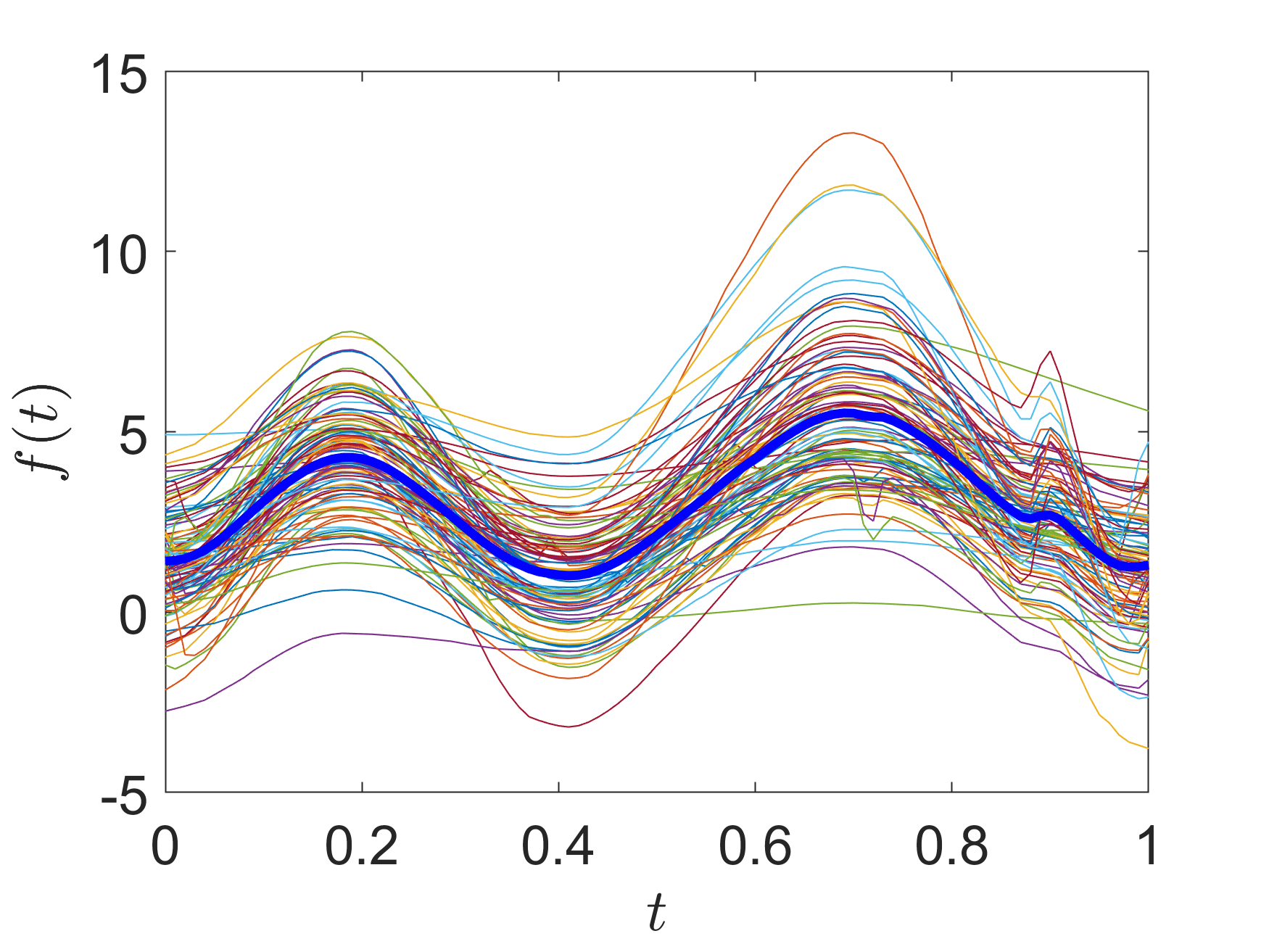}}
    \hspace{-0.2in}
    \subfloat[]{\includegraphics[height = 1.3in]{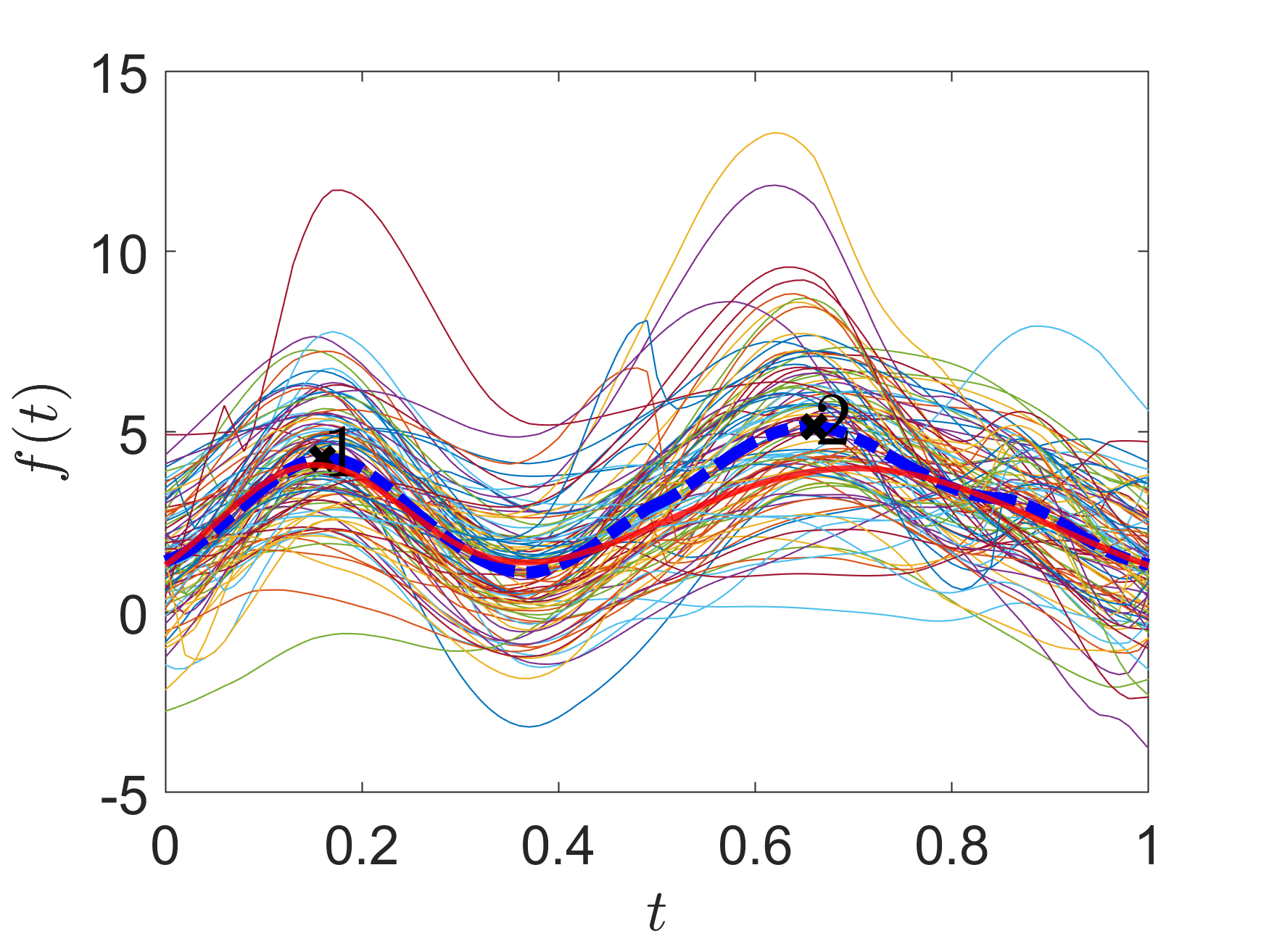}}
    \includegraphics[height = 0.9in]{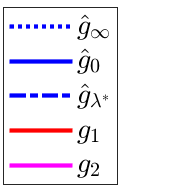}
    \hspace{0in}\\
    \vspace*{-0.05in}
    \hspace{-0.2in}
    \subfloat[]{\includegraphics[height = 1.3in]{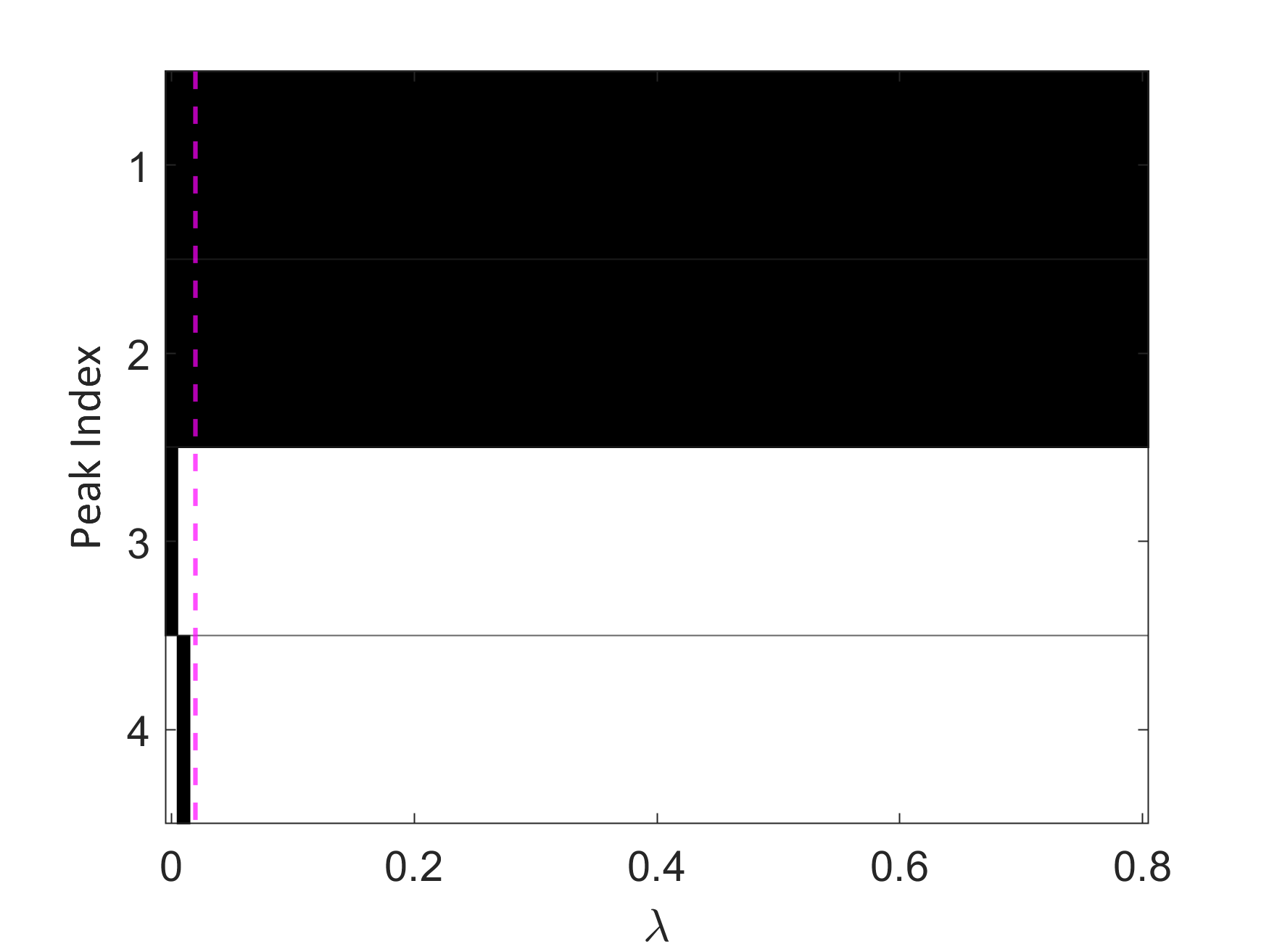}}
    \hspace{-0.2in}
    \subfloat[]{\includegraphics[height = 1.3in]{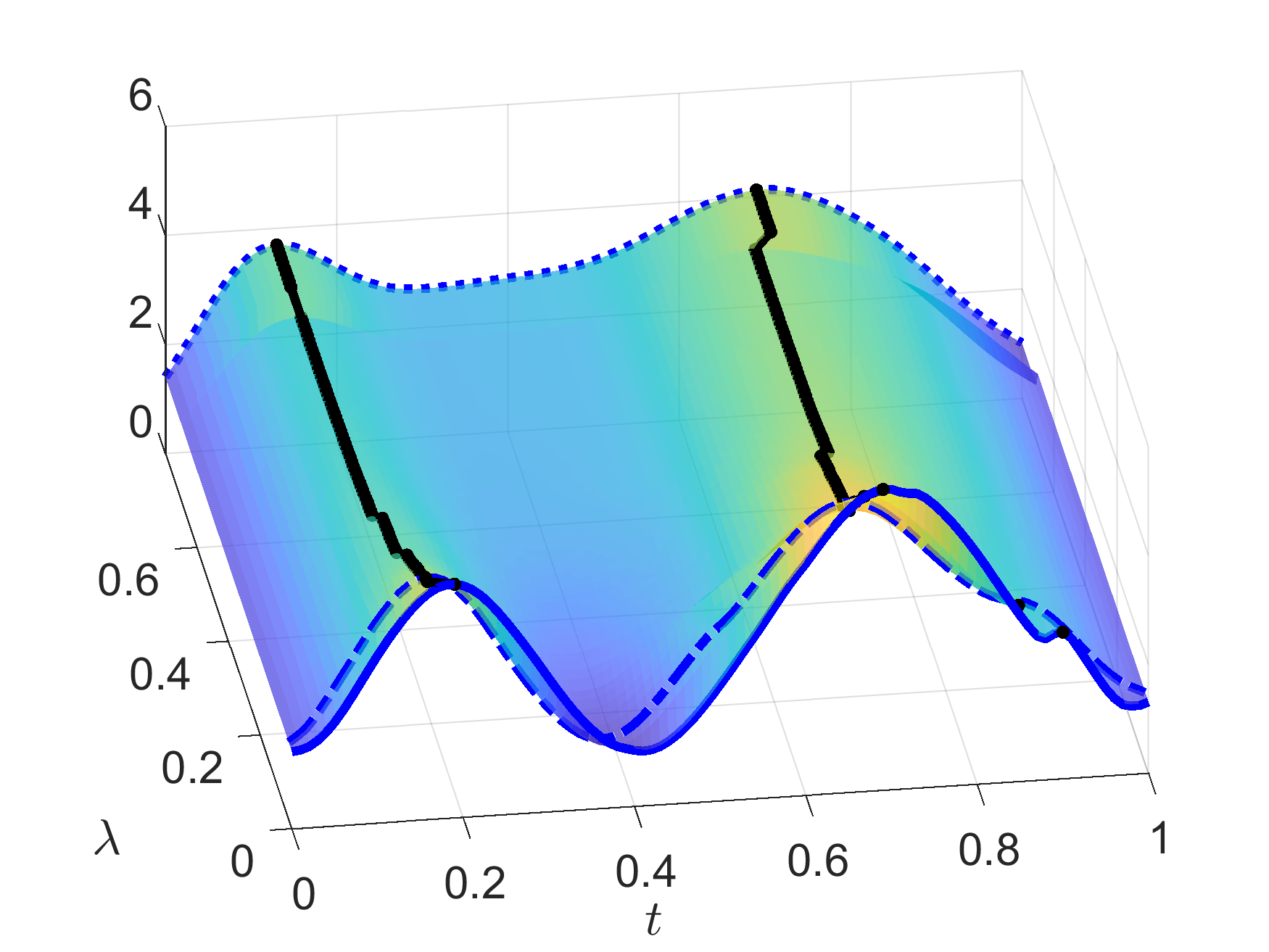}}
    \hspace{-0.2in}
    \subfloat[]{\includegraphics[height = 1.3in]{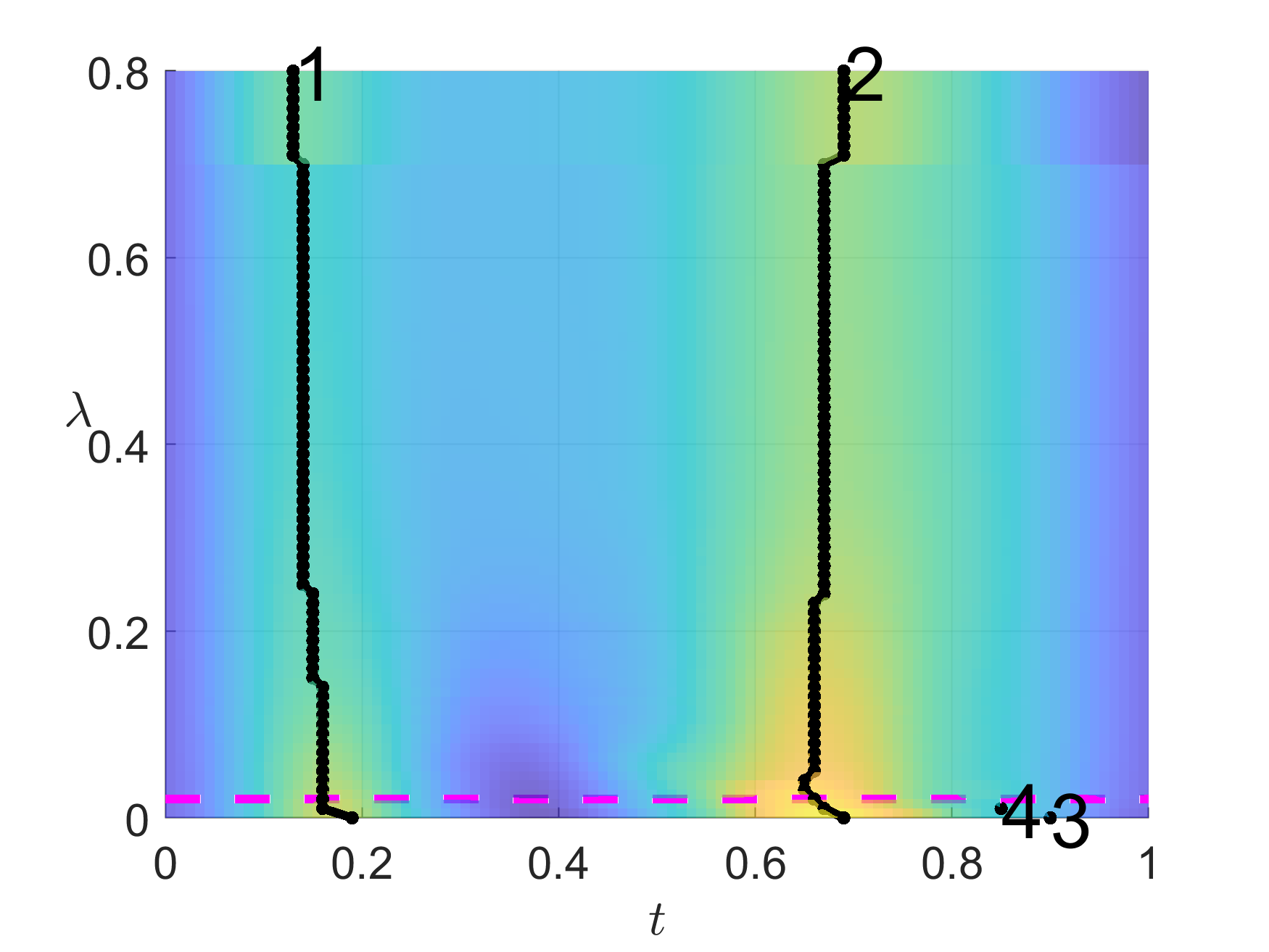}}
    \vspace*{-0.05in}
    \caption{Example 1: Data consists of 80 and 20 functions generated from $g_1$ and $g_2$, respectively. Plots (a), (b), and (c) display $\hat g_\infty$, $\hat g_0$, and $\hat g_{\lambda^*}$. PPDs in (d) through (f) suggest two peaks (1 and 2) are significant and persistent.}
    \label{fig: shape_estimation 1}
\end{figure*}

Now we present examples of PPD-based shape estimation on some simulated and real functional datasets.

\noindent {\bf Example 1}: 
The first simulation generates data from a bimodal function, labeled as $g_1$ (red curve), which we corrupt by adding a few tri-modal curves $g_2$ (magenta curve). The complete set consists of 80 random perturbations of $g_1$ and 20 of $g_2$. The objective is to investigate the estimation of the shape of $g_1$ from this noisy and corrupted data.

Figure \ref{fig: shape_estimation 1} presents the estimation results. Plot (a) shows the original data, and (b) shows the full elastic alignment with the mean, $\hat g_0$, drawn in blue. This mean curve has three peaks, with the third relatively small. Plot (c) shows the outcomes of partial-elastic alignment with $\lambda^* = 0.02$ chosen via PPDs displayed in panels (d), (e), and (f). The bar chart in (d) indicates two significant and persistent peaks (1 and 2), consistent with $g_1$. This example suggests that the PPD method can successfully estimate the shape of the underlying signal even when the data is contaminated with another shape.

\begin{figure*}[htbp]
    \centering
    \subfloat[]{\includegraphics[height = 1.3in]{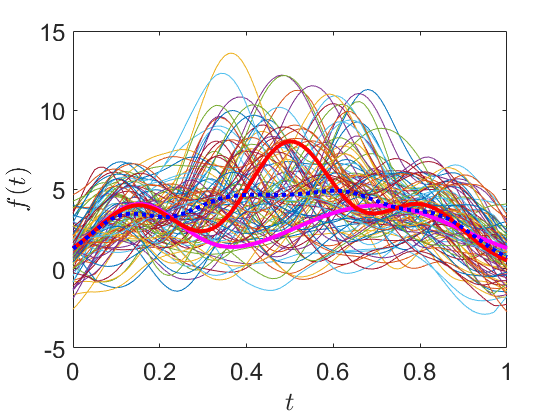}}
    \hspace{-0.2in}
    \subfloat[]{\includegraphics[height = 1.3in]{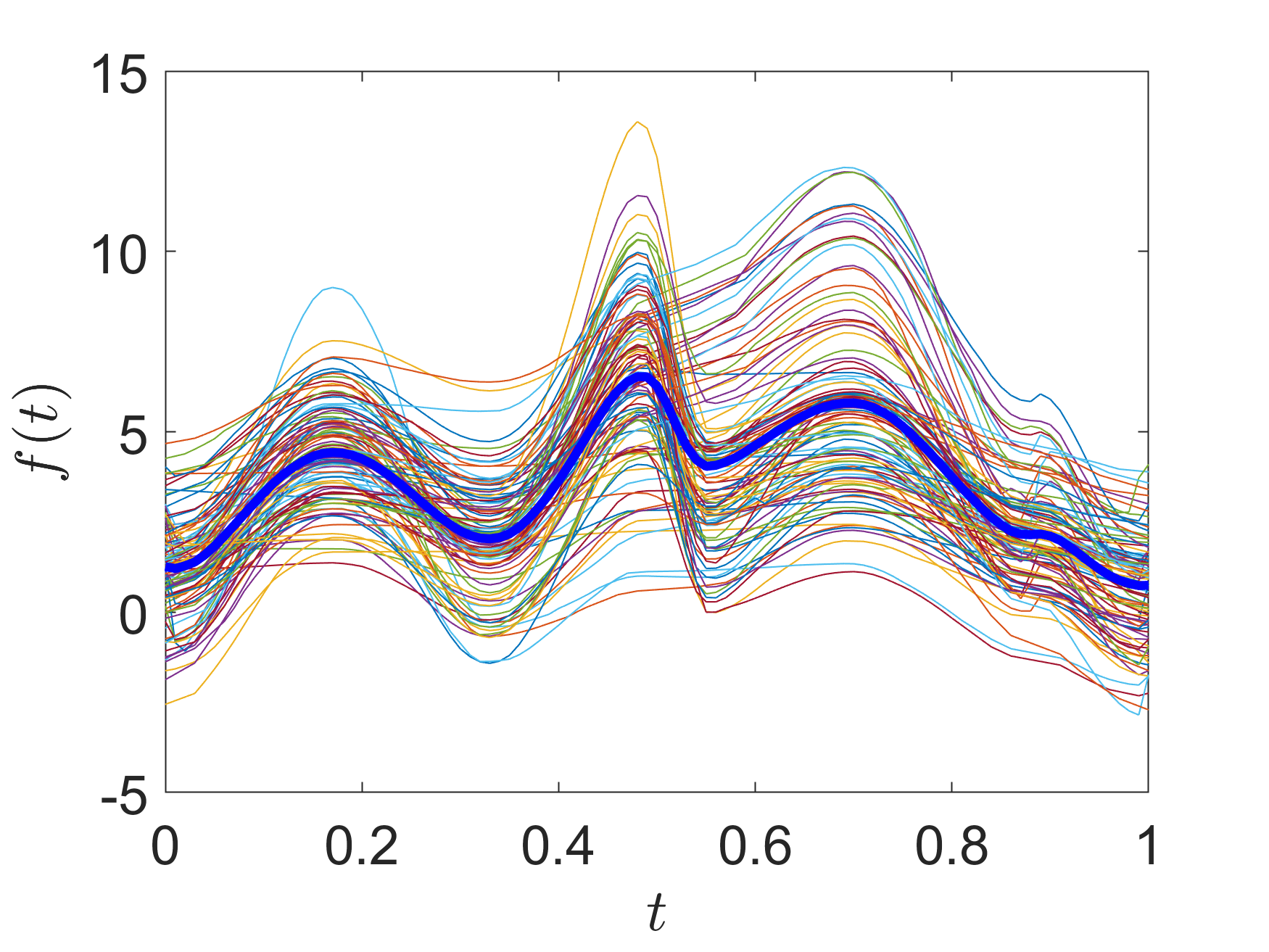}}
    \hspace{-0.2in}
    \subfloat[]{\includegraphics[height = 1.3in]{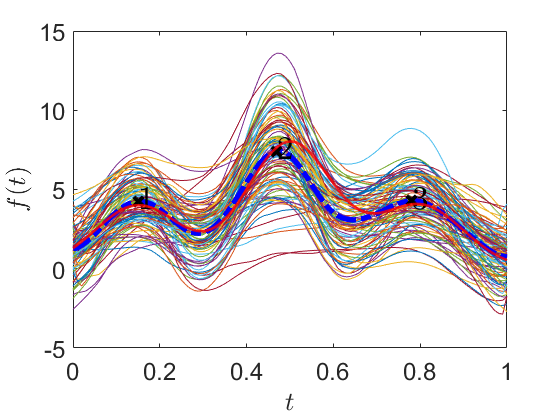}}
    \includegraphics[height = 0.9in]{fig/legend-rd-3.png}
    \hspace{0in}\\
    \vspace*{-0.05in}
    \hspace{-0.2in}
    \subfloat[]{\includegraphics[height = 1.3in]{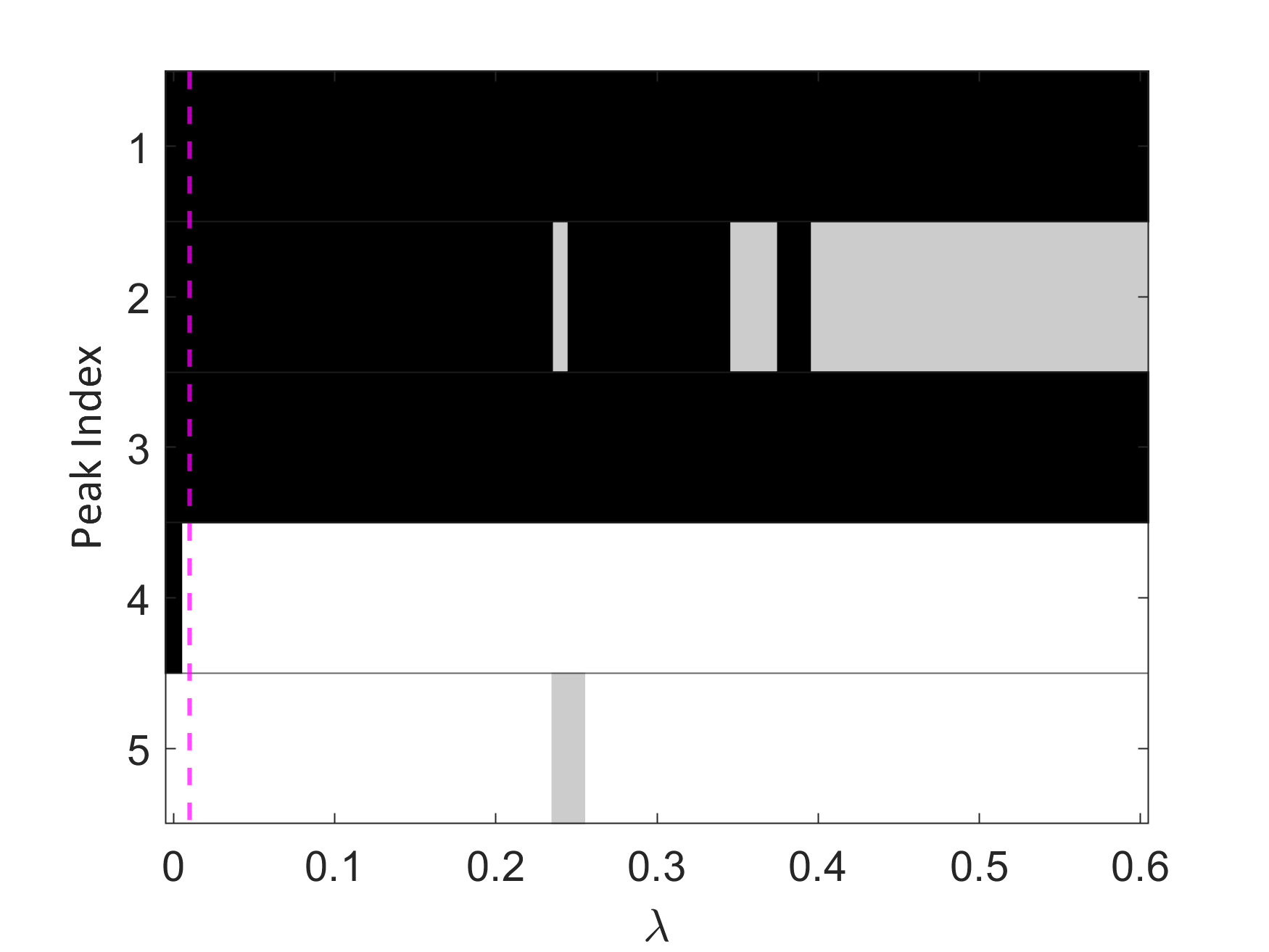}}
    \hspace{-0.2in}
    \subfloat[]{\includegraphics[height = 1.3in]{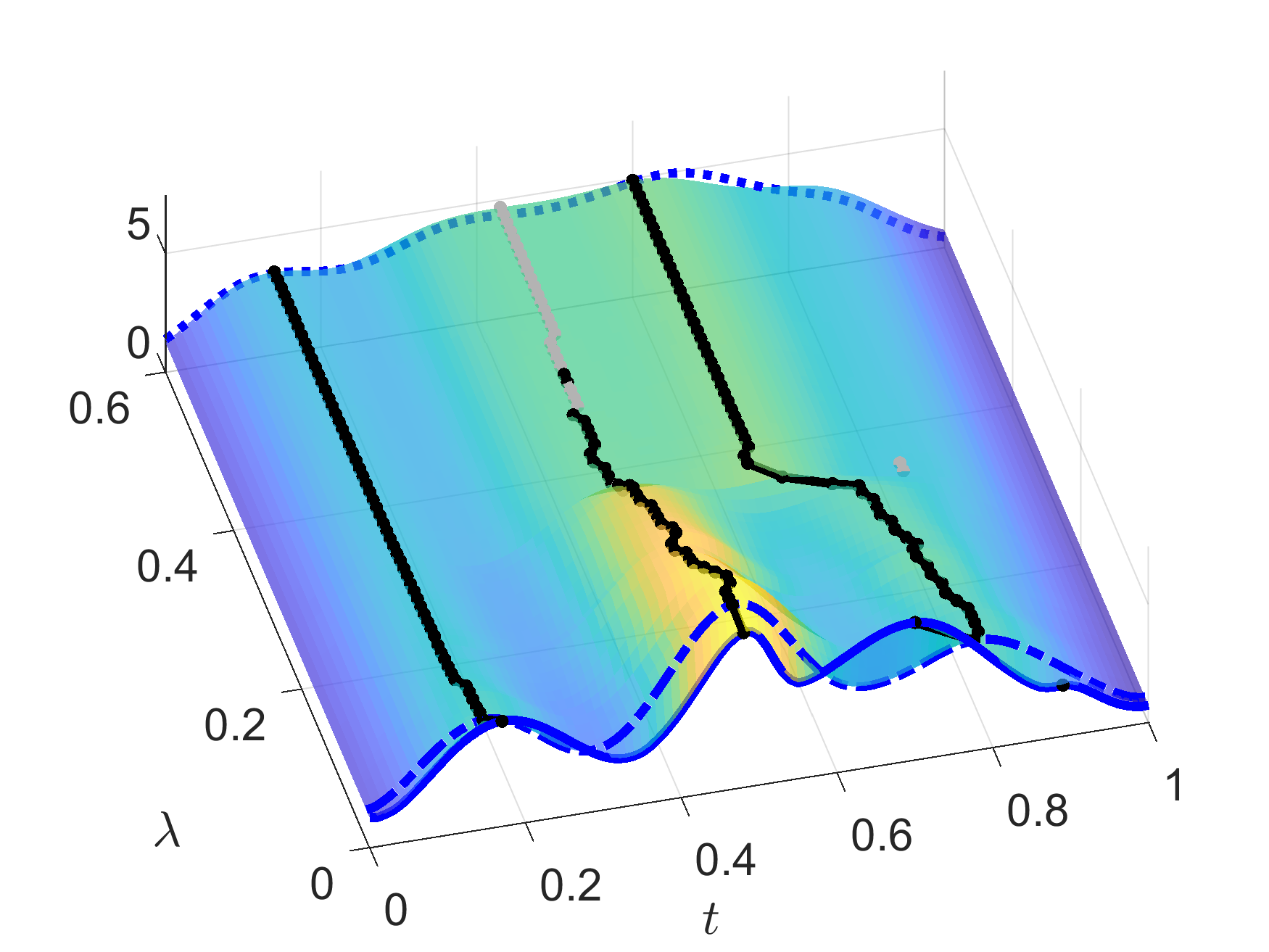}}
    \hspace{-0.2in}
    \subfloat[]{\includegraphics[height = 1.3in]{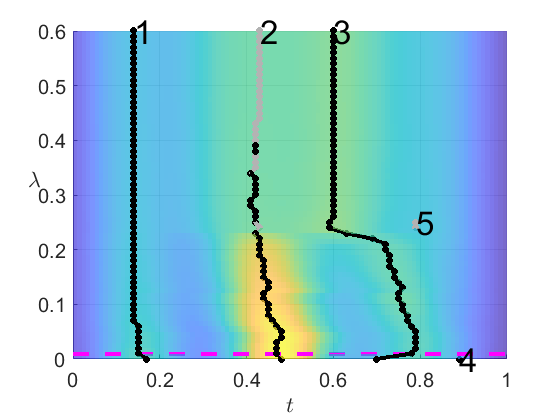}}
    \vspace*{-0.05in}
    \caption{Example 2: This figure is laid out similar to Fig. \ref{fig: shape_estimation 1}. In this case, 80 functions have three peaks, and the other 20 have two. PPDs (d) through (f) indicate three peaks (1,2 and 3) are significant and persistent.}
    \label{fig: shape_estimation 2}
\end{figure*}

\noindent {\bf Example 2}:
In this instance, we have reversed the roles of $g_1$ and $g_2$ in comparison to the first simulation. Specifically, we have 80 samples of the trimodal functions and 20 samples of the bimodal functions. The fourth plot (d) in Fig. 6 demonstrates that three peaks (1, 2, and 3) are significant and persistent, and $\lambda^* = 0.01$ is the optimal alignment. These findings suggest that the PPD method is robust and has the ability to accurately identify the number of peaks in noisy signals.

\begin{figure*}[htbp]
    \centering
    \subfloat[]{\includegraphics[height = 1.25in]{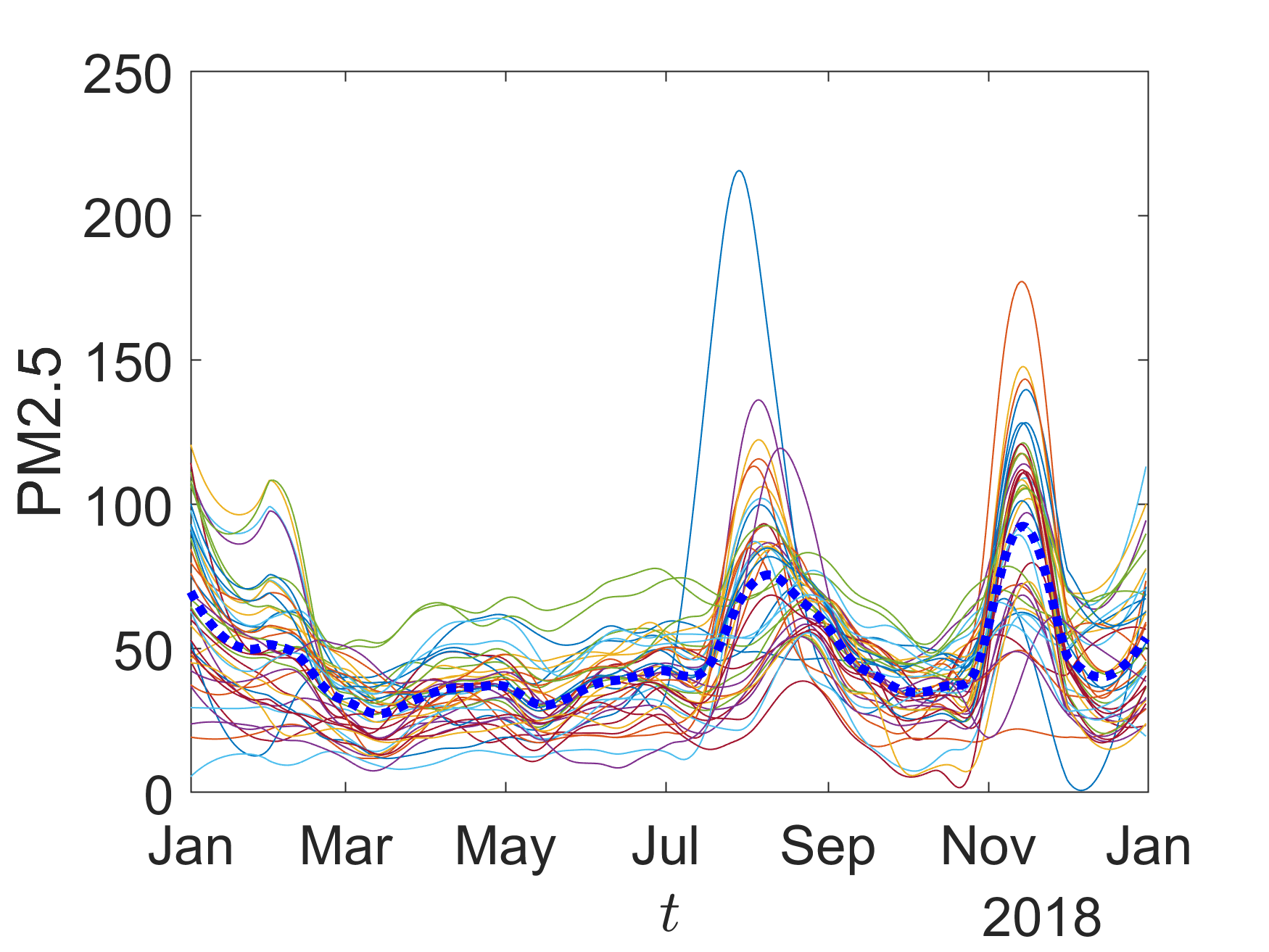}}
    \hspace{-0.2in}
    \subfloat[]{\includegraphics[height = 1.25in]{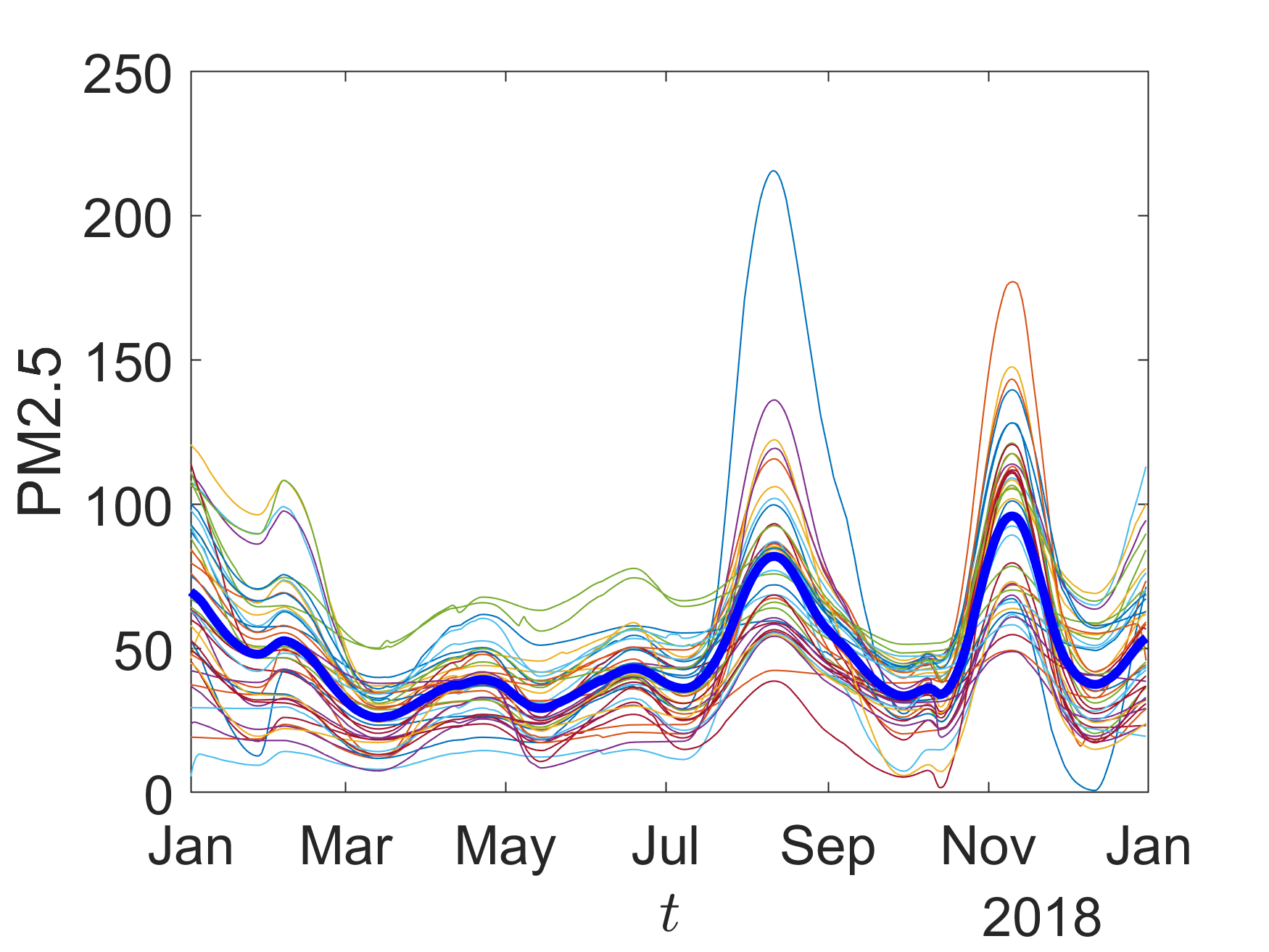}}
    \hspace{-0.2in}
    \subfloat[]{\includegraphics[height = 1.25in]{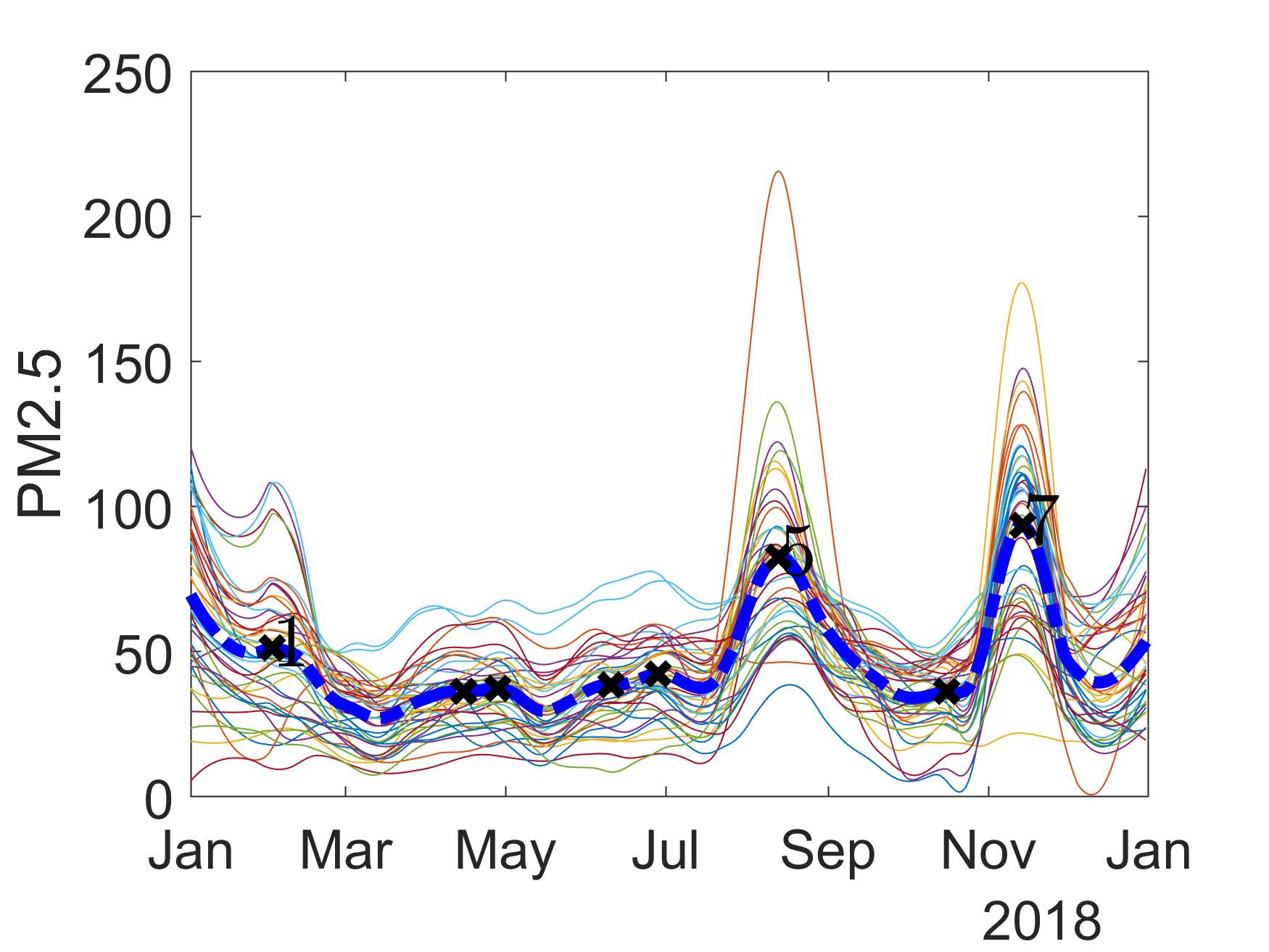}}
    \includegraphics[width = 0.8in]{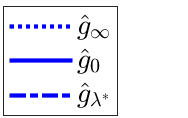}
    \hspace{0in}\\
    \vspace*{-0.05in}
    \hspace{-0.1in}
    \subfloat[]{\includegraphics[height = 1.25in]{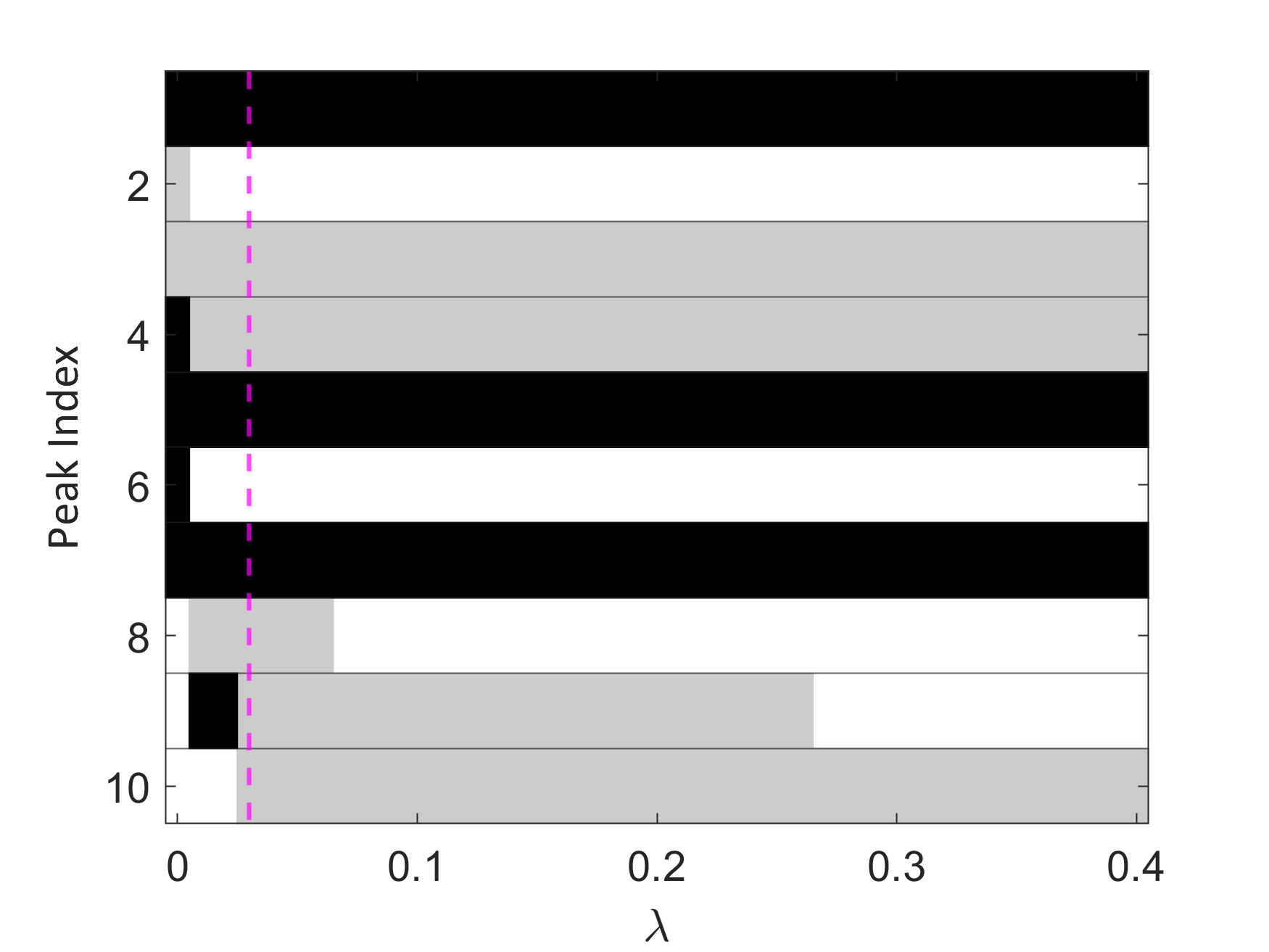}}
    \hspace{-0.1in}
    \subfloat[]{\includegraphics[height = 1.25in]{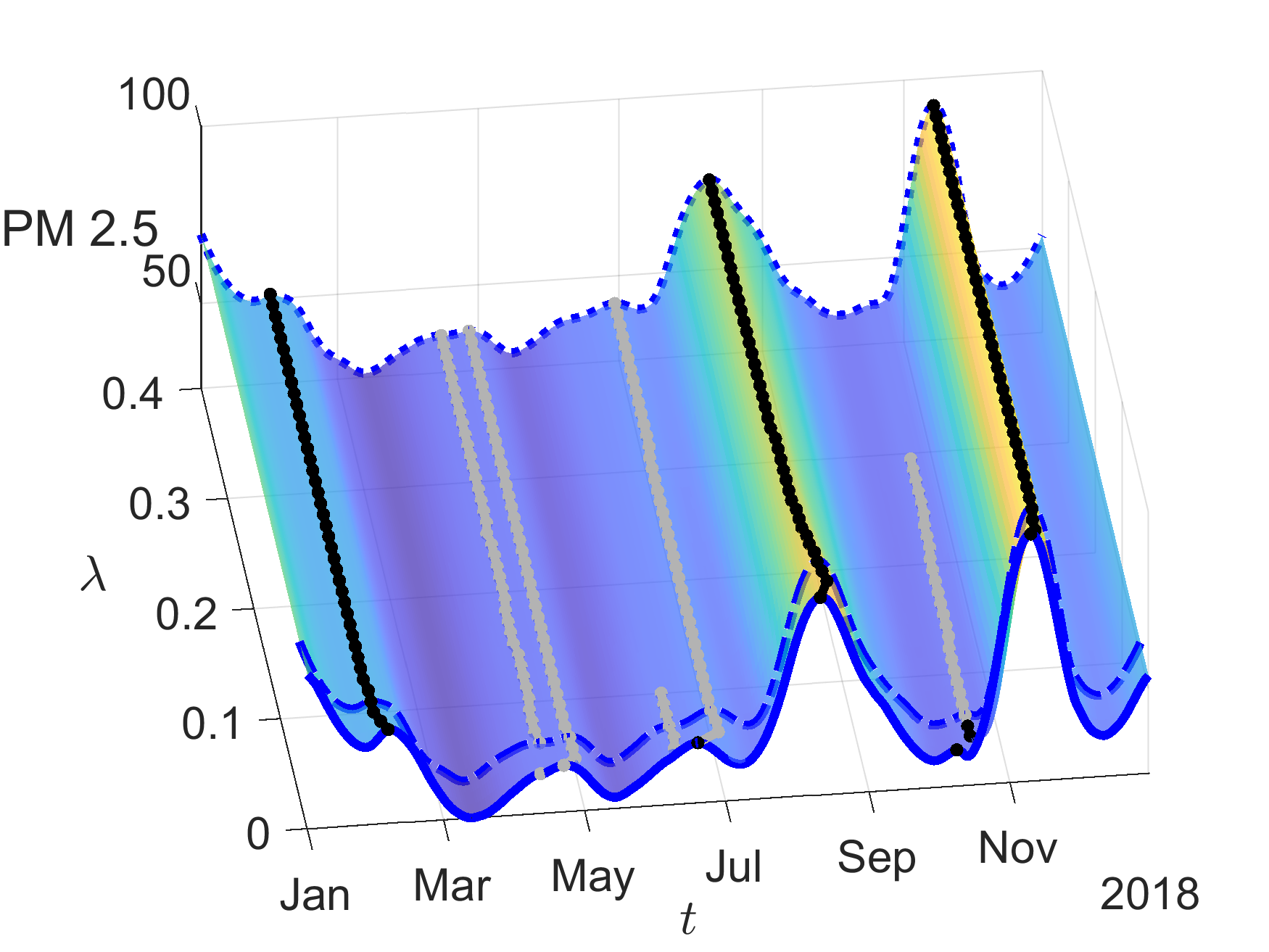}}
    \hspace{-0.1in}
    \subfloat[]{\includegraphics[height = 1.25in]{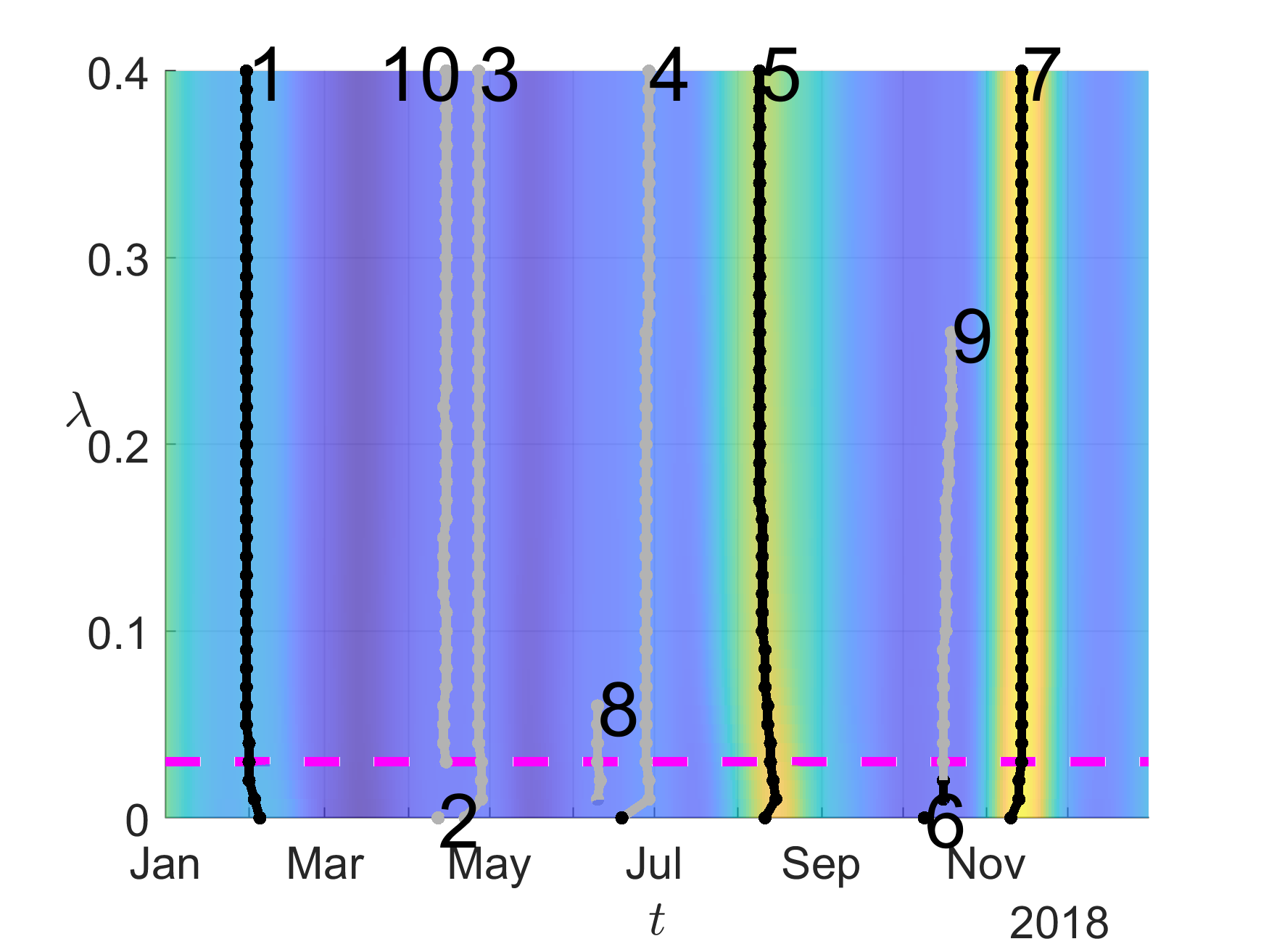}}
    \vspace*{-0.05in}
    \caption{Example 3: This figure is laid out similar to Fig. \ref{fig: shape_estimation 1}. The barchart in (d) indicates that while $\hat g_0$ has seven peaks, but only three peaks (1, 5, and 7) are significant and persistent .}
    \label{fig: shape_estimation Air Data}
\end{figure*}

\noindent {\bf Example 3: Air Quality in California}:
California is frequently plagued by wildfires that significantly impact its air quality.
To study this impact, we analyzed publicly available data from the Environment Protection Agency (EPA) on daily fine particle (PM 2.5) levels in 42 out of 58 counties in California during 2018. Fig. \ref{fig: shape_estimation Air Data} (a) shows the time-indexed data obtained by smoothing the raw measurements. For smoothing, we employed the {\it Lowess} (Locally Weighted Scatterplot Smoothing) method with a window size of 50 days. The PPDs in (d), (e), and (f) reveal that the estimated number of peaks in the ground truth signal, $g$, is three (1, 5, and 7) with $\lambda^* = 0.03$. In addition to Dec-Jan (the boundaries), the internal peak periods are around Feb, late Aug, and late Nov. The Feb peak is small but significant, while the other two internal peaks are visibly dominant. 

\noindent {\bf Example 4: Birth Rate Changes in European Countries}: We analyze a dataset of historical birth rate changes in 50 European countries from 1950 to 2021, collected by \cite{owidworldpopulationgrowth}. The response variable here is the yearly change in birth rates, where the birth rate implies count of live births per 1,000 individuals each year. The raw data is first smoothed using a window size of 15 years. 
The PPD method finds that four peaks are significant and persistent. The optimal parameter for this peak is $\lambda^* = 0.09$. Plot (b) demonstrates that $\hat g_0$ exhibits five peaks, but the fifth peak disappears quickly as the value of $\lambda$ increases. On the other hand, (a) and (d) show that $\hat g_\infty$ has four peaks. However, (c) shows $\hat g_{\lambda^*}$ with four significant peaks, namely peak numbers 1 through 4.

\begin{figure*}[htbp]
    \centering
    \subfloat[]{\includegraphics[height = 1.25in]{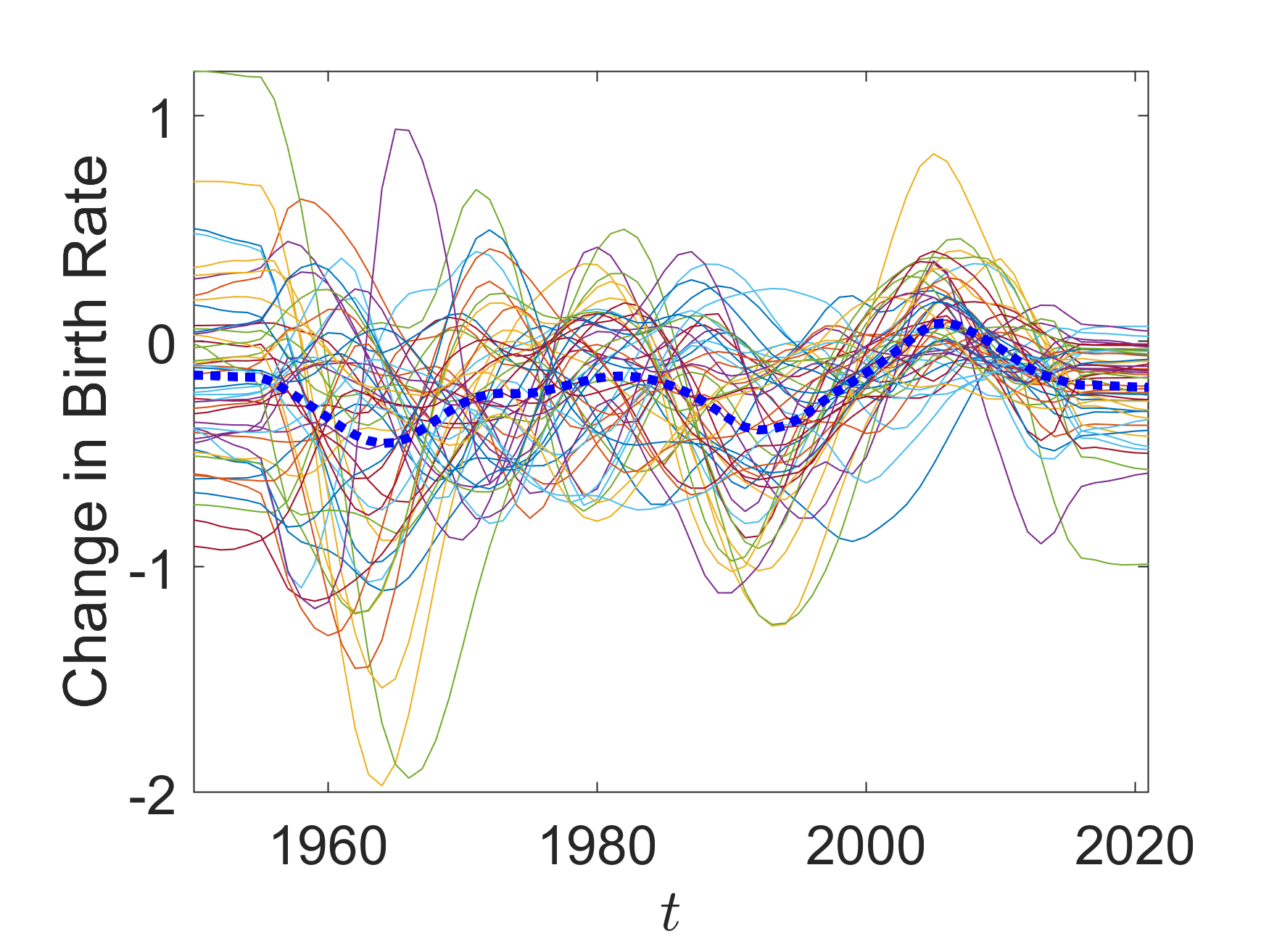}}
    \hspace{-0.2in}
    \subfloat[]{\includegraphics[height = 1.25in]{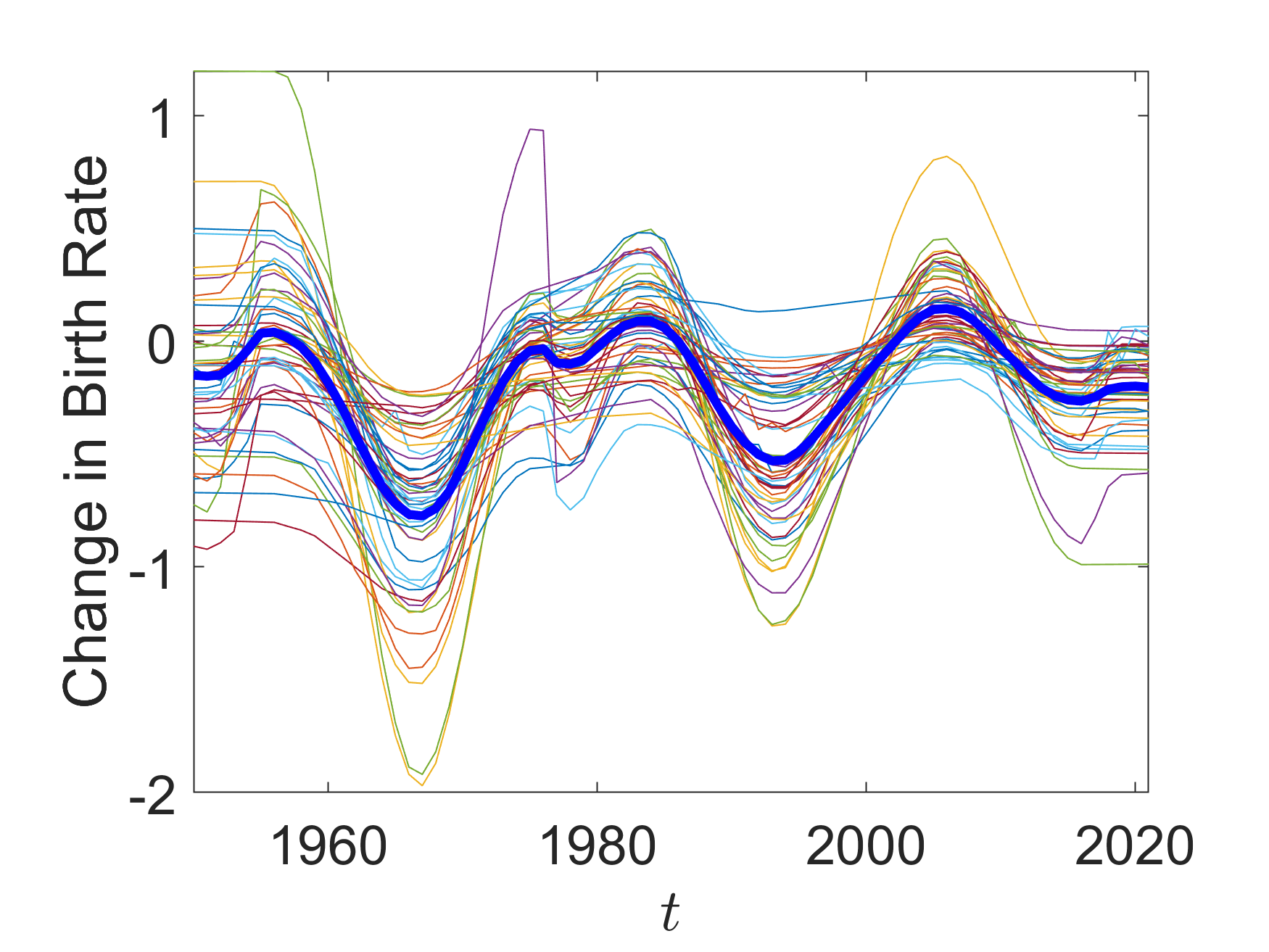}}
    \hspace{-0.1in}
    \subfloat[]{\includegraphics[height = 1.23in]{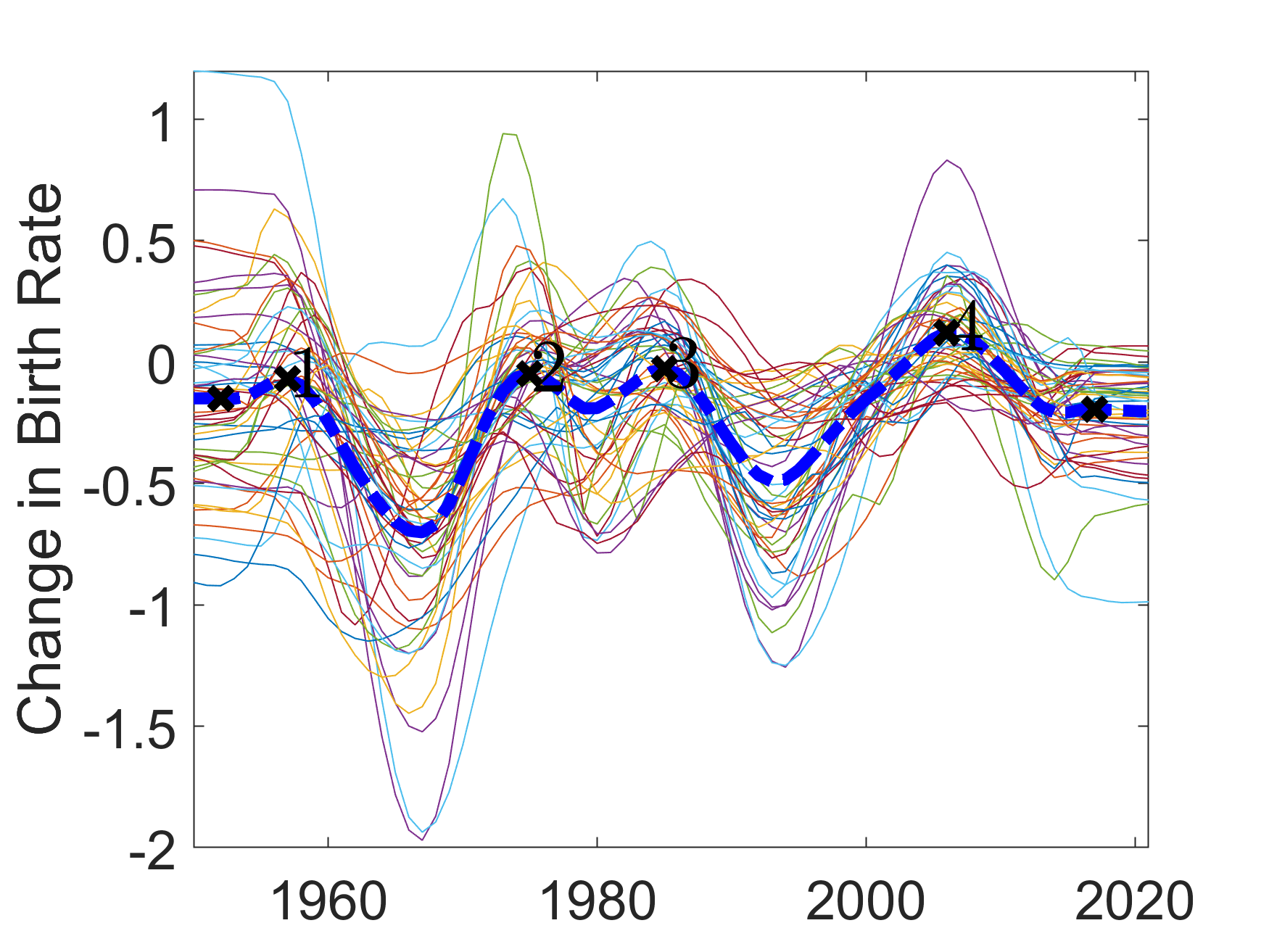}}
    \includegraphics[width = 0.8in]{fig/legend-rd-4.png}
    \hspace{0in}\\
    \vspace*{-0.05in}
    \hspace{-0.1in}
    \subfloat[]{\includegraphics[height = 1.25in]{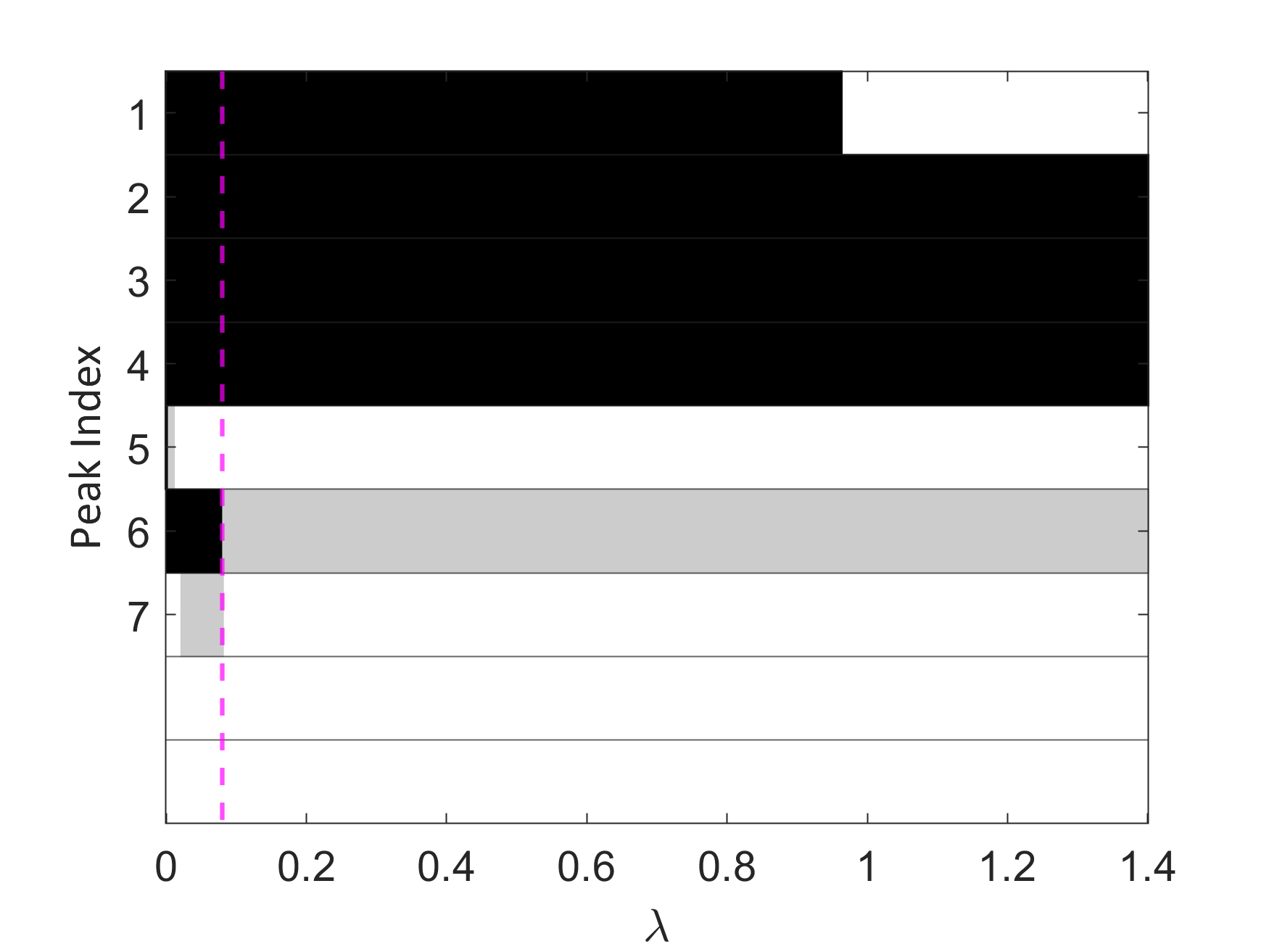}}
    \hspace{-0.1in}
    \subfloat[]{\includegraphics[height = 1.25in]{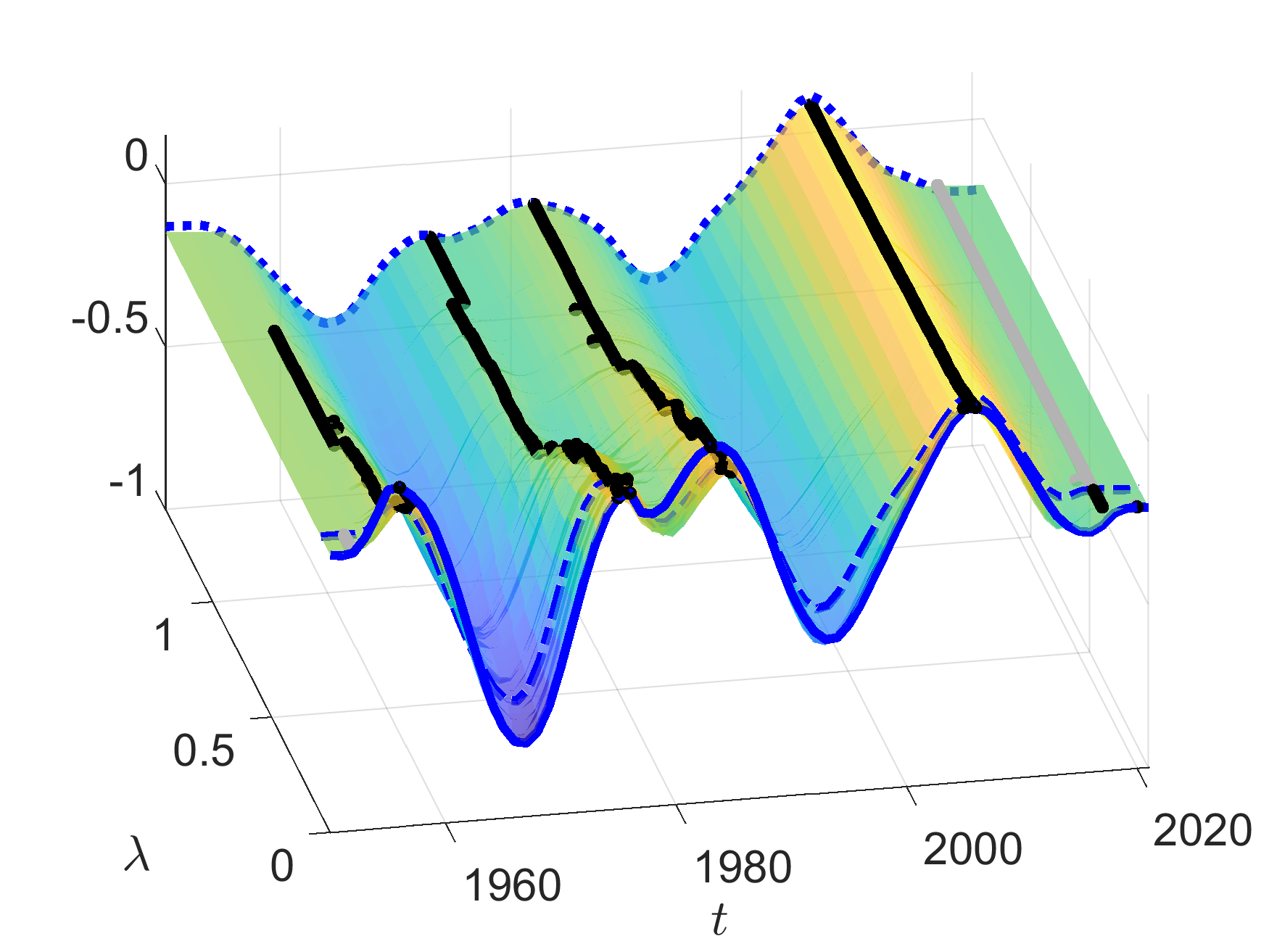}}
    \hspace{-0.1in}
    \subfloat[]{\includegraphics[height = 1.25in]{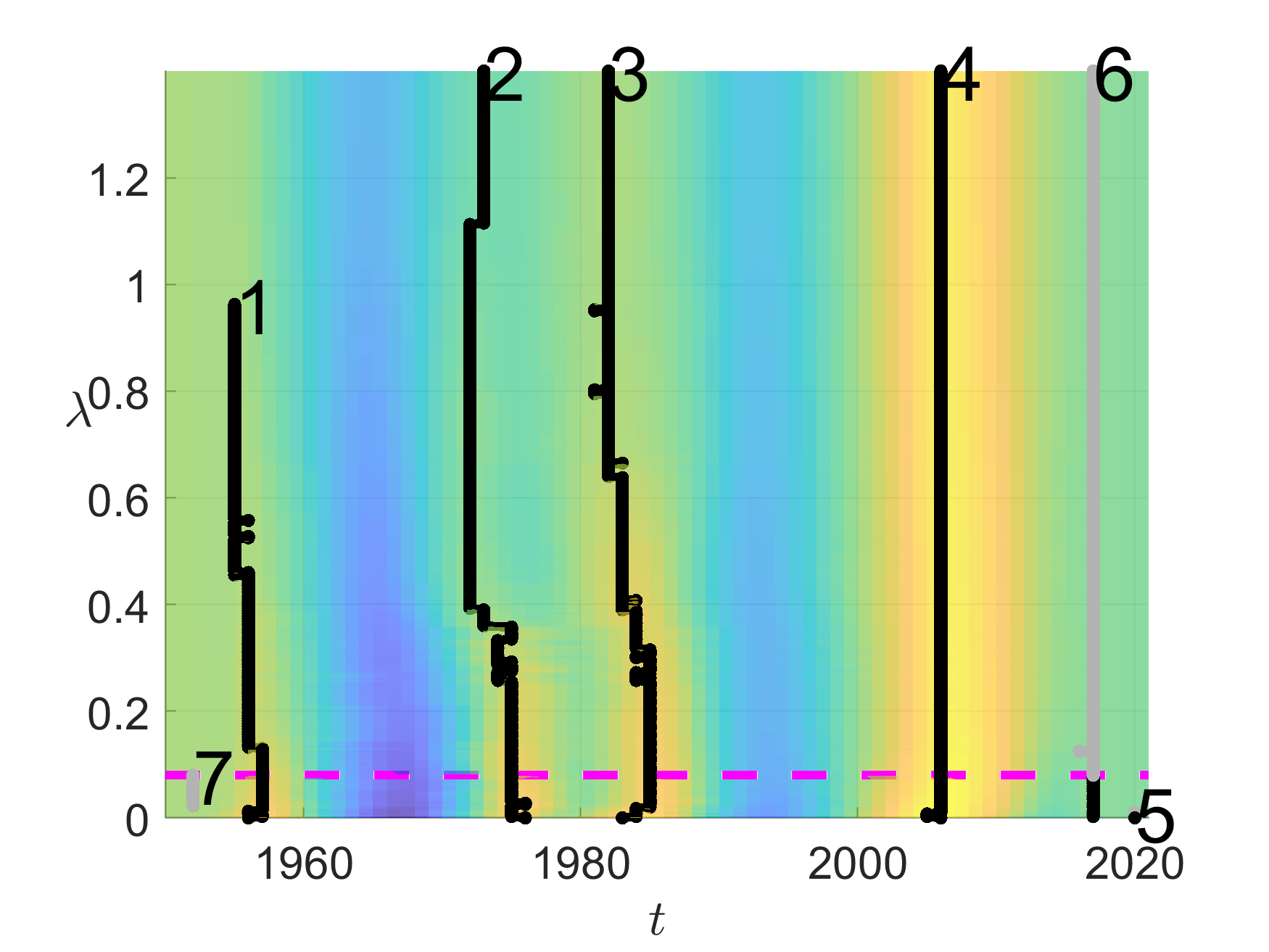}}
    \vspace*{-0.05in}
    \caption{Example 4:  The estimated number of peaks by PPDs is four (peaks 1,2, 3, and 4).}
    \label{fig: shape_estimation Birth Rate}
\end{figure*}

\noindent {\bf Example 5: Consumer Confidence Index of OCED Countries}:
This example analyzes the monthly consumer confidence index (CCI) of 38 OECD countries from 2008 to 2022, made public by the Organisation for Economic Co-operation and Development (OECD). The CCI is based on surveys of households' financial expectations and economic sentiments. A value above 100 indicates increased confidence and a willingness to spend, while a value below 100 signals pessimism and a tendency to save more and spend less. As pre-processing, we performed Lowess smoothing method with a window size of 24 months.

\begin{figure*}[htbp]
    \centering
    \subfloat[]{\includegraphics[height = 1.25in]{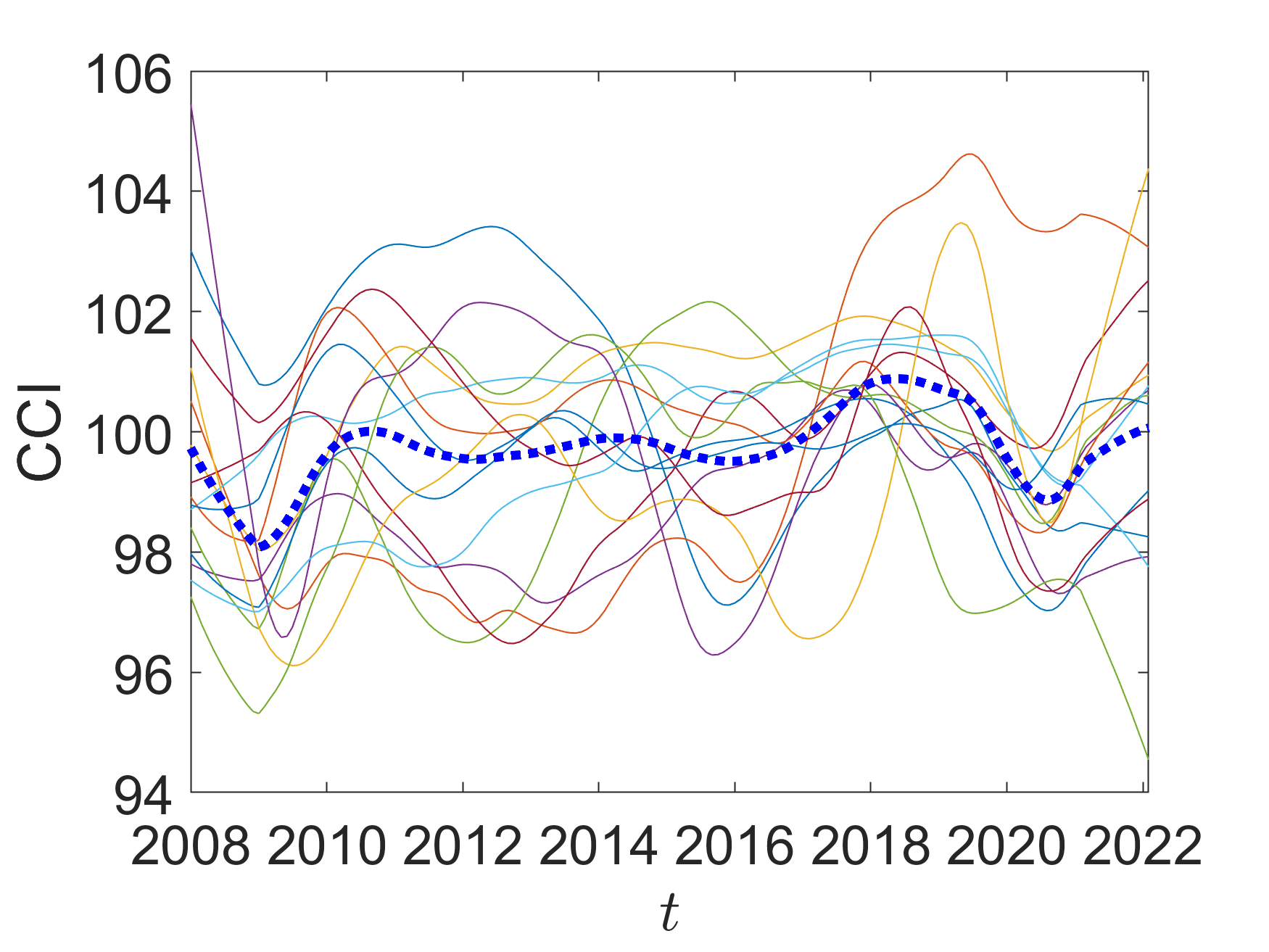}}
    \hspace{-0.2in}
    \subfloat[]{\includegraphics[height = 1.25in]{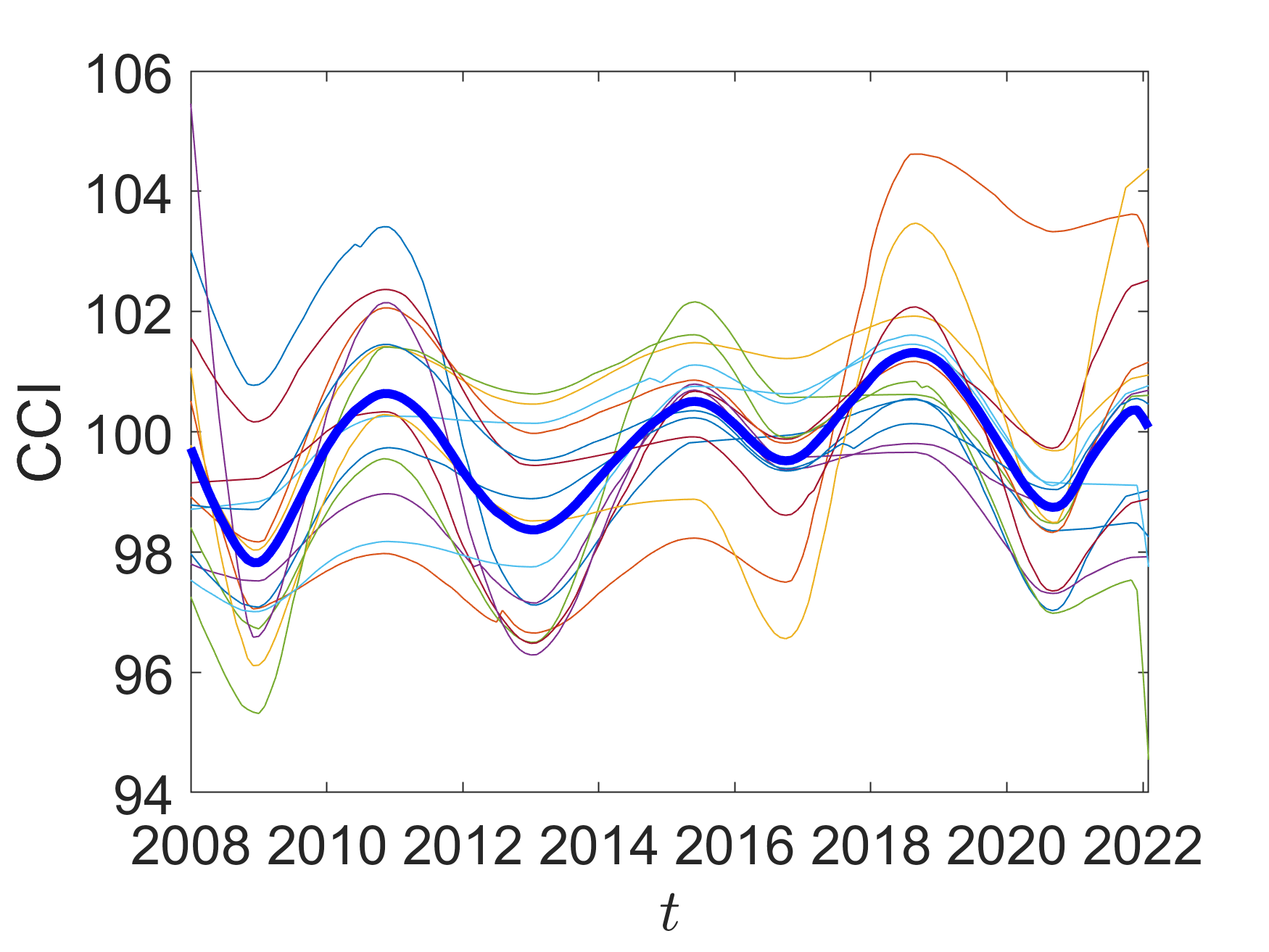}}
    \hspace{-0.2in}
    \subfloat[]{\includegraphics[height = 1.25in]{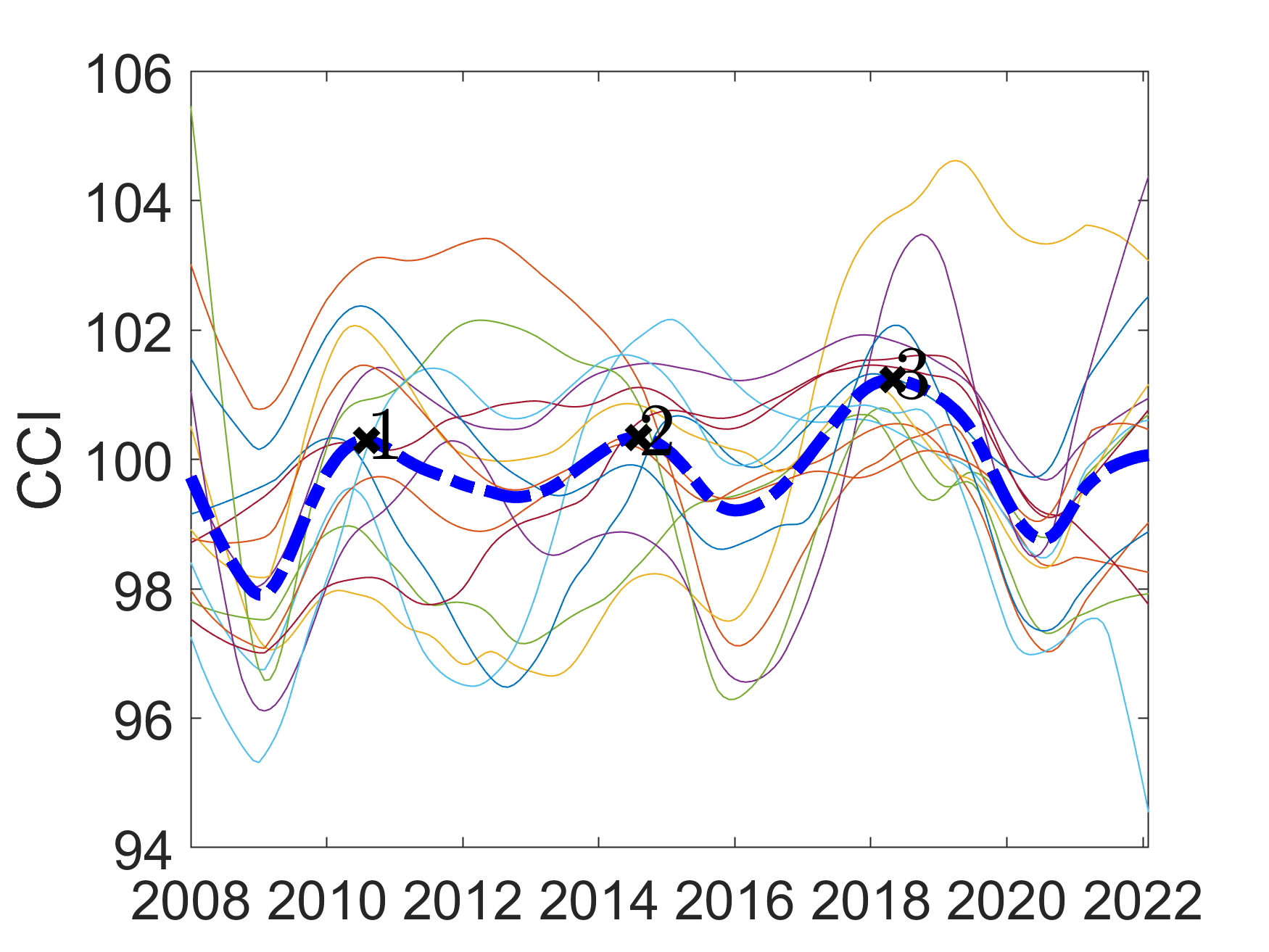}}
    \includegraphics[width = 0.8in]{fig/legend-rd-4.png}
    \hspace{0in}\\
    \vspace*{-0.05in}
    \hspace{-0.1in}
    \subfloat[]{\includegraphics[height = 1.25in]{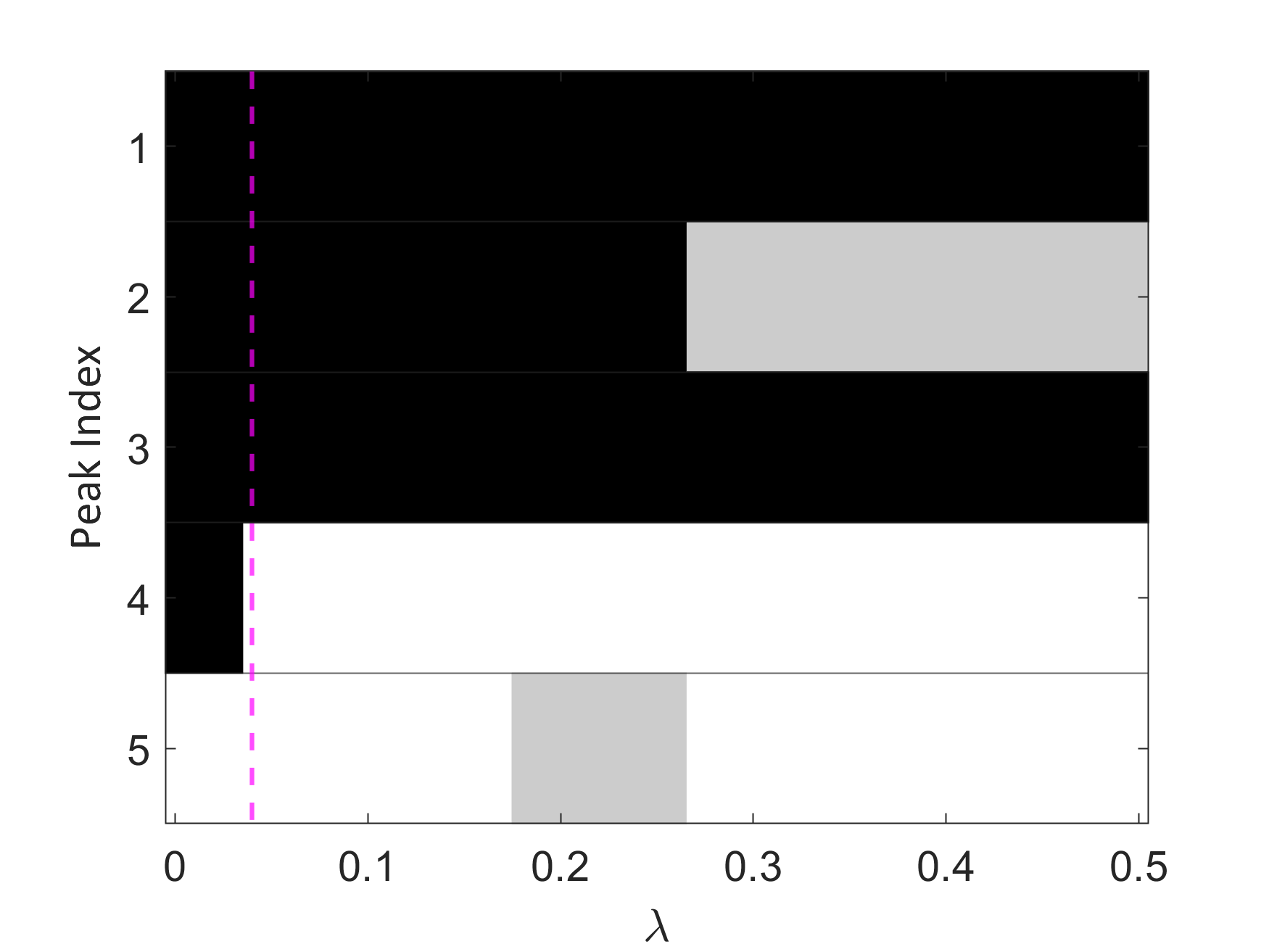}}
    \hspace{-0.1in}
    \subfloat[]{\includegraphics[height = 1.25in]{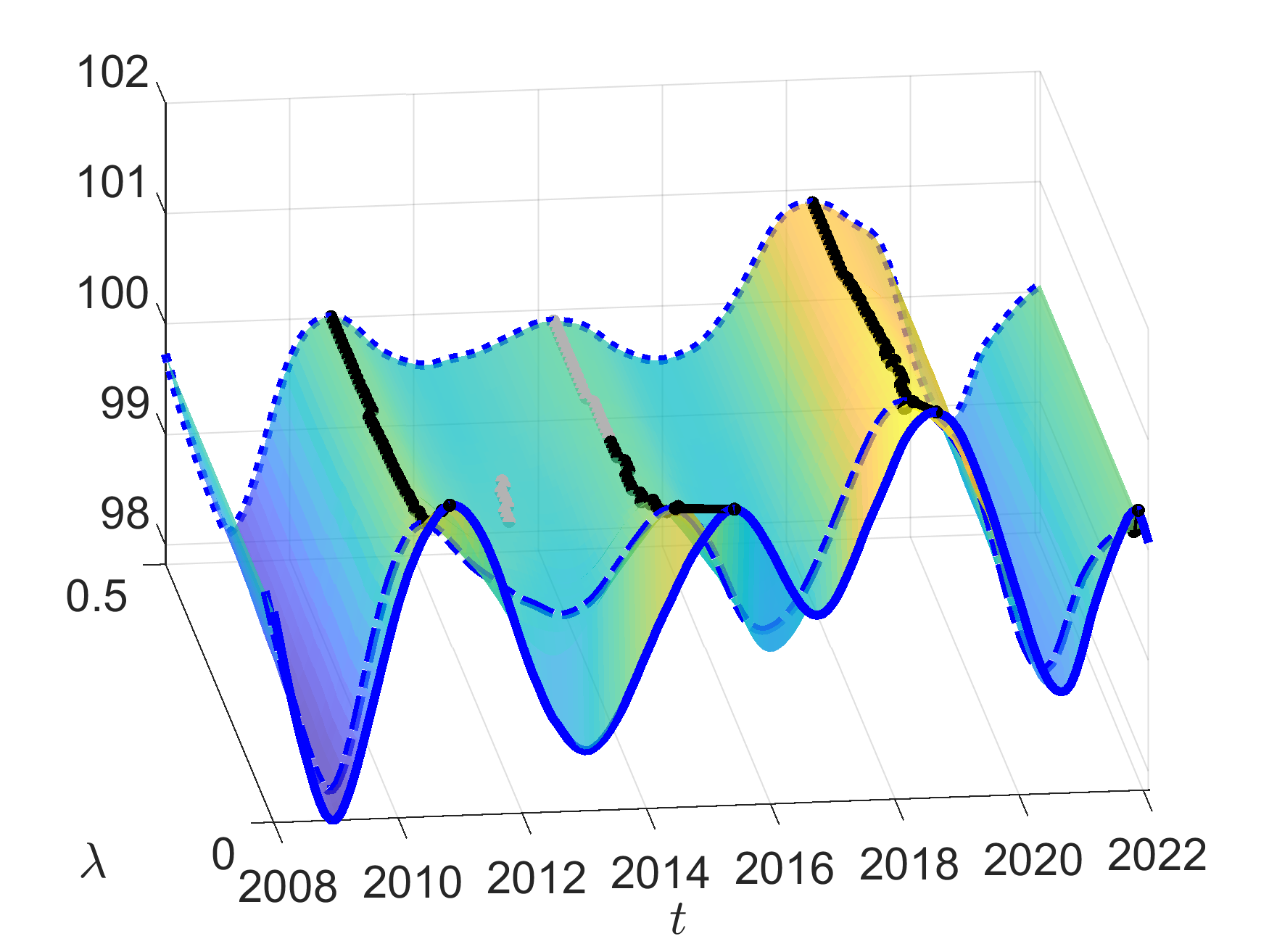}}
    \hspace{-0.1in}
    \subfloat[]{\includegraphics[height = 1.25in]{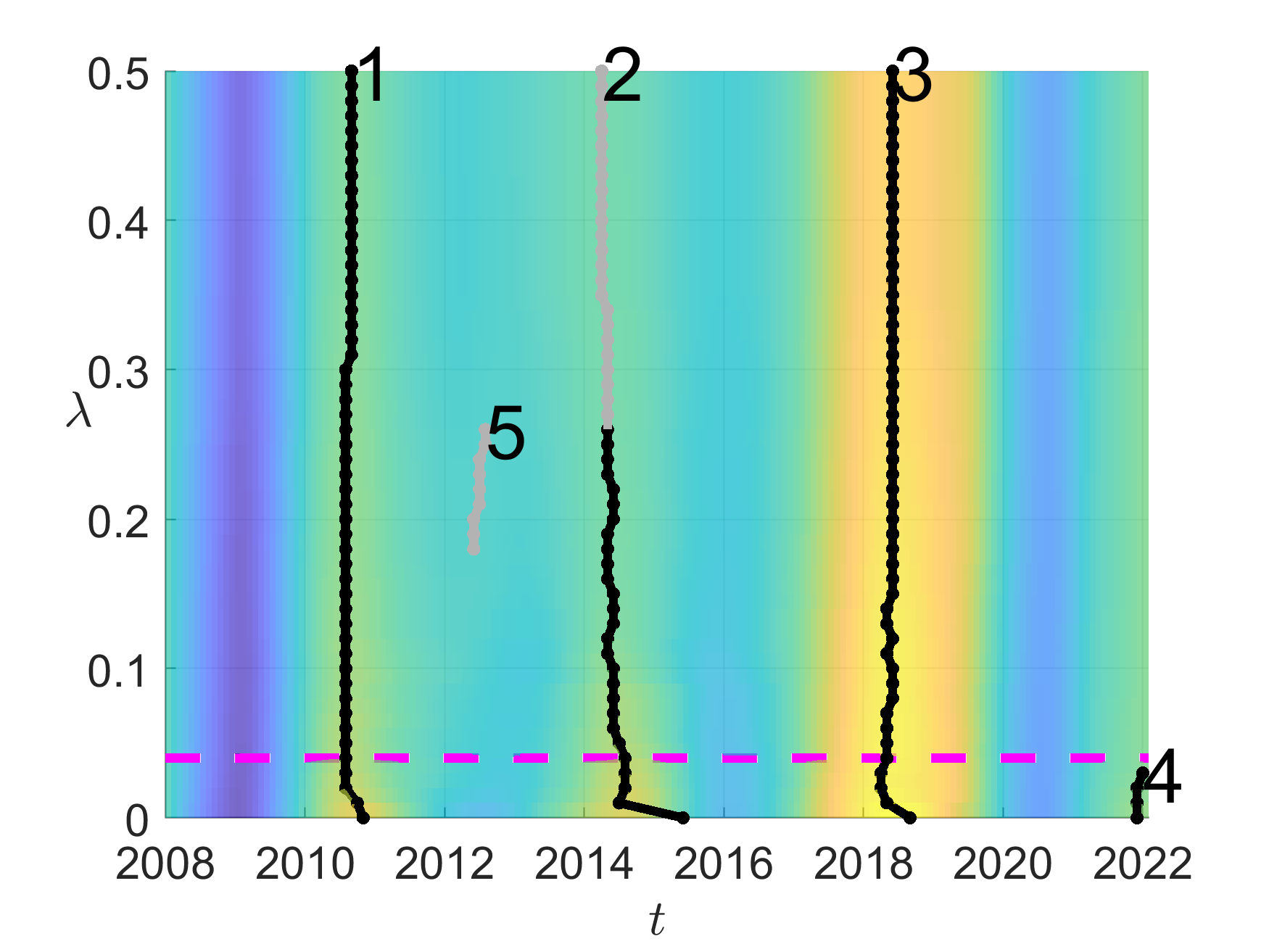}}
    \vspace*{-0.05in}
    \caption{(Shape Estimation of Consumer Confidence Index) This figure is laid out similar to Fig. \ref{fig: shape_estimation 1}. The estimated number of peaks by PPDs is three (1, 2, and 3.)}
    \label{fig: shape_estimation Consumer Confidence Index}
\end{figure*}

Analysis of the dataset using the PPD method, as shown in Fig. \ref{fig: shape_estimation Consumer Confidence Index}, reveals that (b) illustrates $\hat g_0$ has four peaks. However, PPDs in plots (d) through (f) suggest that $\lambda^*=0.04$, with three peaks (1, 2, and 3) identified as significant in $\hat g_{\lambda^*}$ in (c).

\section{Step 2: Peak-constrained Curve Estimation} \label{sec:Peak-constrained Curve Estimation}
The next step involves estimating $g$ by utilizing a pre-defined shape class and employing a penalized maximum-likelihood criterion. The objective here is to limit the exploration to the desired shape class and determine the optimal element of that class through geometric search. This technique is an adaptation of the shape-constrained density estimation method suggested in \citet{dasgupta2021} and has been customized for the functional estimation problem. It involves altering an initial estimate $\hat{g}_{init}$  while simultaneously maximizing a penalized log-likelihood function. (We use the peaks and valleys, including locations and heights, in $\hat{g}_{\lambda^*}$  and smooth interpolations between them to form an initial estimate, $\hat{g}_{init}$.) The optimization is conducted across the complete range of functions having the designated shape.


\subsection{Penalized-MLE Approach}
 Let $M$ be the number of {\it{extrema points}} of $\hat{g}_{init}$, which comprises interior peaks, valleys, and the two endpoints.
$M$ lies between $2m+1$ to $2m+3$ (inclusive), where $m$ is the estimated number of interior peaks. For instance, if both endpoints are minima and there are $m-1$ interior valleys, then $M$ equals $2m+1$. 
\\

\noindent
{\bf Set of Shape-Constrained Functions}:
Let a function $f$ have $M$ extrema points in $I$.
We define ${\mathcal F}_m \subset {\mathcal F}$ as the set of all functions with $m$ internal peaks on $I$. Any two elements of ${\mathcal F}_m$ differ in the locations and heights of their extrema points. Correspondingly, we are going to define two sets of variables. 

\begin{figure}[ht]
    \centering
    \begin{tabular}{ccc}
    \includegraphics[width=1in]{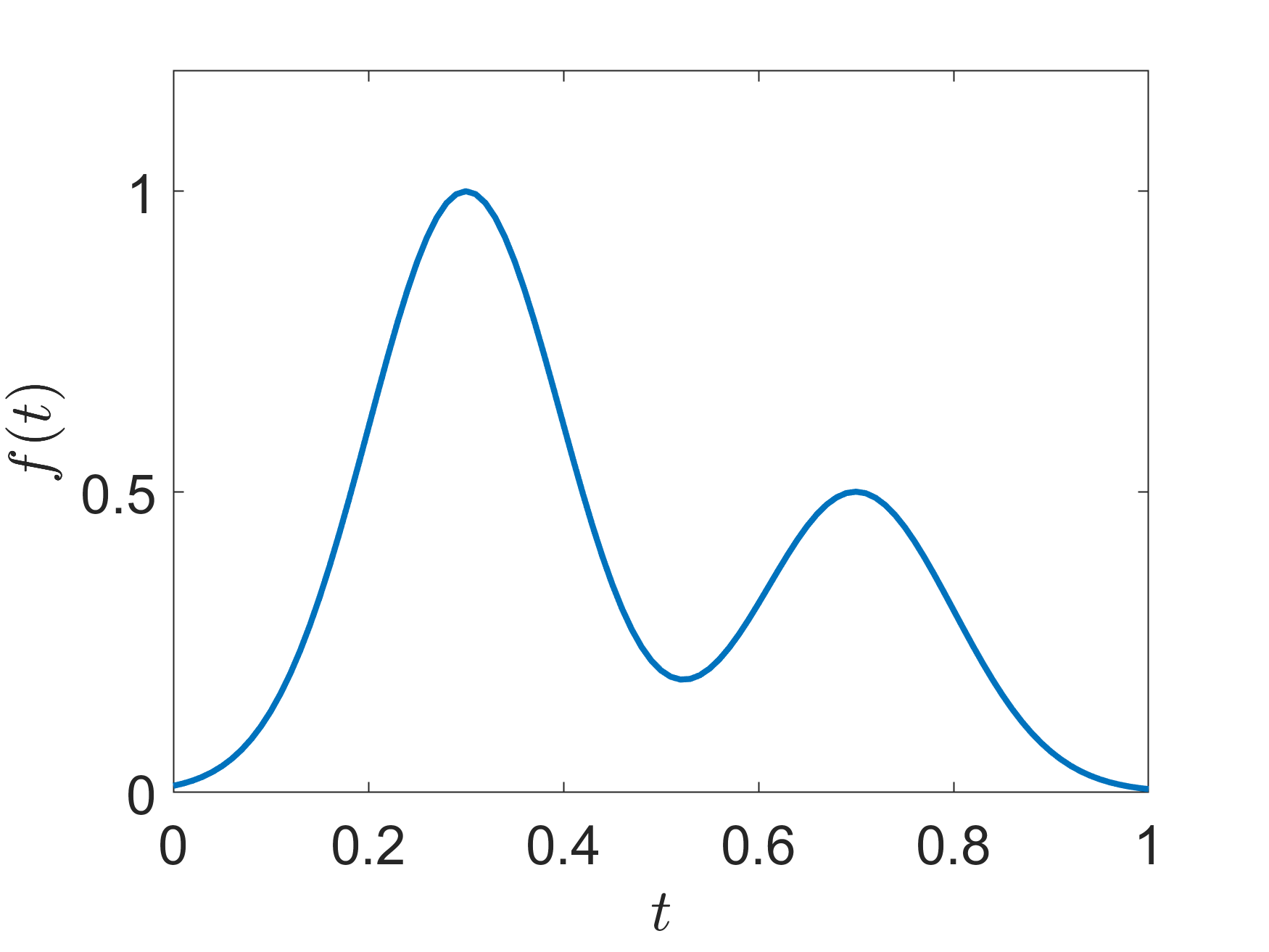} &
    \hspace*{-0.3in}  \includegraphics[width=1in]{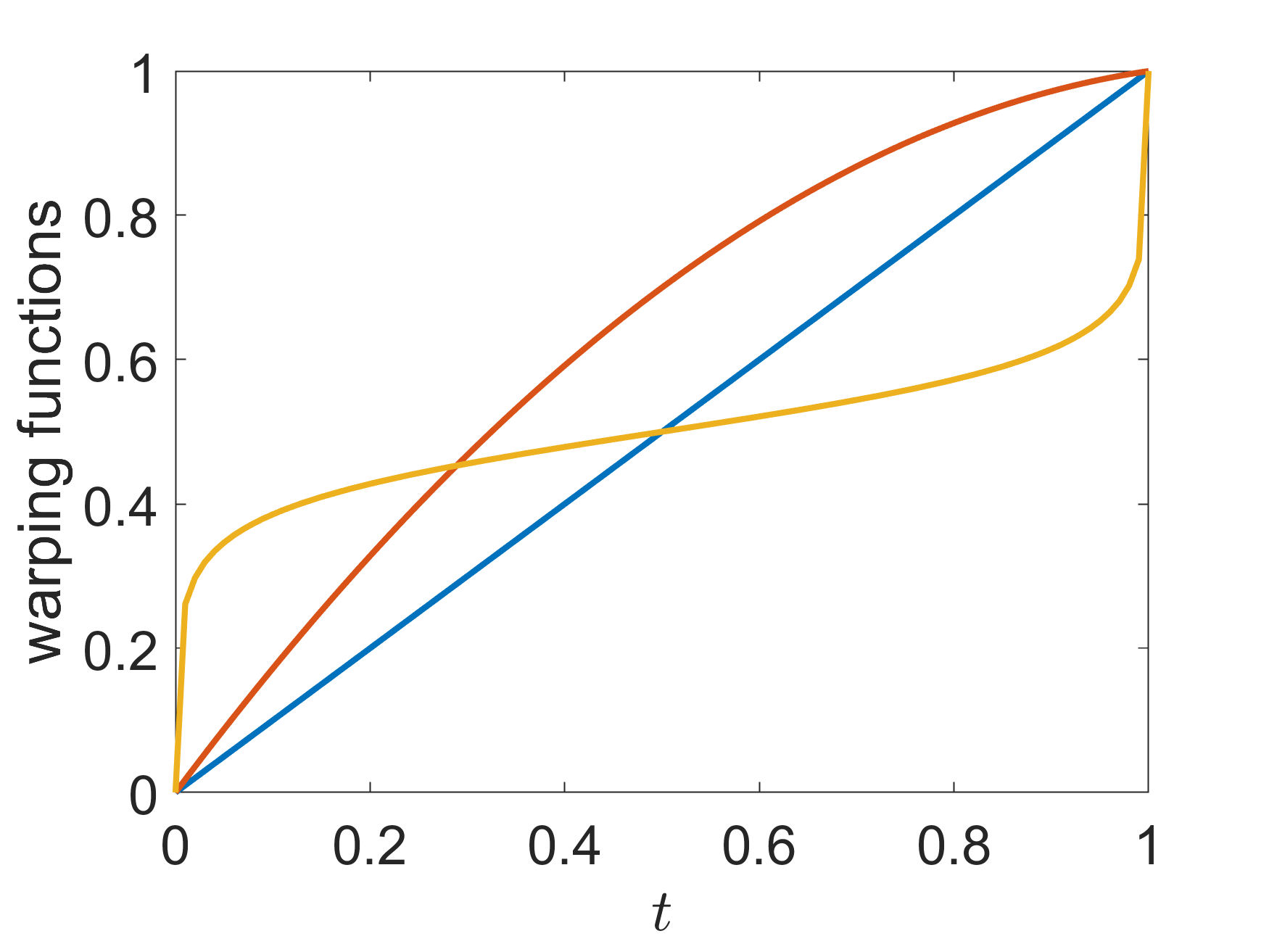} &
    \hspace*{-0.3in} \includegraphics[width=1in]{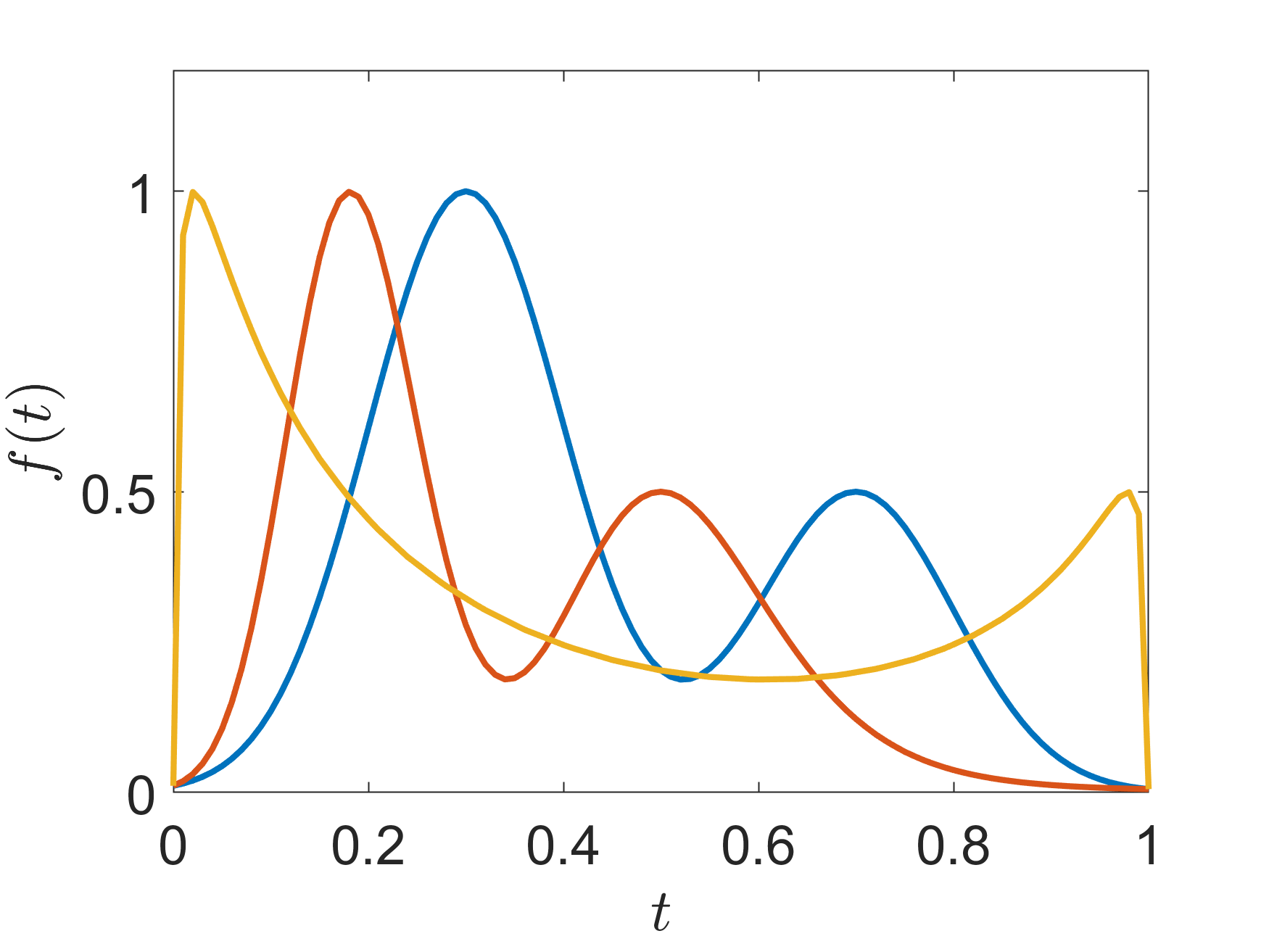} \\
    (a) & (b) & (c) \\
    \end{tabular}
    \caption{Time-warping of functions. (a) is the original function, (b) shows three warping functions, (c) is the function warped by corresponding $\gamma$'s in (b).}
    \label{fig: Diffeo}
\end{figure}

\begin{itemize}
\item {\bf Locations}: 
We will use the time-warping functions -- elements of $\Gamma$  -- to vary the locations of the extrema points. That is, the composition $\hat{g}_{init} \circ \gamma$, for a $\gamma \in \Gamma$, changes the locations of the extrema points while maintaining their heights and order. Fig. \ref{fig: Diffeo} illustrates how the number and the heights of the peaks of a function are invariant to time-warping.

\item {\bf Heights}: We also define ${\bf s} =\{s_1, s_2, \ldots, s_{M}\}$ as the heights of the extremal points in $\hat{g}_{init}$. There are some natural constraints on the values of ${\bf s}$. A valley's height should be less than the heights of its neighboring peaks. Let ${\mathcal S}$ be the set of all vectors ${\bf s}$ that satisfy these constraints. We will use $f_{\bf s} \in {\mathcal F}_m$ to denote a function that has ${\bf s}$ as the height vector of its extrema.
\end{itemize}

Starting with the initial estimate $\hat{g}_{init}$, we adjust the locations and the heights of the extrema points in order to explore the set ${\mathcal F}_m$. The final estimate is given by: $\hat{g} = (f_{\mathbf{s}^*} \circ \gamma^*)$, where 
\begin{align}\label{eqn:final-optimization}
\begin{split}
(\gamma^*, \mathbf{s}^*) = \underset{\gamma \in \Gamma,\mathbf{s} \in {\mathcal S}}{\mbox{argmin}} \Bigg( \sum_{i=1}^n \int_0^1 (f_{\mathbf{s}}(\gamma(t)) - \tilde{f}_{\lambda^*,i}(t))^2 dt \\
+ \rho \int_0^1 \ddot{f}_{\mathbf{s}}(\gamma(t))^2dt \Bigg)
\end{split}
\end{align}
Here $\rho> 0$ is an infinitesimal 
weight ($\sim 10^{-8}$) for favoring smooth functions. 
This solution differs from Eqn.~\ref{eq:penalized-align-L2} in several ways. Here we optimize over only one $\gamma$ while Eqn.~\ref{eq:penalized-align-L2} uses a $\gamma_i$ for each observation. Also, we use the partially aligned data that favors $m$ peaks rather than using the original data. 

\begin{figure}[ht]
    \centering
    \begin{tabular}{cc}
    \hspace*{-0.4in}  \includegraphics[width=1.7in]{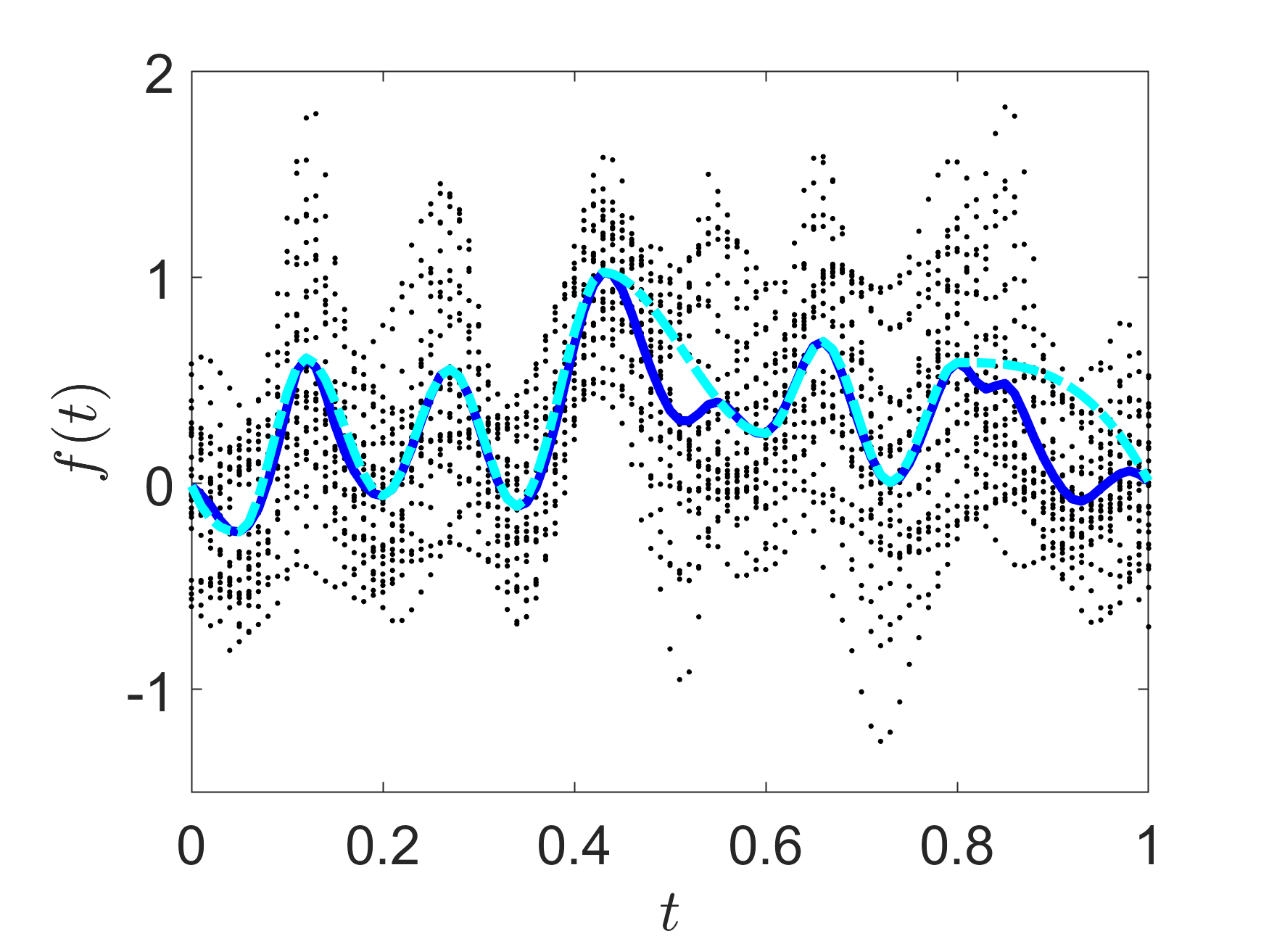} &
    \hspace*{-0.3in}  \includegraphics[width=1.7in]{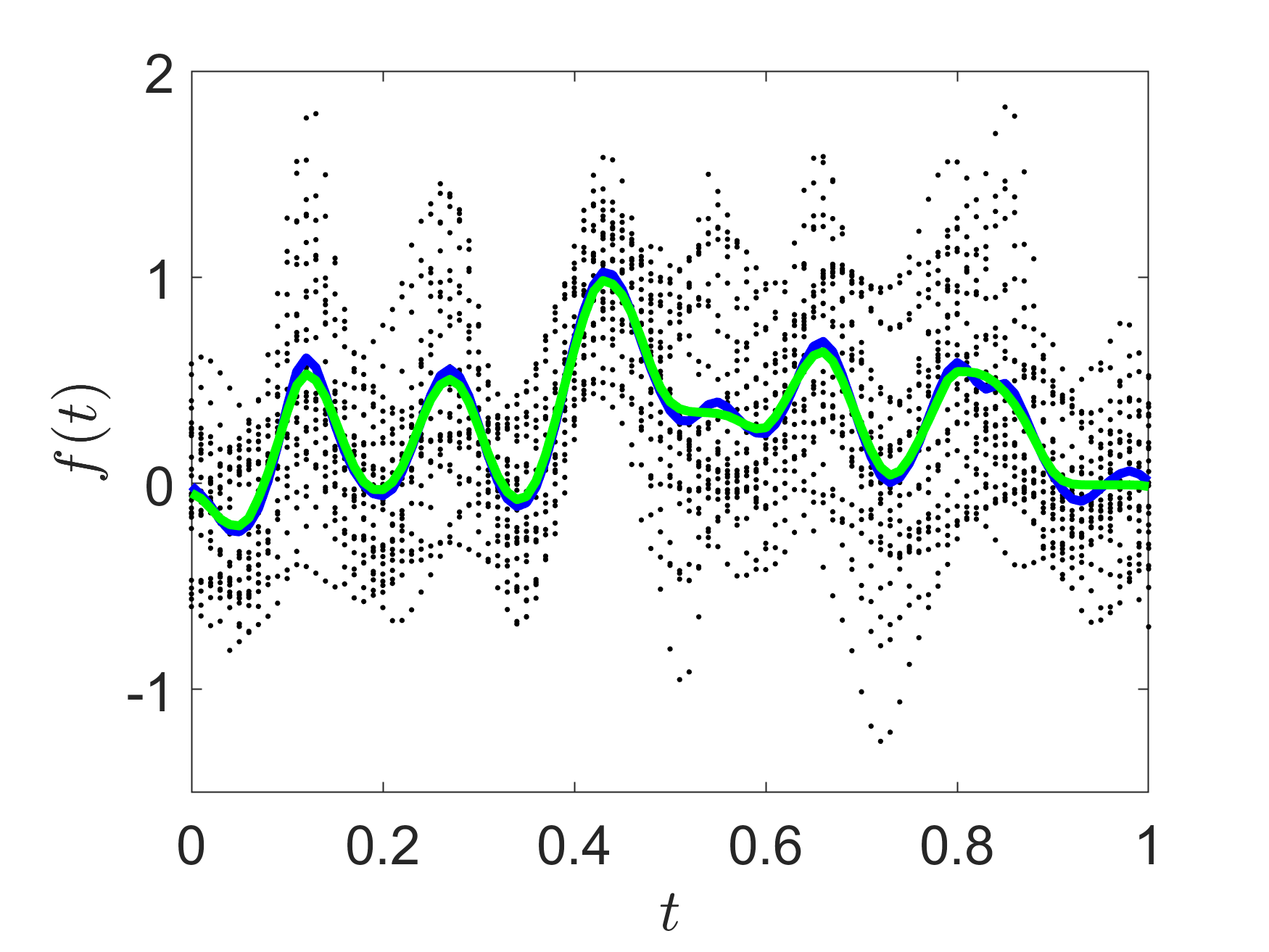}\\
    (a) & (b)
    \end{tabular}
    \caption{(Shape Constrained Function Estimation) he dotted points represent the partially-aligned functions $\{ \tilde{f}_{\lambda^*,i}\}$ and blue lines are their cross sectional mean. The cyan and the green lines are the initial and final function estimates of $g$.}
    \label{fig: curve_est}
\end{figure}

The optimization problem in Eqn.~\ref{eqn:final-optimization} is solved using the \texttt{fmincon} function in Matlab. However, since $\Gamma$ is a nonlinear manifold of infinite dimension, direct optimization poses a challenge. To overcome this, we employ an SRVF map followed by an inverse exponential map to represent the warping functions in a vector space, and use an orthogonal basis to represent $\gamma$ by its coefficients. Interested readers can find the optimization details in \cite{dasgupta2017twostep}. Algorithm \ref{alg: Peak-constrained Curve Estimation} outlines the steps for shape-constrained estimation of $g$. Fig. \ref{fig: curve_est} presents an example. In Plot (a), we show partially-aligned functions ${ \tilde{f}_{\lambda^*,i}}$ as dotted points, along with the initial estimate $\hat{g}_{init}$ in cyan color. Plot (b) shows the result of Algorithm~2 in green color. The dark blue line represents the ground truth $g$.

\begin{algorithm}[ht]
\begin{algorithmic}[1]
\Require Data: $\{\tilde{f}_{\lambda^*,i}\}_{i=1}^{n}$, initial shape estimate $\hat{g}_{init}(t)$


\State Initialize the heights of the extrema points of $\hat{g}_{init}$: $\{s_1,s_2,...,s_{M}\}$, 
\State Initialize $\gamma \leftarrow \gamma_{id}$.
\State Obtain $\hat{\gamma}$ and $\hat{\mathbf{s}}$ by solving Eqn.~\ref{eqn:final-optimization} by \texttt{fmincon}.

    
\State \Return $\hat f(t) = f_{{\hat{\gamma}},\mathbf{\hat{s}}}(t)$
\end{algorithmic}
\caption{Peak-constrained Curve Estimation} \label{alg: Peak-constrained Curve Estimation}
\end{algorithm}

\subsection{Bootstrapping for Estimating Confidence Bands}

To evaluate the performance of the estimate $\hat{g}$, we compute pointwise confidence bands for our estimator and display them with the estimation results. In the case of simulated data, we check if these estimated confidence bands contain the actual signal and validate our results. Algorithm~\ref{alg: Constructing Pointwise Functional Confidence Band} lays out the steps for computing these confidence bands.

\begin{algorithm}[ht] 


\begin{algorithmic}[1]
\Require $\{f_i\}_{i=1}^n$, Initial shape estimate, $\hat{g}_{init}$, Estimated smoothing parameter, $\lambda^*$
\For{$j = 1,2,\dots,B$}
    \State Resample $n$ functions with replacement to obtain a bootstrapped set of functions, $\{f_i^{(j)}\}_{i=1}^{n}$.
    \State Obtain partially aligned functions with smoothing parameter, $\lambda^*$,  $\{\tilde{f}_{\lambda^*,i}^{(j)}\}_{i=1}^{n}$
    \State Given $\hat{g}_{init}$ and $\{\tilde{f}_{\lambda^*,i}^{(j)}\}_{i=1}^{n}$, obtain the best estimate of $g$ by Algorithm \ref{alg: Peak-constrained Curve Estimation}: $\hat g^{(j)}$
\EndFor
\State With $\{\hat g^{(j)}\}_{j=1}^{B}$, \Return $(\alpha/2)$ and $(1-\alpha/2)$ quantiles to set the lower and upper bound of the confidence bound point-wisely. 
\end{algorithmic}

\caption{Constructing Pointwise Functional Confidence Band}\label{alg: Constructing Pointwise Functional Confidence Band}
\end{algorithm}

\subsection{Function Estimator Properties}

The basic approach for shape-constrained function estimation follows that presented in \cite{dasgupta2021}, albeit with a different action of the group $\Gamma$ since that previous paper focused on estimating {\it pdf}s. The asymptotic properties of the current estimator $\hat{g}$ remain similar to that of earlier work, and we refer the reader to that paper for details.

\section{Experimental Results: Function Estimation} \label{sec: result}
This section evaluates the performance of our proposed method using a range of simulated and real datasets. The goal is to compare our estimate $\hat{g}$ with two prior solutions, which include: (1) the mean of unaligned functions, $\bar{f} = \hat{g}_\infty$, and (2) the fully-elastic mean, $\hat{g}_0$. In addition, we study the results obtained using Eqn.~\ref{eq:penalized-align-L2} in the case of simulated data. However, due to the unknown optimal $\kappa$ value (as discussed in Section \ref{sec: l2 alignment}), we present results for several $\kappa$ values. We use RMSE as the metric for comparing the estimation errors and provide 95\% pointwise bootstrap confidence bands to visualize the variability of our estimator.

\subsection{Simulation Studies} \label{sec:Simulation Studies}

We try four different simulation scenarios on the domain $I = [0,1]$. In each experiment, we choose a different $g$ and generate $n = 100$ samples according to Eqn.~\ref{eq: model specification}. To compute and compare different solutions, we use $50$ independent simulations of each scenario and compute statistics of RMSEs of the estimators.

\begin{figure*}[htbp]
    \centering
    \subfloat[]{\includegraphics[height = 1.2in]{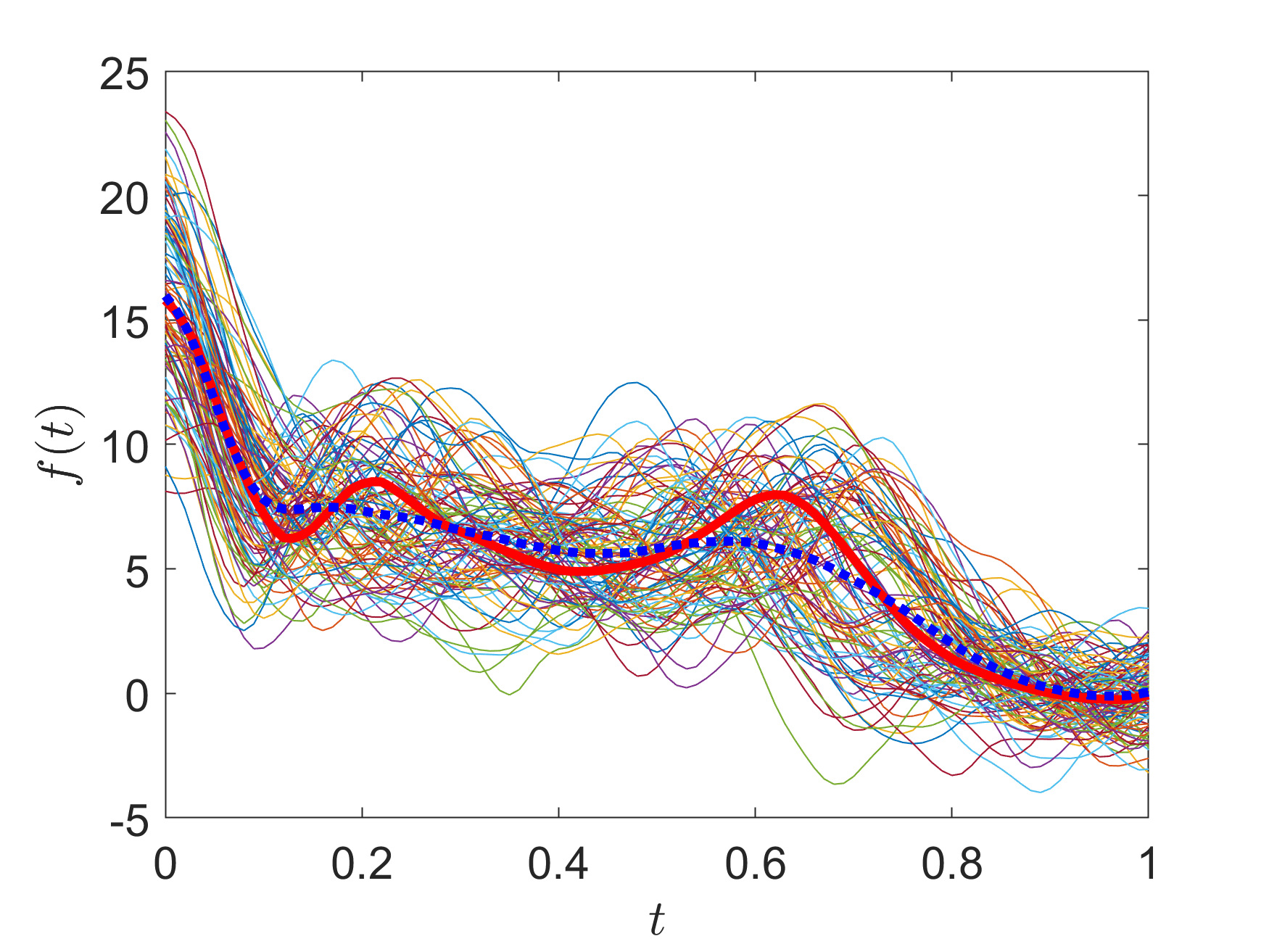}}
    \hspace{-0.2in}
    \subfloat[]{\includegraphics[height = 1.2in]{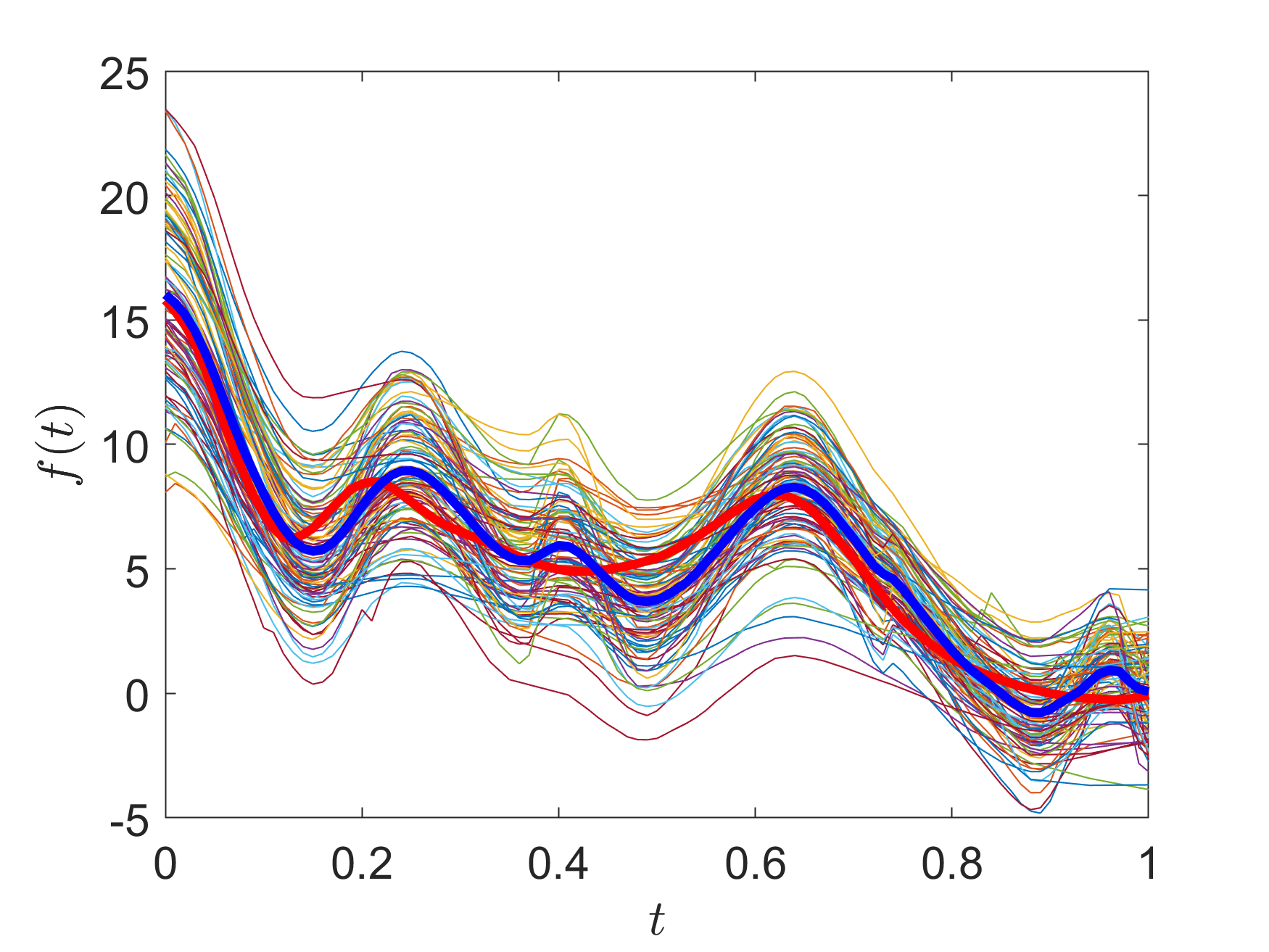}}
    \hspace{-0.2in}
    \subfloat[]{\includegraphics[height = 1.2in]{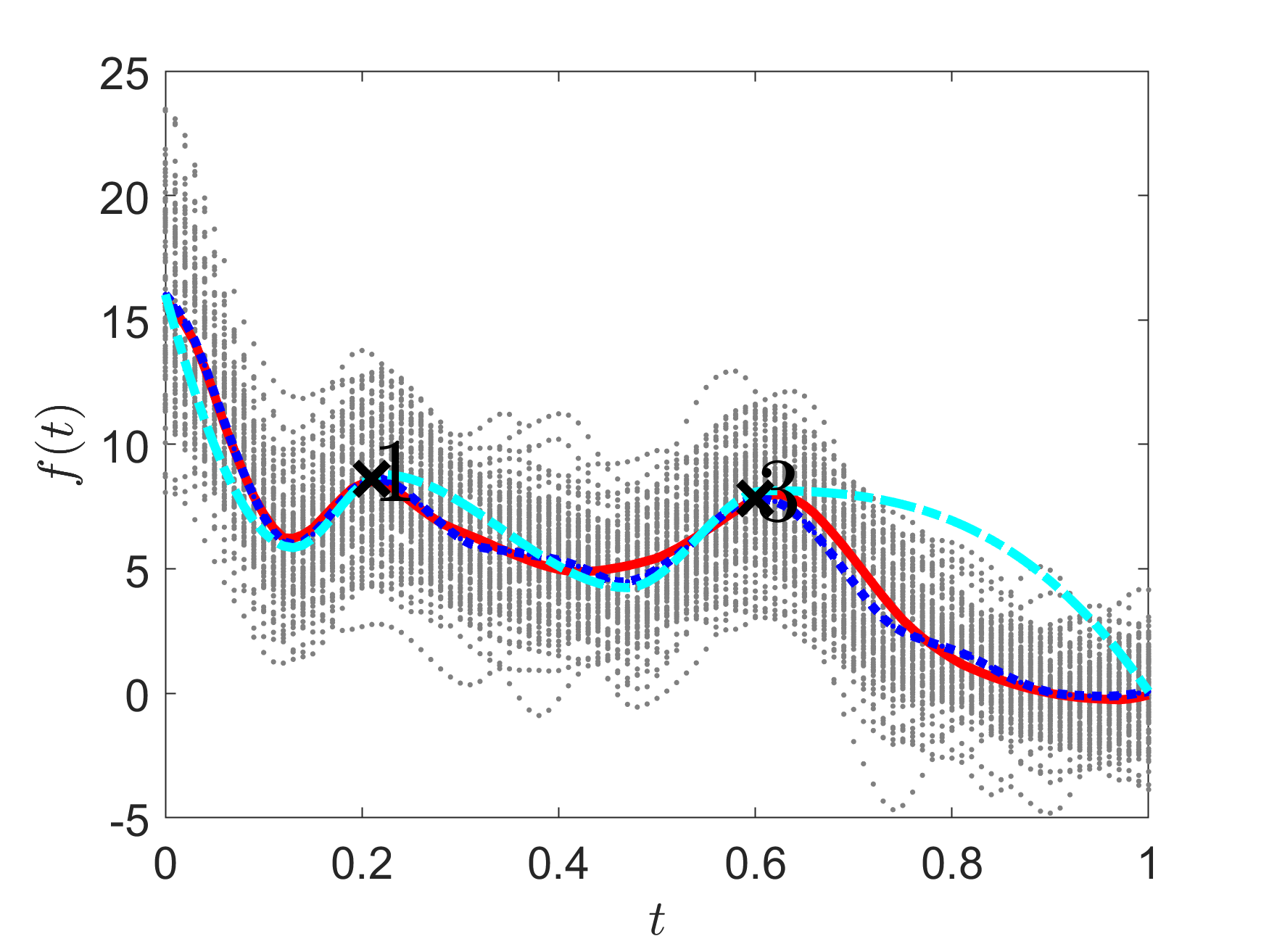}}
    \includegraphics[height = 1in]{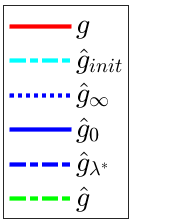}
    \hspace{0in}\\
    \vspace*{-0.05in}
    \hspace{-0.2in}
    \subfloat[]{\includegraphics[height = 1.2in]{fig/ex1_3.png}}
    \hspace{-0.2in}
    \subfloat[]{\includegraphics[height = 1.2in]{fig/ex1_ppd3_1.png}}
    \hspace{-0.2in}
    \subfloat[]{\includegraphics[height = 1.2in]{fig/ex1_ppd3_2.png}}
    \hspace{-0.2in}
    \subfloat[]{\includegraphics[height = 1.2in]{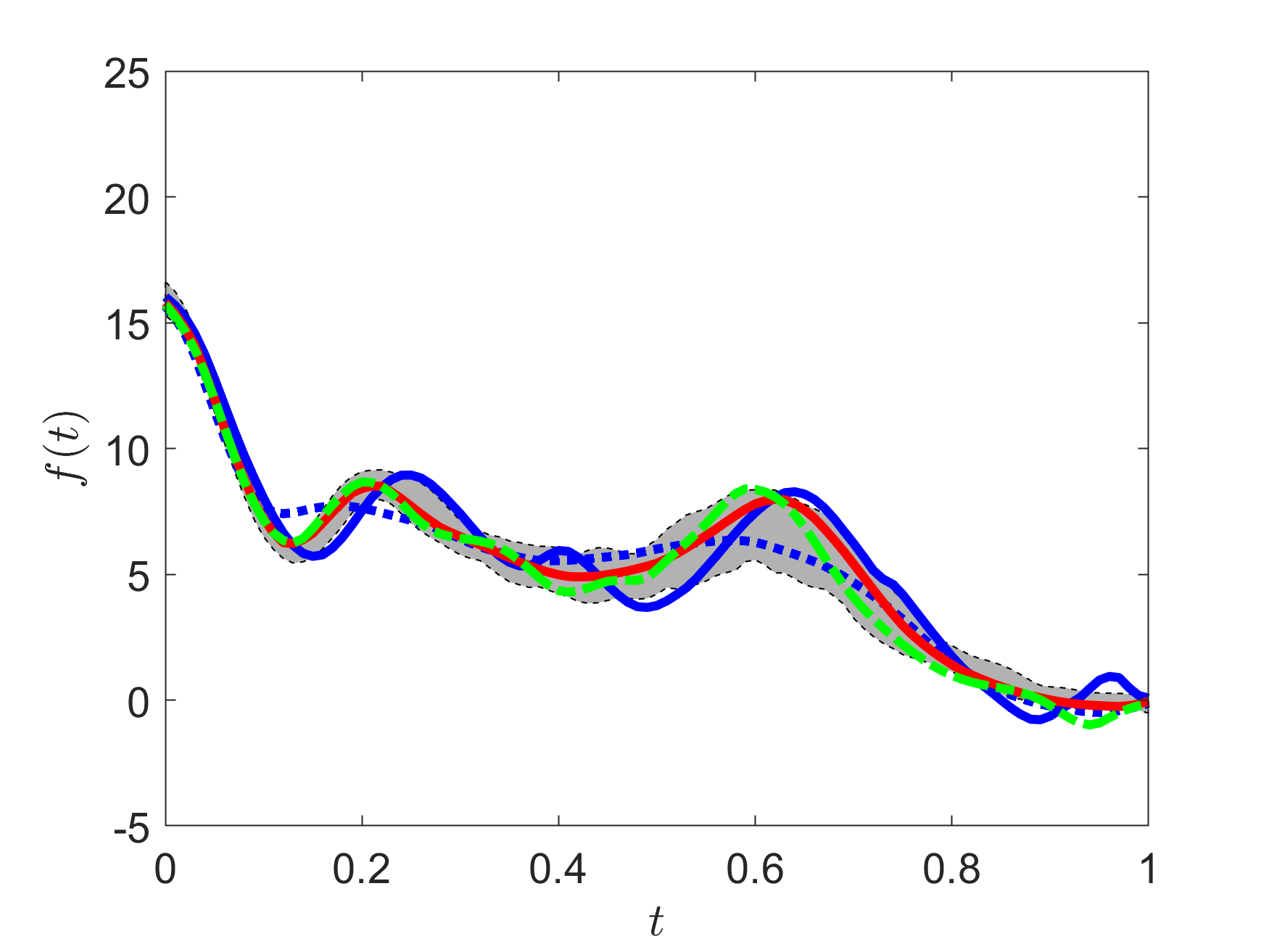}}
    \vspace*{-0.05in}
    \caption{(Dataset 1) Plots (a) and (b) are aligned functions at $\lambda = \infty$ and $0$. Thes figures also show $\hat{g}_0$ (bold blue), $\hat{g}_{0.05}$ (bold dotted-line blue) and $\hat{g}_{\infty}$ (bold dotted blue). The ground truth $g$, initial estimate $\hat{g}_{init}$, and the final estimate $\hat{g}$ are drawn in red, cyan, and green, respectively. Plot (d) is the PPD, with $\lambda^*=0.05$. Plot (e) and (f) show the evolution of $\hat{g}_{\lambda}$ versus $\lambda$. The black lines on the surface indicate the interior peaks, and (f) is the top view of (e). Plot (c) contains the aligned functions at $\lambda^*$. Panel (g) compares the unaligned mean, fully elastic mean, our estimate, and the ground truth, $g$. A 95\% bootstrap confidence band is drawn using gray regions.}
    \label{fig: simulation_data 1}
\end{figure*}

\begin{itemize}
  \item {\bf{\emph{Simulated Dataset 1}}}
\end{itemize}
Fig. \ref{fig: simulation_data 1} shows results from experiments on the first dataset. The original data is shown in panel (a), with the true function $g$ drawn in red. 
Plots (a) and (b) show the alignment of functions under the extreme values of $\lambda$: $\infty$ and $0$, respectively. The cross-sectional means are shown in blue in each case. The PPD bar chart for this data is shown in (d), which estimates two significant internal peaks. The optimal parameter $\lambda^* = 0.05$ is shown in the magenta, dotted vertical line in (d). Plots (e) and (f) are PPD surfaces displaying gradual changes in $\hat{g}_\lambda$ when $\lambda$ increases. The black lines on this surface are traces of significant peaks. Plot (f) shows the top view of the surface.  
For this $\lambda^*$, (c) shows the partially aligned functions, $\{\tilde{f}_{\lambda^*,i}\}_{i=1}^{n}$ (as a point cloud to help visibility). Their mean $\hat{g}_{\lambda^*}$ is shown using a dotted-line blue curve and captures the geometric features of $g$ better than $\hat g_0$ and $\hat g_\infty$. The dashed cyan line is the initial estimate, $\hat{g}_{init}$.

Next we use these quantities to estimate $g$ as described in Section \ref{sec:Peak-constrained Curve Estimation}. The
green dashed curve in (g) is the final estimate $\hat{g}$. We also display a pointwise bootstrap confidence band using a gray region. We can see that both the function estimate (in green) and the band (in gray) are distinguishable from $\hat{g}_{0}$ and $\hat{g}_{\infty}$ over large intervals. Furthermore, $\hat g$ is the closest function to the ground truth $g$ in terms of the $\mathbb L^2$ distance. A detailed quantitative evaluation of the results is presented later. 

\begin{figure*}[htbp]
    \centering
    \subfloat[]{\includegraphics[height = 1.2in]{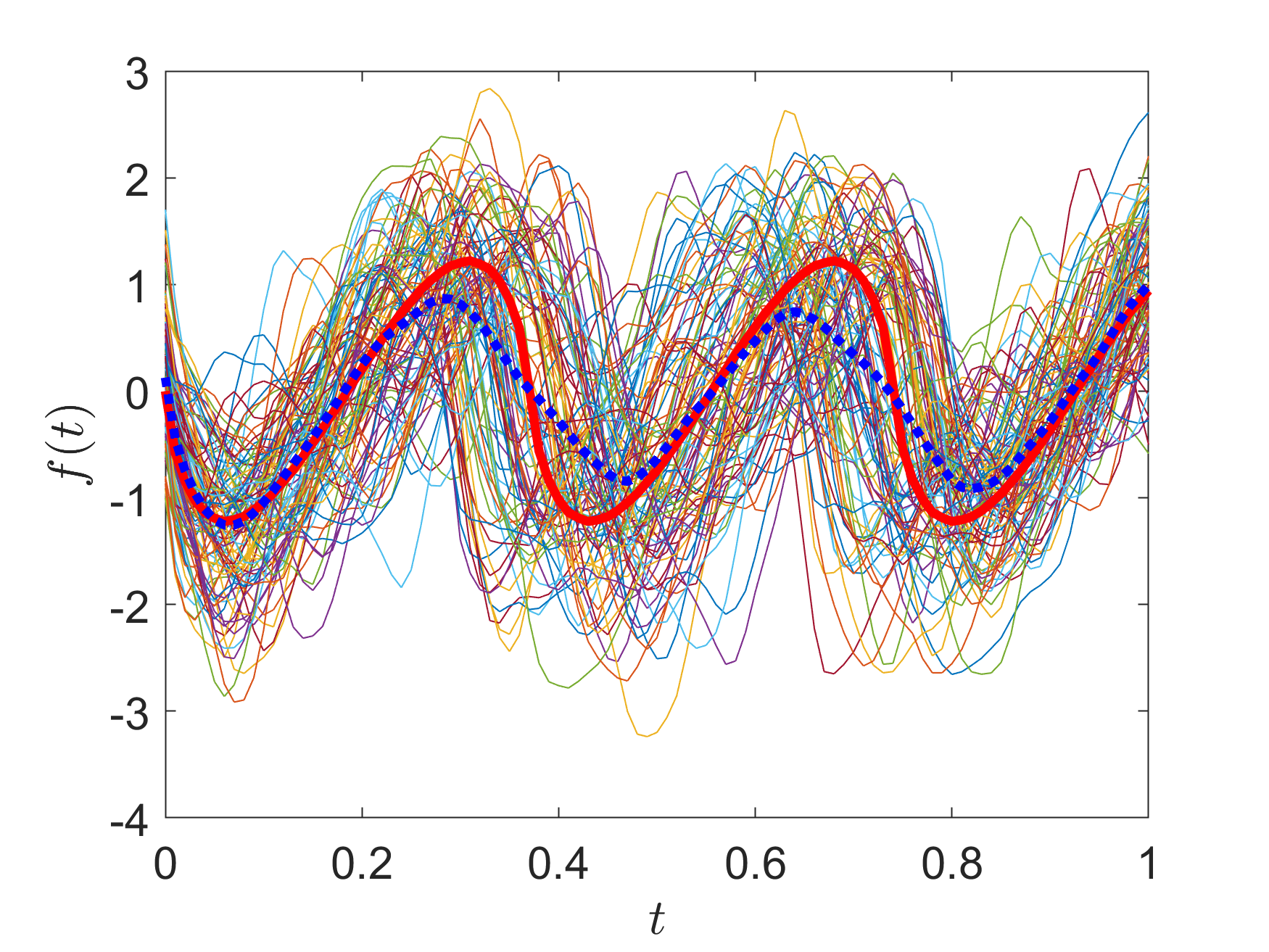}}
    \hspace{-0.2in}
    \subfloat[]{\includegraphics[height = 1.2in]{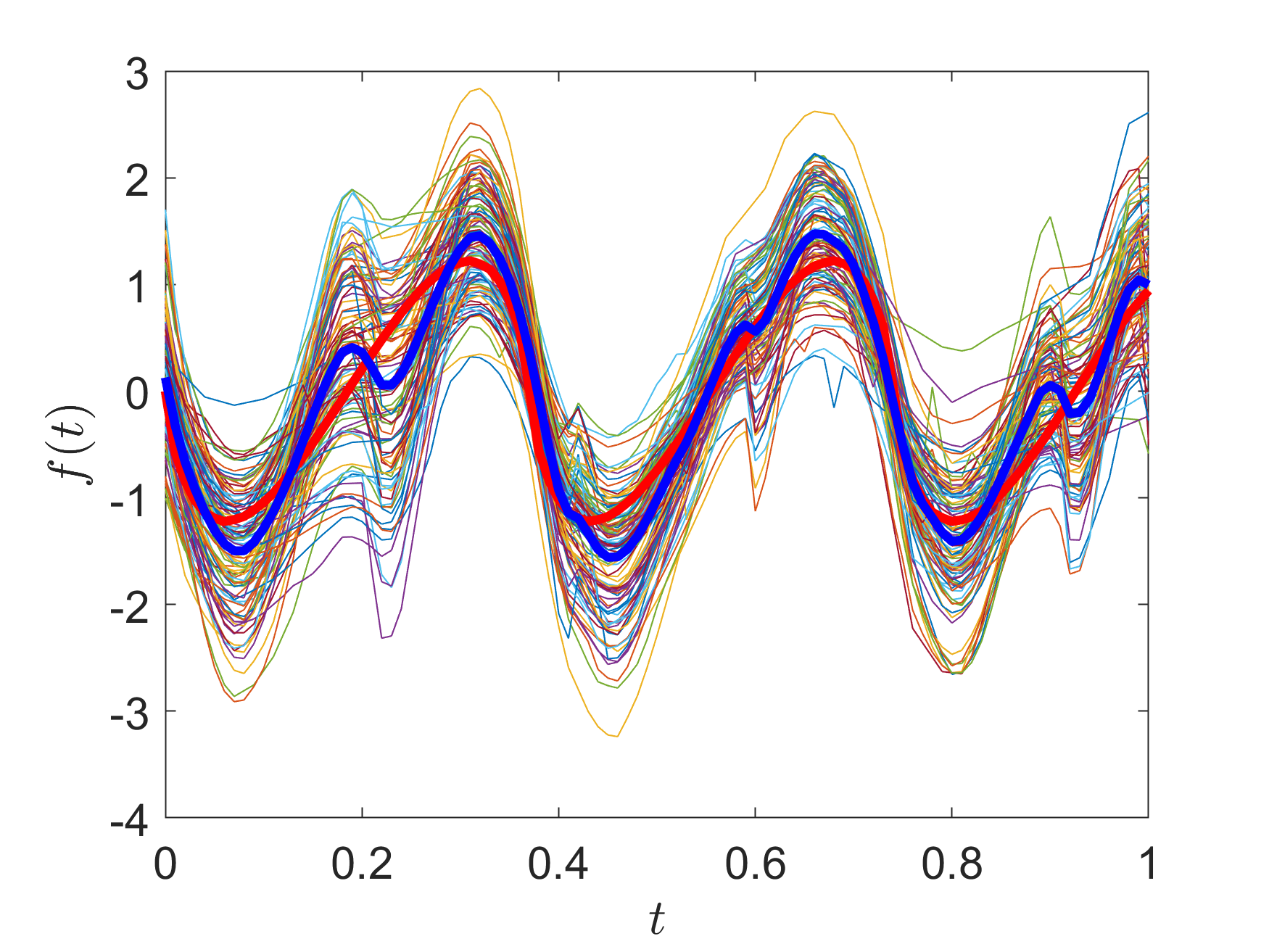}}
    \hspace{-0.2in}
    \subfloat[]{\includegraphics[height = 1.2in]{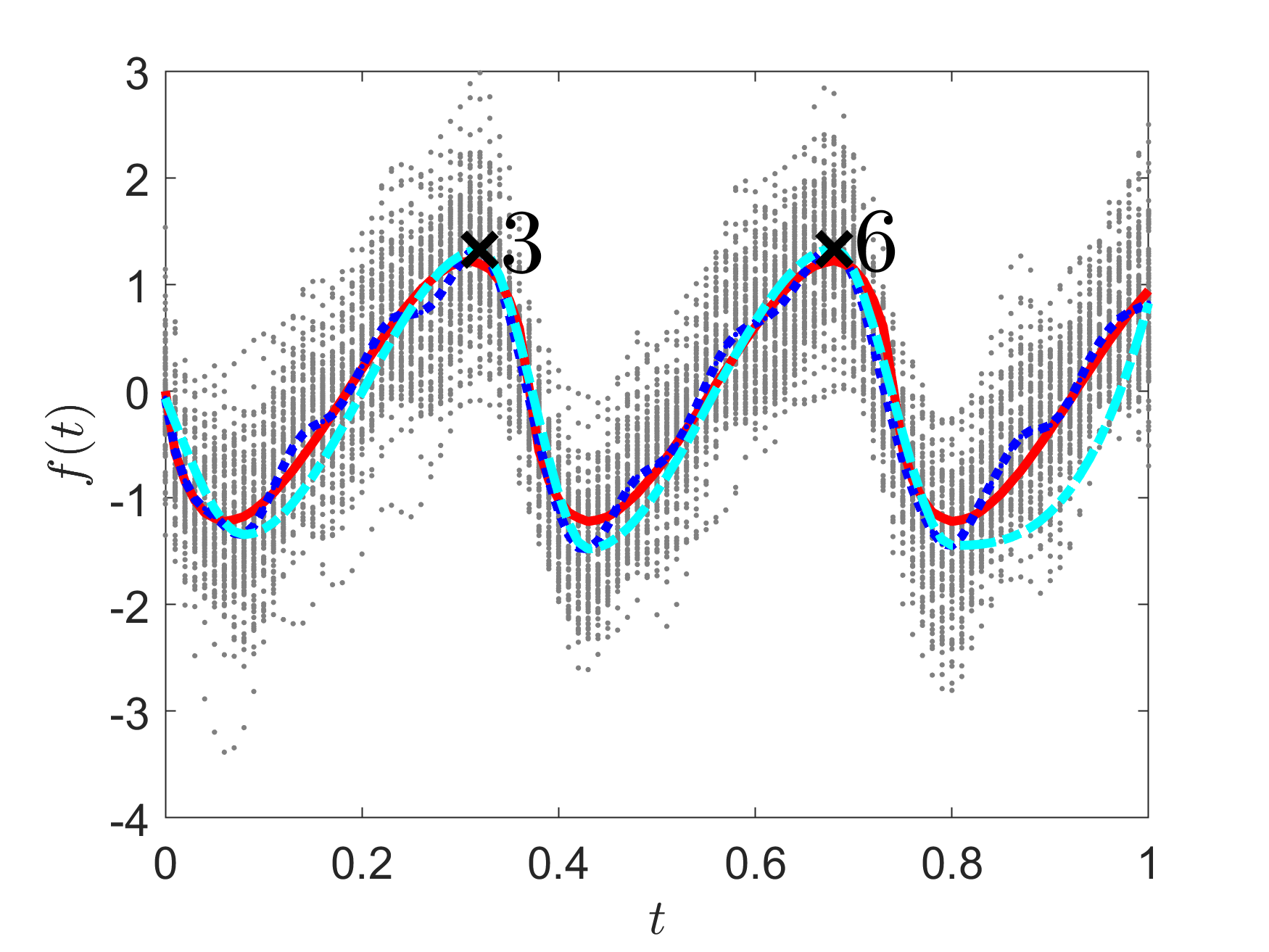}}
    \includegraphics[height = 1in]{fig/legend-sim-1.png}
    \hspace{0in}\\
    \vspace*{-0.05in}
    \hspace{-0.2in}
    \subfloat[]{\includegraphics[height = 1.2in]{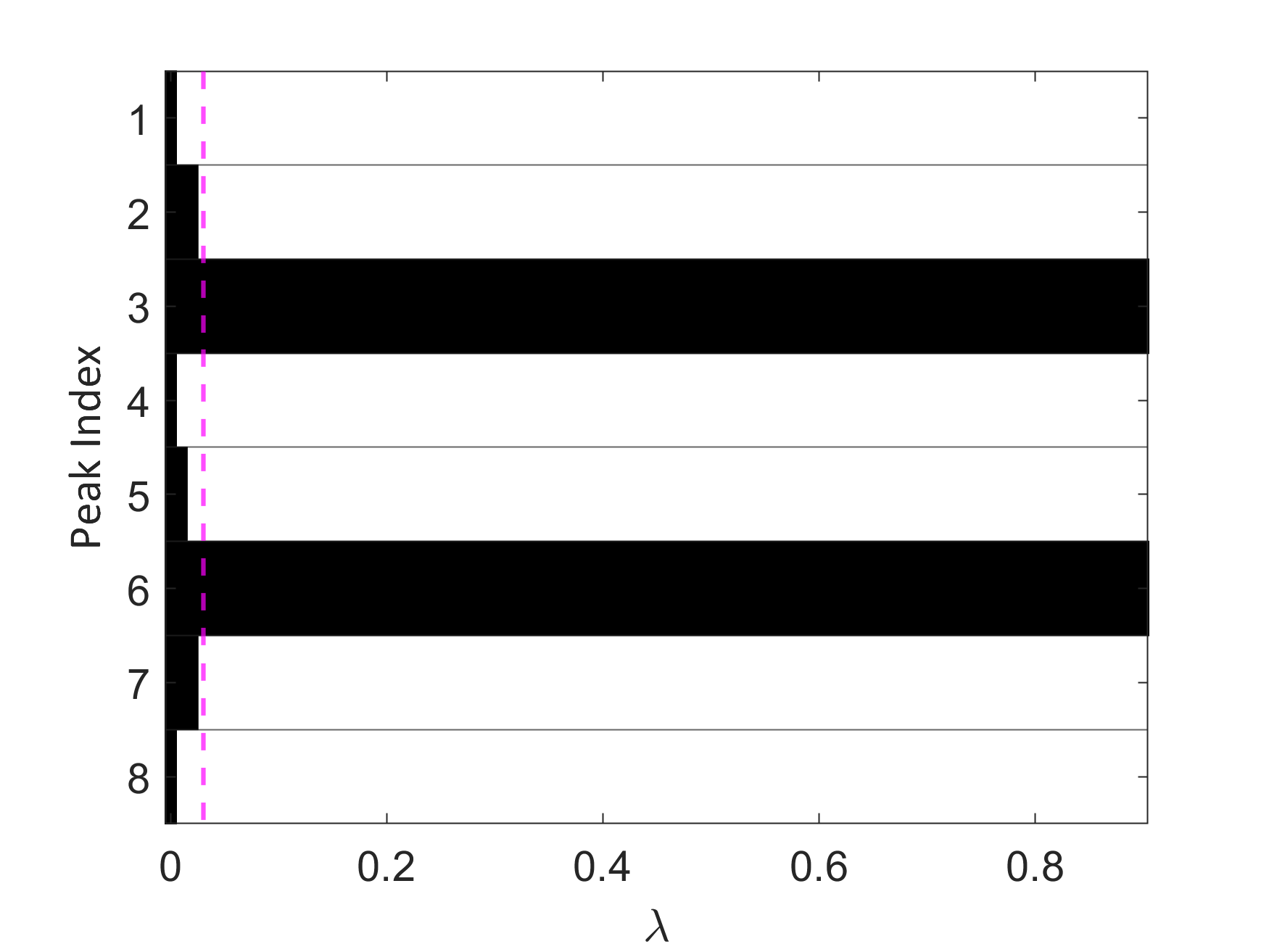}}
    \hspace{-0.2in}
    \subfloat[]{\includegraphics[height = 1.2in]{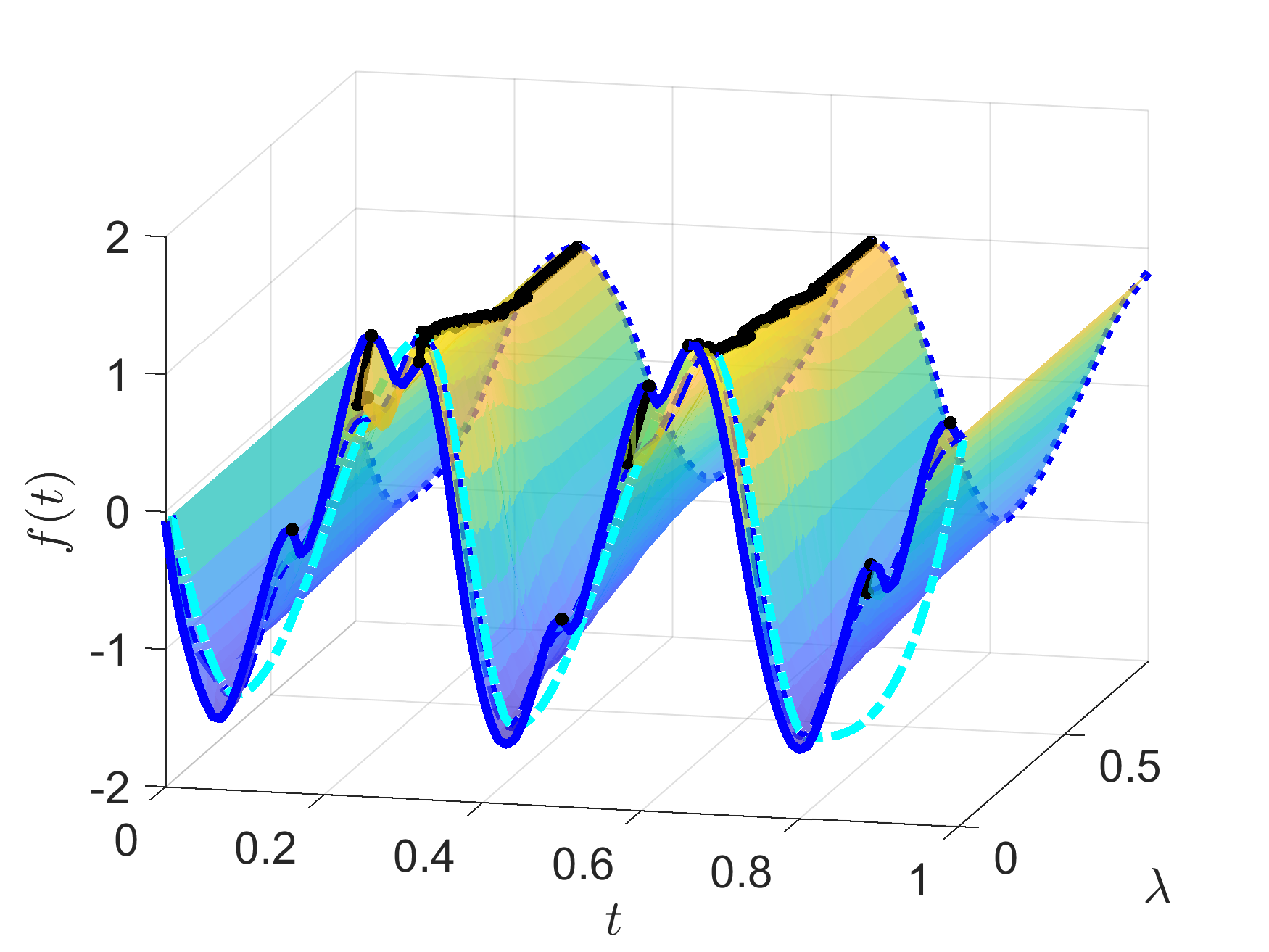}}
    \hspace{-0.2in}
    \subfloat[]{\includegraphics[height = 1.2in]{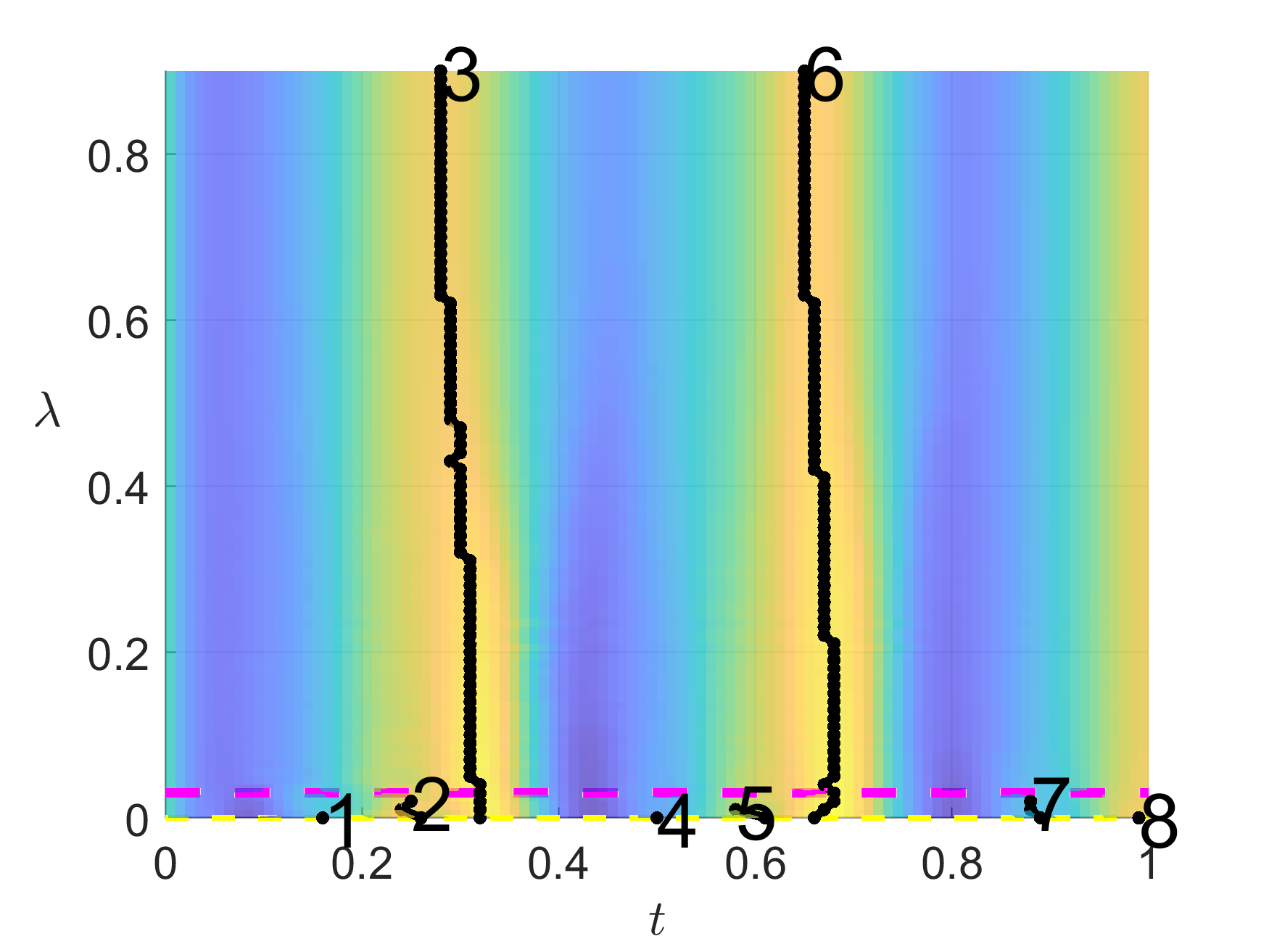}}
    \hspace{-0.2in}
    \subfloat[]{\includegraphics[height = 1.2in]{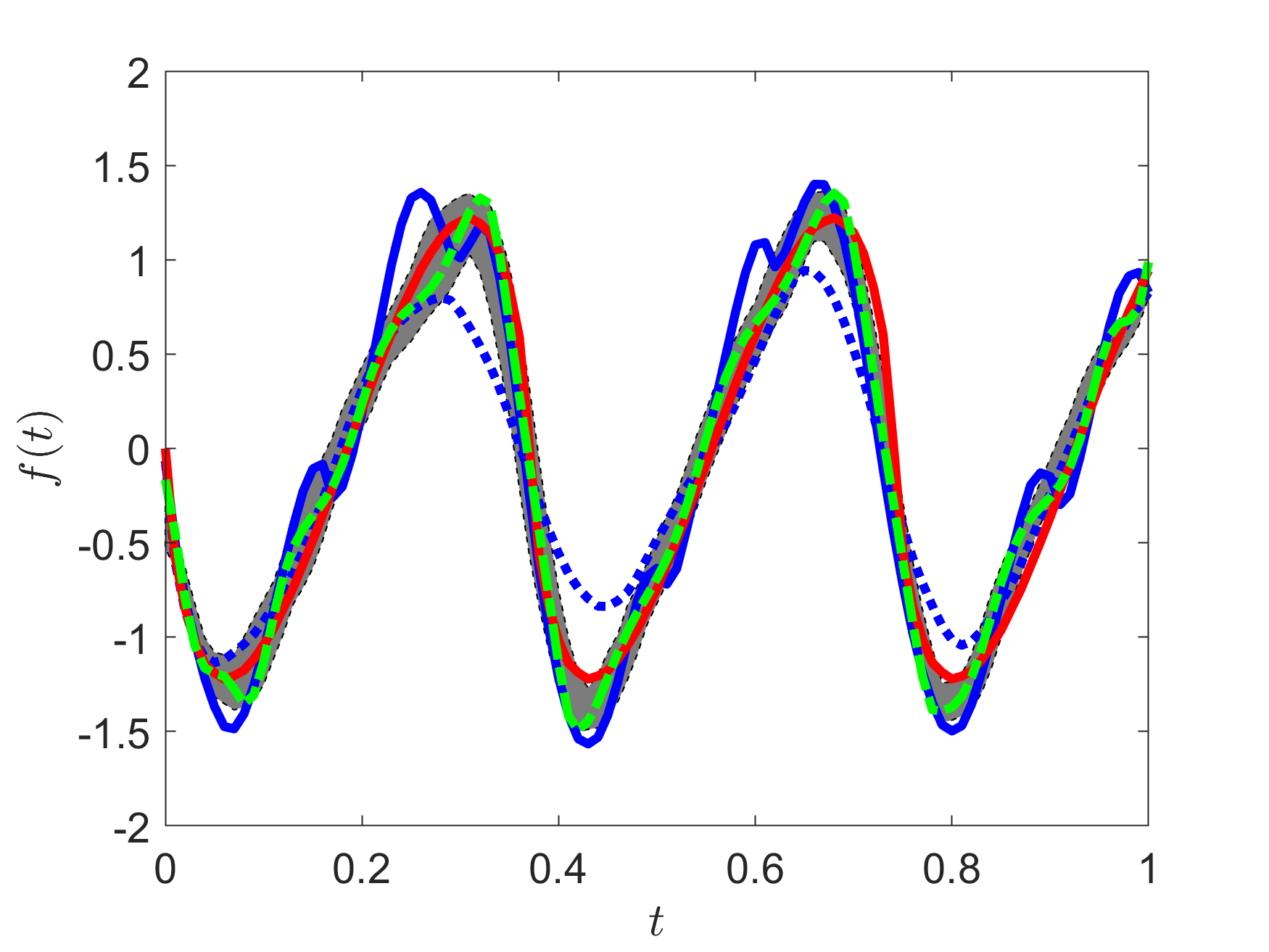}}
    \vspace*{-0.05in}
    \caption{(Simulated Dataset 2)
    This figure is laid out similar to Fig.~\ref{fig: simulation_data 1}. Our estimate$\hat{g}$ is the closest $g$ and the 95\% confidence band contains $g$ entirely.}
    \label{fig: simulation_data 2}
\end{figure*}

\begin{itemize}
  \item {\bf{\emph{Simulated Dataset 2}}}
\end{itemize}
This experiment uses a sawtooth wave function for $g$, and the data is shown in Fig. \ref{fig: simulation_data 2} (a). The rest of this figure is laid out similarly to Fig.~\ref{fig: simulation_data 1}. 
The PPD bar chart in Plot (d) successfully detects two significant internal peaks, labeled $3$ and $6$, and estimates $\lambda^* = 0.04$. The evolution of peaks is seen in (e) and (f) with PPD surface plots. Plot (c) shows the partially-aligned functions, $\{\tilde{f}_{\lambda^*,i}\}_{i=1}^n$, their mean, $\hat{g}_{\lambda^*}$, and the initial estimate, $\hat{g}_{init}$ at $\lambda^*=0.04$. 
Plot (g) shows the final estimate $\hat{g}$ (in green) with a bootstrap confidence band in gray. As shown, the elastic mean, $\hat{g}_0$ (solid blue), has spurious peaks while  $\hat{g}_\infty$ (dotted blue) underestimates the heights of peaks and valleys. Our estimate $\hat g$ (in green) provides an excellent estimate of $g$.

\begin{figure*}[htbp]
    \centering
    \subfloat[]{\includegraphics[height = 1.2in]{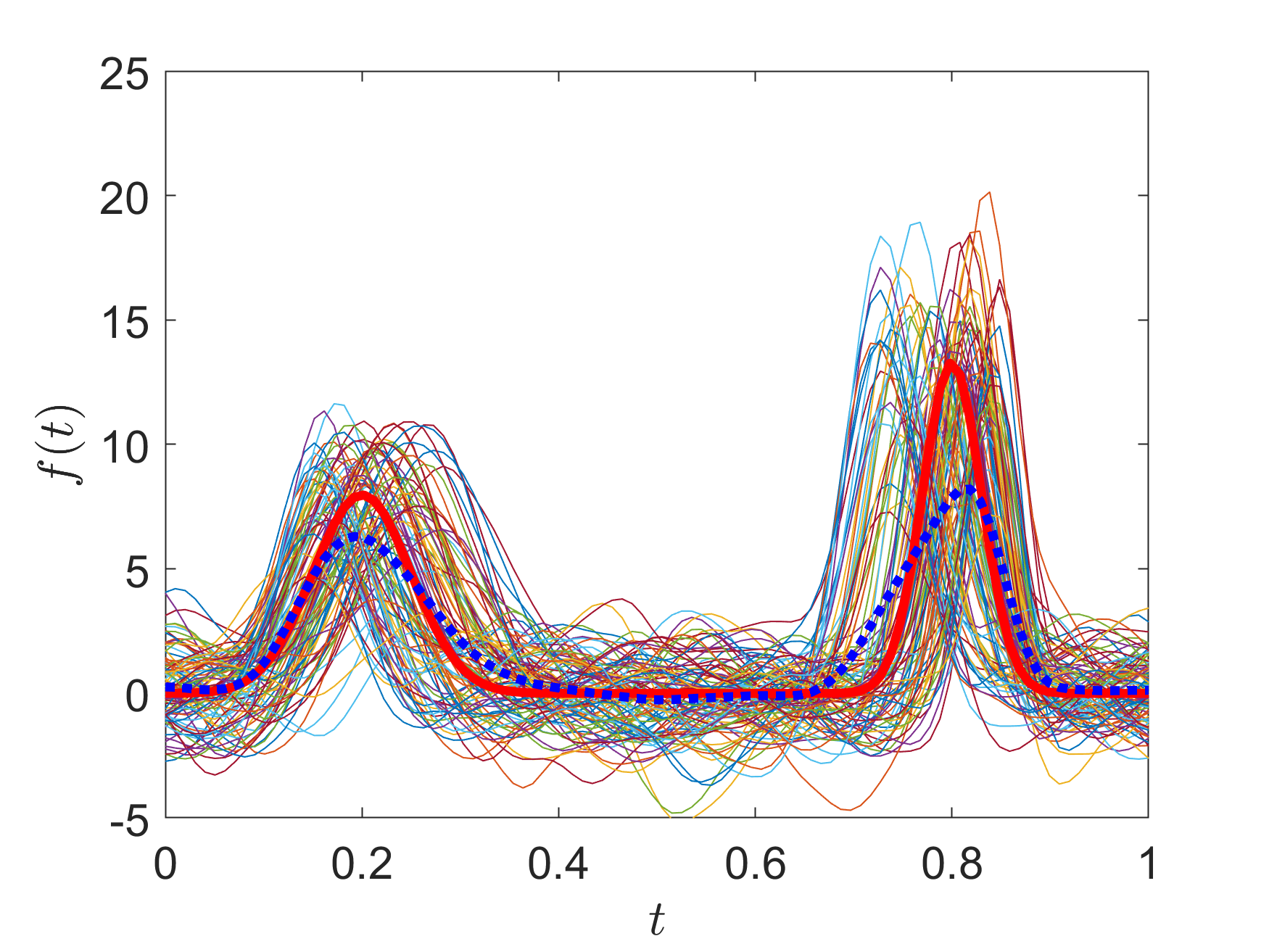}}
    \hspace{-0.2in}
    \subfloat[]{\includegraphics[height = 1.2in]{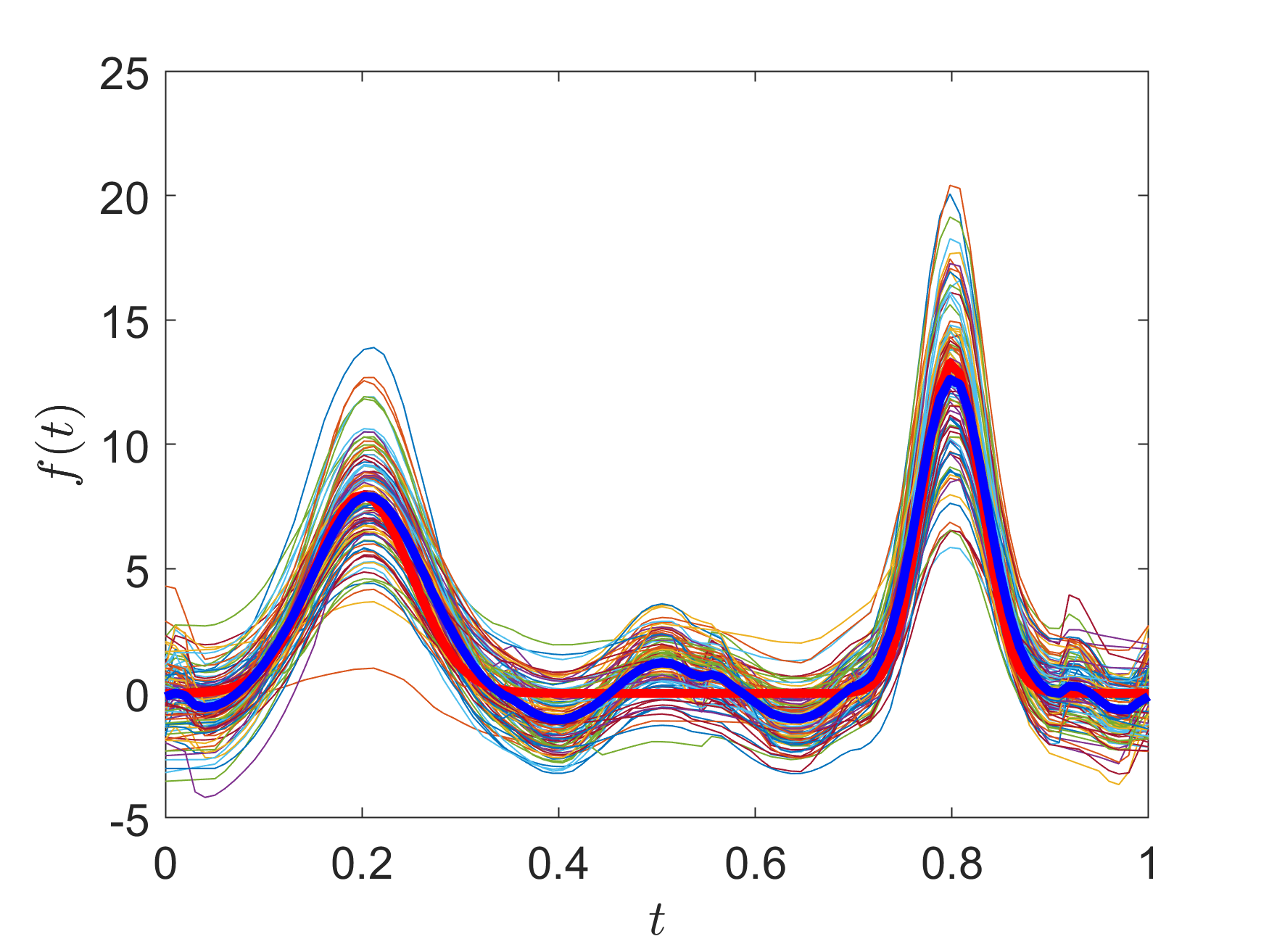}}
    \hspace{-0.2in}
    \subfloat[]{\includegraphics[height = 1.2in]{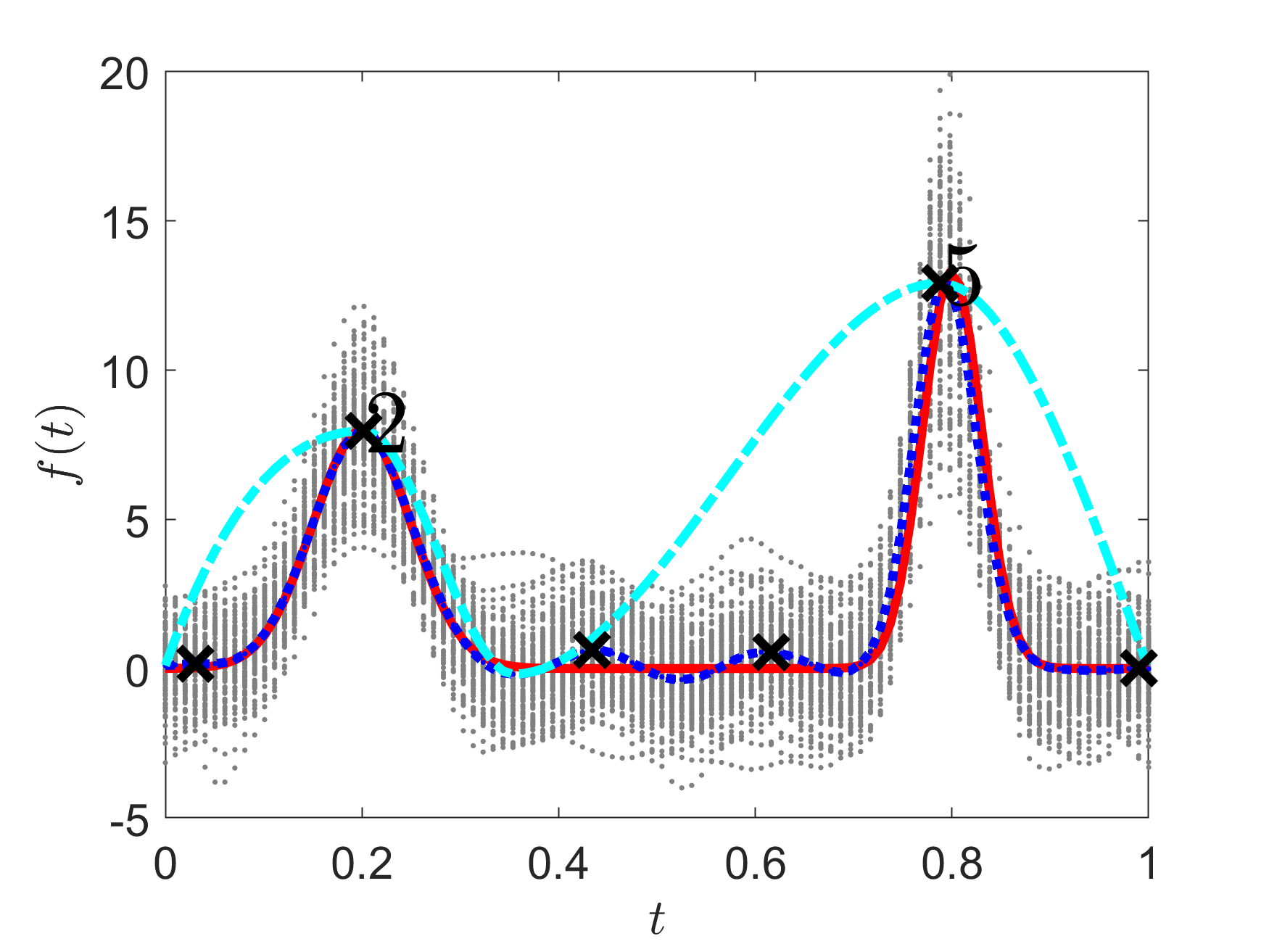}}
    \includegraphics[height = 1in]{fig/legend-sim-1.png}
    \hspace{0in}\\
    \vspace*{-0.05in}
    \hspace{-0.2in}
    \subfloat[]{\includegraphics[height = 1.2in]{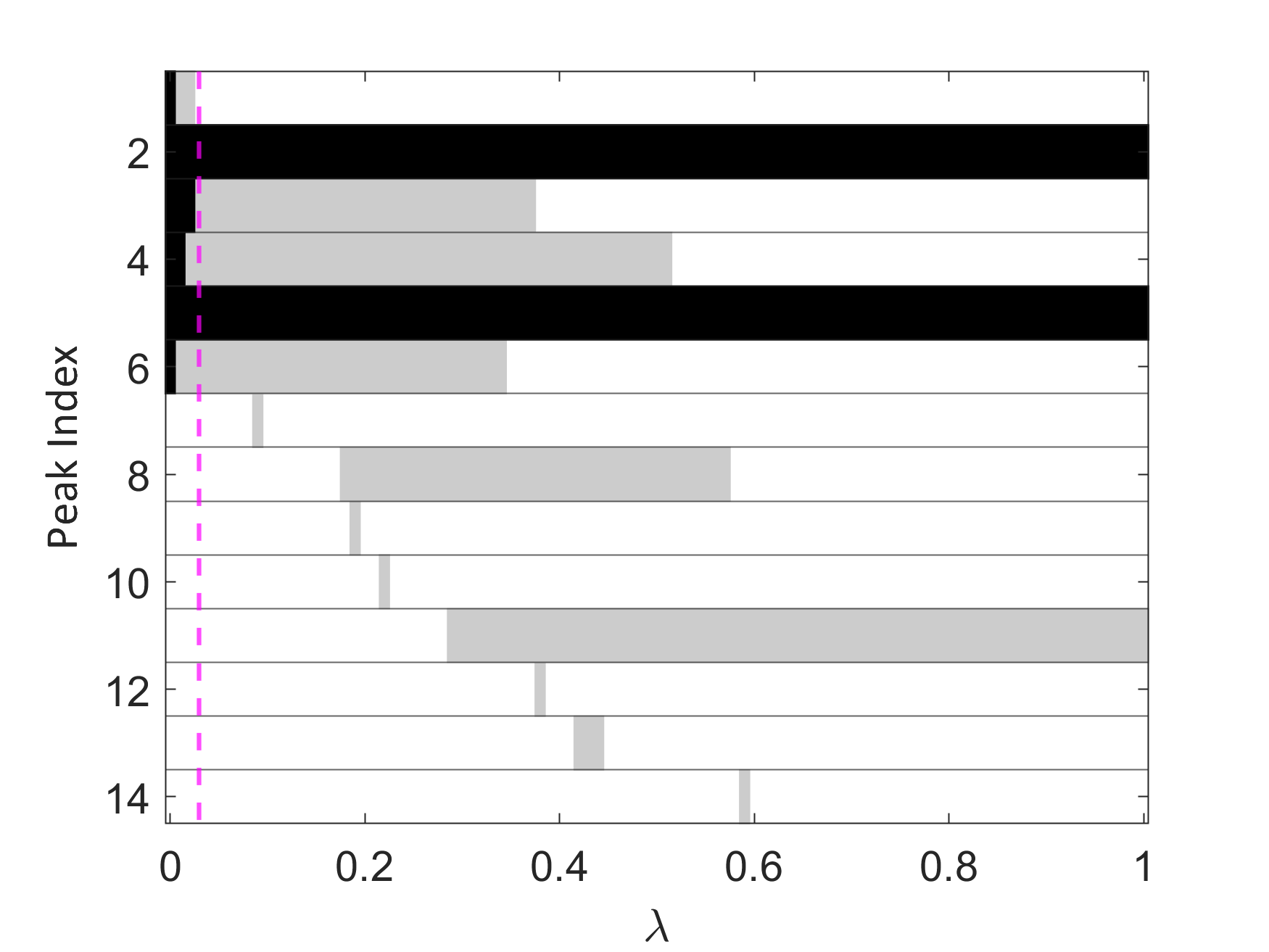}}
    \hspace{-0.2in}
    \subfloat[]{\includegraphics[height = 1.2in]{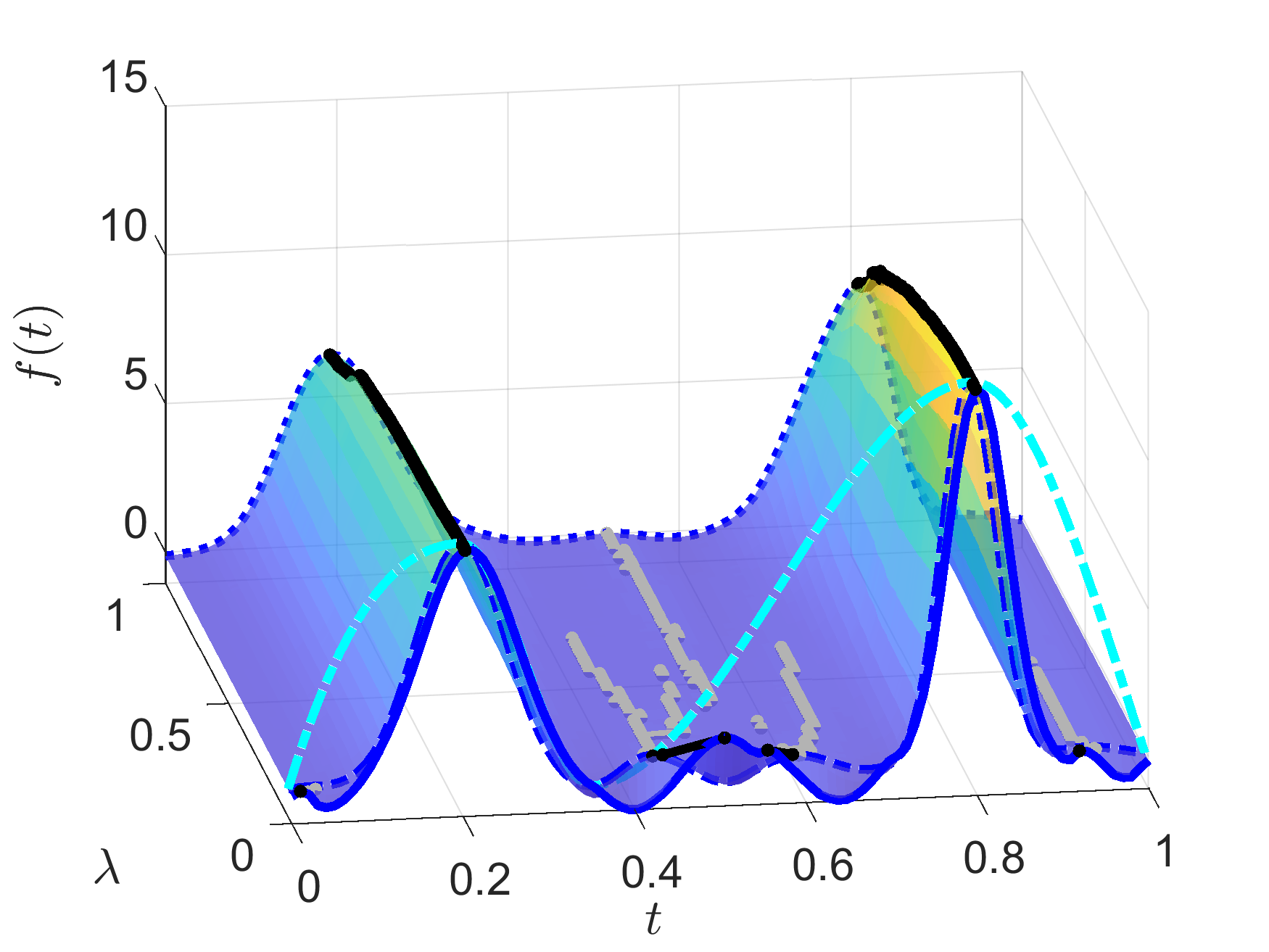}}
    \hspace{-0.2in}
    \subfloat[]{\includegraphics[height = 1.2in]{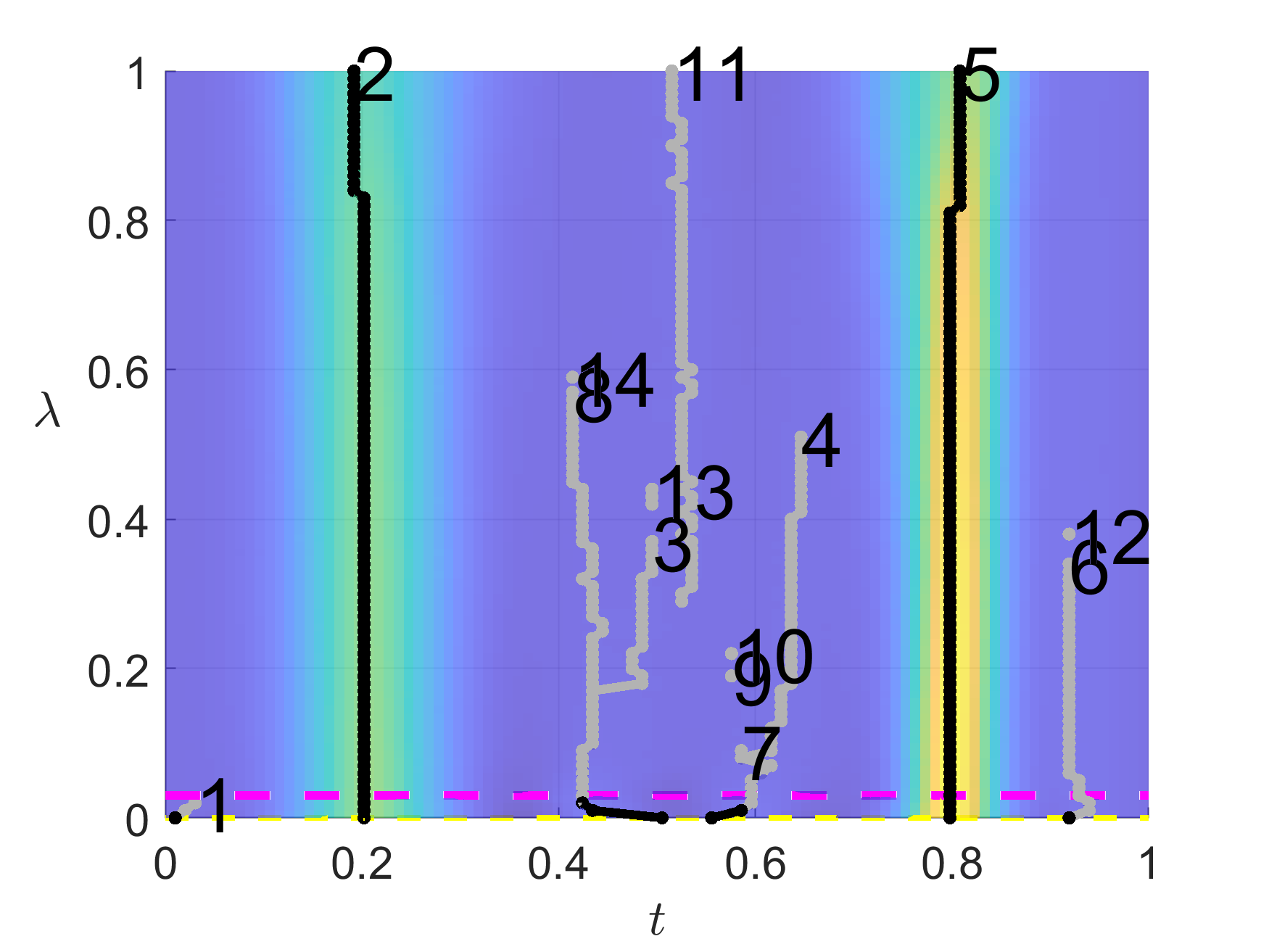}}
    \hspace{-0.2in}
    \subfloat[]{\includegraphics[height = 1.2in]{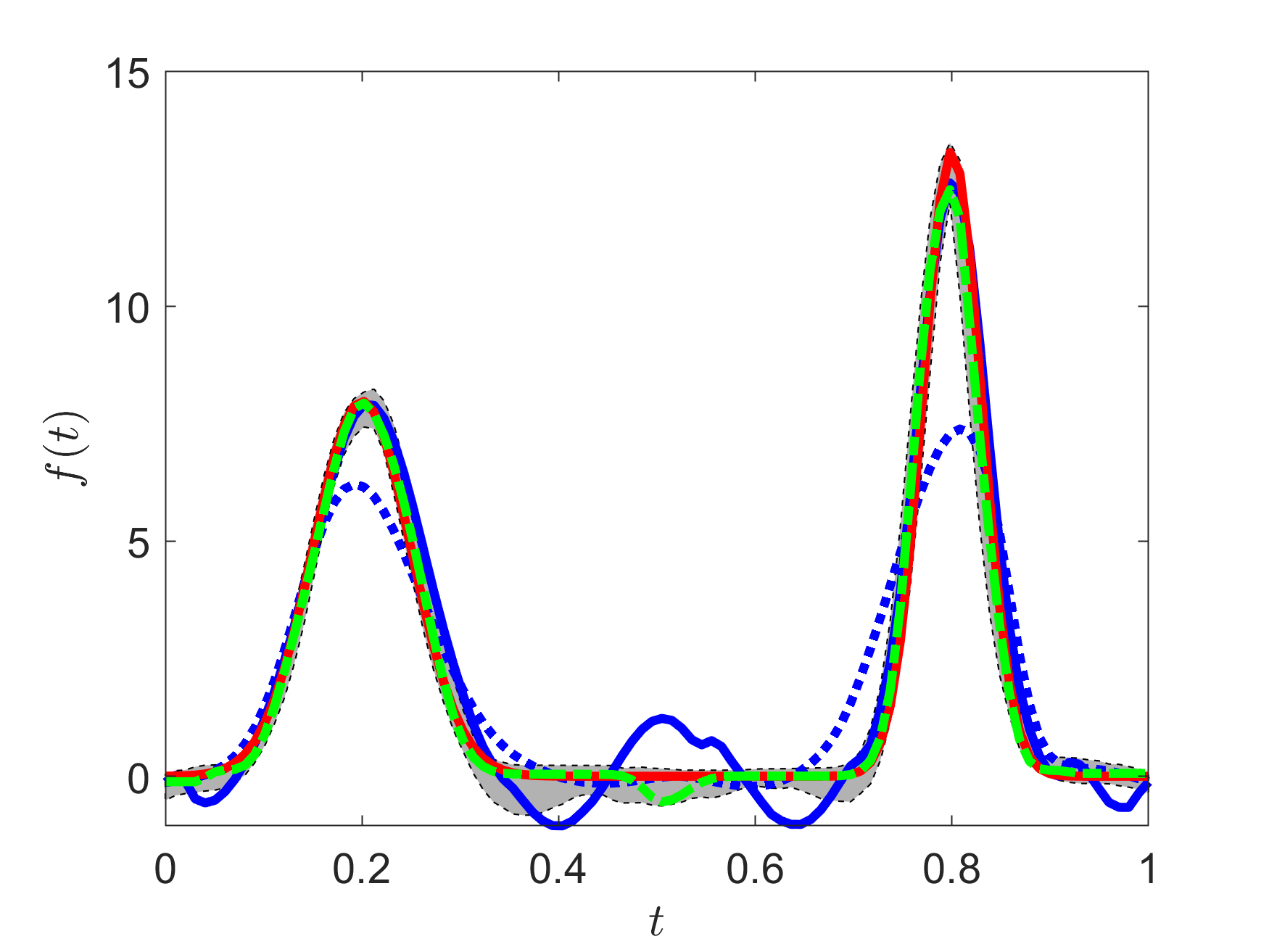}}
    \vspace*{-0.05in}
    \caption{(Simulated Dataset 3)
    The PPD barchart in (d) indicates that two peaks numbered 2 and 5. Accordingly, the dotted cyan lines in (c) and (e) are the estimated shape of $g$. The final estimate $\hat{g}$ is shown in (g).}
    \label{fig: simulation_data 3}
\end{figure*}

\begin{itemize}
  \item{\bf{\emph{Simulated Dataset 3}}}
\end{itemize}
In this case, we select a function $g$ with two peaks and a large constant region in between peaks. The constant interval is difficult to estimate as noise can erroneously introduce spurious peaks in a flat region, as shown in Plot (b) of Fig. \ref{fig: simulation_data 3}. This data has some short-lived peaks and others that do not pass the significance test. The PPD bar chart detects peaks 2 and 5 as significant and persistent, with the resulting $\lambda^* = 0.03$. The PPD surface plots in (e) and (f) show the evolution of these peaks. 
Plot (g) presents the final estimate (green) with a point-wise confidence band (gray) which is considerably different from $\hat{g}_0$ (solid blue) and $\hat{g}_{\infty}$ (dotted blue).


\begin{figure*}[htbp]
    \centering
    \subfloat[]{\includegraphics[height = 1.2in]{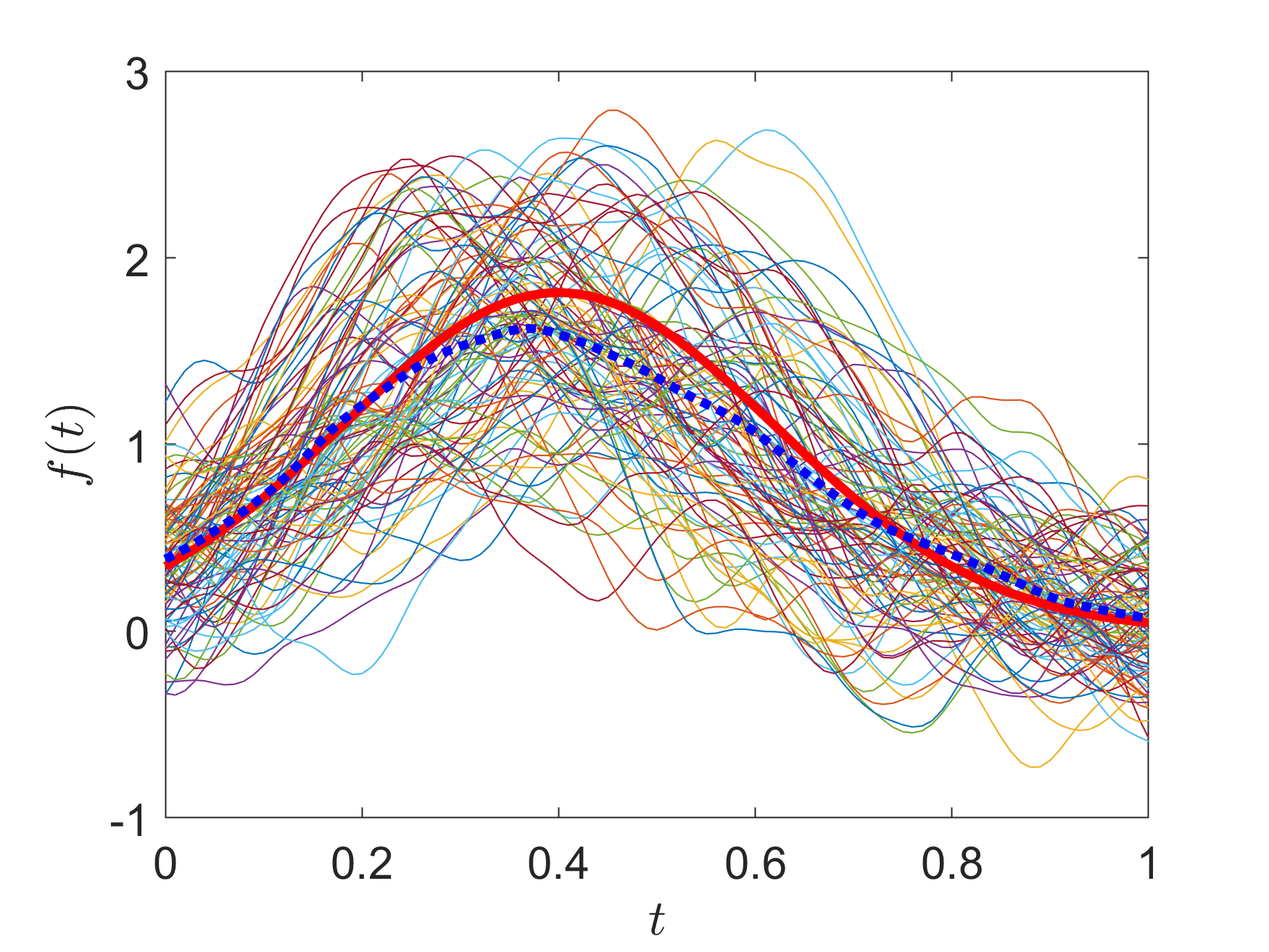}}
    \hspace{-0.2in}
    \subfloat[]{\includegraphics[height = 1.2in]{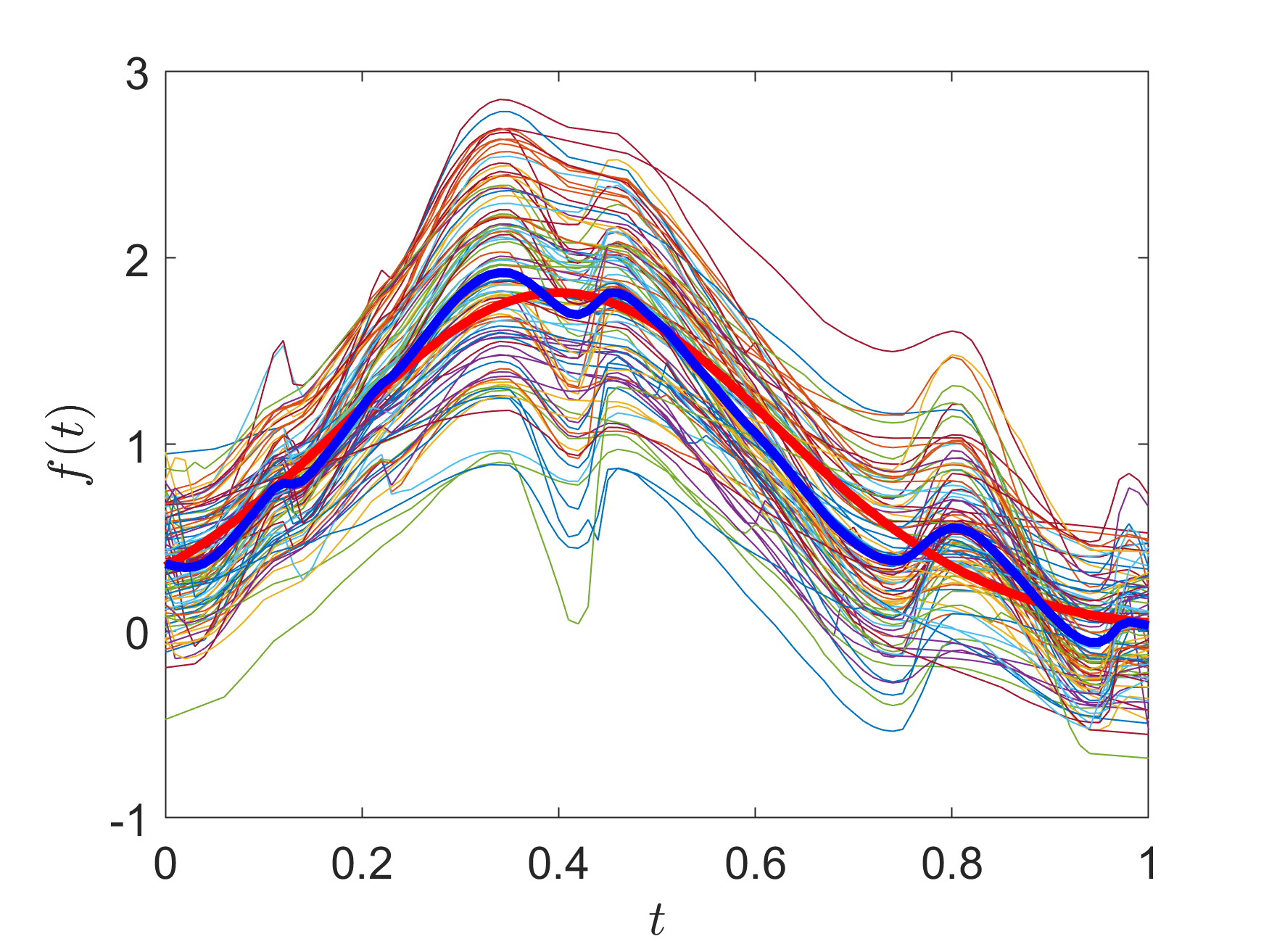}}
    \hspace{-0.2in}
    \subfloat[]{\includegraphics[height = 1.2in]{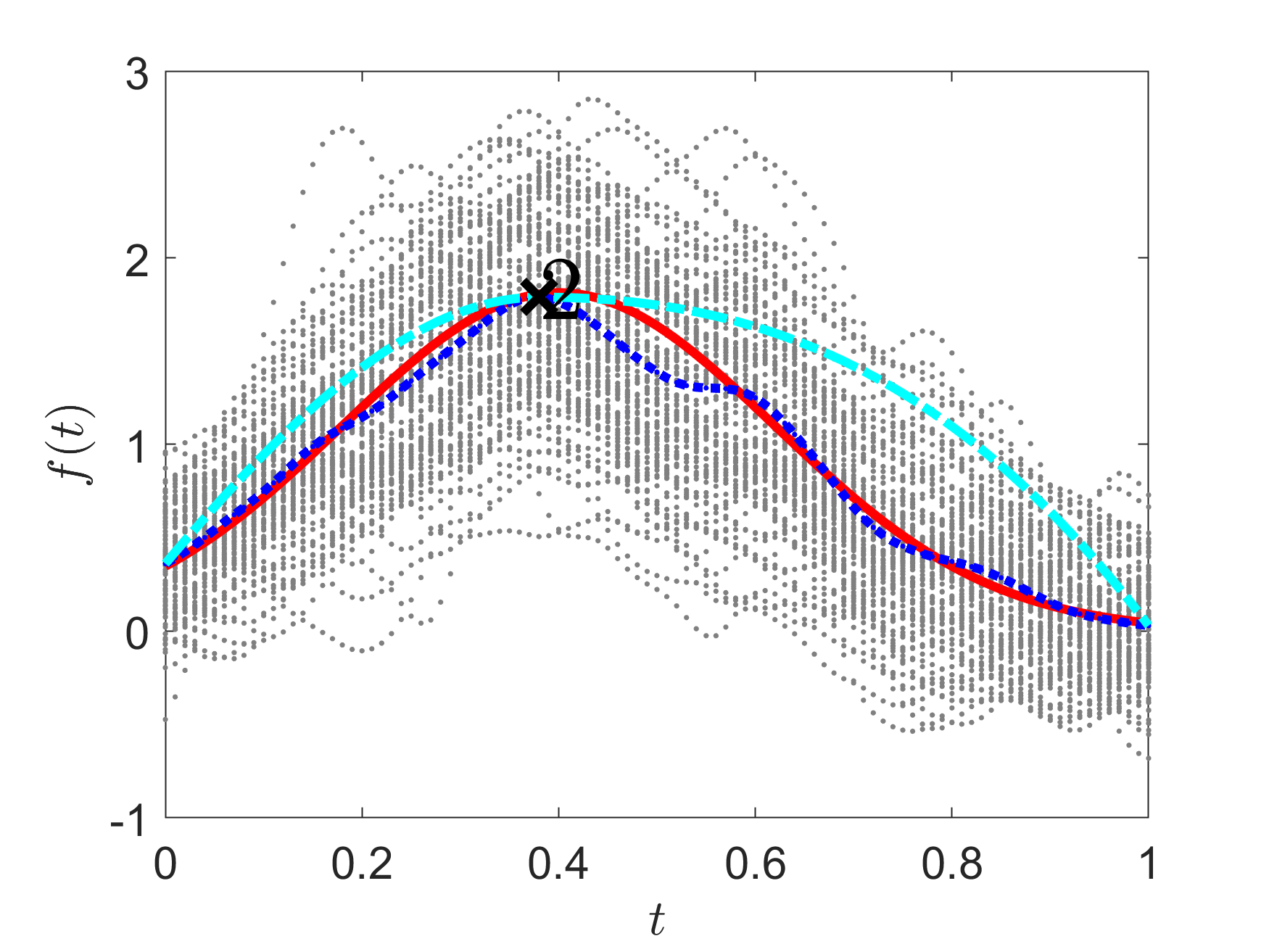}}
    \includegraphics[height = 1in]{fig/legend-sim-1.png}
    \hspace{0in}\\
    \vspace*{-0.05in}
    \hspace{-0.2in}
    \subfloat[]{\includegraphics[height = 1.2in]{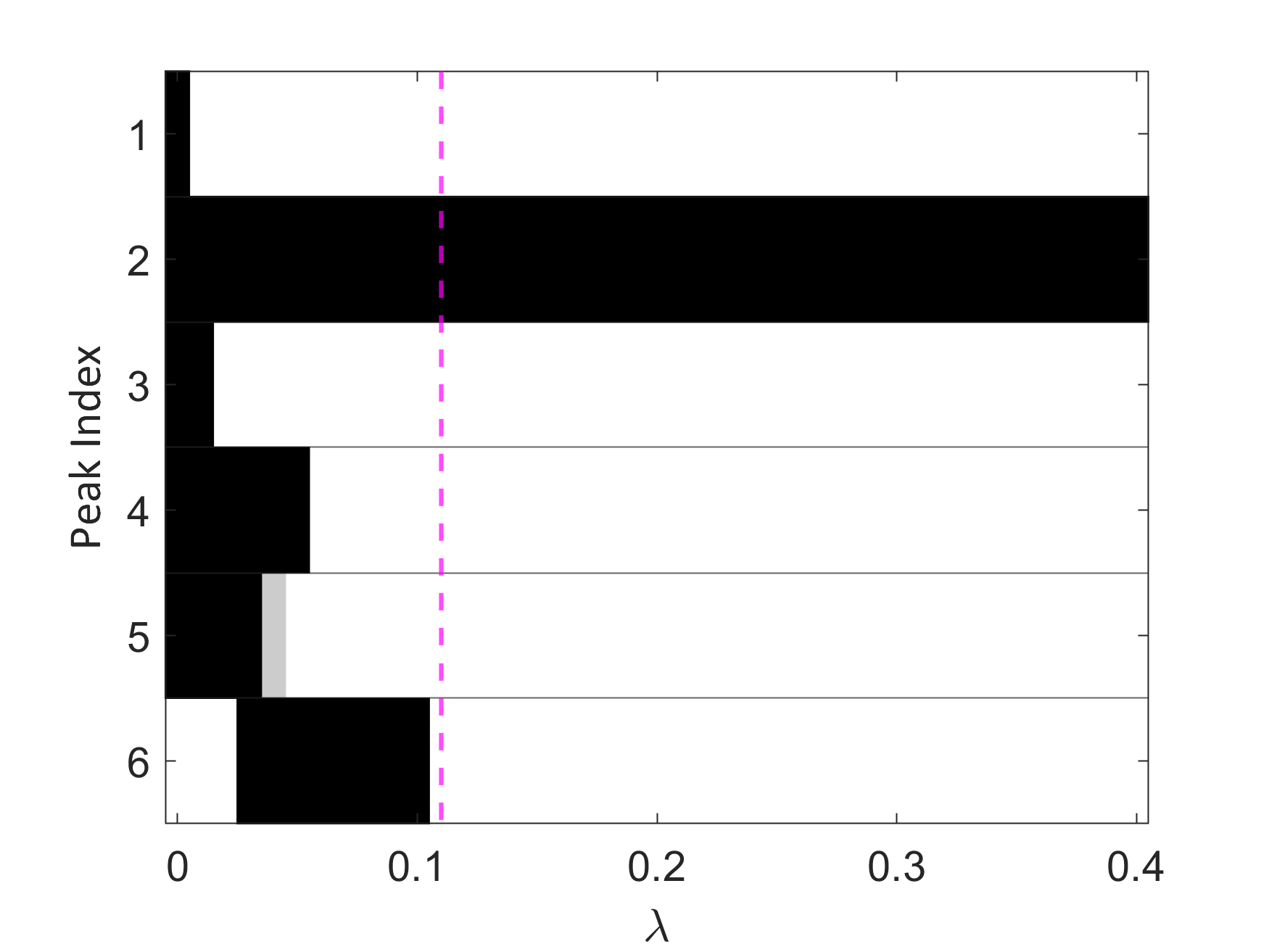}}
    \hspace{-0.2in}
    \subfloat[]{\includegraphics[height = 1.2in]{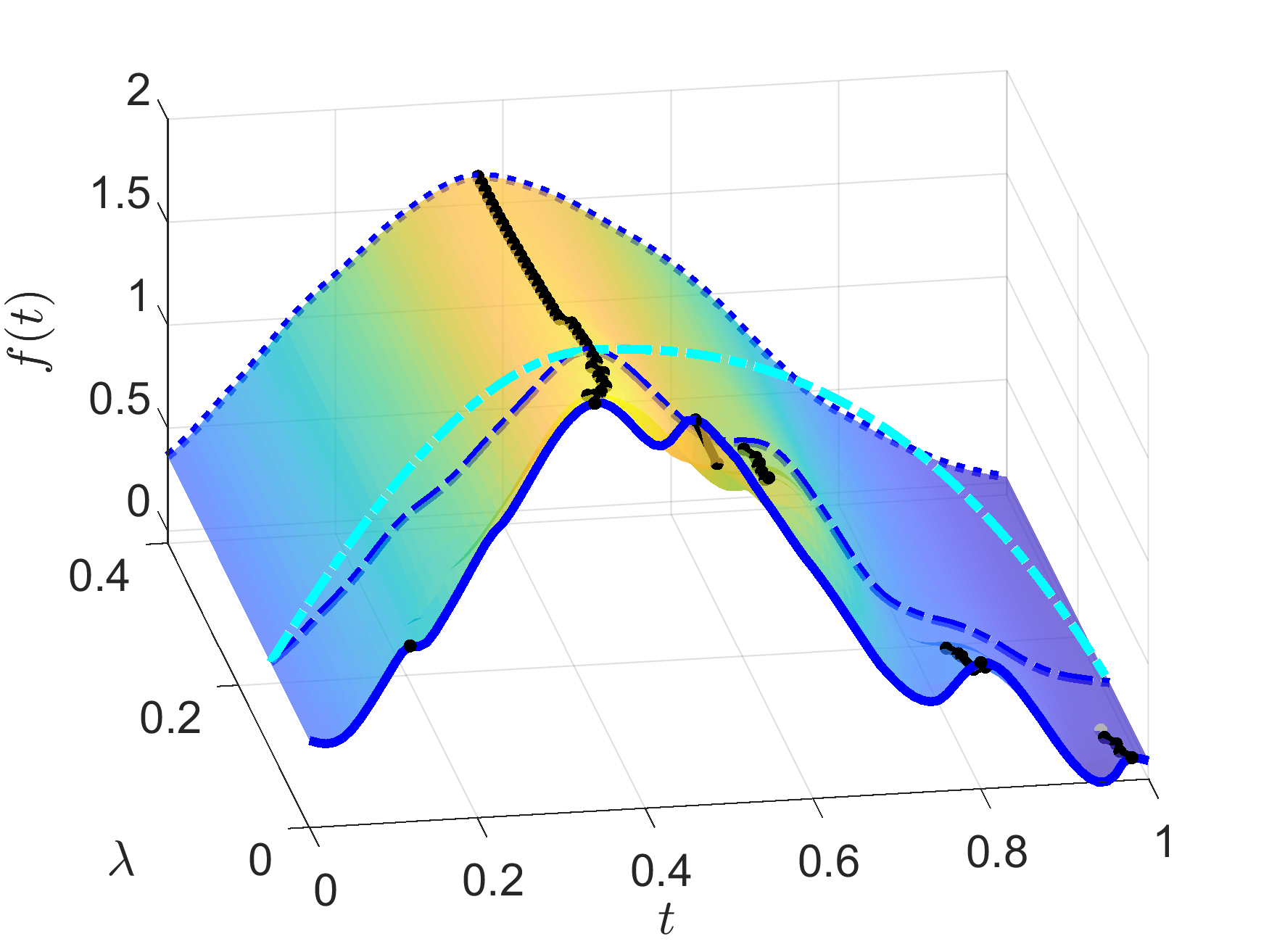}}
    \hspace{-0.2in}
    \subfloat[]{\includegraphics[height = 1.2in]{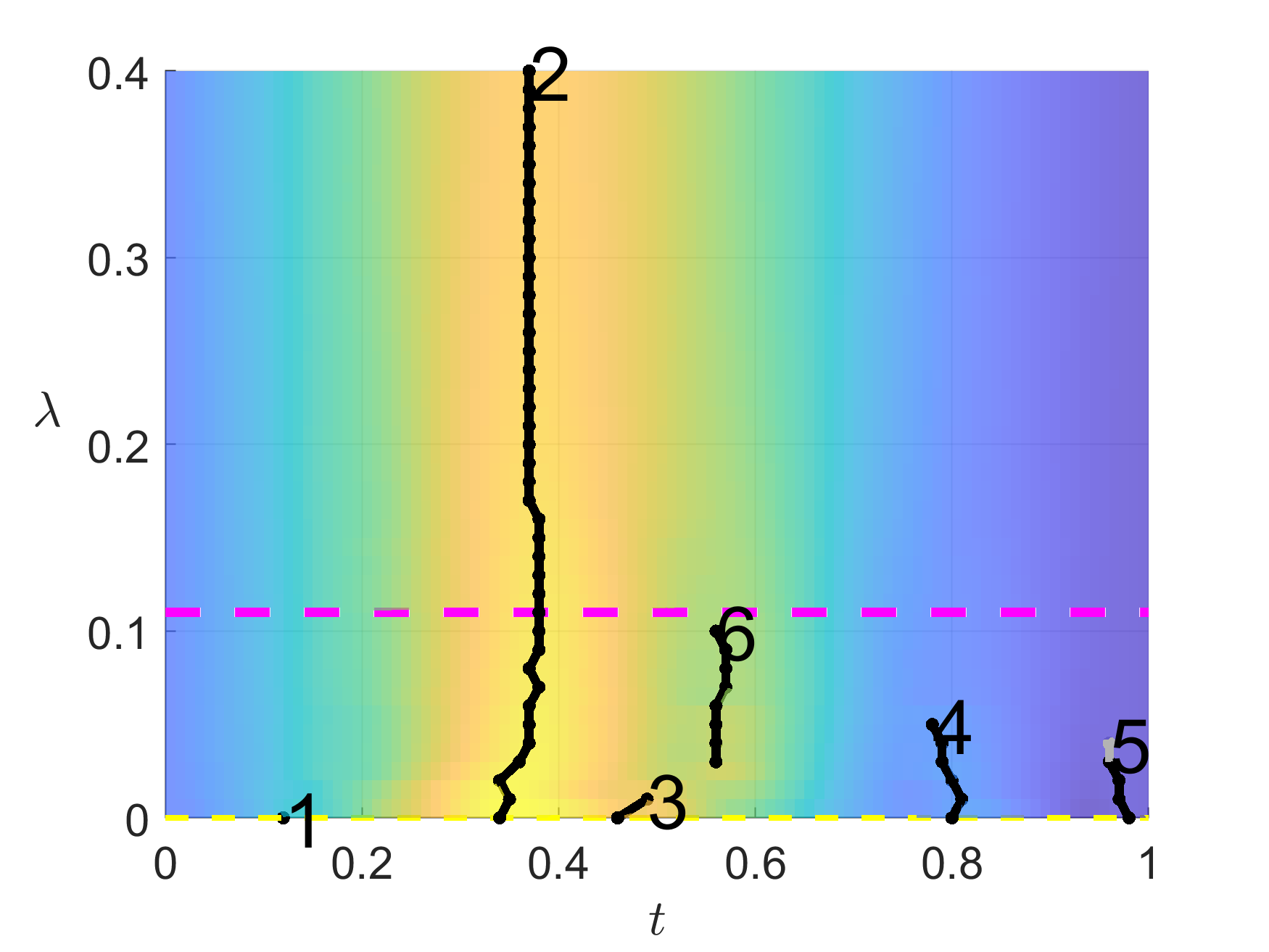}}
    \hspace{-0.2in}
    \subfloat[]{\includegraphics[height = 1.2in]{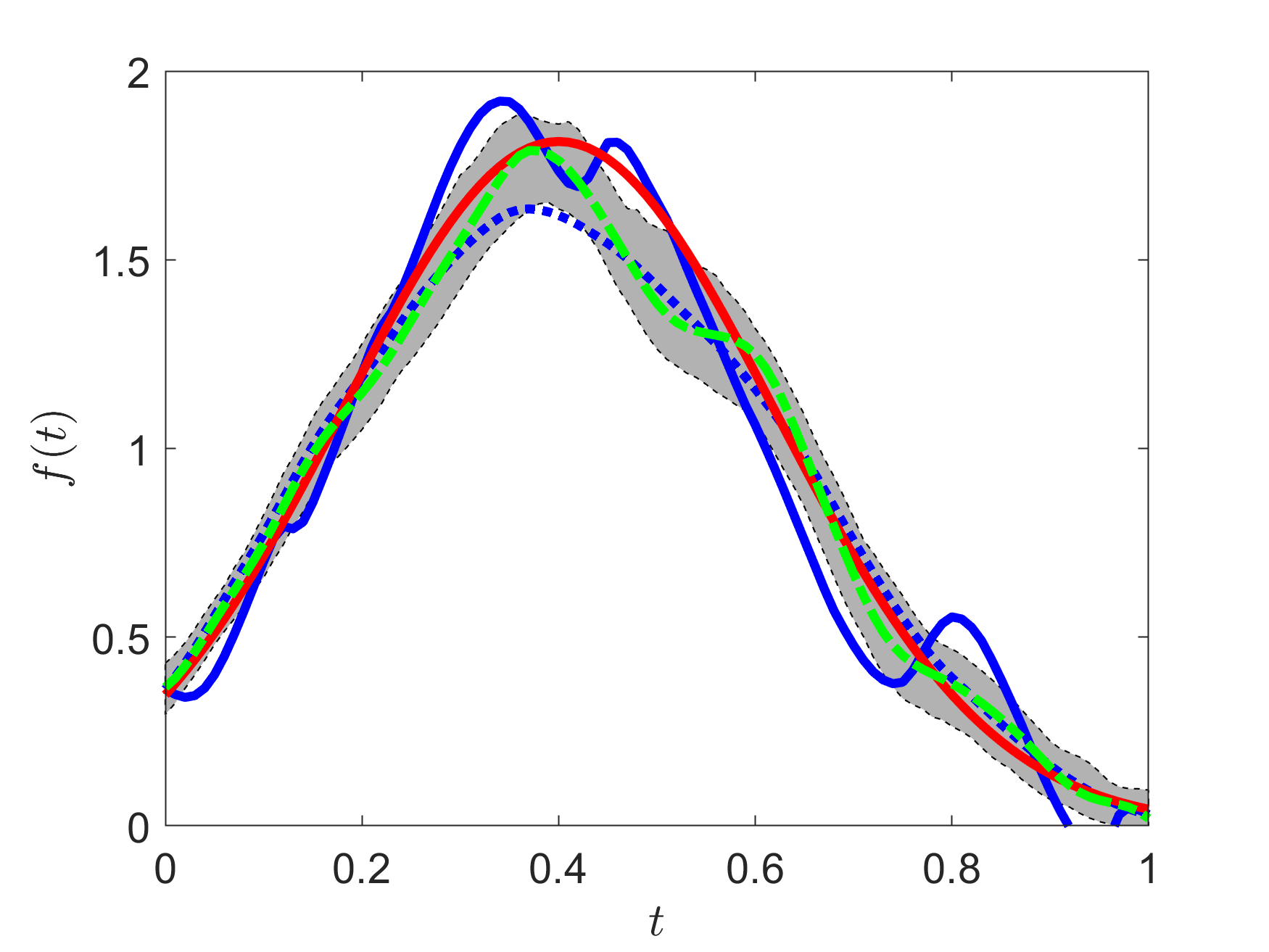}}
    \vspace*{-0.05in}
    \caption{(Simulated Dataset 4): Details similar to Fig.~\ref{fig: simulation_data 1}.}
    \label{fig: simulation_data 4}
\end{figure*}

\begin{itemize}
  \item{\bf{\emph{Simulated Dataset 4}}}
\end{itemize}
In this example, we use a broad unimodal function with moderate slope and large phase variability in the data.  Fig. \ref{fig: simulation_data 4} (a) shows the original data.  The PPD barchart in (d) successfully screens spurious peaks at $\lambda^*=0.11$. 
Plot (g) displays the final estimate, $\hat g$. We find that $\hat g$ provides a good estimate by capturing the location and the height of the peak of $g$.

\begin{figure*}[ht]
    \centering
    \begin{tabular}{cccc}
    \includegraphics[width=1.5in]{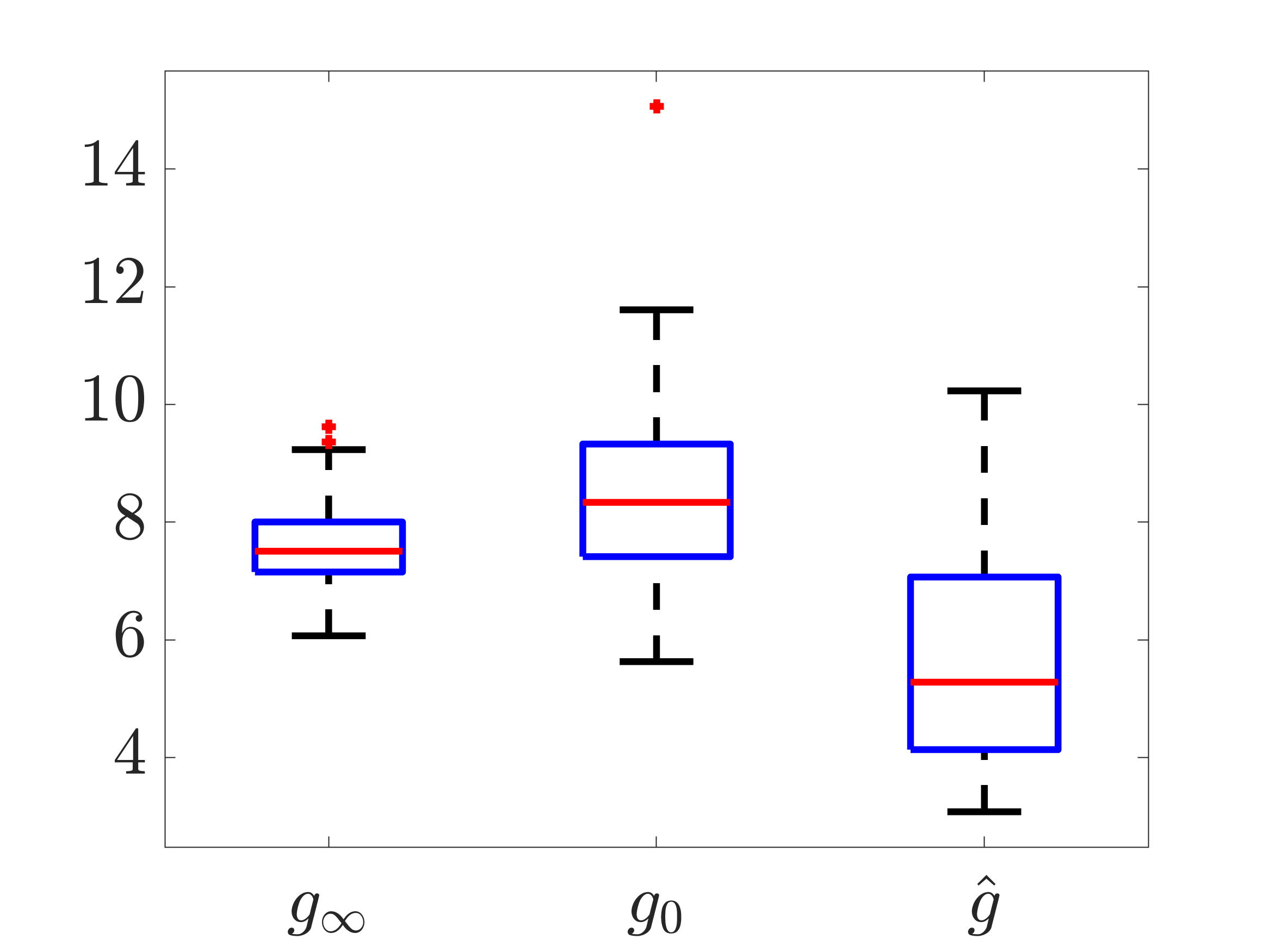} &
  \hspace*{-0.22in}  \includegraphics[width=1.5in] {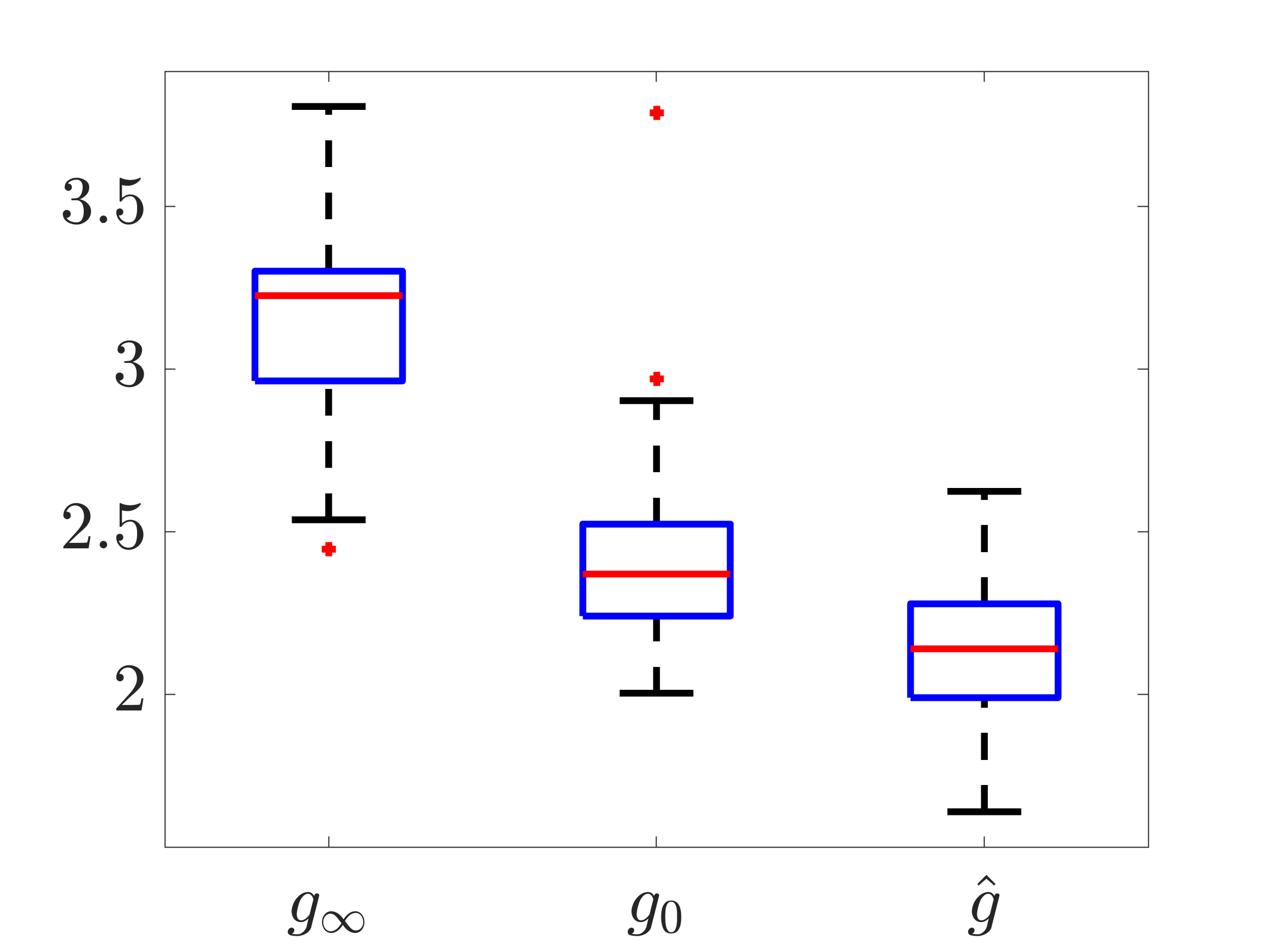} &
   \hspace*{-0.22in} \includegraphics[width=1.5in]{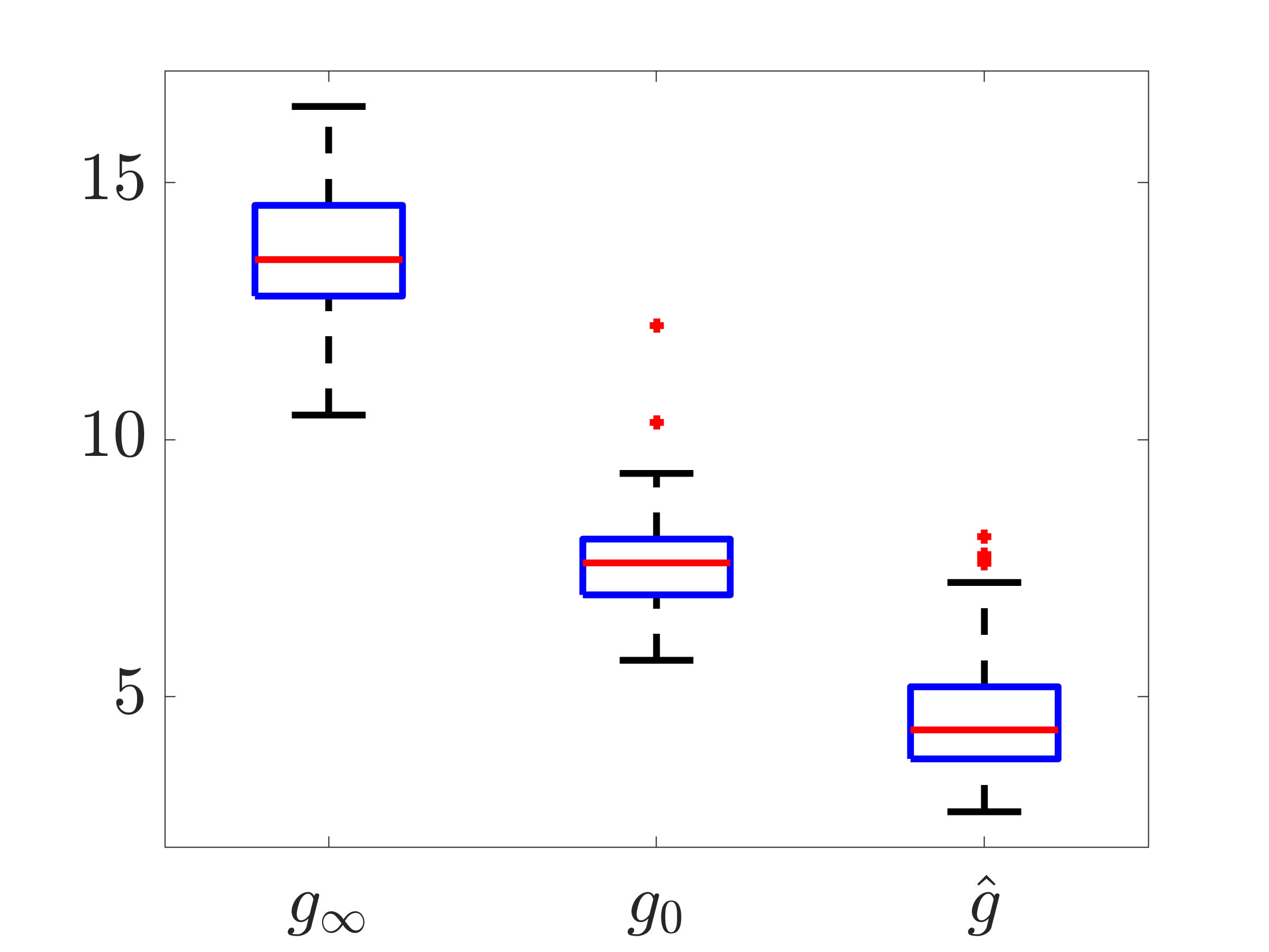} &
   \hspace*{-0.22in} \includegraphics[width=1.5in]{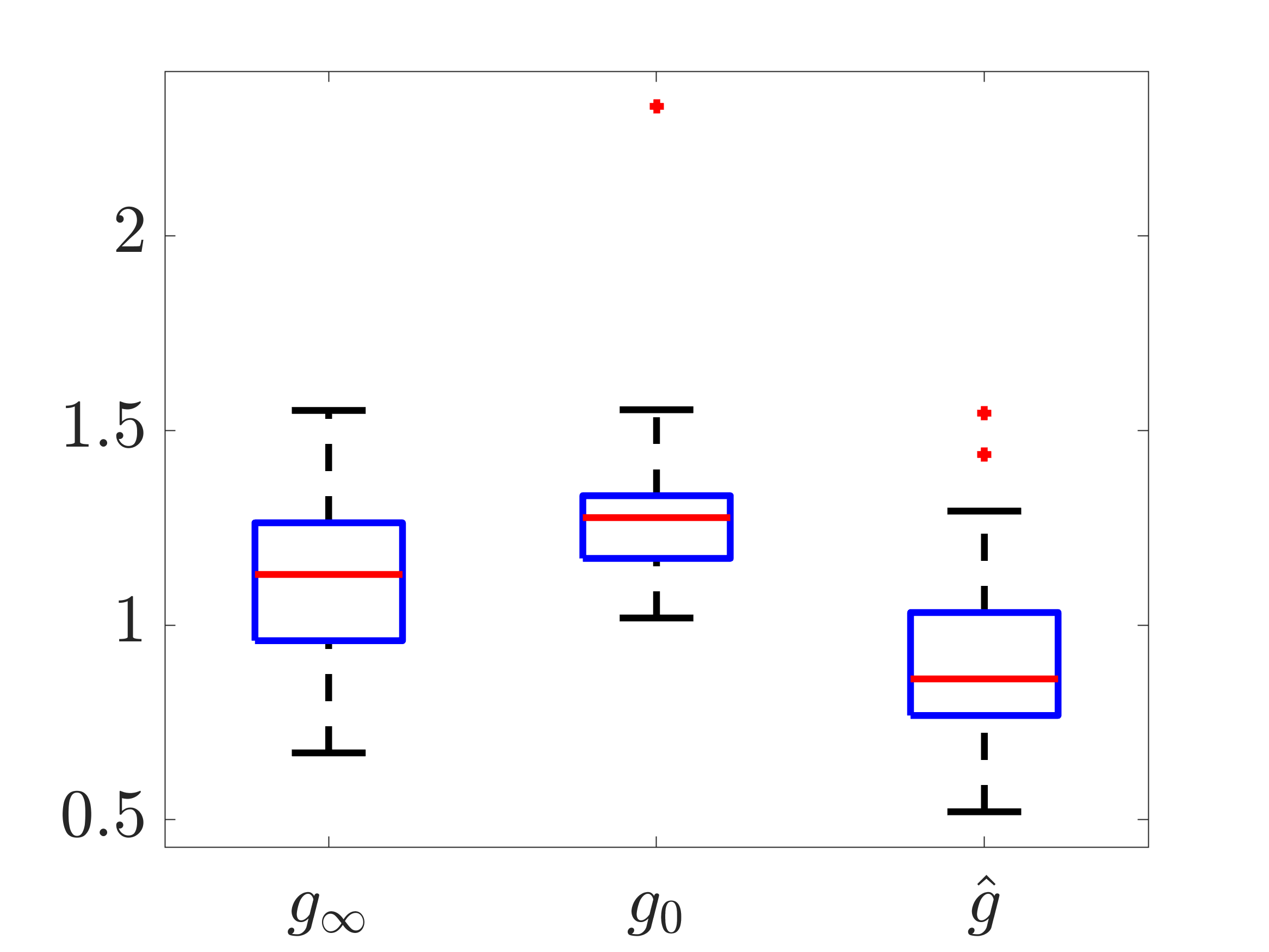} \\
    Simulation 1& Simulation 2& Simulation 3& Simulation 4\\
    \includegraphics[width=1.5in]{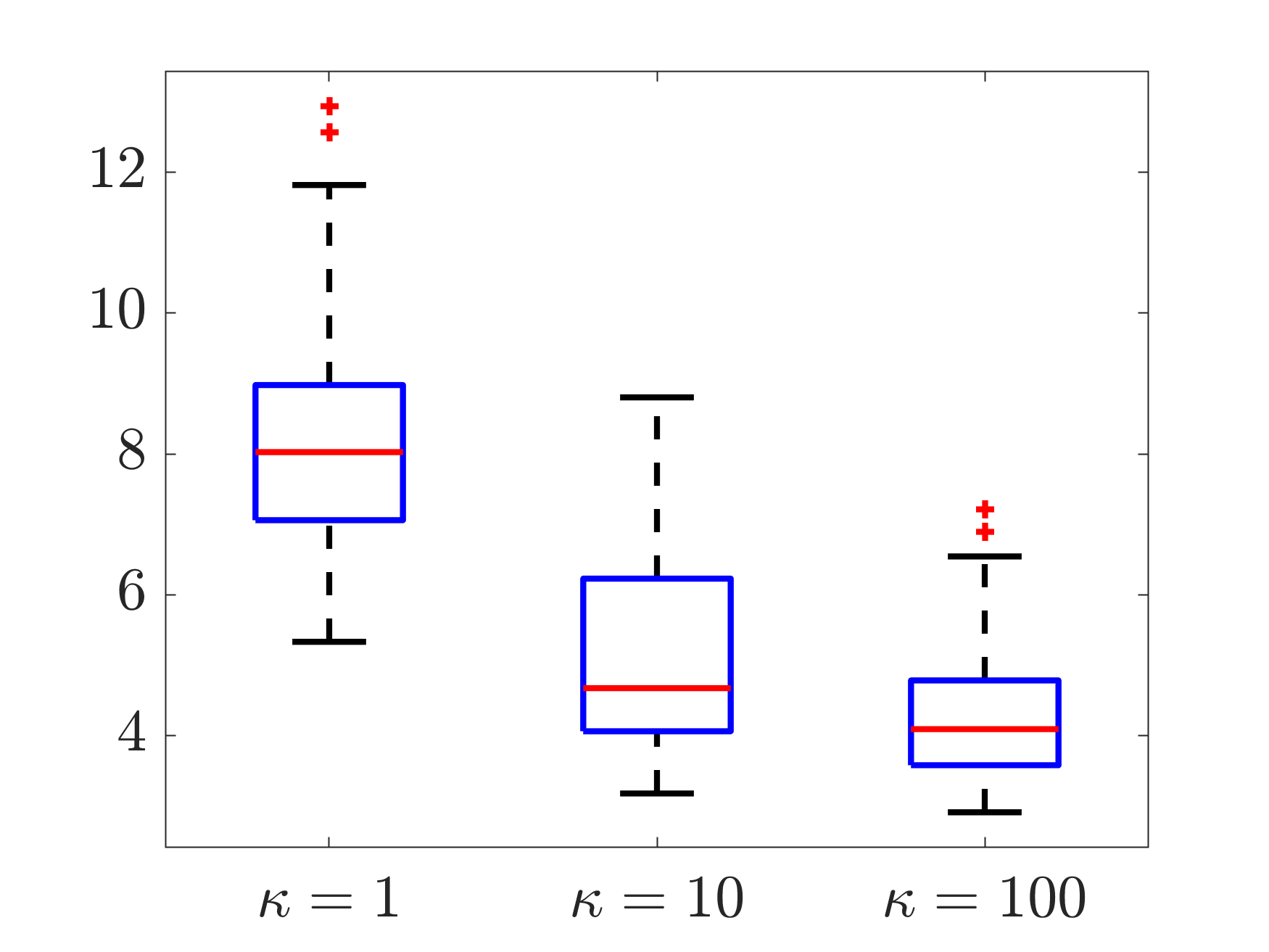} &
  \hspace*{-0.22in}  \includegraphics[width=1.5in] {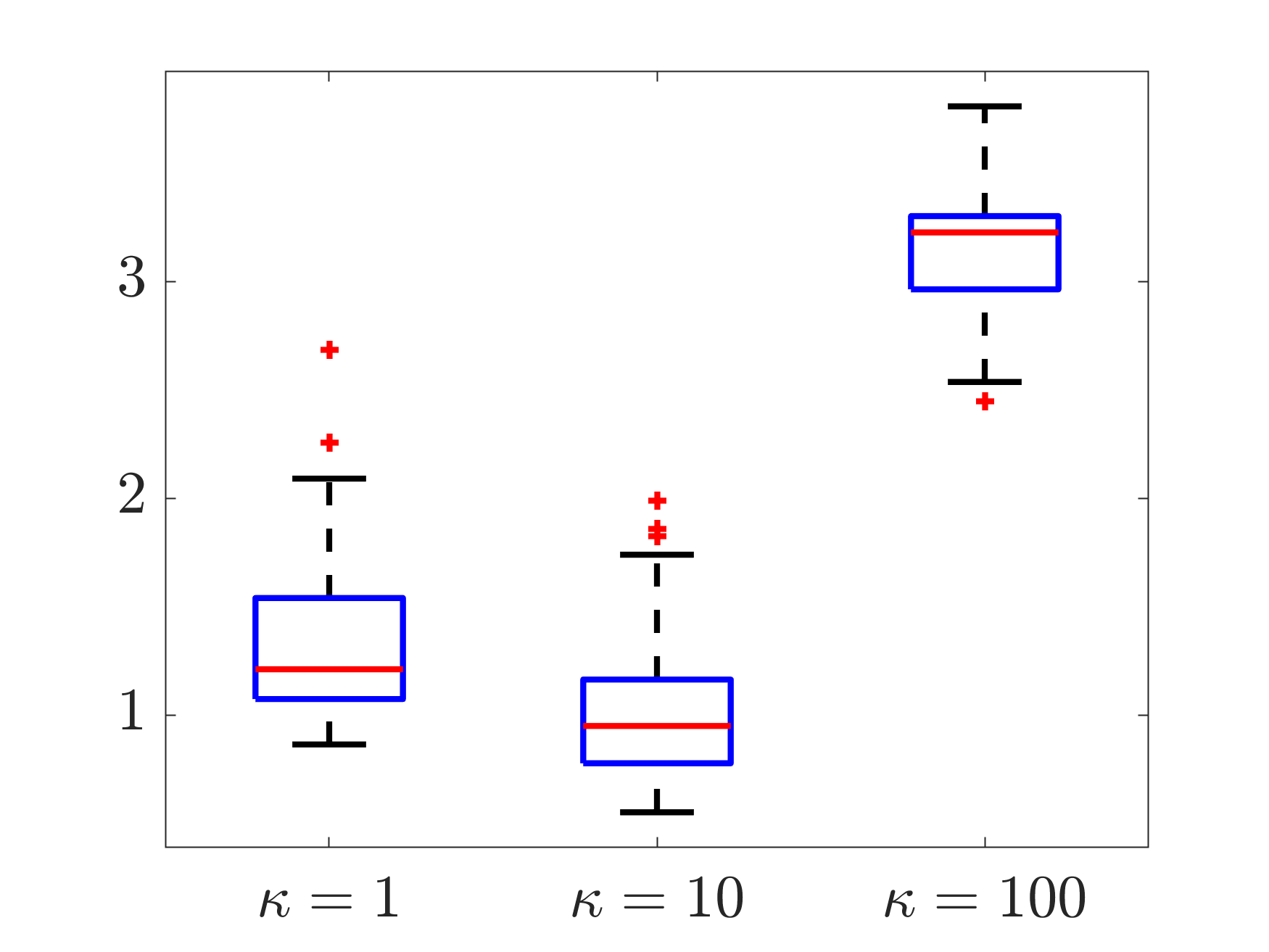} &
   \hspace*{-0.22in} \includegraphics[width=1.5in]{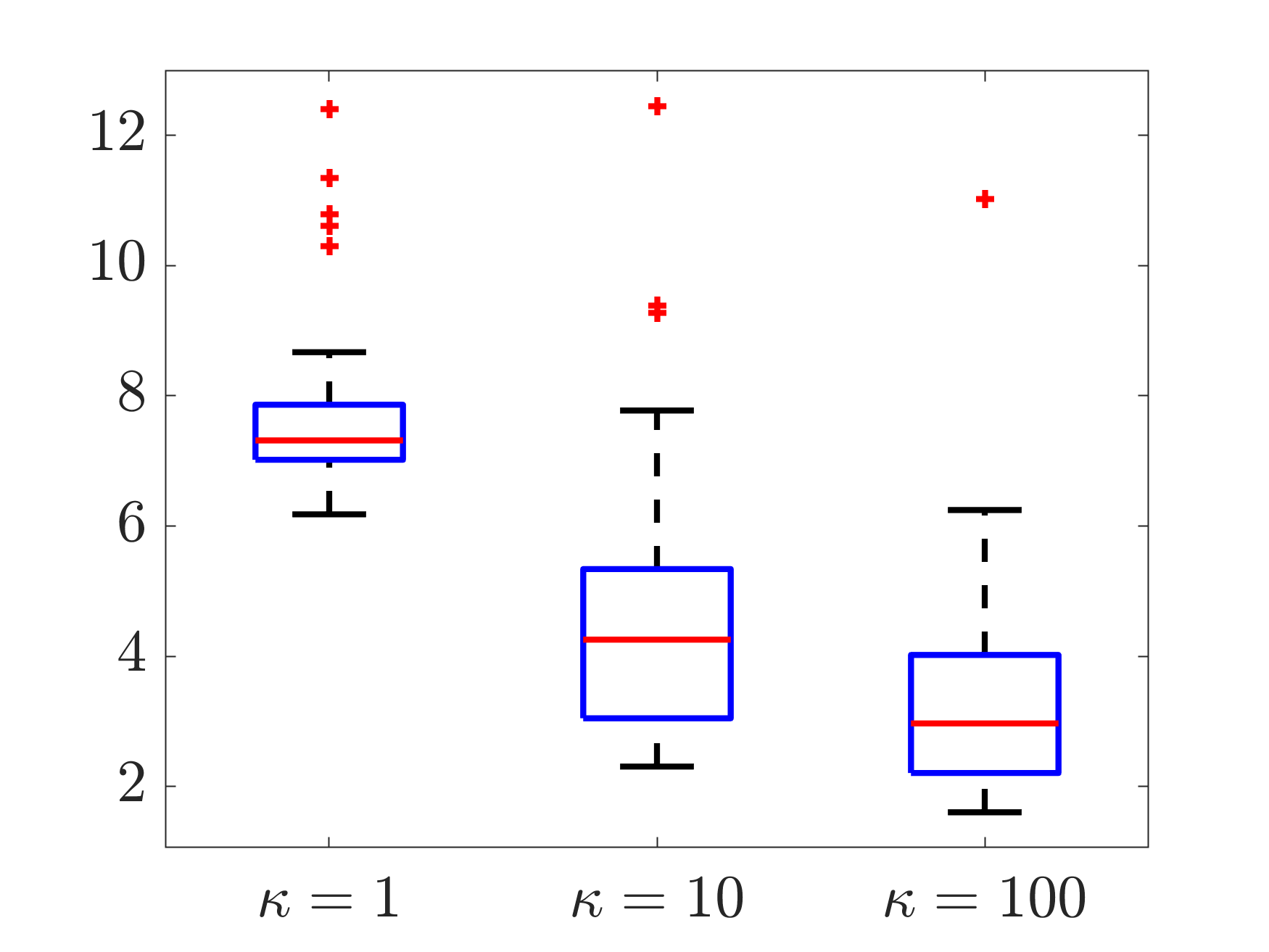} &
   \hspace*{-0.22in} \includegraphics[width=1.5in]{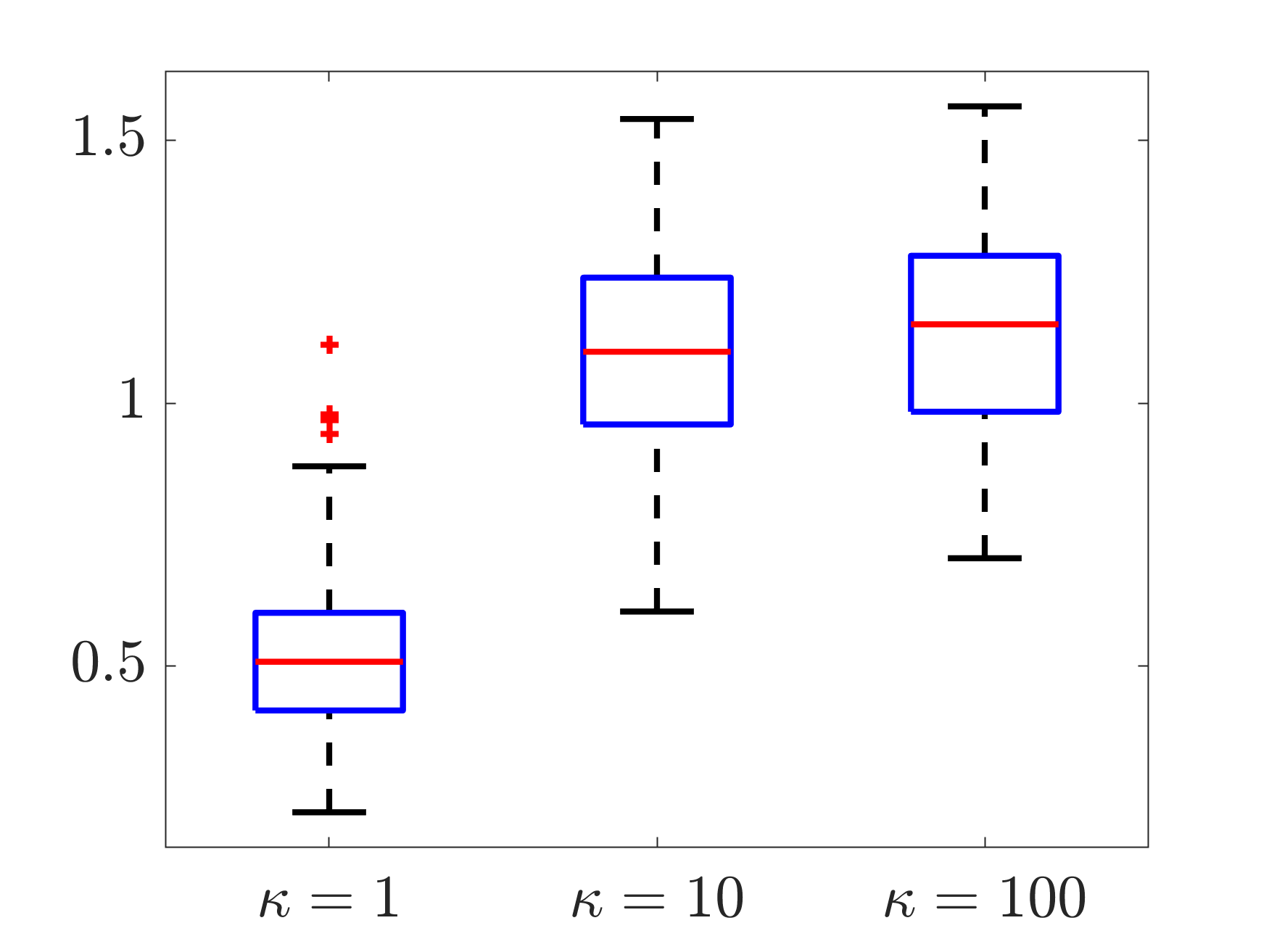} \\
    Simulation 1& Simulation 2& Simulation 3& Simulation 4\\
    \end{tabular}
    \caption{Top: The boxplots of the $\ltwo$ errors of three different estimators on four simulation scenarios. Bottom: The boxplots for the penalized-$\ltwo$ estimators for different values of $\kappa$.}
    \label{fig:boxplots}
\end{figure*}

\begin{figure*}[ht]
  \captionsetup[subfigure]{labelformat=empty}
  \centering
  \raisebox{45pt}{\parbox[b]{.07\textwidth}{$\hat g_{\lambda^*}$}}%
  \vspace*{-0.05in}\hspace{-0.2in}
  \subfloat[][]{\includegraphics[width=.275\textwidth]{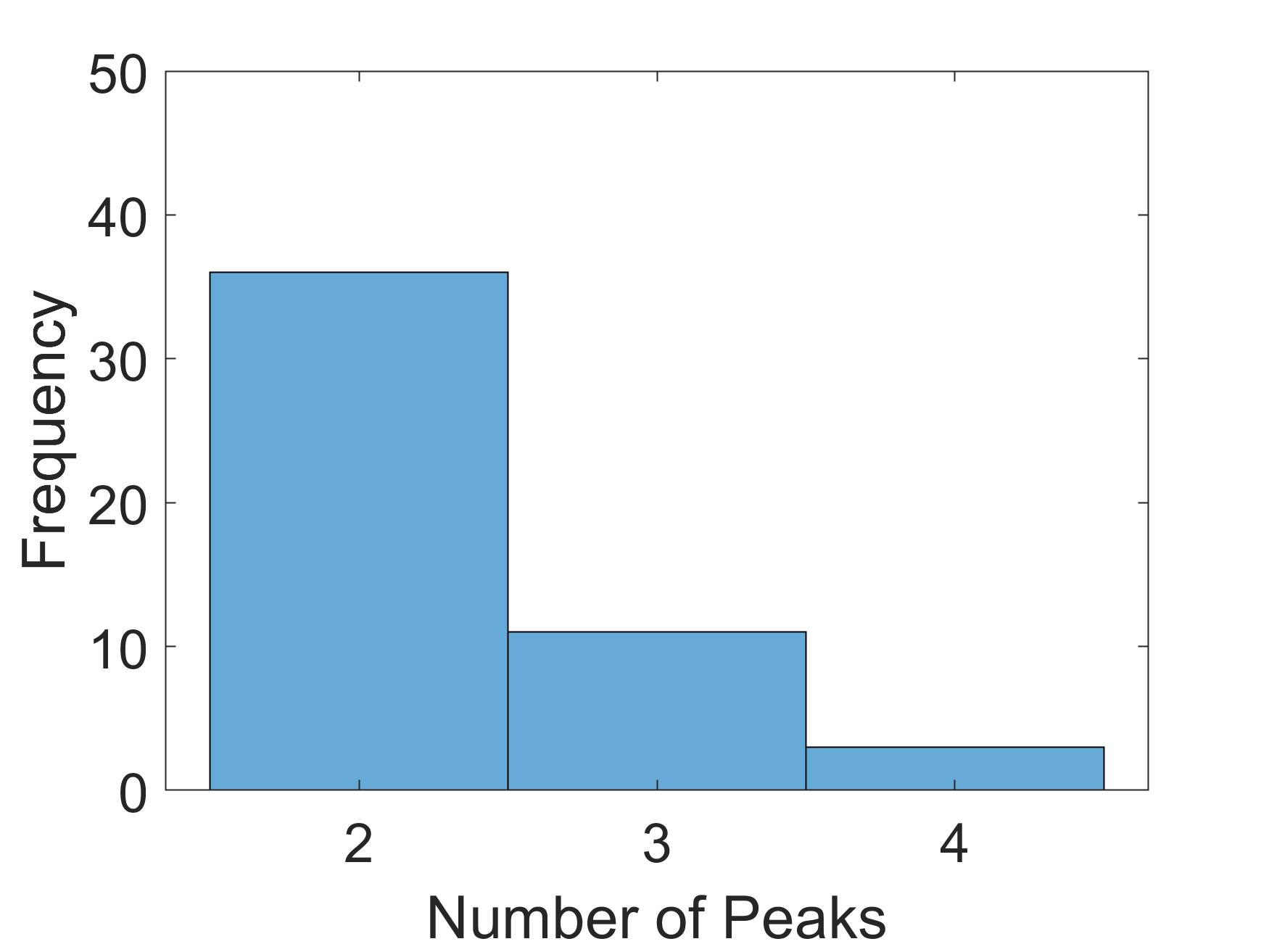}}\hfill
  \hspace{-0.32in}
  \subfloat[][]{\includegraphics[width=.275\textwidth]{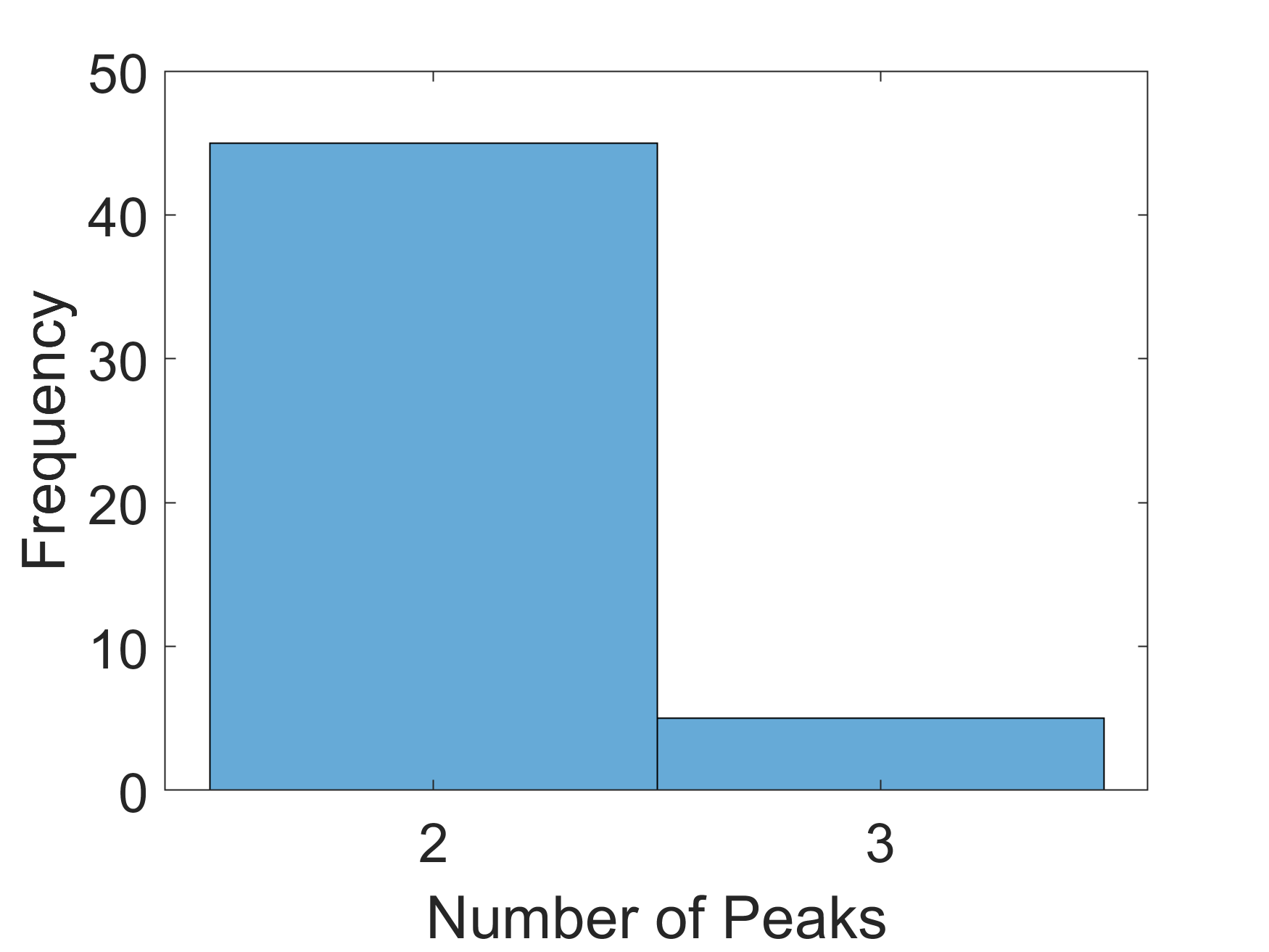}}\hfill
  \hspace{-0.32in}
  \subfloat[][]{\includegraphics[width=.275\textwidth]{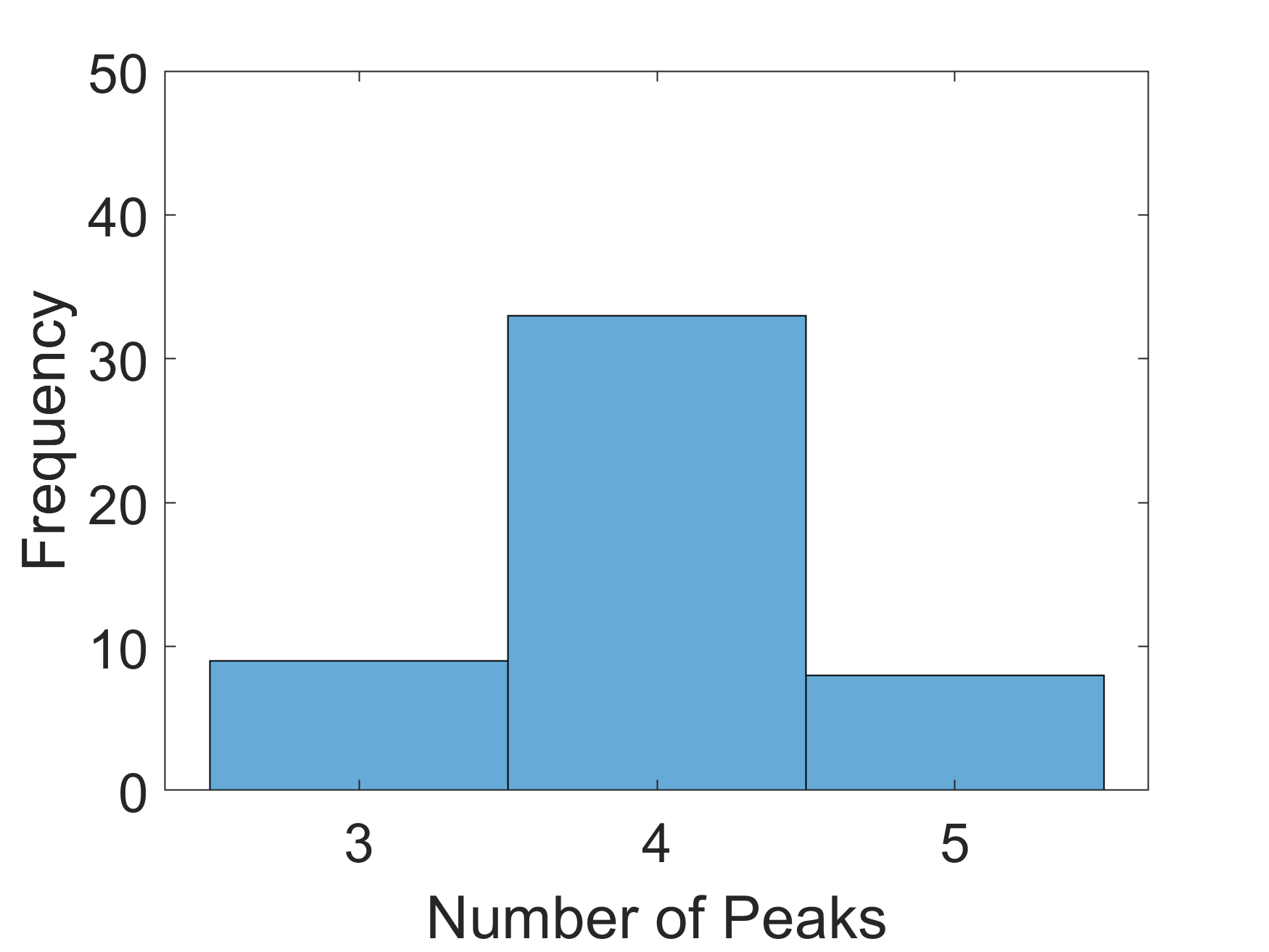}}\hfill
  \hspace{-0.32in}
  \subfloat[][]{\includegraphics[width=.275\textwidth]{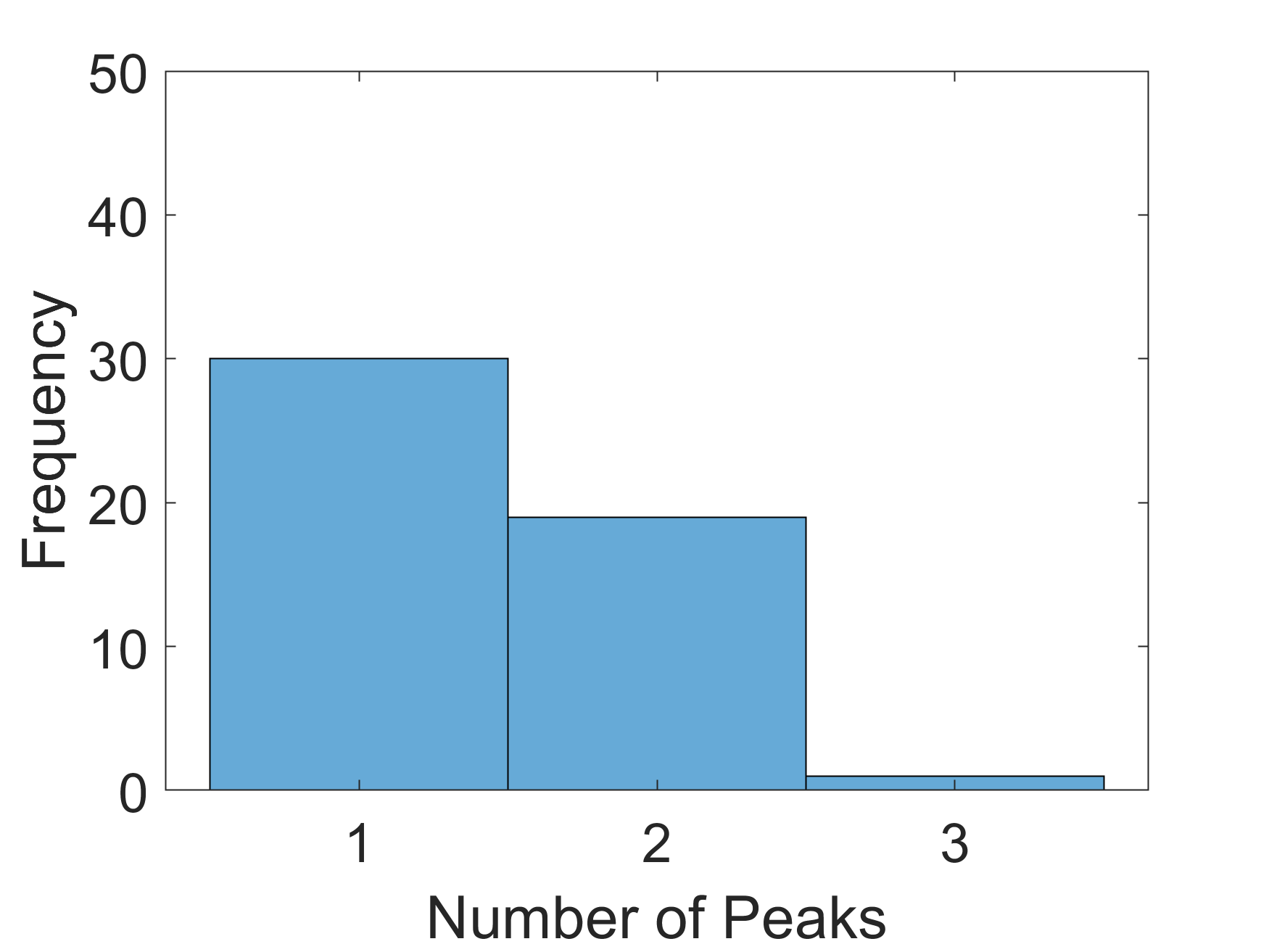}}\hfill\\
  \raisebox{45pt}{\parbox[b]{.07\textwidth}{$\hat g_{init}$}}%
  \vspace*{-0.05in}\hspace{-0.2in}
  \subfloat[][Sim1 (True = 2)]{\includegraphics[width=.275\textwidth]{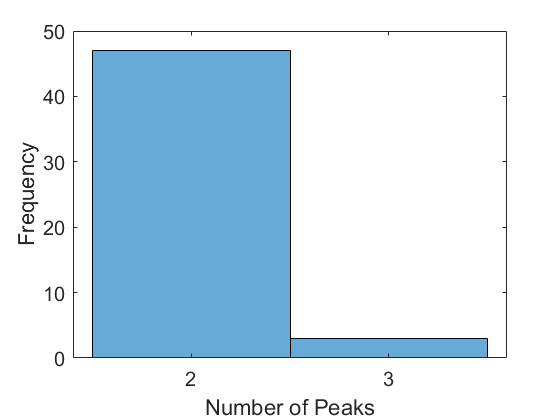}}\hfill
  \hspace{-0.32in}
  \subfloat[][Sim2 (True = 2)]{\includegraphics[width=.275\textwidth]{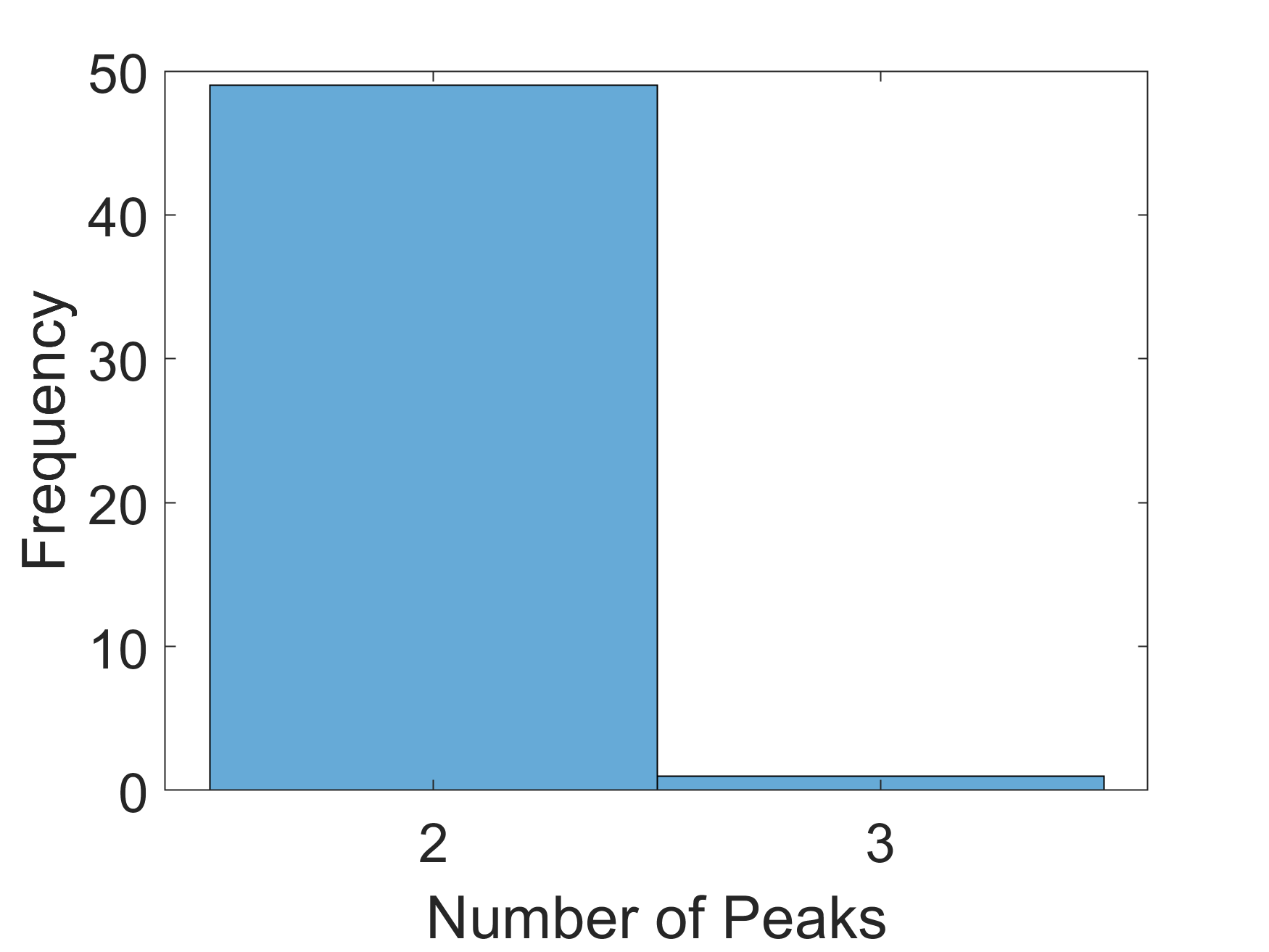}}\hfill
  \hspace{-0.32in}
  \subfloat[][Sim3 (True = 2)]{\includegraphics[width=.275\textwidth]{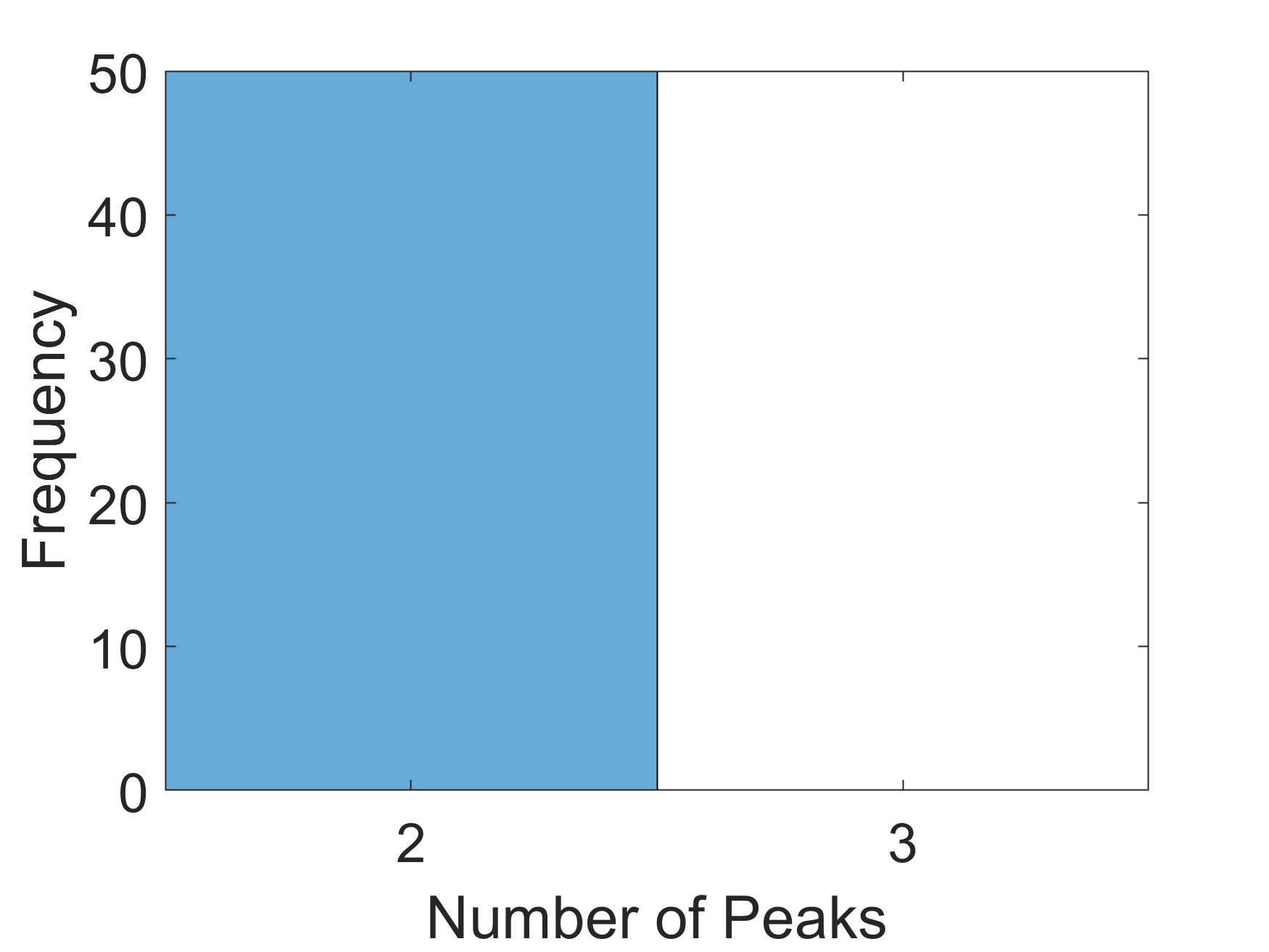}}\hfill
  \hspace{-0.32in}
  \subfloat[][Sim4 (True = 1)]{\includegraphics[width=.275\textwidth]{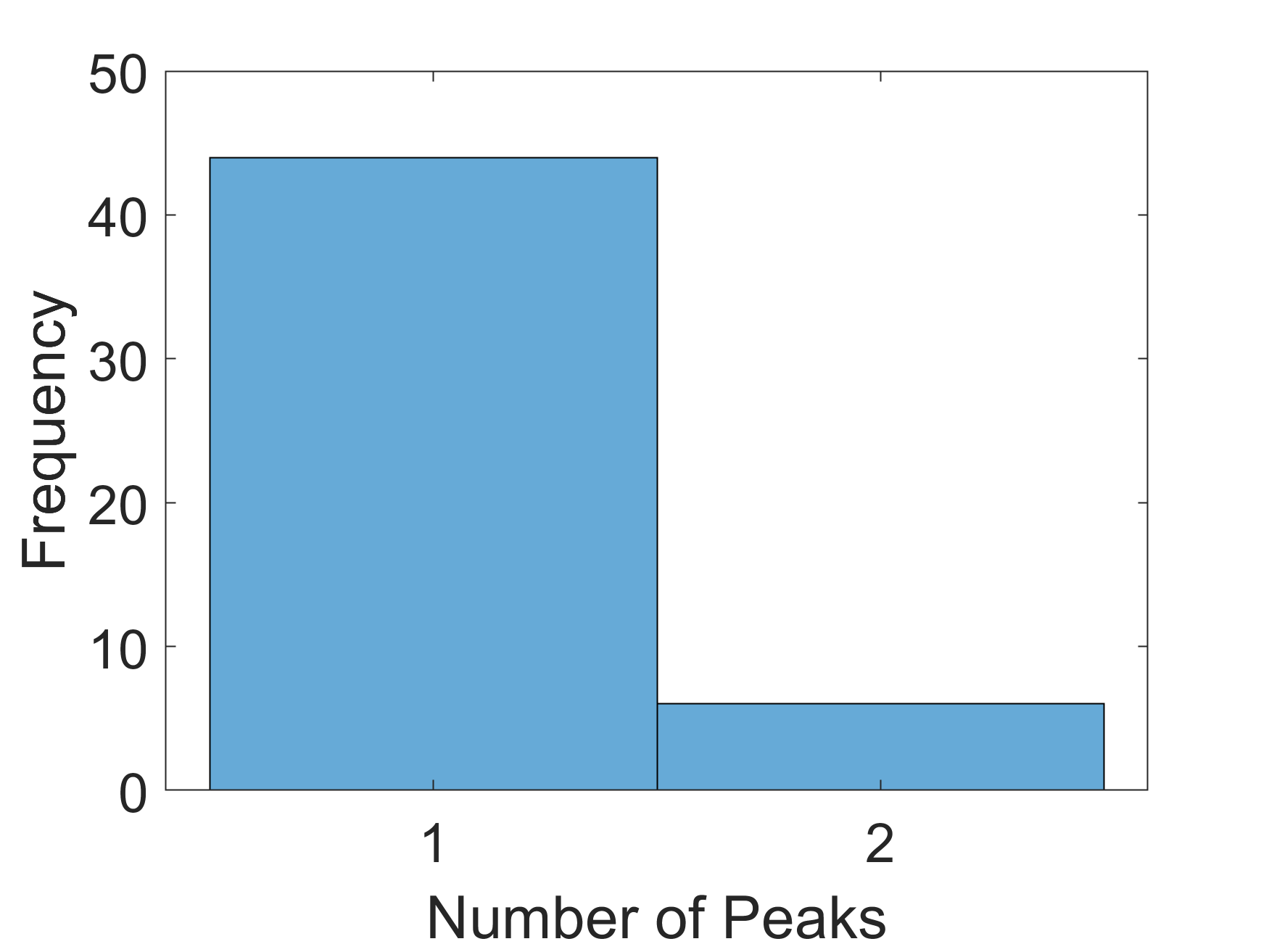}}\hfill
    \caption{Top: Histograms of estimated number of peaks (insignificant and significant) in $\hat{g}_{\lambda^*}$ for 50 repetitions. Bottom: Histograms of numbers of significant peaks estimated using PPD.}
  \label{fig: PPD_Histogram}
\end{figure*}

\begin{itemize}
  \item {\bf{\emph{Quantification of Estimation Performance}}}
\end{itemize}
So far we have pictorially analyzed estimates from different methods. To quantify accuracy of these estimates, we conduct 50 replications of each experiment and calculated the estimation errors. Fig. \ref{fig:boxplots} displays boxplots of the mean and standard deviations of the $\ltwo$ errors for the four simulation experiments. It shows that $\hat g$, the third box in the boxplot, provides the lowest RMSE. In comparisons, the unaligned mean ($\hat{g}_\infty$) fairs poorly in estimating the heights of peaks and valleys, and the fully aligned mean ($\hat{g}_0$) overestimates the number of peaks.  
To emphasize the role of shape in estimation, Fig. \ref{fig: PPD_Histogram} shows the histograms of the estimated number of peaks for different simulation datasets. The upper row histograms the number of peaks in the cross-sectional mean of the partially aligned mean, $\hat g_{\lambda^*}$. These include both significant and insignificant peaks. The bottom row shows histograms of the number of the significant peaks selected by PPD. As these results show, the PPD approach is able to estimate the correct number of peaks in $g$ most of the time. This contributes to the improved performance of the proposed shape-constrained function estimator.

\tabcolsep 1pt

Finally, we analyze the computational cost of our estimation procedure relative to the past methods. Table~\ref{tab: computational time} lists the time to perform different algorithmic steps. This cost is computed on a CPU (Intel(R)) i7-11800H @ 2.30GHz. 
We omit the cost of unaligned mean as it does not require any functional alignment. The cost for computing the fully elastic mean averaged over $50$ replications is in the first column. 
Lastly, the costs of different steps of our method -- costs for partial alignment, computing PPD, and function estimation --  are listed in the three right columns. Most of the time is consumed in the partial alignment of functions for different values of $\lambda$.

\begin{table*}
\centering
\caption{Computational time (in seconds) for various steps in our procedure.}
\label{tab: computational time}
\begin{tabular}{|l|c||ccc|}
\hline
 &
  Fully Elastic &
  \multicolumn{3}{c|}{Partial Alignment} \\ \hline \hline
 &
  Aligning Functions &
  \multicolumn{1}{c|}{\begin{tabular}[c]{@{}c@{}}Aligning Functions\\ (1st Step)\end{tabular}} &
  \multicolumn{1}{c|}{\begin{tabular}[c]{@{}c@{}}Computing PPDs\\ (1st Step)\end{tabular}} &
  \begin{tabular}[c]{@{}c@{}}Function Estimation\\ (2nd Step)\end{tabular} \\ \hline
\multicolumn{1}{|c|}{Simulation 1} &
  \begin{tabular}[c]{@{}c@{}}15.29 \end{tabular} &
  \multicolumn{1}{c|}{\begin{tabular}[c]{@{}c@{}}1868.4 \end{tabular}} &
  \multicolumn{1}{c|}{\begin{tabular}[c]{@{}c@{}}0.13\end{tabular}} &
  \begin{tabular}[c]{@{}c@{}}17.87 
  \end{tabular} \\ \hline
Simulation 2 &
  \begin{tabular}[c]{@{}c@{}}17.12 \end{tabular} &
  \multicolumn{1}{c|}{\begin{tabular}[c]{@{}c@{}}2021.6\end{tabular}} &
  \multicolumn{1}{c|}{\begin{tabular}[c]{@{}c@{}}0.12\end{tabular}} &
  \begin{tabular}[c]{@{}c@{}}18.92\end{tabular} \\ \hline
Simulation 3 &
  \begin{tabular}[c]{@{}c@{}}14.59\end{tabular} &
  \multicolumn{1}{c|}{\begin{tabular}[c]{@{}c@{}}2205.0\end{tabular}} &
  \multicolumn{1}{c|}{\begin{tabular}[c]{@{}c@{}}0.14\end{tabular}} &
  \begin{tabular}[c]{@{}c@{}}38.16\end{tabular} \\ \hline
Simulation 4 &
  \begin{tabular}[c]{@{}c@{}}15.83\end{tabular} &
  \multicolumn{1}{c|}{\begin{tabular}[c]{@{}c@{}}1564.8\end{tabular}} &
  \multicolumn{1}{c|}{\begin{tabular}[c]{@{}c@{}}0.09\end{tabular}} &
  \begin{tabular}[c]{@{}c@{}}17.01\end{tabular} \\ \hline
\end{tabular}
\end{table*}

\subsection{Real Data Studies} \label{sec:Real data}
In this section, we apply our method to estimate functions $g$ underlying some real datasets. However, in these cases, we are unable to compare our estimates to a ground truth function to evaluate their accuracy. Therefore, the results must be interpreted and compared to other estimators in order to assess their performance.
\\

\begin{figure*}[ht]
    \centering
    \subfloat[]{\includegraphics[height = 1.2in]{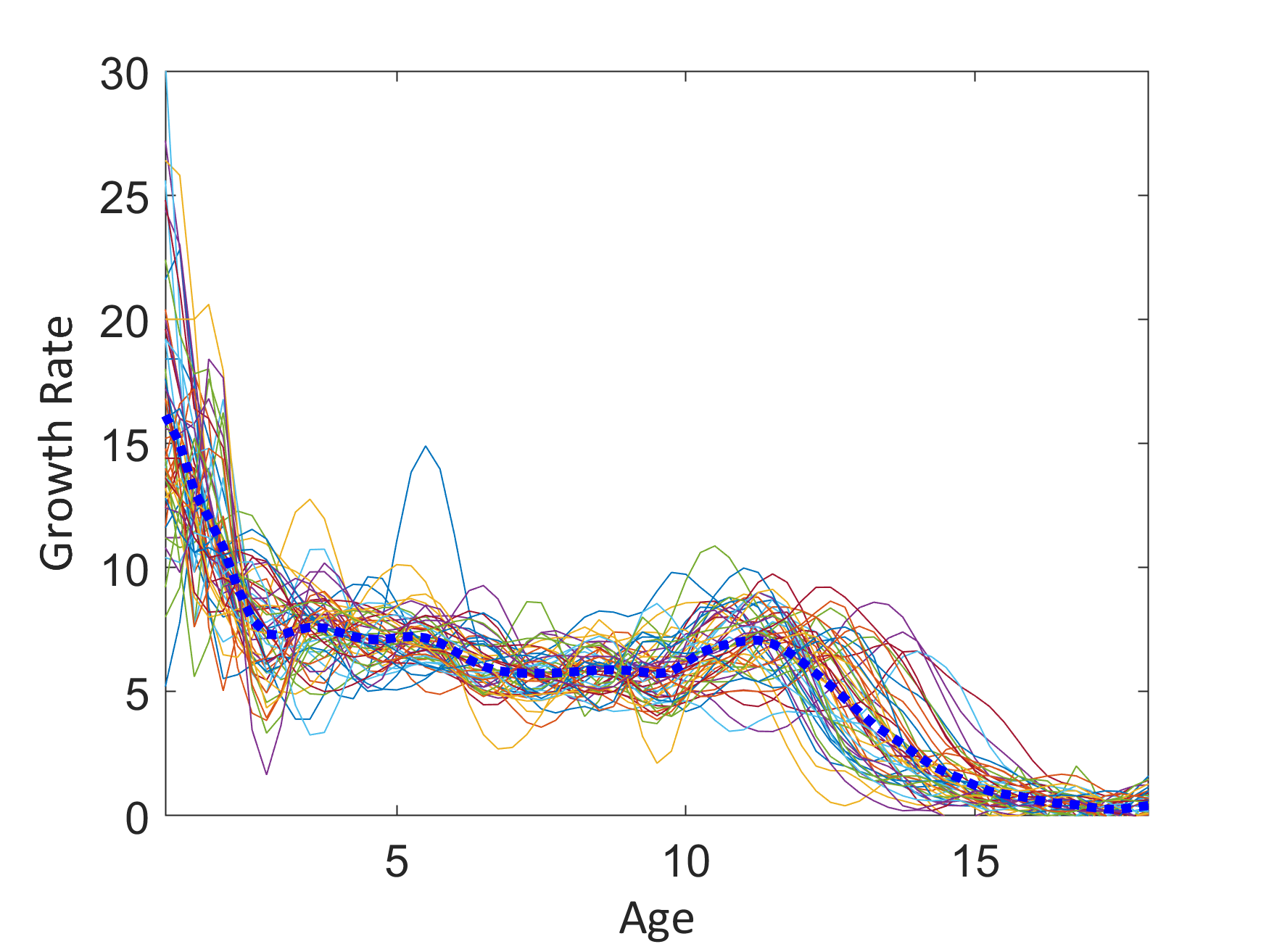}}
    \hspace{-0.2in}
    \subfloat[]{\includegraphics[height = 1.2in]{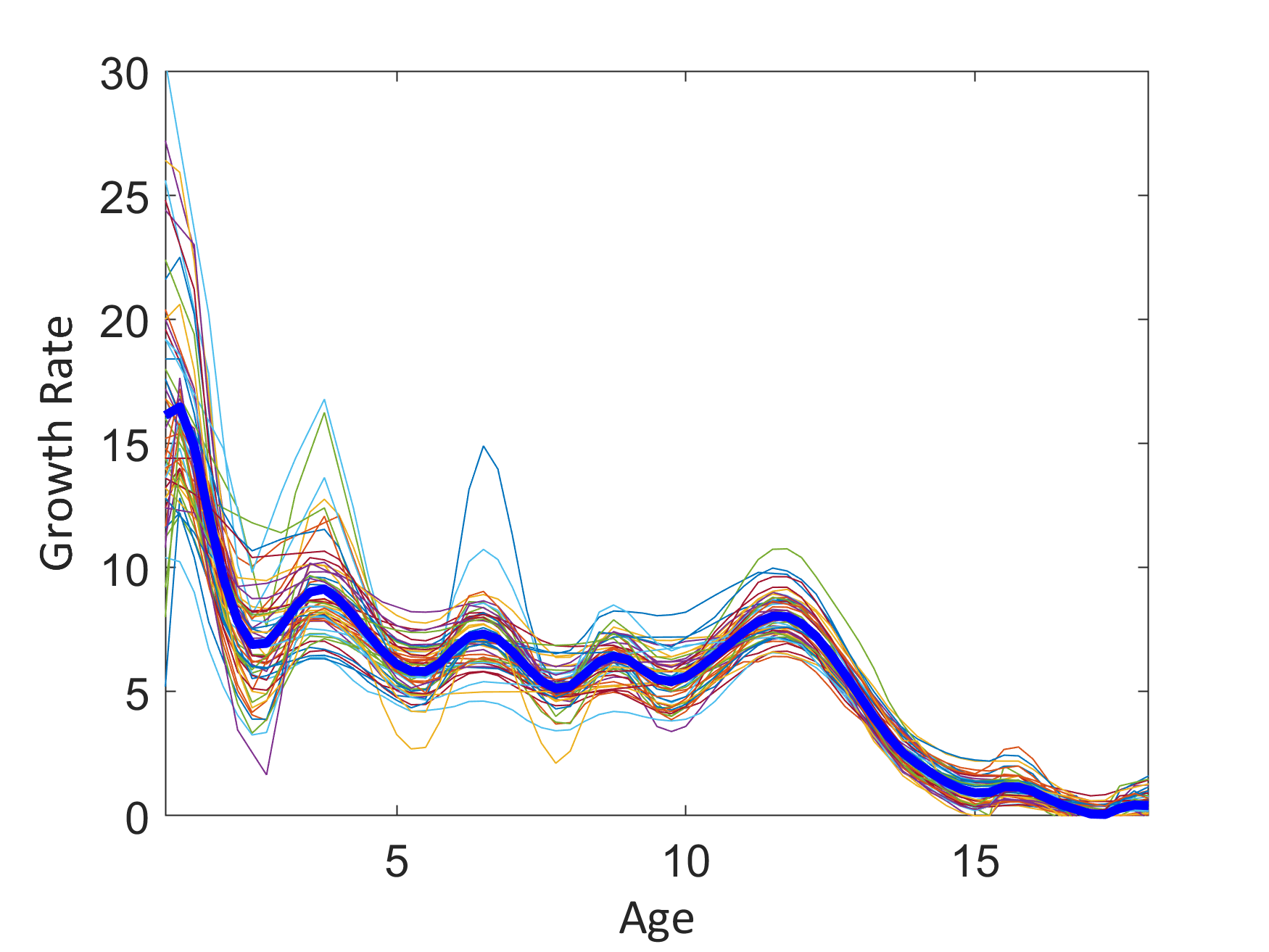}}
    \hspace{-0.2in}
    \subfloat[]{\includegraphics[height = 1.2in]{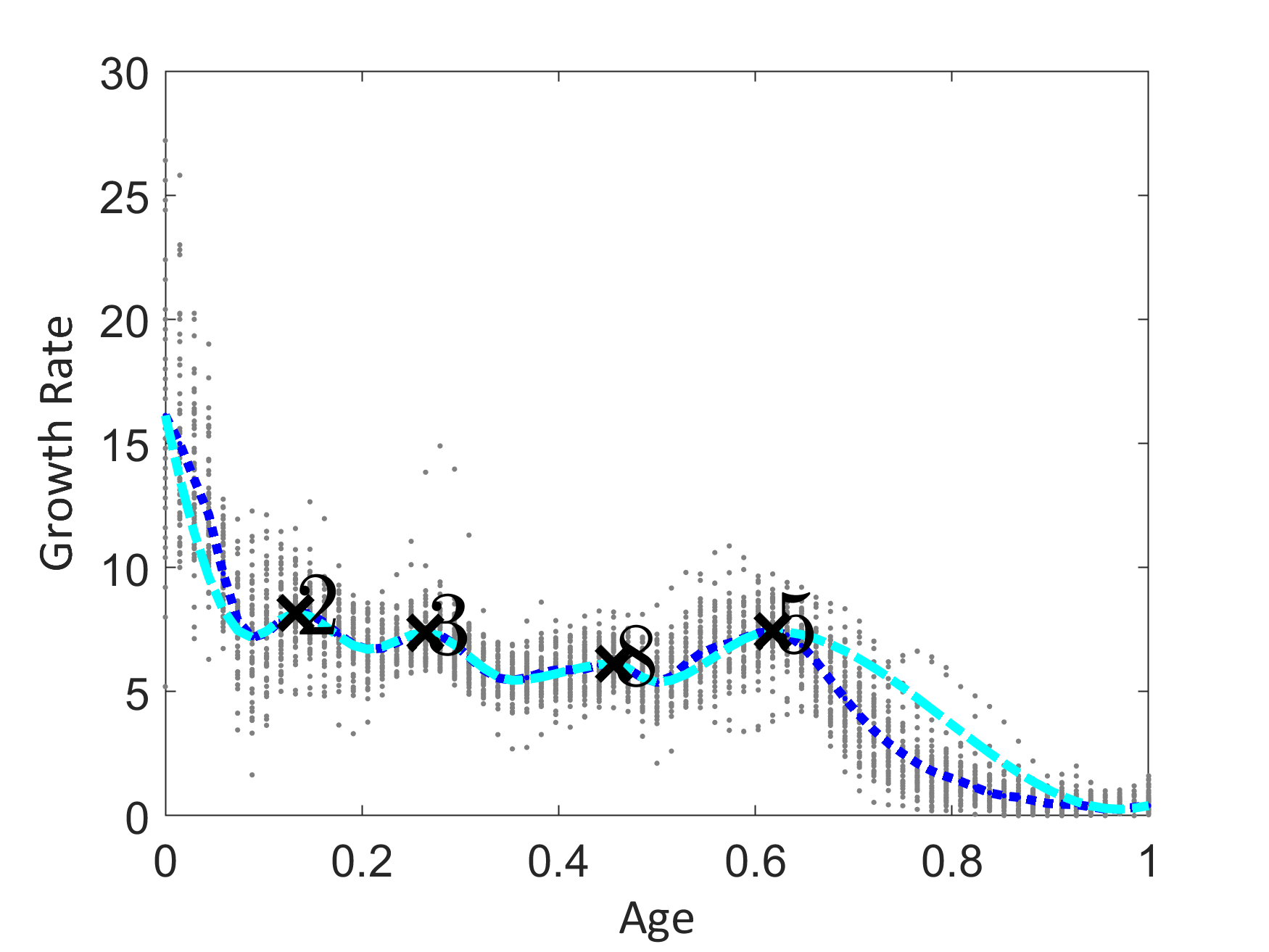}}
    \includegraphics[height = 1in]{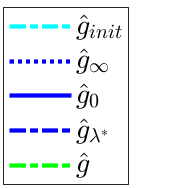}
    \hspace{0in}\\
    \hspace{-0.2in}
    \subfloat[]{\includegraphics[height = 1.2in]{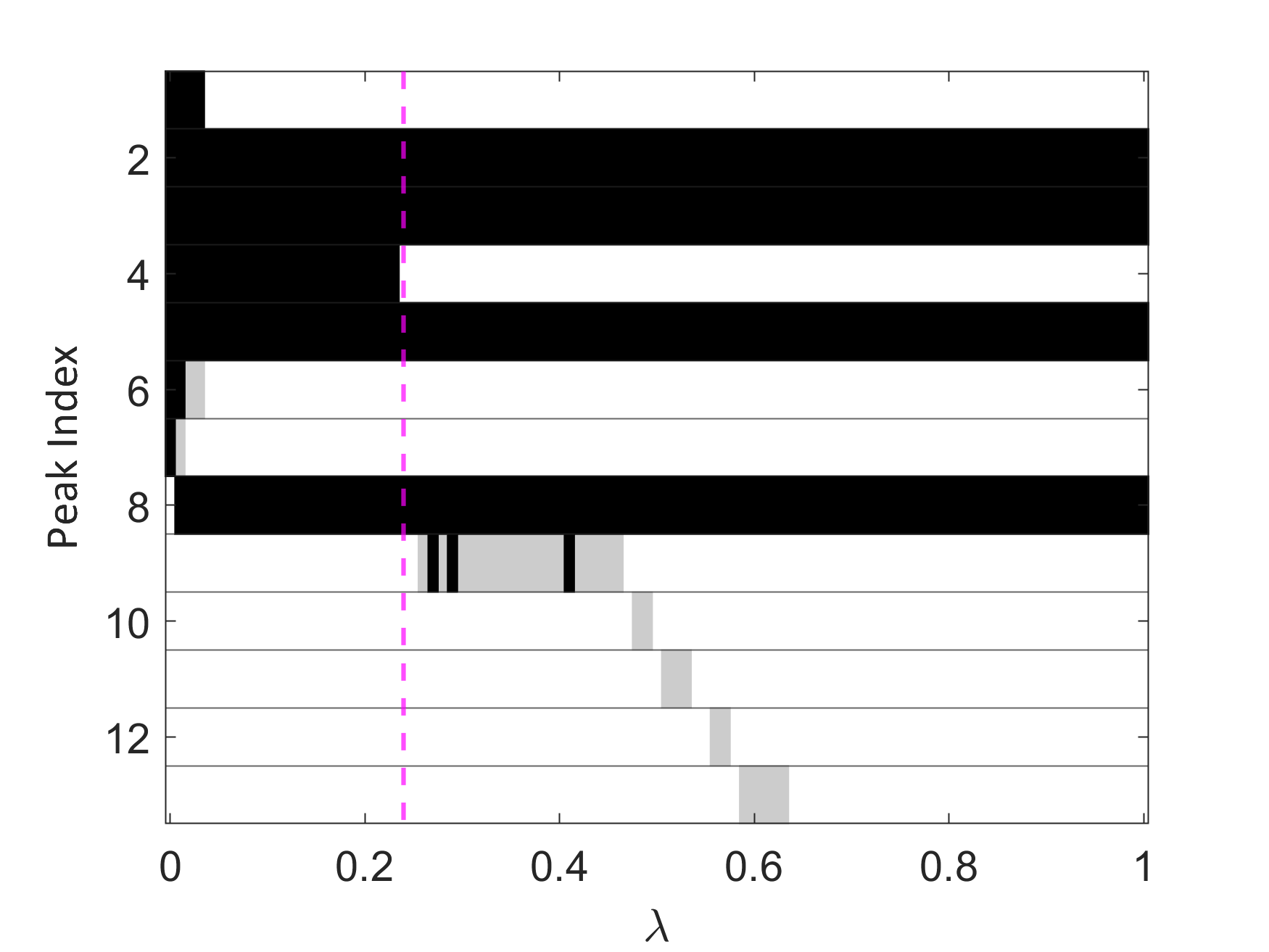}}
    \hspace{-0.2in}
    \subfloat[]{\includegraphics[height = 1.2in]{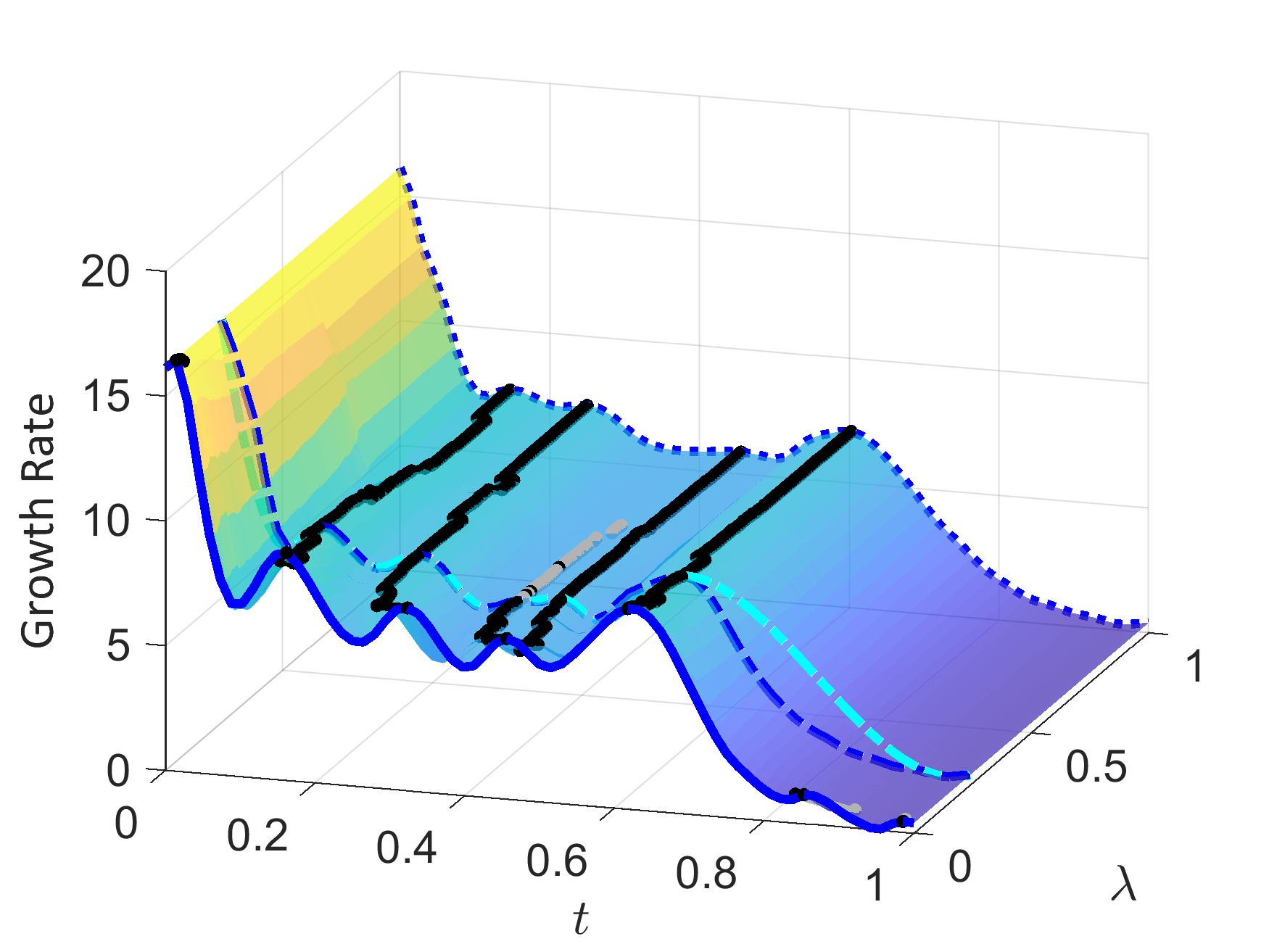}}
    \hspace{-0.2in}
    \subfloat[]{\includegraphics[height = 1.2in]{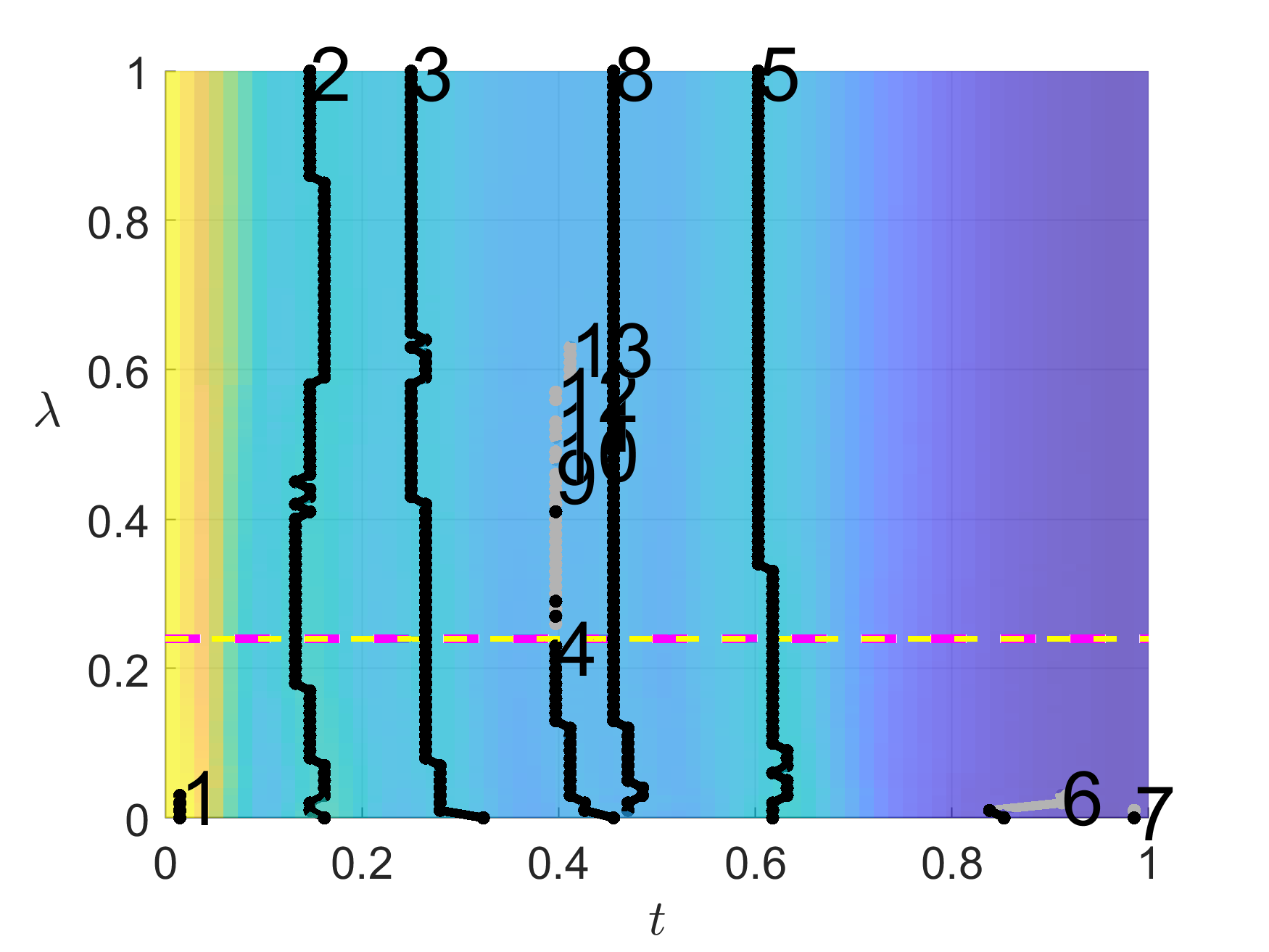}}
    \hspace{-0.2in}
    \subfloat[]{\includegraphics[height = 1.2in]{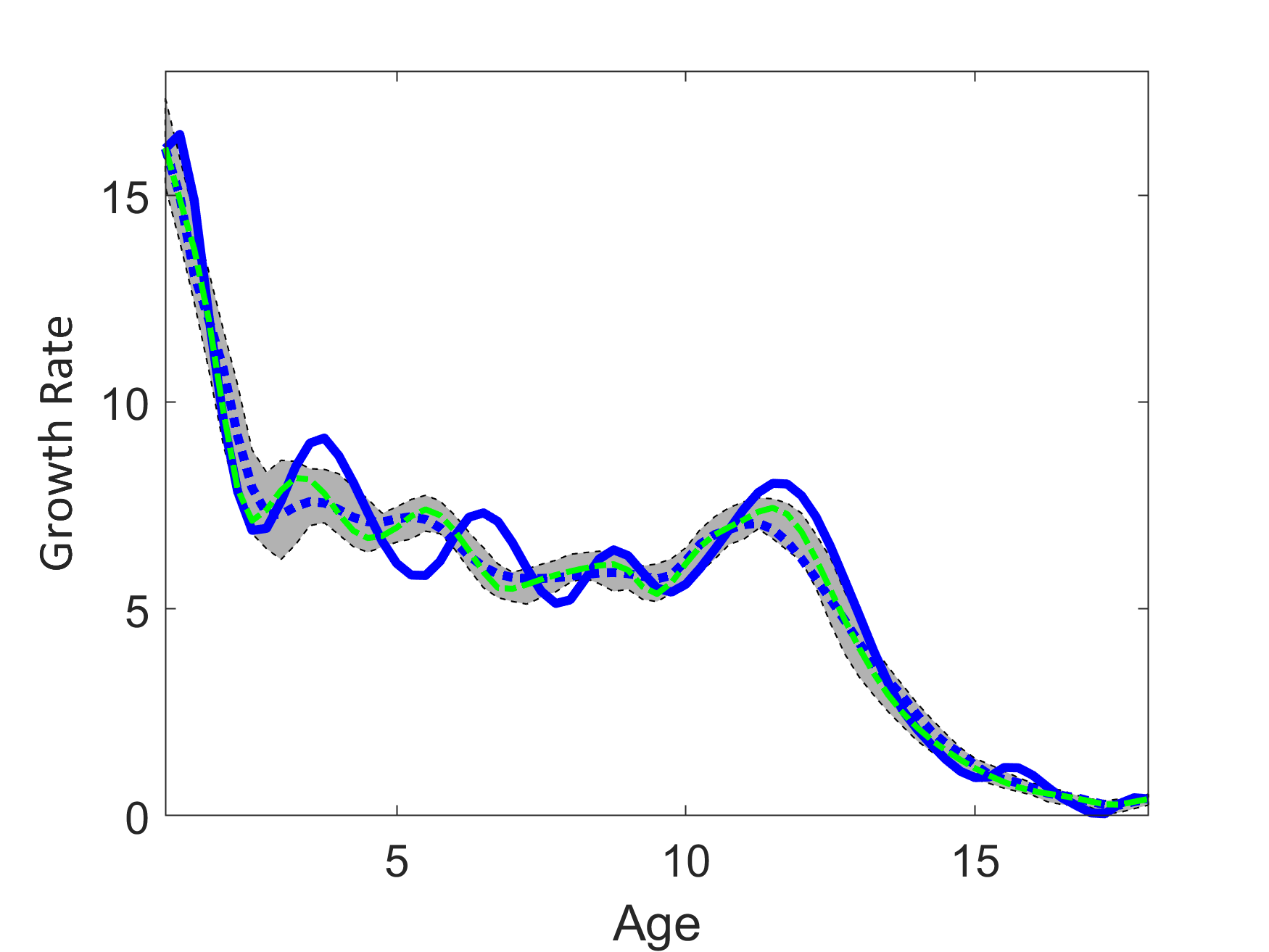}}
    \caption{(Female Growth Rate Data) Unaligned raw data is in (a), and fully-elastic alignment is in (b). PPD in (d) denotes that five peaks (2, 3, 5, and 8) are persistent. (c) shows the partially aligned point cloud and initial shape estimate of $g$ (cyan curve). Plots (e) and (f) demonstrate the gradual changes of aligned functions and their peak process behaviors as $\lambda$ increases. (g) shows the estimate underlying function.}
    \label{fig: real1(2)}
\end{figure*}

\noindent {\bf Berkeley Growth Rate Data}: \\
First, we study the classical Berkeley Growth Study data (\citet{ramsay1997functional}) often used to demonstrate the phase variability in functional data. The female datasets contains the rates of growth in heights of 54 girls from ages 1 to 18. The peaks in growth curves are called the {\it growth spurts} and of great interest. Scientists and physicians use a growth chart to analyze individual growths and ask the question: How many growth spurts to the individuals (male or female) have on average? What is the average growth profile for a human being in that category? 
Fig.~\ref{fig: real1(2)} (a) shows the raw data for female subjects and (b) shows fully aligned functions. The unaligned mean (blue curve in (a)) smooths out almost all the growth spurts and can be a bad representative of the population. The full aligned mean (blue curve in (b)) shows several strong spurts but also some relatively weak spurts. Among the $13$ peaks generated, the PPD chart selects four internal peaks ($2$, $3$, $5$, and $8$) as significant and persistent, concluding thet female growth has three main growth spurts. The plot on (g) shows the estimated function $\hat{g}$ in green color.  
\\

\noindent {\bf COVID-19 Data Analysis for Europe}\\
There is a great interest in analyzing COVID pandemic data and associated surges in infection rates. While different communities experienced COVID peaks (or waves) at asynchronous times, there are common patterns underlying COVID incidences and outcomes. For instance, if we focus on COVID outcomes for different countries in the region, there is arguably an overall pattern of waves during the pandemic period.

\begin{itemize}
\item {\bf Daily Hospitalization Rate Curves}: 
We consider the daily hospital-occupancy counts associated with COVID-19 in 25 European countries during the period from April 2020 to July 2021 (\citet{owid}).
\begin{figure*}[htbp]
    \centering
    \subfloat[]{\includegraphics[height = 1.2in]{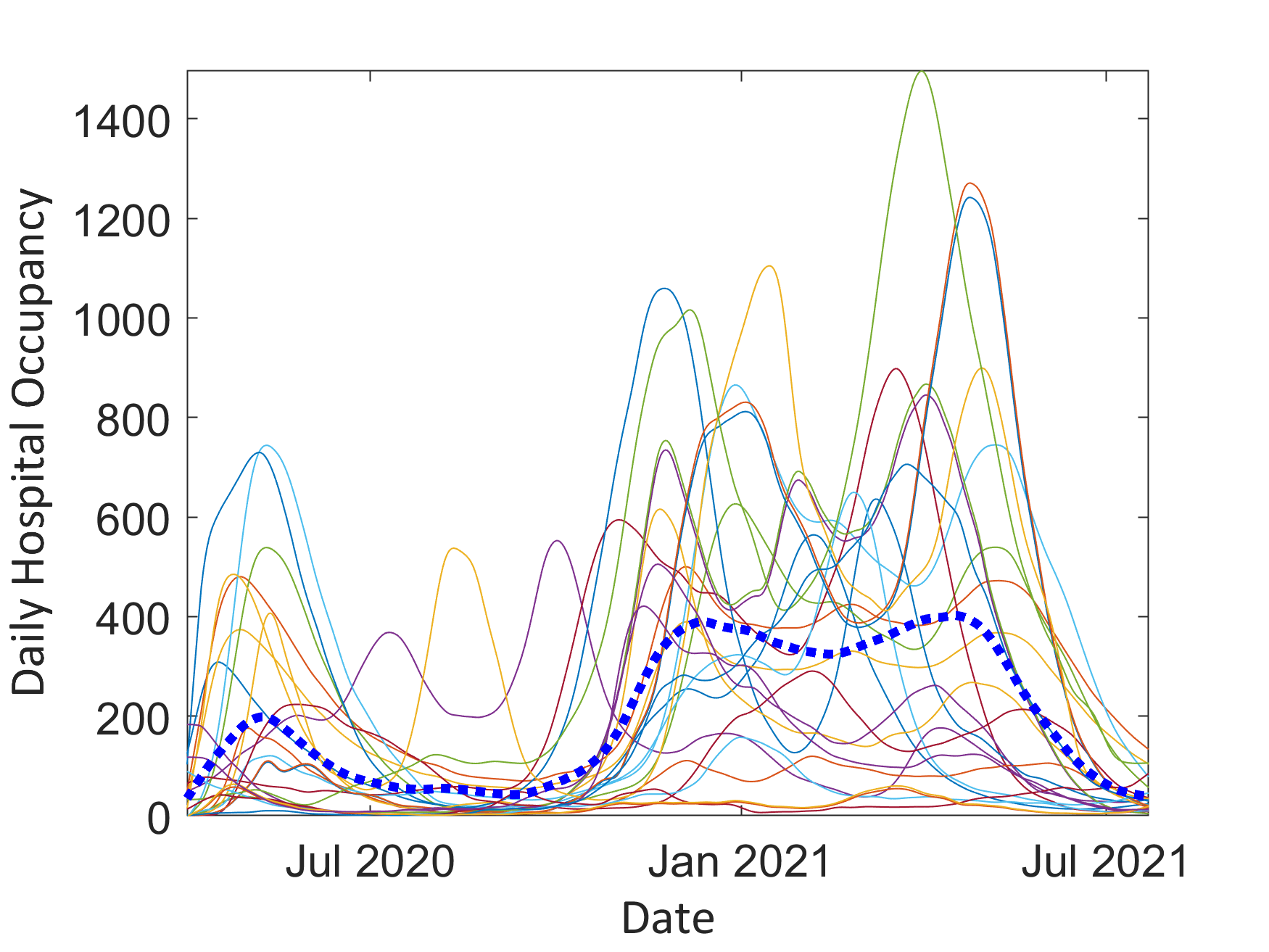}}
    \hspace{-0.2in}
    \subfloat[]{\includegraphics[height = 1.2in]{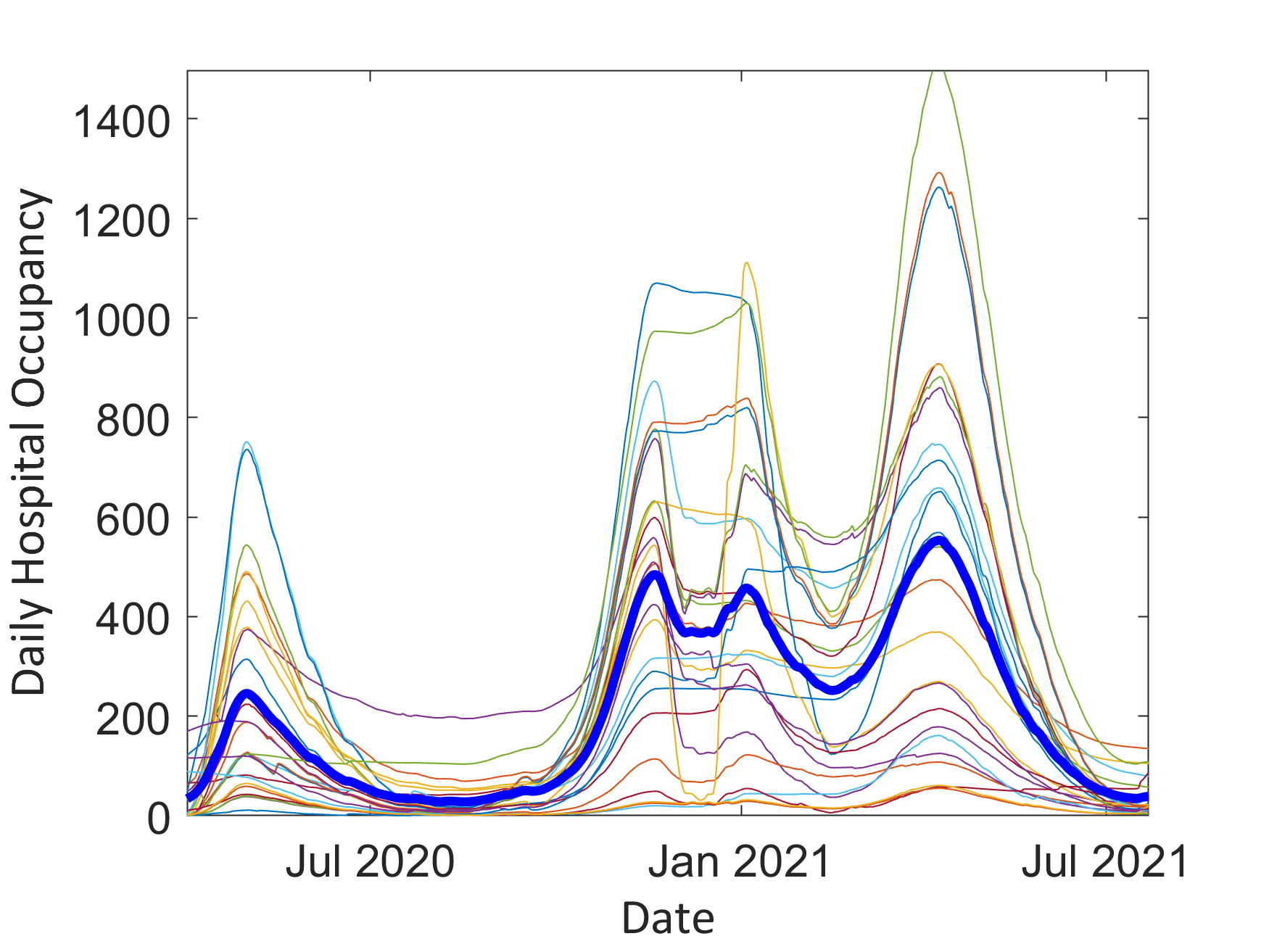}}
    \hspace{-0.2in}
    \subfloat[]{\includegraphics[height = 1.2in]{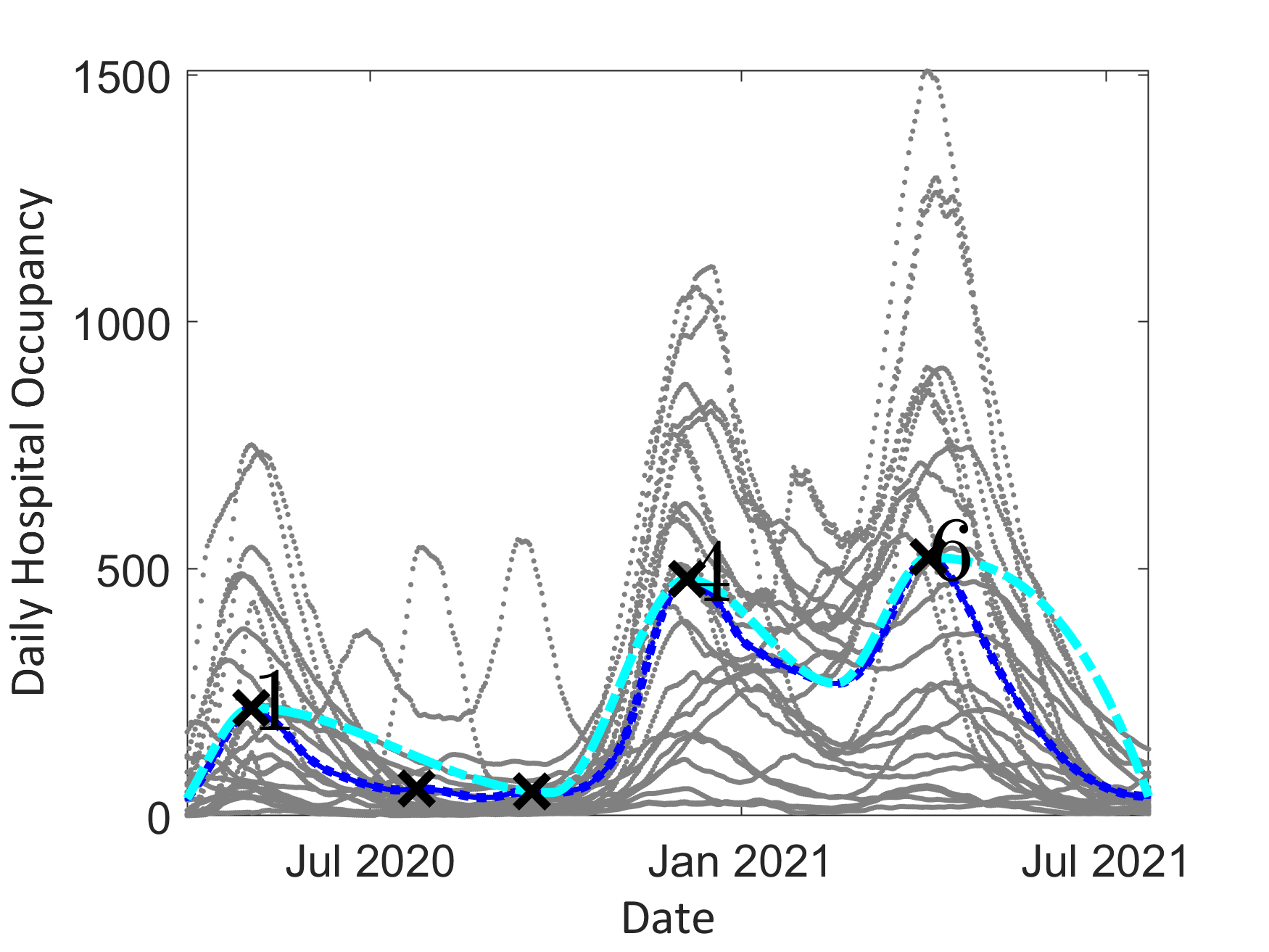}}
    \includegraphics[height = 1in]{fig/legend-rd-1.png}
    \hspace{0in}\\
    \hspace{-0.2in}
    \subfloat[]{\includegraphics[height = 1.2in]{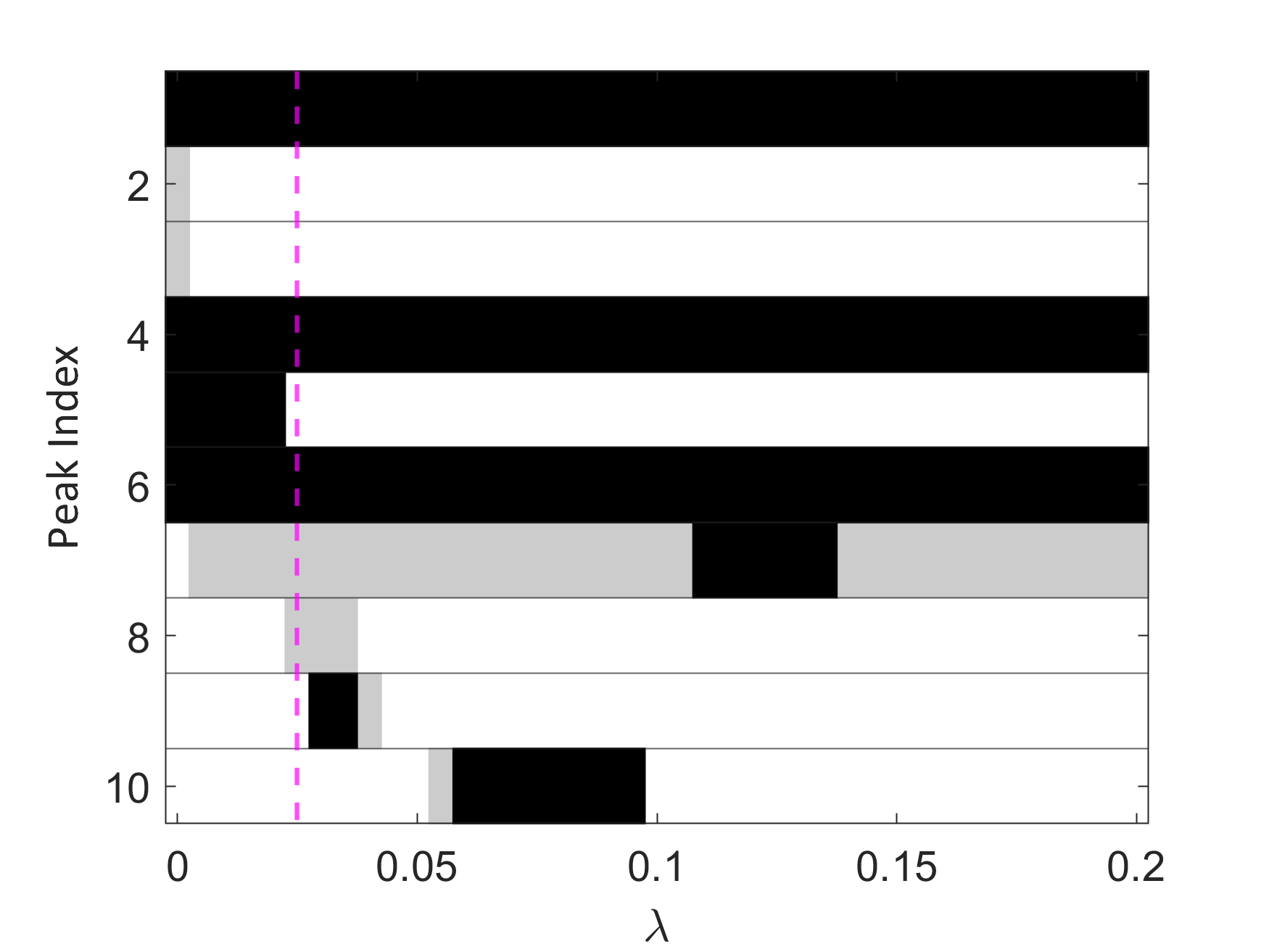}}
    \hspace{-0.2in}
    \subfloat[]{\includegraphics[height = 1.2in]{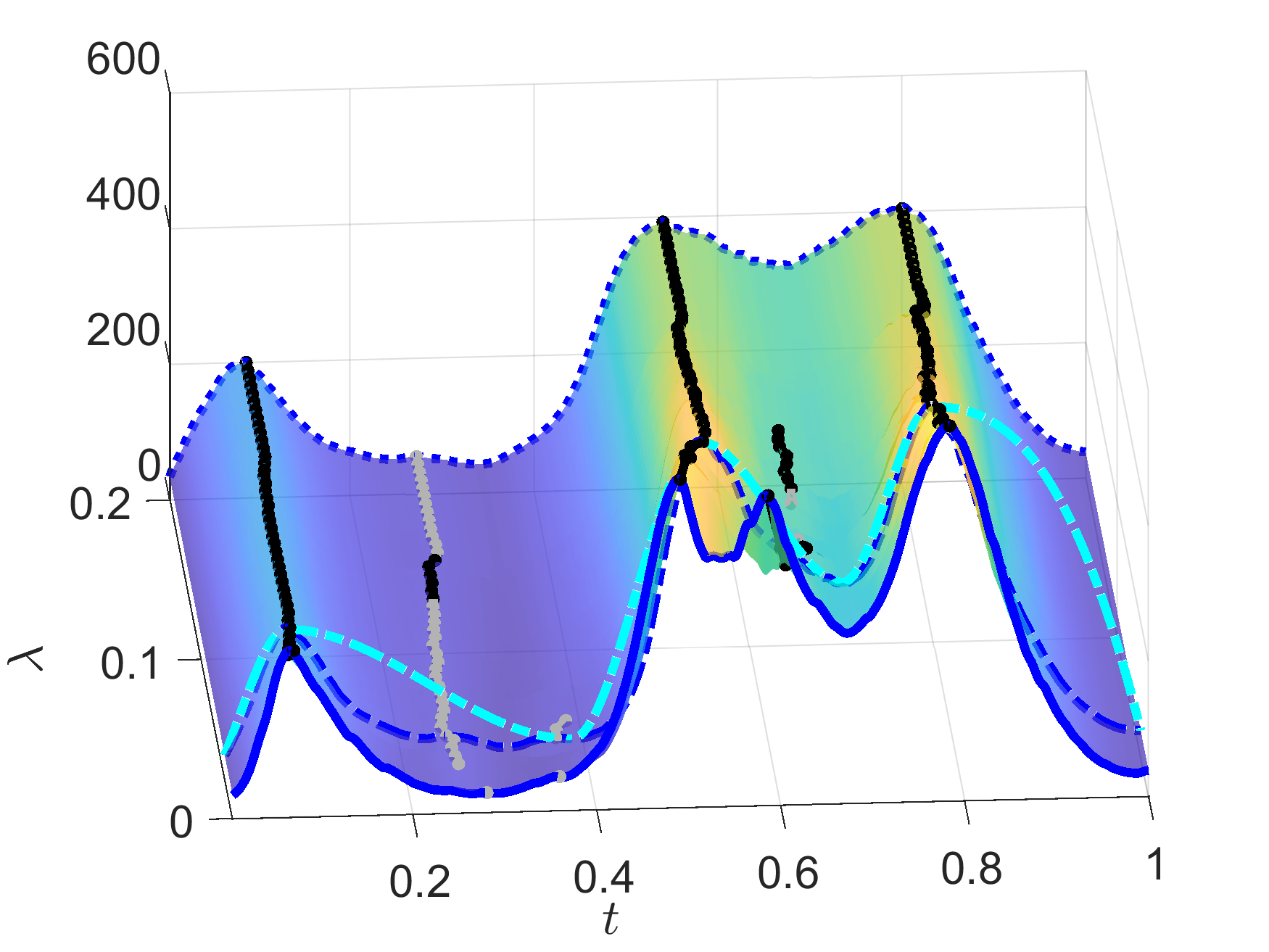}}
    \hspace{-0.2in}
    \subfloat[]{\includegraphics[height = 1.2in]{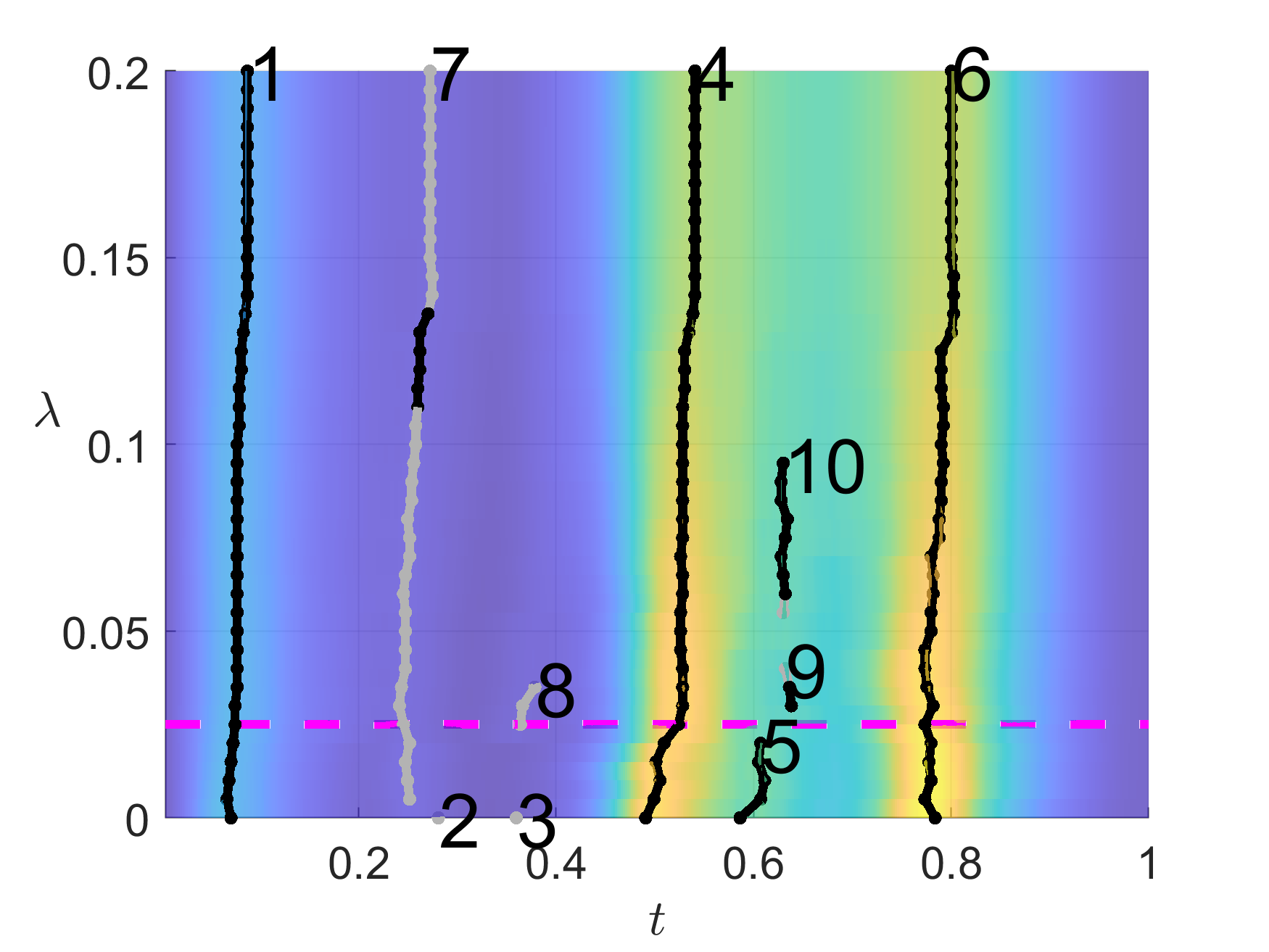}}
    \hspace{-0.2in}
    \subfloat[]{\includegraphics[height = 1.2in]{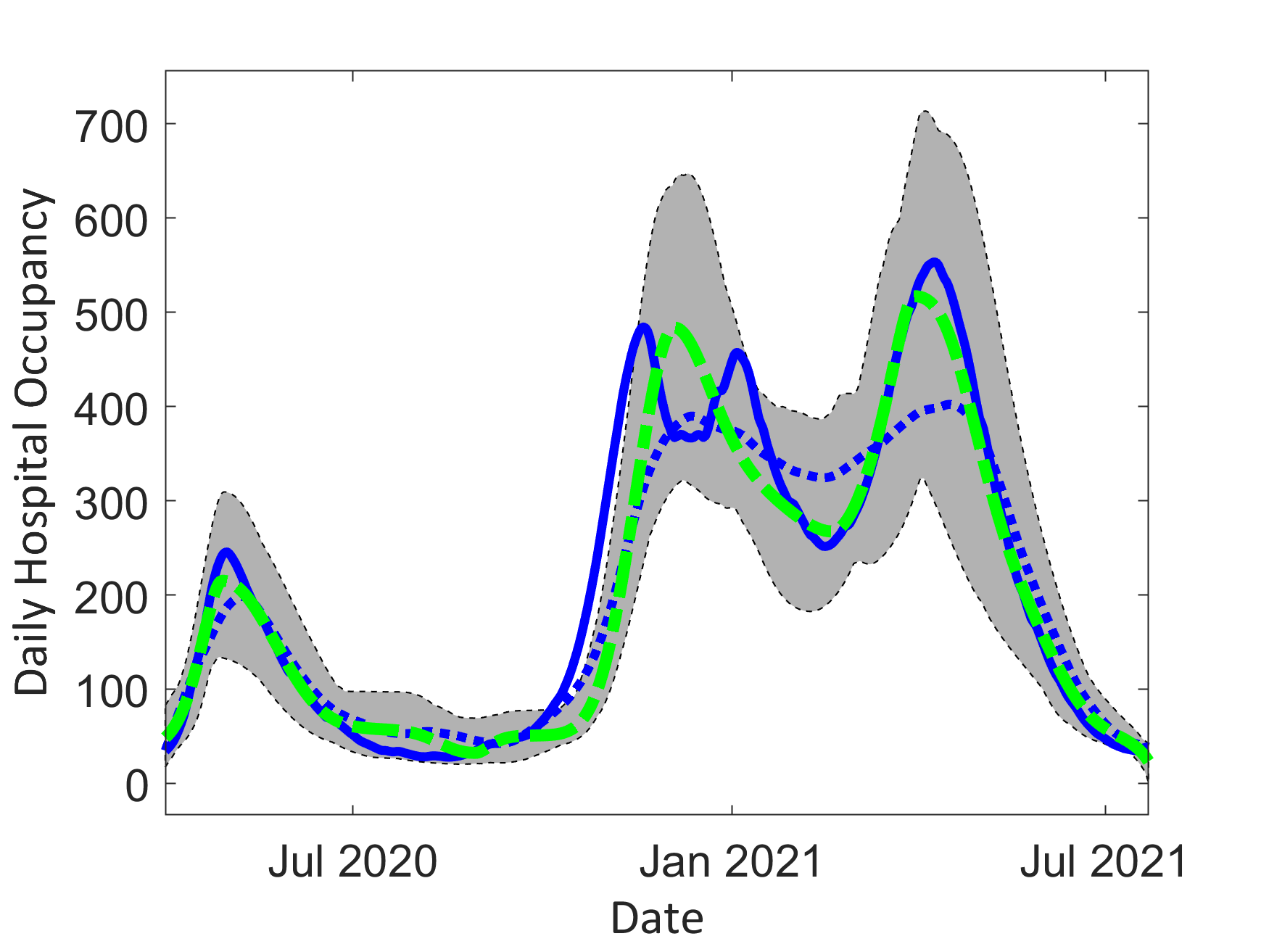}}
    \caption{(Daily Hospital Occupancy in Europe) PPD in (d) suggests that three peaks (1,4 and 6) are real features of the ground truth. In Plot (b), four peaks exist, and peak number 5 is a notable difference between $\hat g_\infty$ and $\hat g_0$. With $\lambda^* = 0.025$ from (d), the final estimation (in green) in Plot (g) is distinguishable with the others, $\hat g_\infty$ and $\hat g_0$. The 95\% bootstrapped confidence band seems wide due to the small sample size.}
    \label{fig: real2}
\end{figure*}
Fig. \ref{fig: real2} (a) and (b) display the functional data for infection rates and fully-aligned functions with their means, $\hat g_{\infty}$ and $\hat g_0$, respectively. As shown, $\hat g_{\infty}$ and $\hat g_0$ differ a lot in terms of the shapes and the heights of peaks. PPD surface plots in (e) and (f) show the gradual changes in $\hat g_{\lambda}$ with respect to $\lambda \in [0,0.2]$, and the barchart in (d) selects $\lambda^* = 0.025$ with three persistent peaks (1, 4 and 6). Our final estimate (in green) in panel (g) highlights this estimated shape. The 95\% pointwise bootstrapped confidence band (in gray) in (g) has a wide range, and this can be attributed to a small sample size $n=25$.
\\

\item {\bf Daily Infection Rate Curves}: The daily infection rate curves of COVID-19 can be used to study rate of spread of a variant or the effectiveness of a vaccine. We collected the infection rate curves in 25 European countries from OWID (\cite{owid}) for the period April 2020 to March 2022. 
Fig. \ref{fig: real3} (a) and (b) show the functional data and the fully-aligned functions, respectively. Although the functions in (b) are well aligned, some of the peaks in the first half of $I$ appear weak. Indeed, the PPD method rejects all peaks prior to July 2021 as insignificant and keeps only three later peaks. Plot (g) shows our final estimate (in green) and it shows a long flat region in the first half. This suggests that significant waves occurred during the outbreak of the delta variant in the Fall 2021. It also discovers a prominent peak around January 2022, which is not present in $\hat{g}_{\infty}$.
\\

\begin{figure*}[htbp]
    \centering
    \subfloat[]{\includegraphics[height = 1.2in]{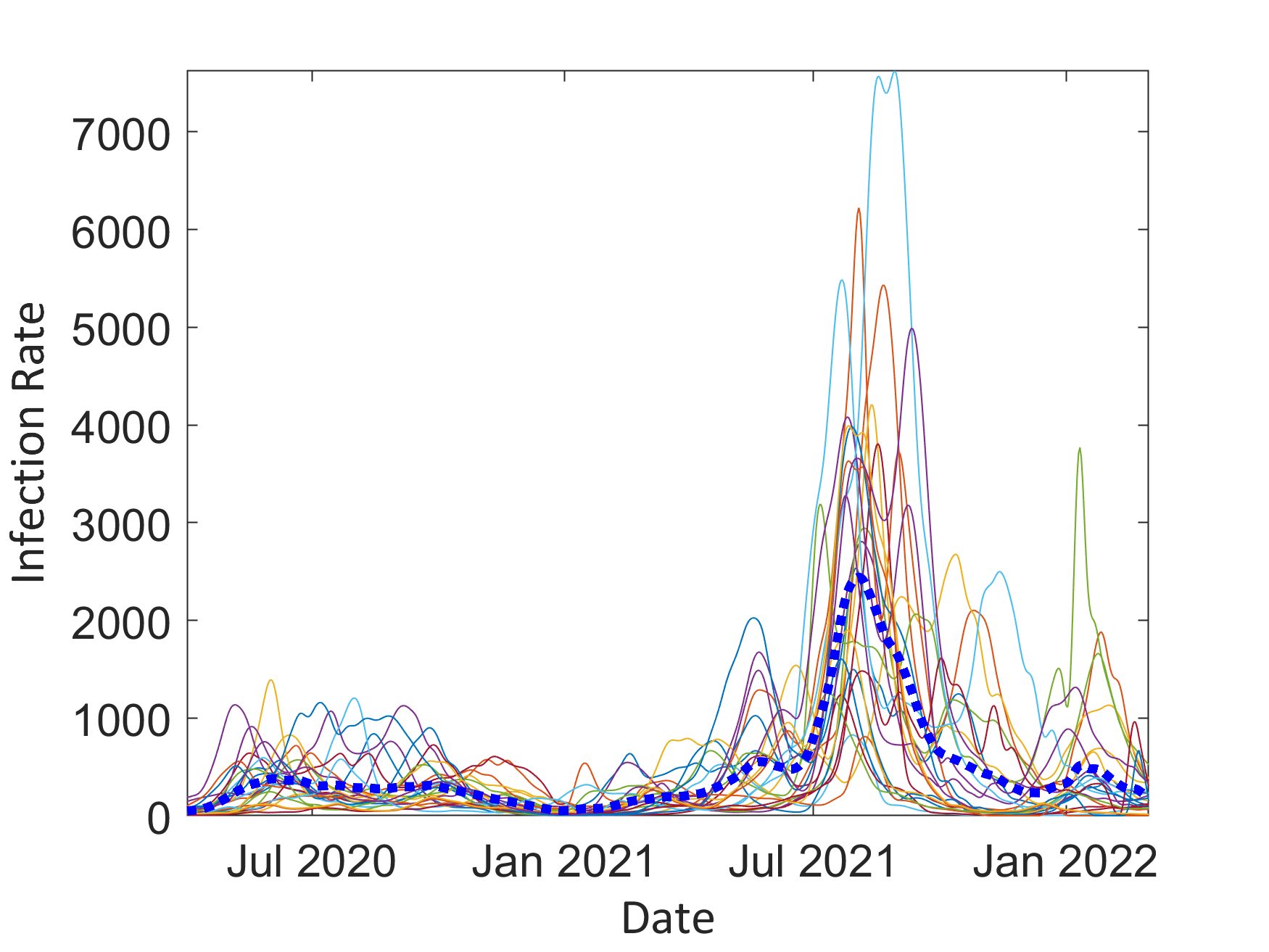}}
    \hspace{-0.2in}
    \subfloat[]{\includegraphics[height = 1.2in]{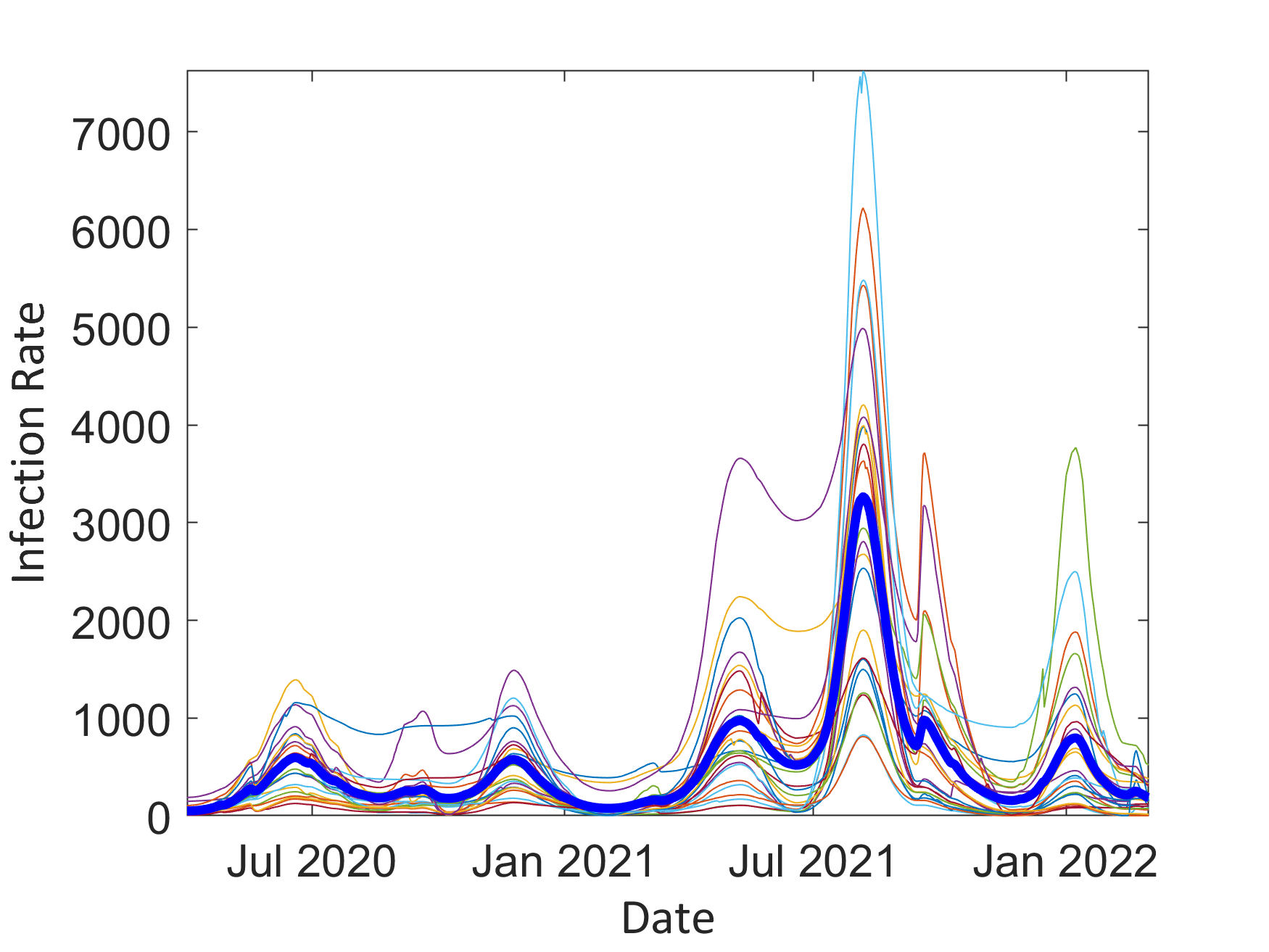}}
    \hspace{-0.2in}
    \subfloat[]{\includegraphics[height = 1.2in]{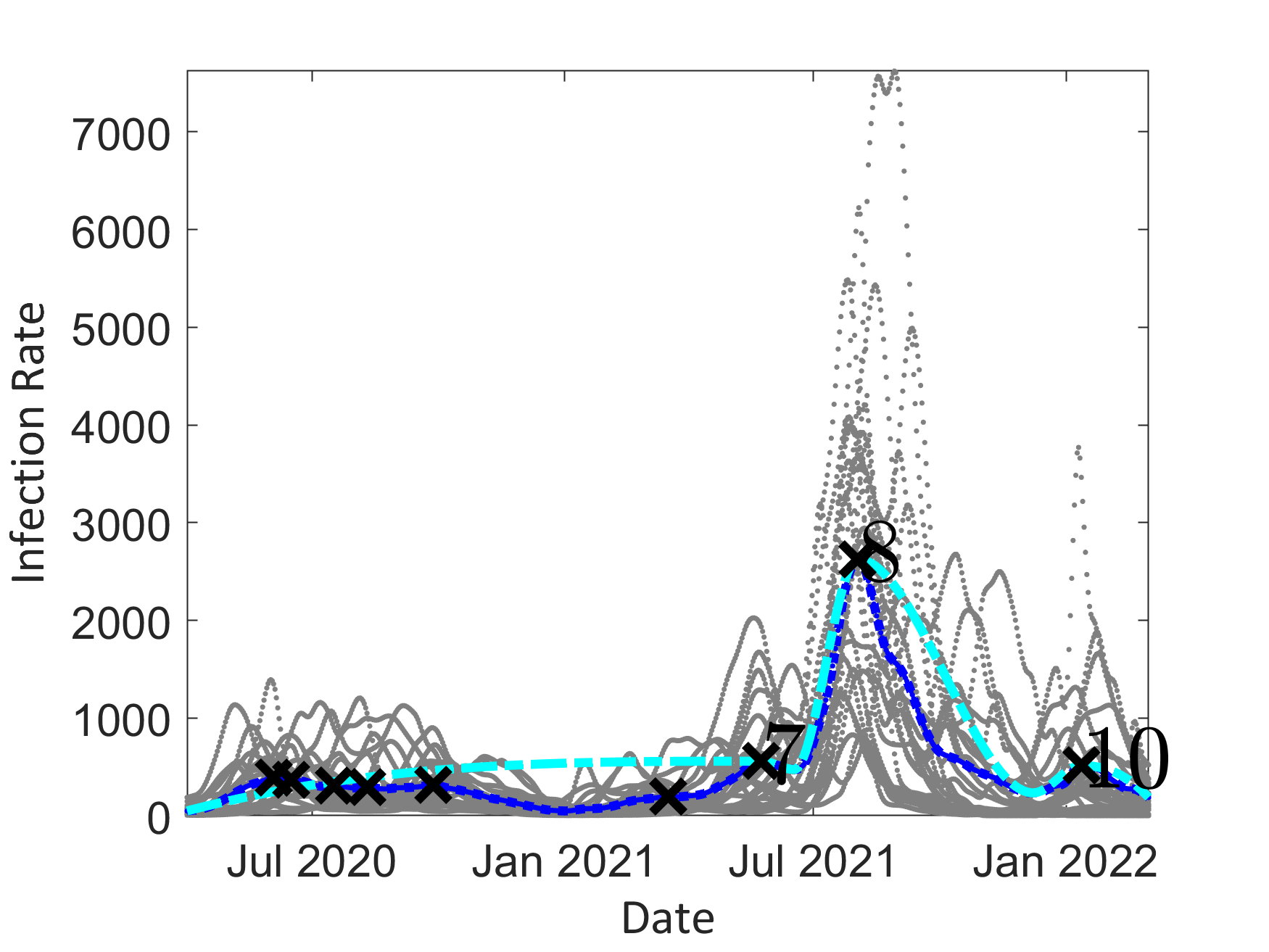}}
    \includegraphics[height = 1in]{fig/legend-rd-1.png}
    \hspace{0in}\\
    \hspace{-0.2in}
    \subfloat[]{\includegraphics[height = 1.2in]{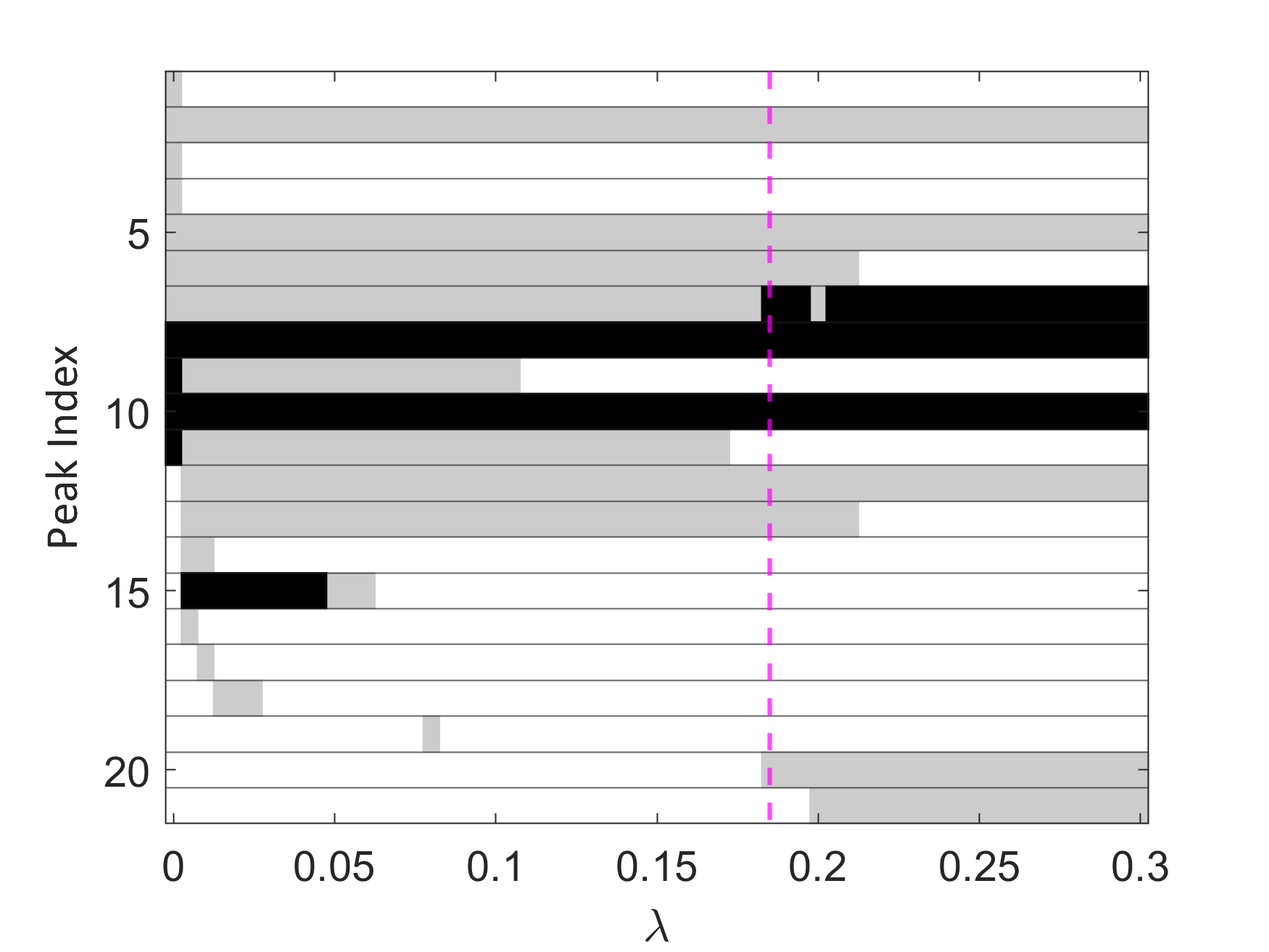}}
    \hspace{-0.2in}
    \subfloat[]{\includegraphics[height = 1.2in]{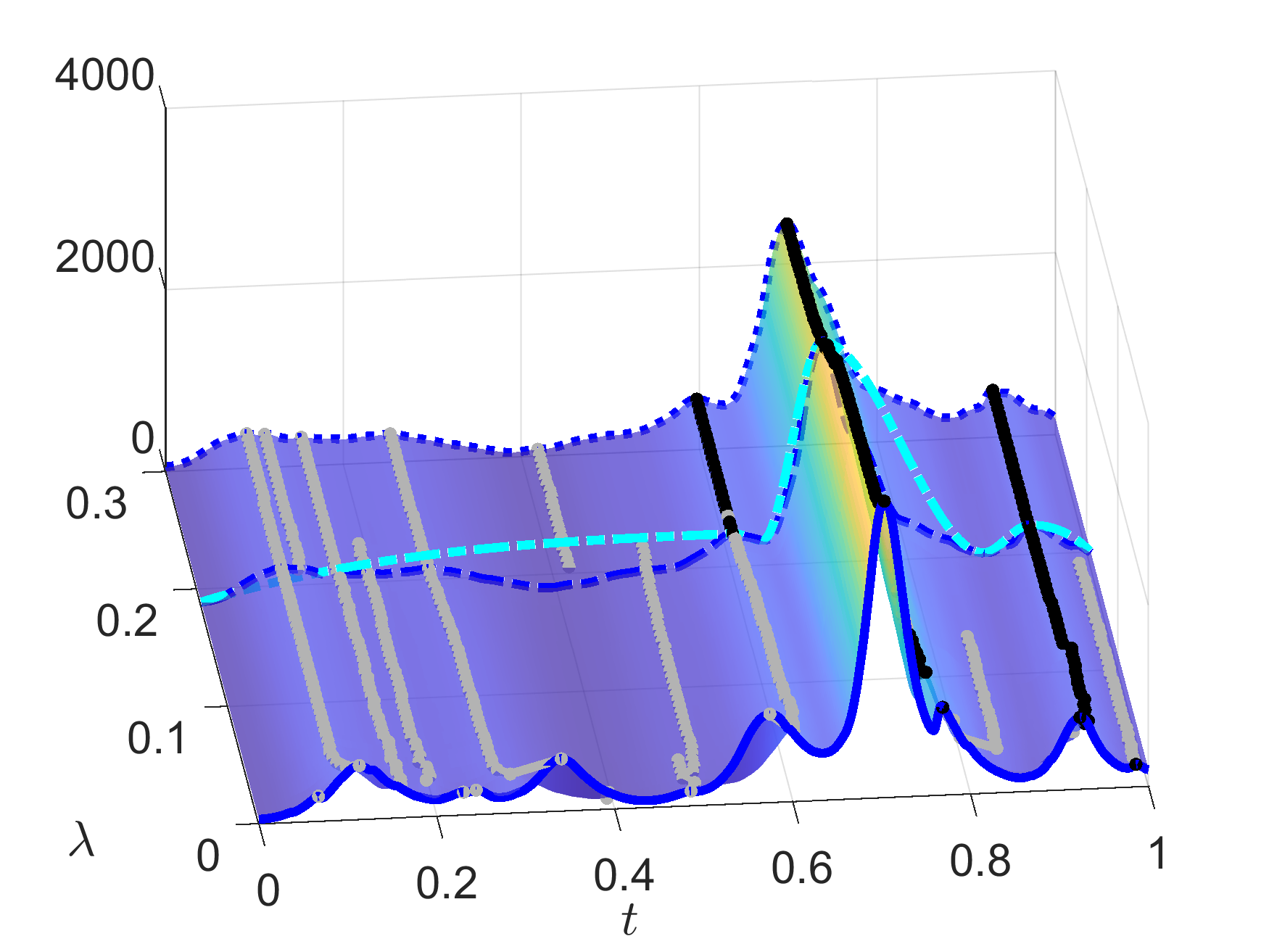}}
    \hspace{-0.2in}
    \subfloat[]{\includegraphics[height = 1.2in]{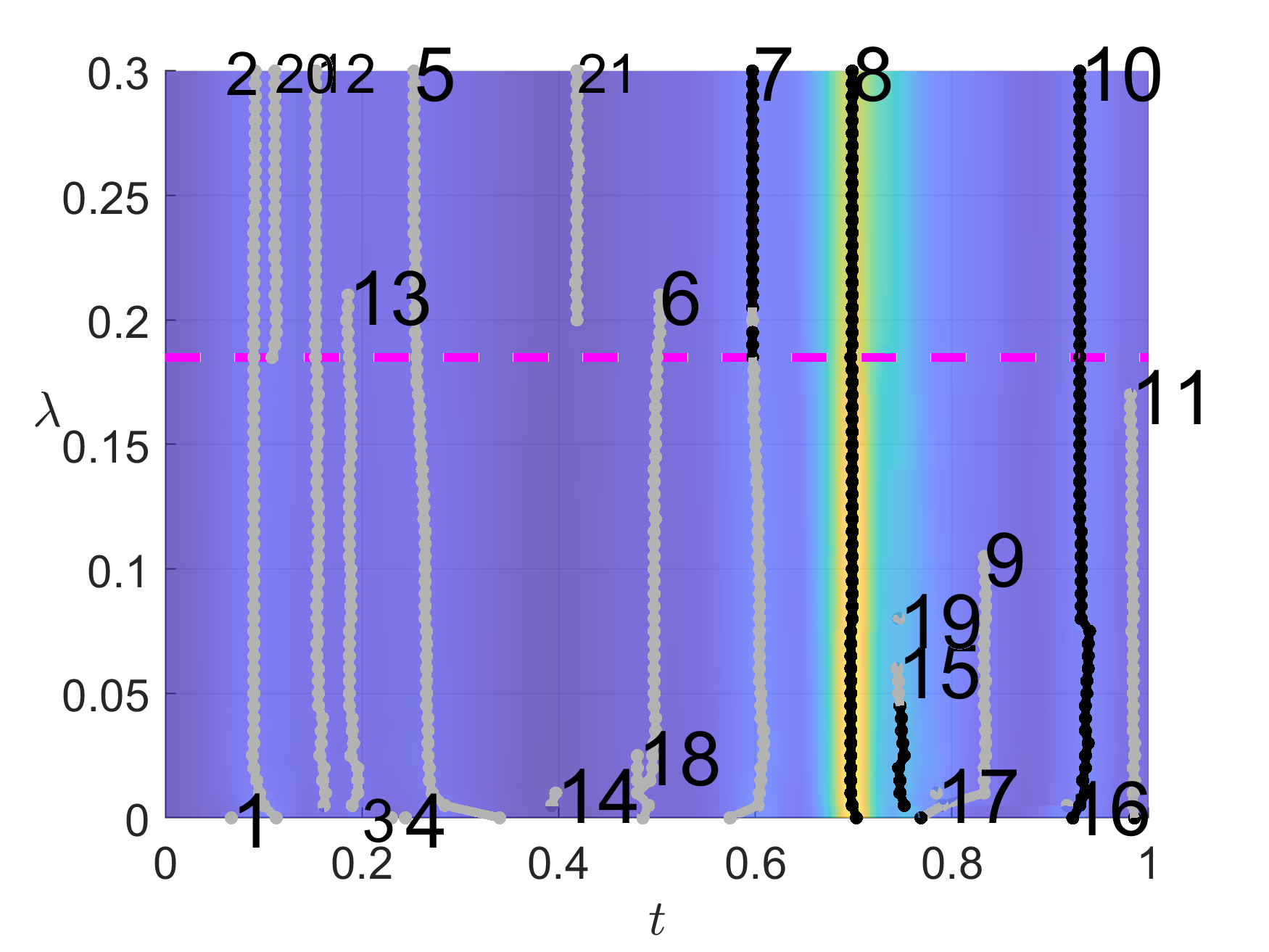}}
    \hspace{-0.2in}
    \subfloat[]{\includegraphics[height = 1.2in]{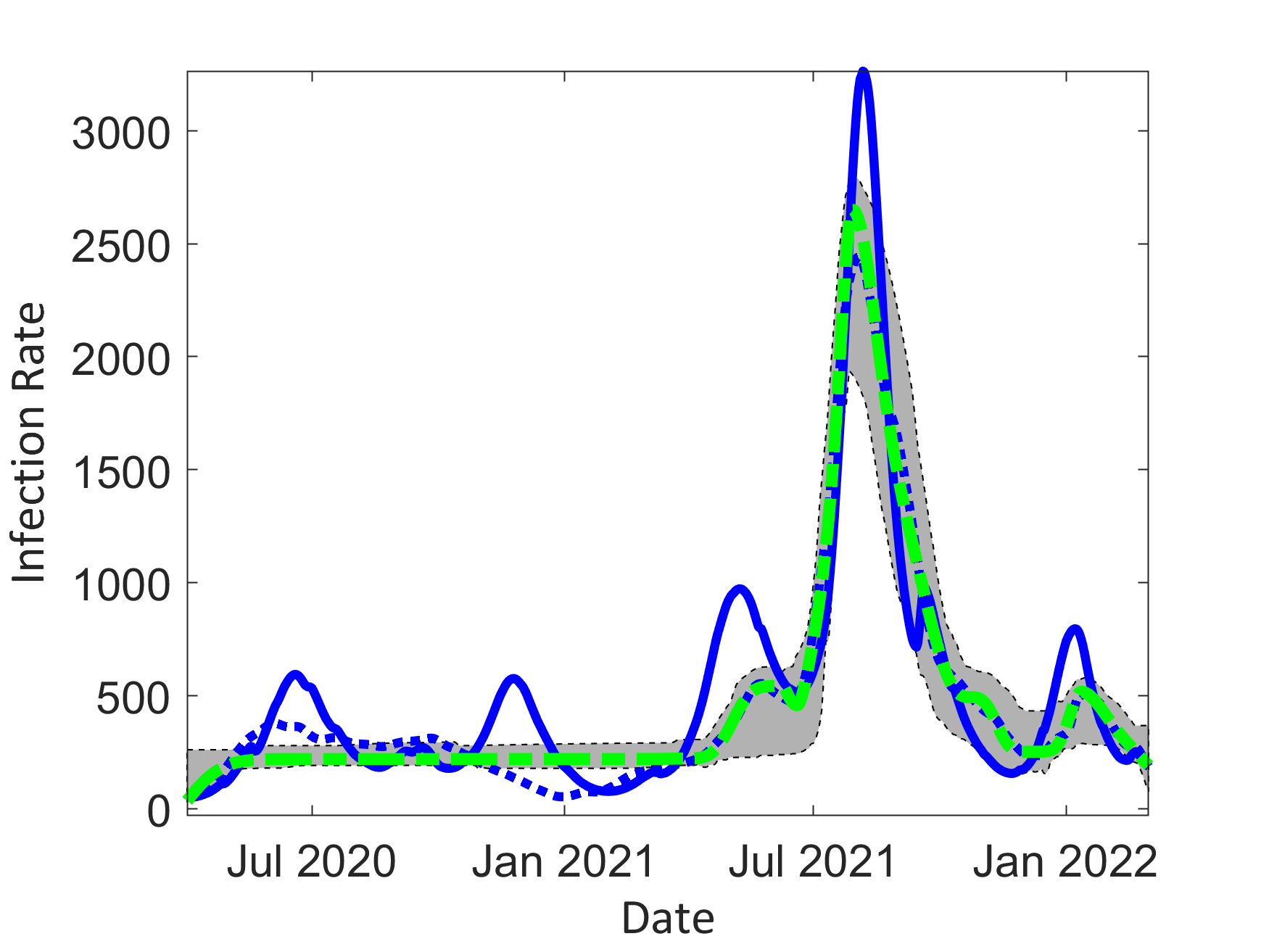}}
    \caption{(Daily Infection Rate in Europe) PPD in (d) labels three peaks (7, 8, and 10) as significant. Our final estimation (green curve in (g)) removes all small peaks in the early stages and select the three most persistent waves.}
    \label{fig: real3}
\end{figure*}

\item {\bf Daily Death Rate Curves}: We also studied the daily death rates of COVID in these 25 European countries for April 2020 to March 2022. 
The standard estimates $\hat g_{\infty}$ (dotted blue) and $\hat g_0$ (solid blue) in Fig. \ref{fig: real4} (a) and (b) differ considerably: $\hat g_0$ has several prominent peaks whereas $\hat g_{\infty}$ has fewer. Specifically, the peak around January 2021 in $\hat {g}_0$ seems artificial as no country has a peak in the original data, and there is no reason to align the peak at the specific time. The PPDs in (d), (e), and (f) estimate $\lambda^* = 0.025$ and the peak labeled 4 disappears as $\lambda$ increases. As shown in Plot (g), the final estimation, $\hat g$ (in green), is different from both $\hat g_\infty$ and $\hat g_0$. Although the partial alignment of functions is not too different from the original data, our estimate $\hat g$ picks up three distinct peaks in the early stage of COVID before January 2021.

\begin{figure*}[htbp]
    \centering
    \subfloat[]{\includegraphics[height = 1.2in]{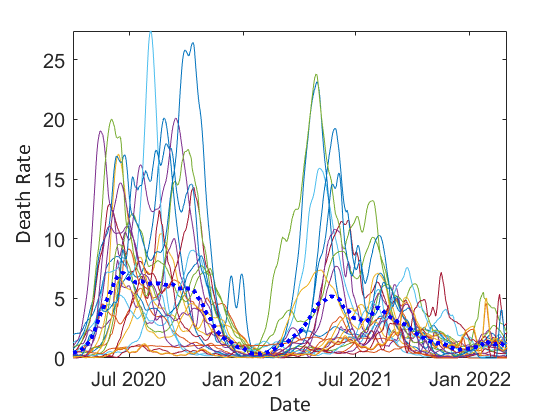}}
    \hspace{-0.2in}
    \subfloat[]{\includegraphics[height = 1.2in]{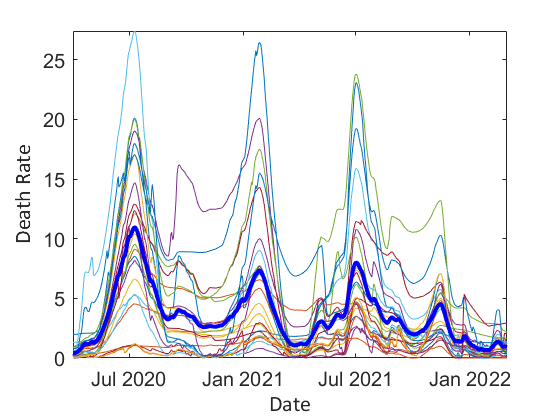}}
    \hspace{-0.2in}
    \subfloat[]{\includegraphics[height = 1.2in]{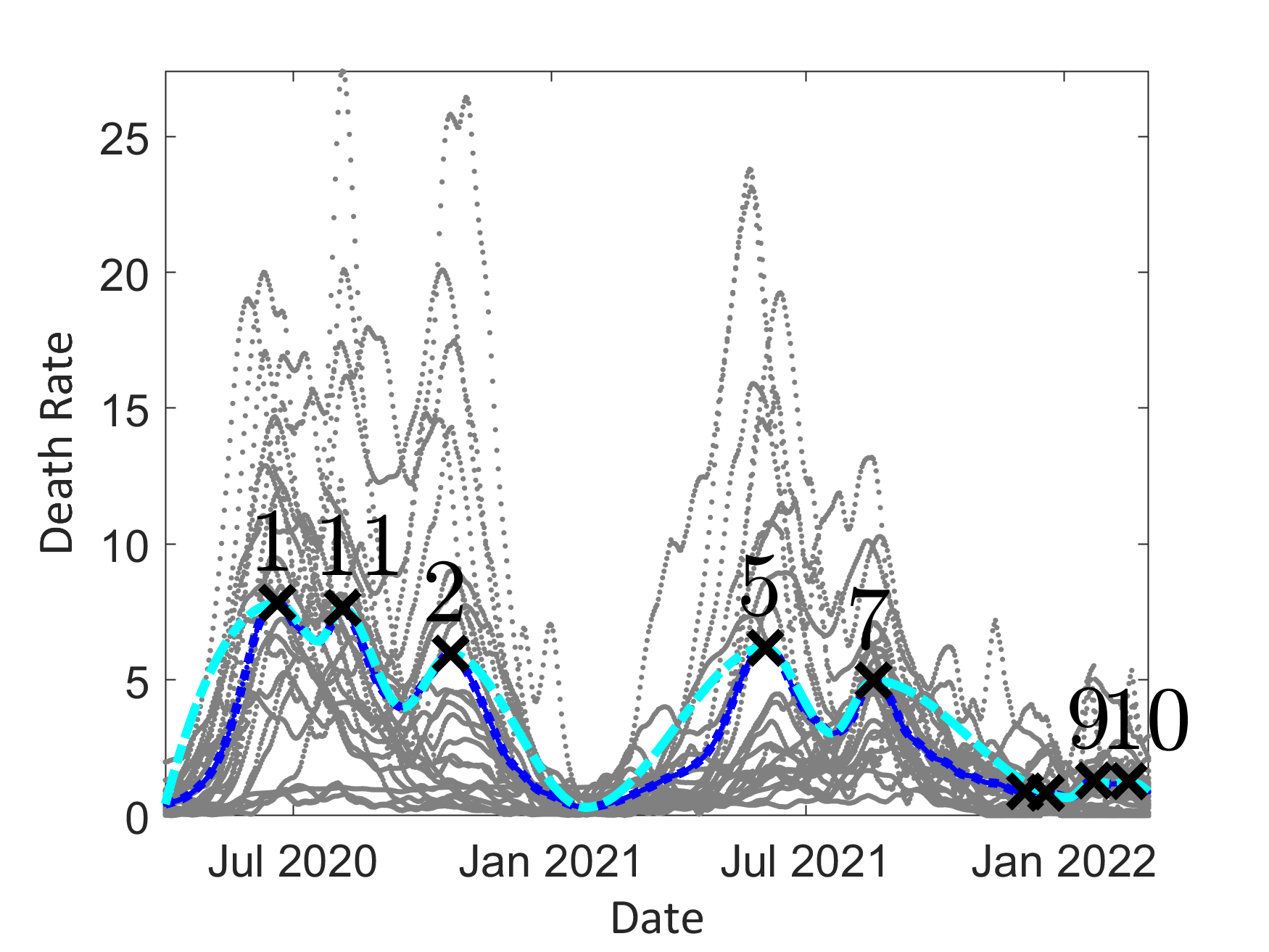}}
    \includegraphics[height = 1in]{fig/legend-rd-1.png}
    \hspace{0in}\\
    \hspace{-0.2in}
    \subfloat[]{\includegraphics[height = 1.2in]{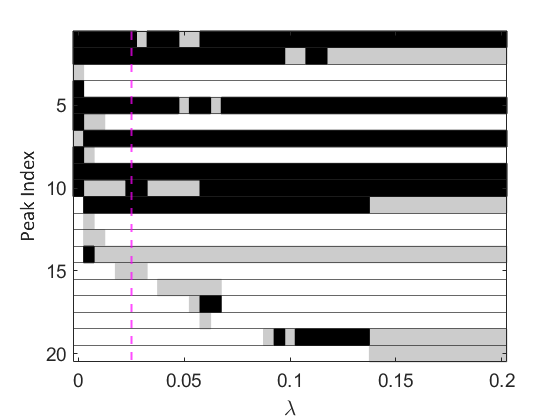}}
    \hspace{-0.2in}
    \subfloat[]{\includegraphics[height = 1.2in]{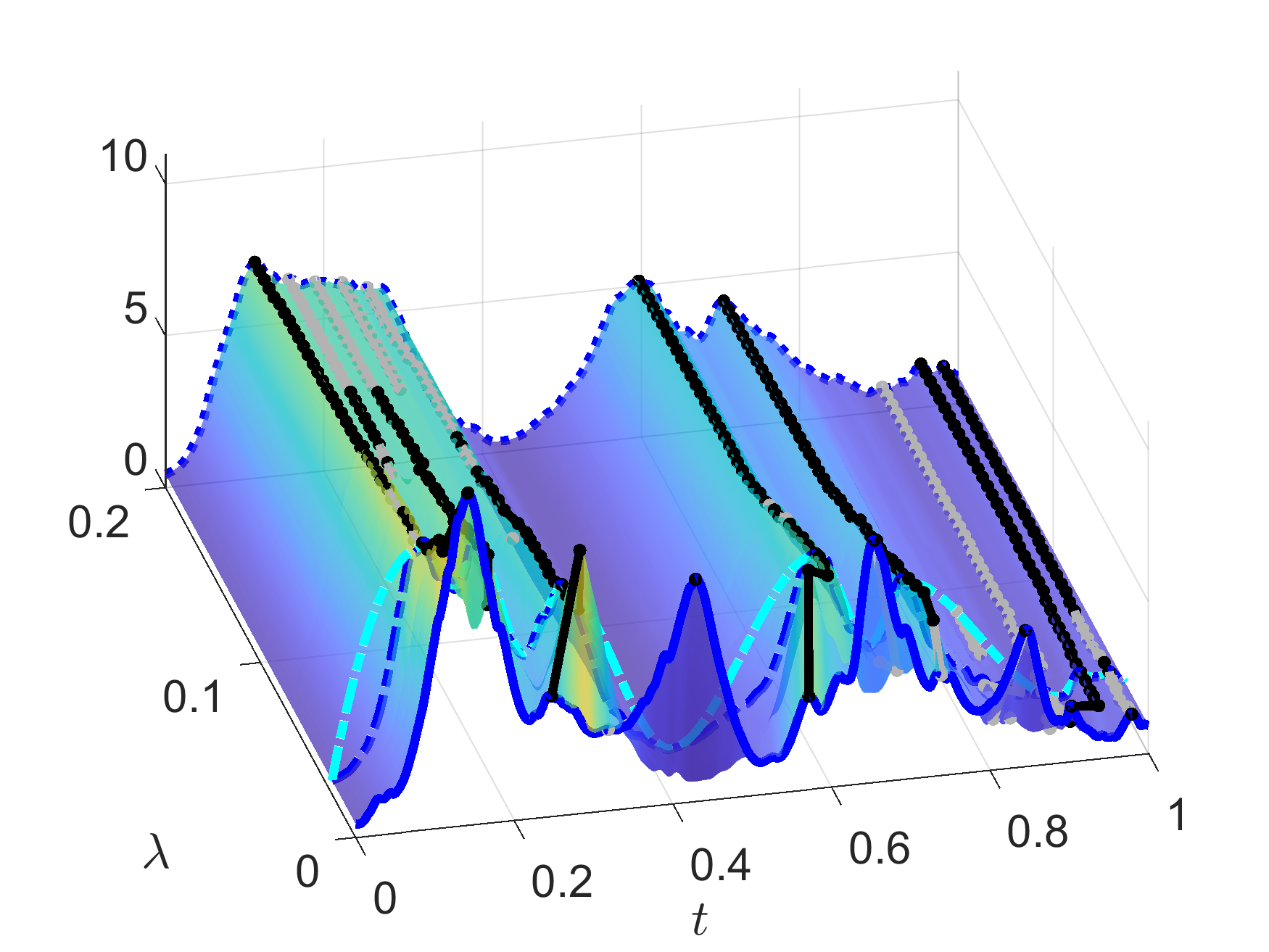}}
    \hspace{-0.2in}
    \subfloat[]{\includegraphics[height = 1.2in]{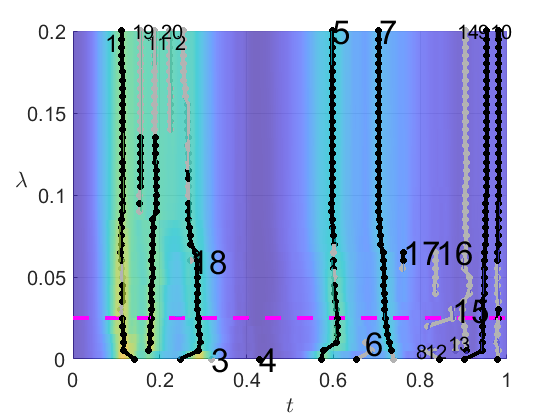}}
    \hspace{-0.2in}
    \subfloat[]{\includegraphics[height = 1.2in]{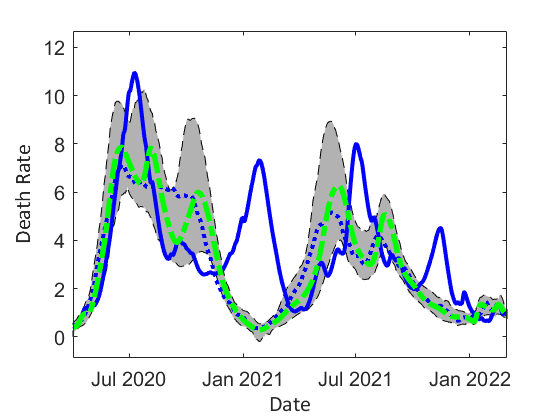}}
    \caption{(Death Rate in Europe) PPD in (d) suggests that seven peaks (1, 2, 5, 7, 9, 10, and 11) are significant. Our final estimation (green curve in (g)) indicates five major peaks and two minor ones with a big valley in the middle.}
    \label{fig: real4}
\end{figure*}


\end{itemize}

\noindent {\bf Household Electricity Consumption Data}\label{sec: electricity consumption}
\begin{figure*}[htbp]
    \centering
    \subfloat[]{\includegraphics[height = 1.2in]{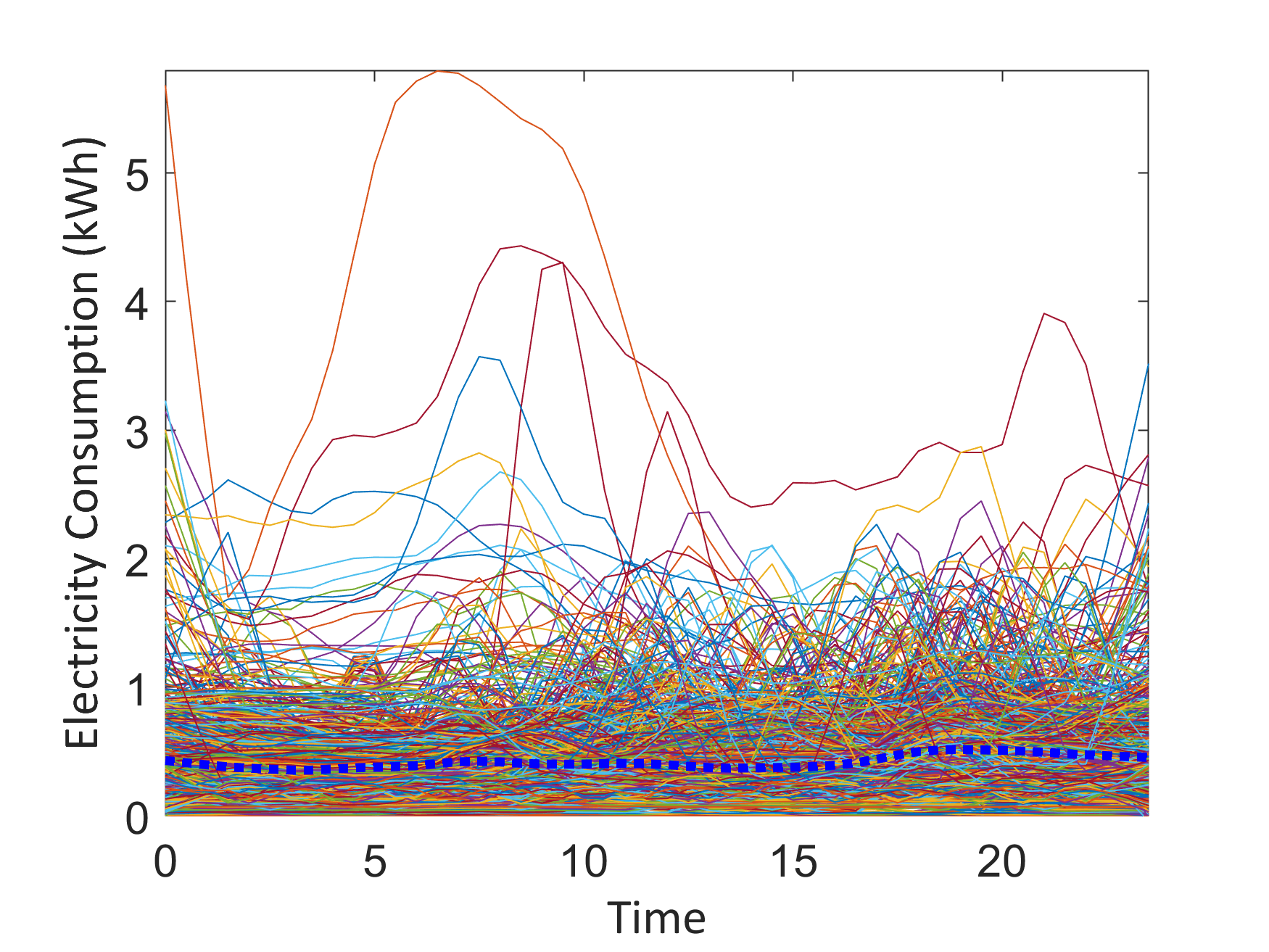}}
    \hspace{-0.2in}
    \subfloat[]{\includegraphics[height = 1.2in]{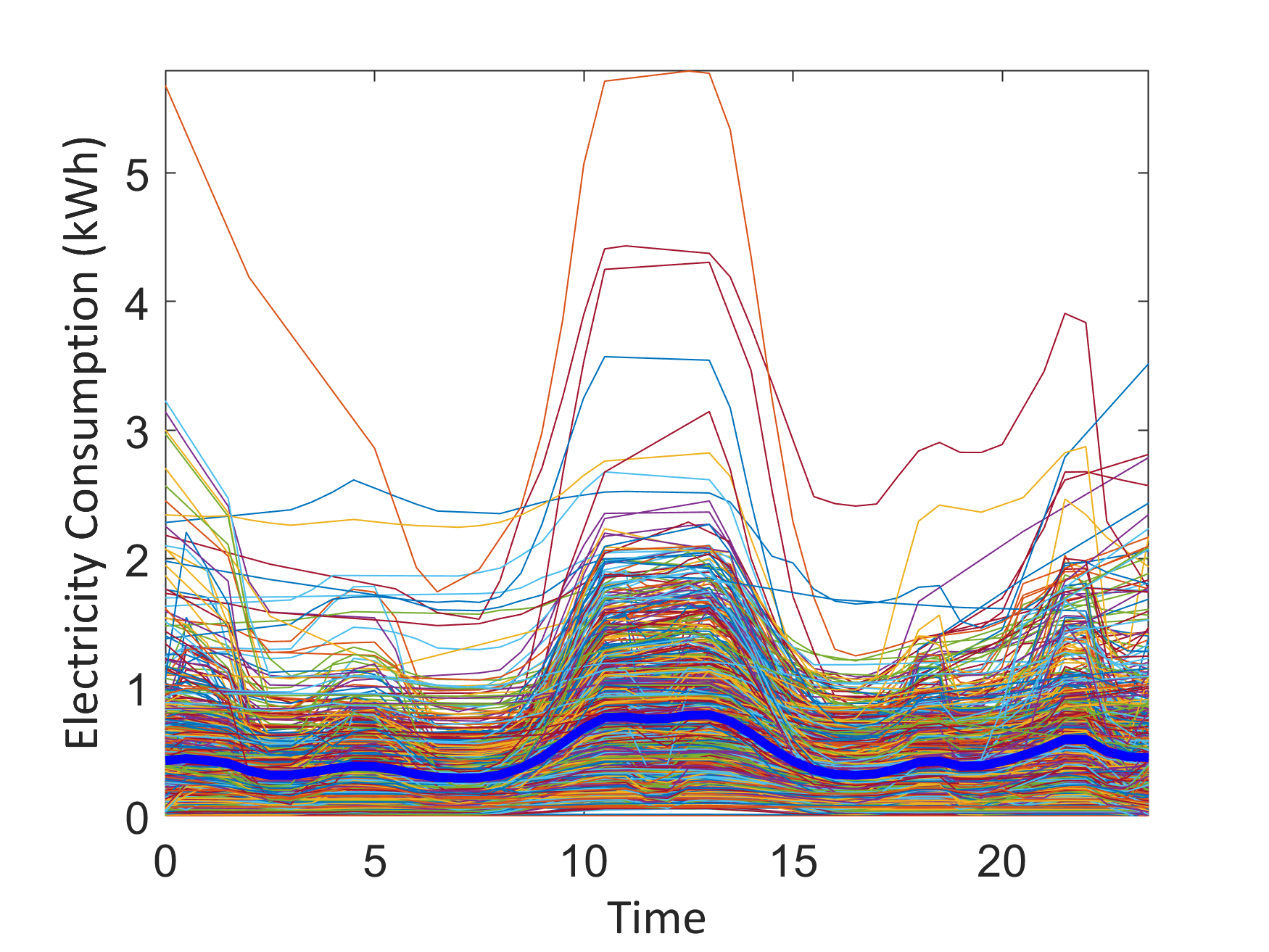}}
    \hspace{-0.2in}
    \subfloat[]{\includegraphics[height = 1.2in]{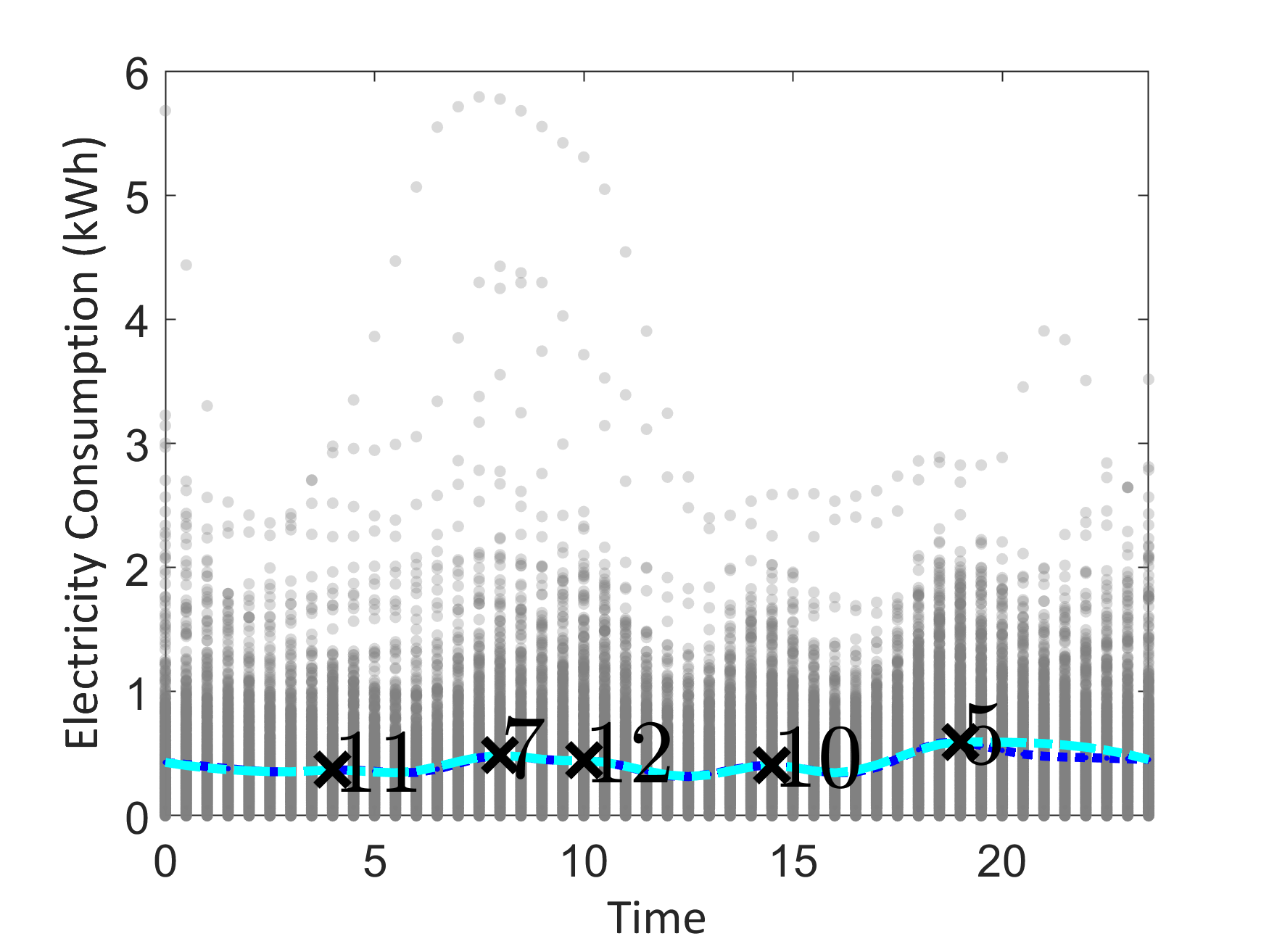}}
    \includegraphics[height = 1in]{fig/legend-rd-1.png}
    \hspace{0in}\\
    \hspace{-0.2in}
    \subfloat[]{\includegraphics[height = 1.2in]{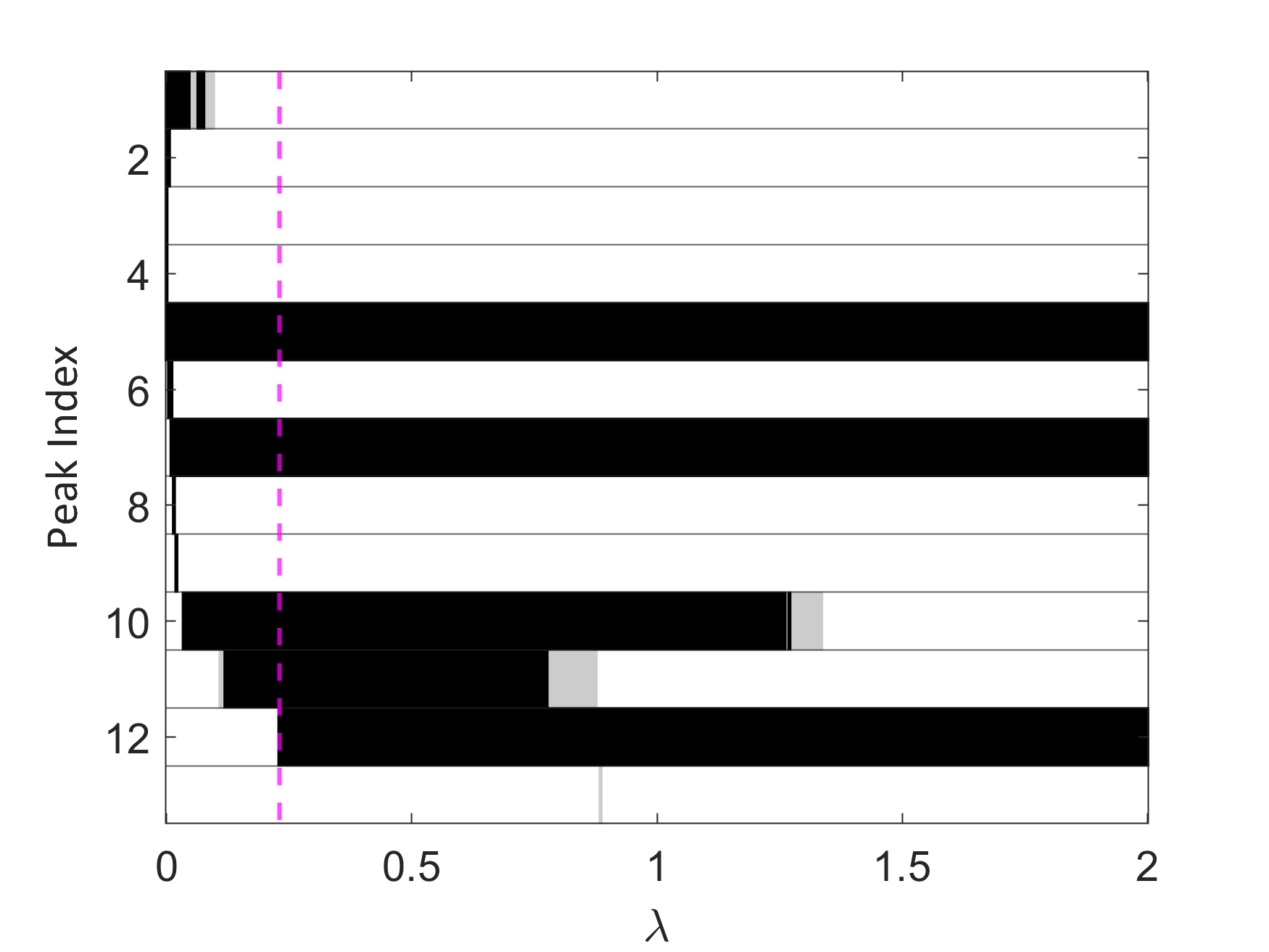}}
    \hspace{-0.2in}
    \subfloat[]{\includegraphics[height = 1.2in]{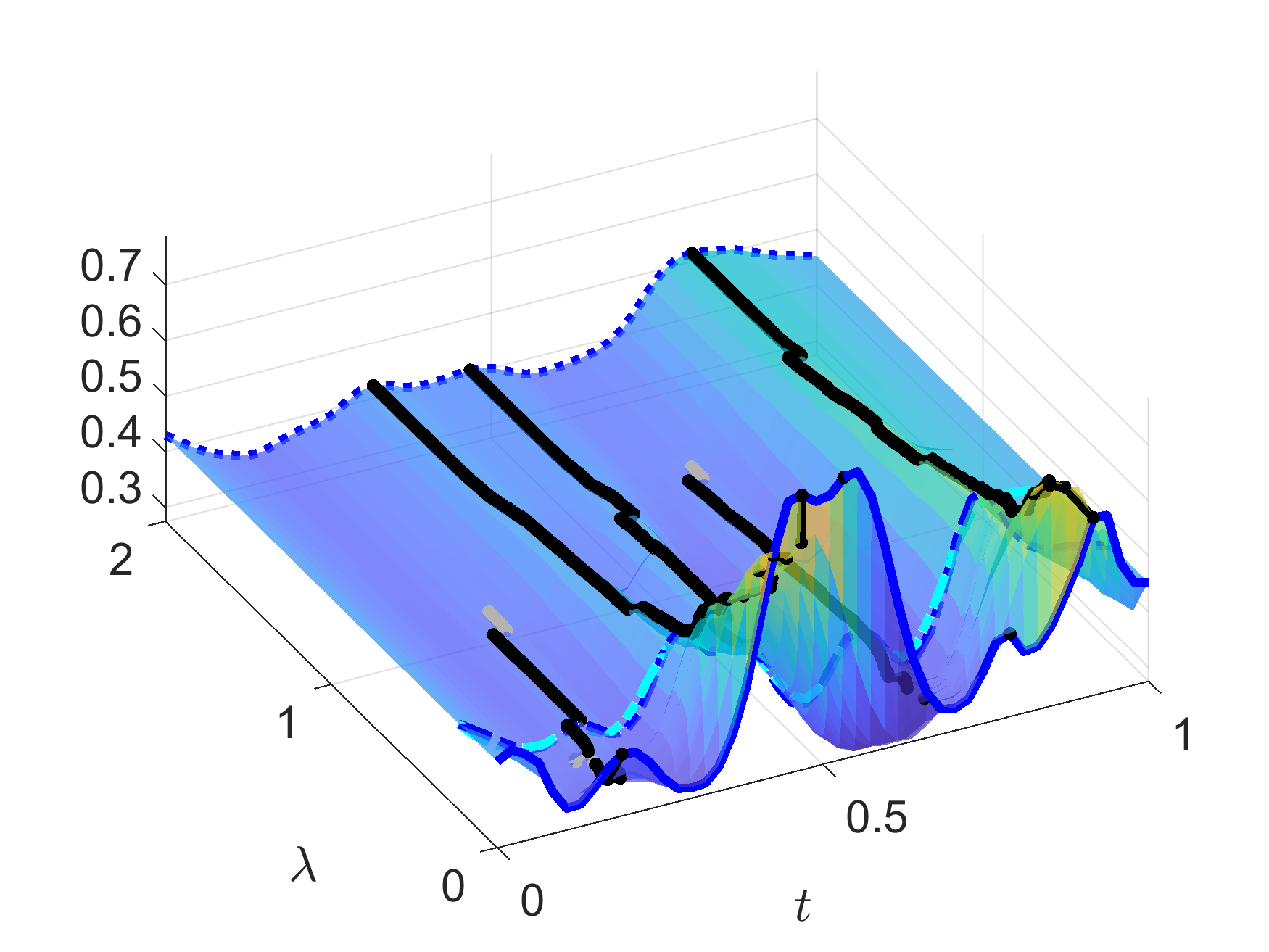}}
    \hspace{-0.2in}
    \subfloat[]{\includegraphics[height = 1.2in]{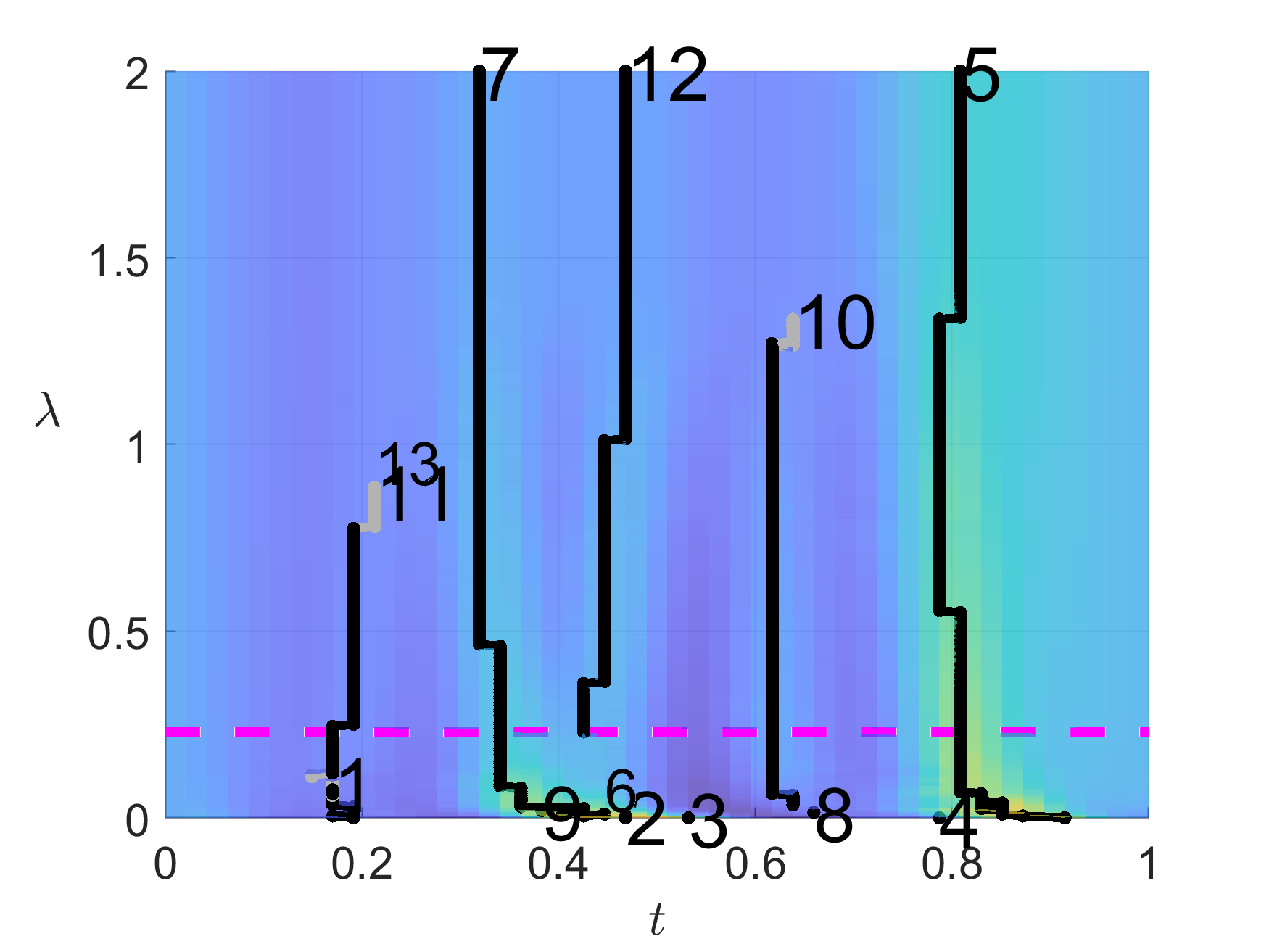}}
    \hspace{-0.2in}
    \subfloat[]{\includegraphics[height = 1.2in]{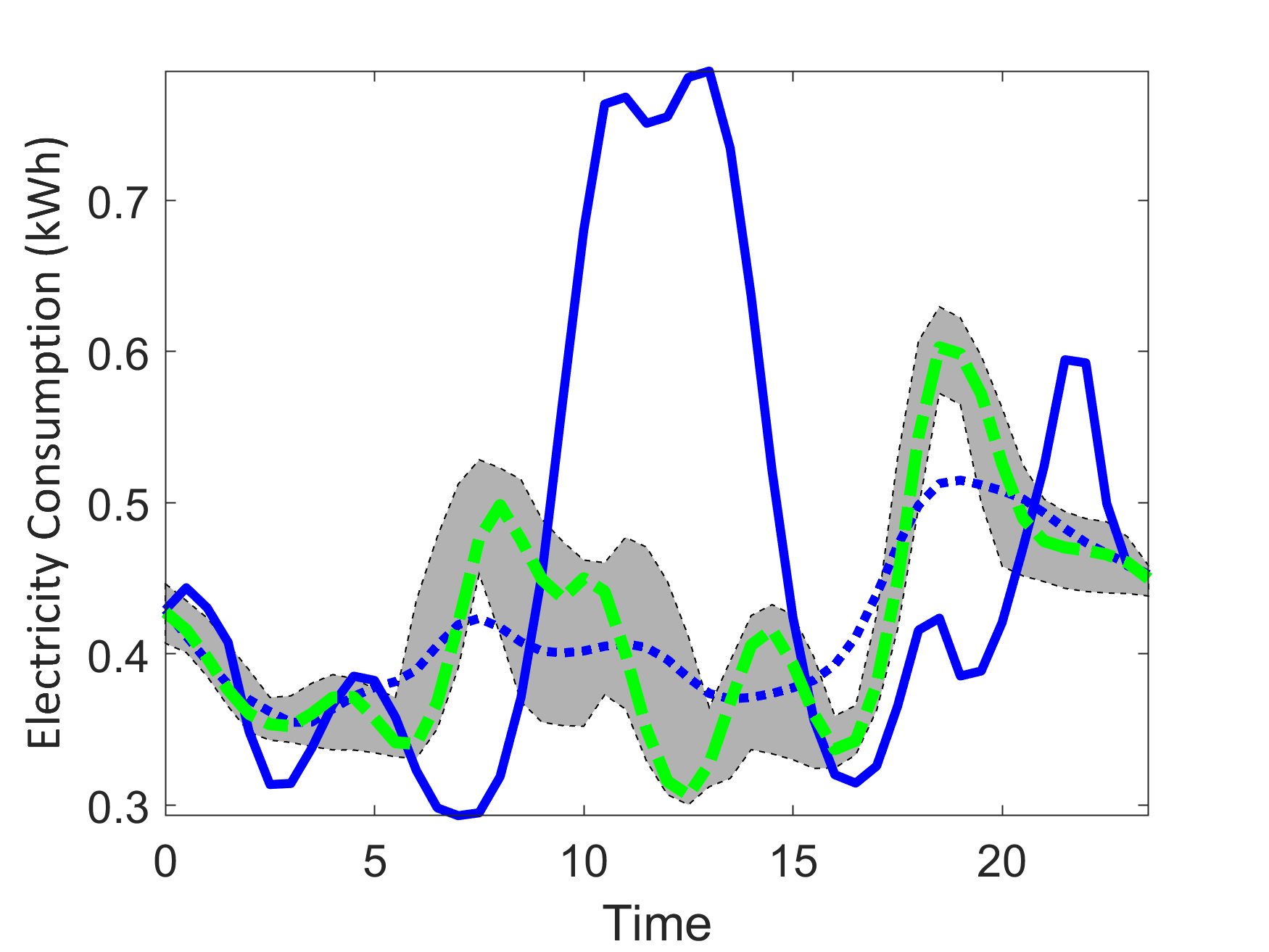}}
    \caption{(Electric Consumption in Tallahassee, FL,  in January 2015) Plots (a) and (b) show household data and fully-aligned functions, respectively. PPD in (d) provides $\lambda^* = 0.235$, and five significant peaks (5, 7 10, 11 12). The final estimation, $\hat g$, (in green)in (g), is distinctive from the others, $\hat g_\infty$, and $\hat g_0$.}
    \label{fig: real5(1)}
\end{figure*}

\begin{figure*}[htbp]
    \centering
    \subfloat[]{\includegraphics[height = 1.2in]{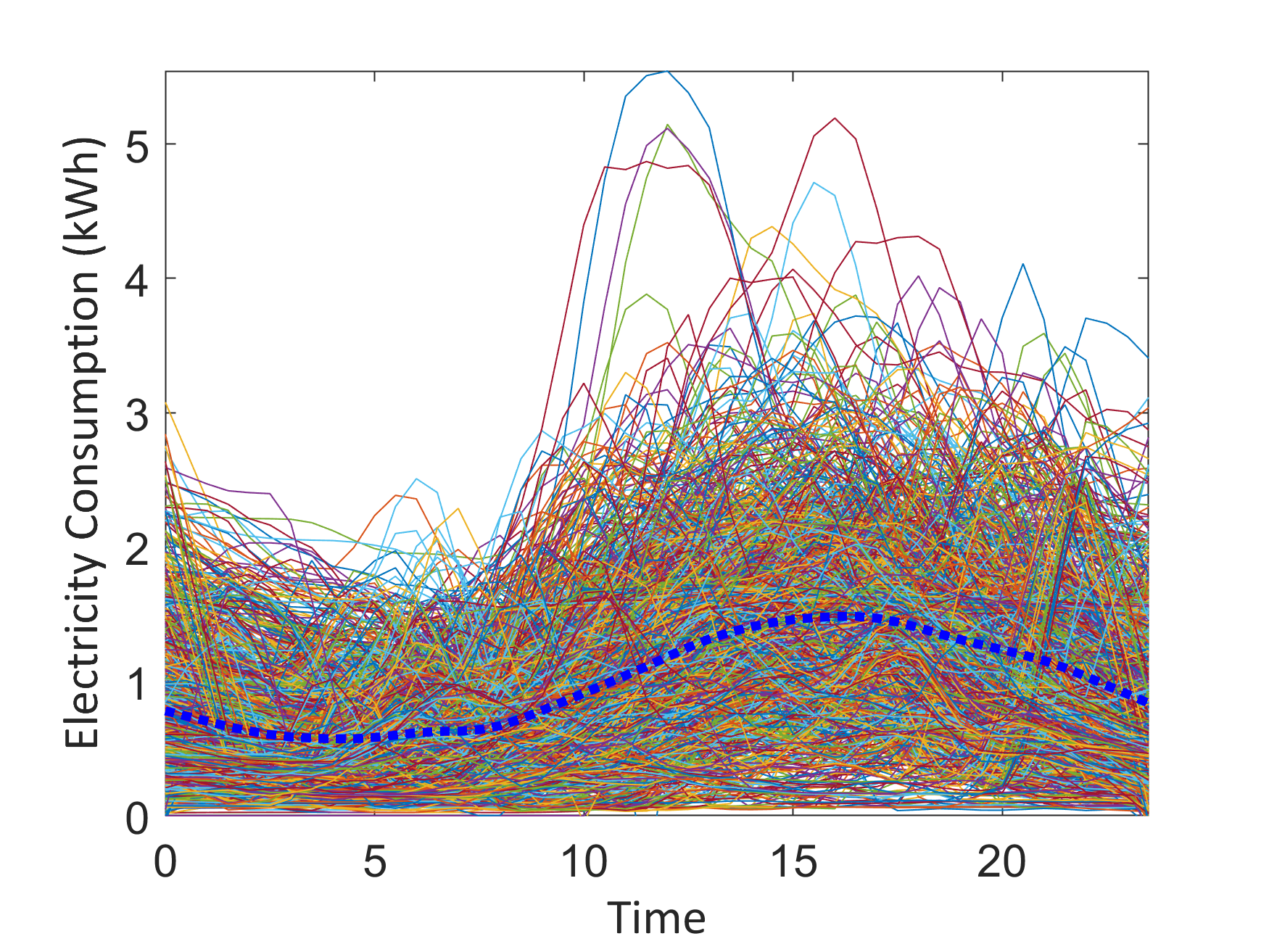}}
    \hspace{-0.2in}
    \subfloat[]{\includegraphics[height = 1.2in]{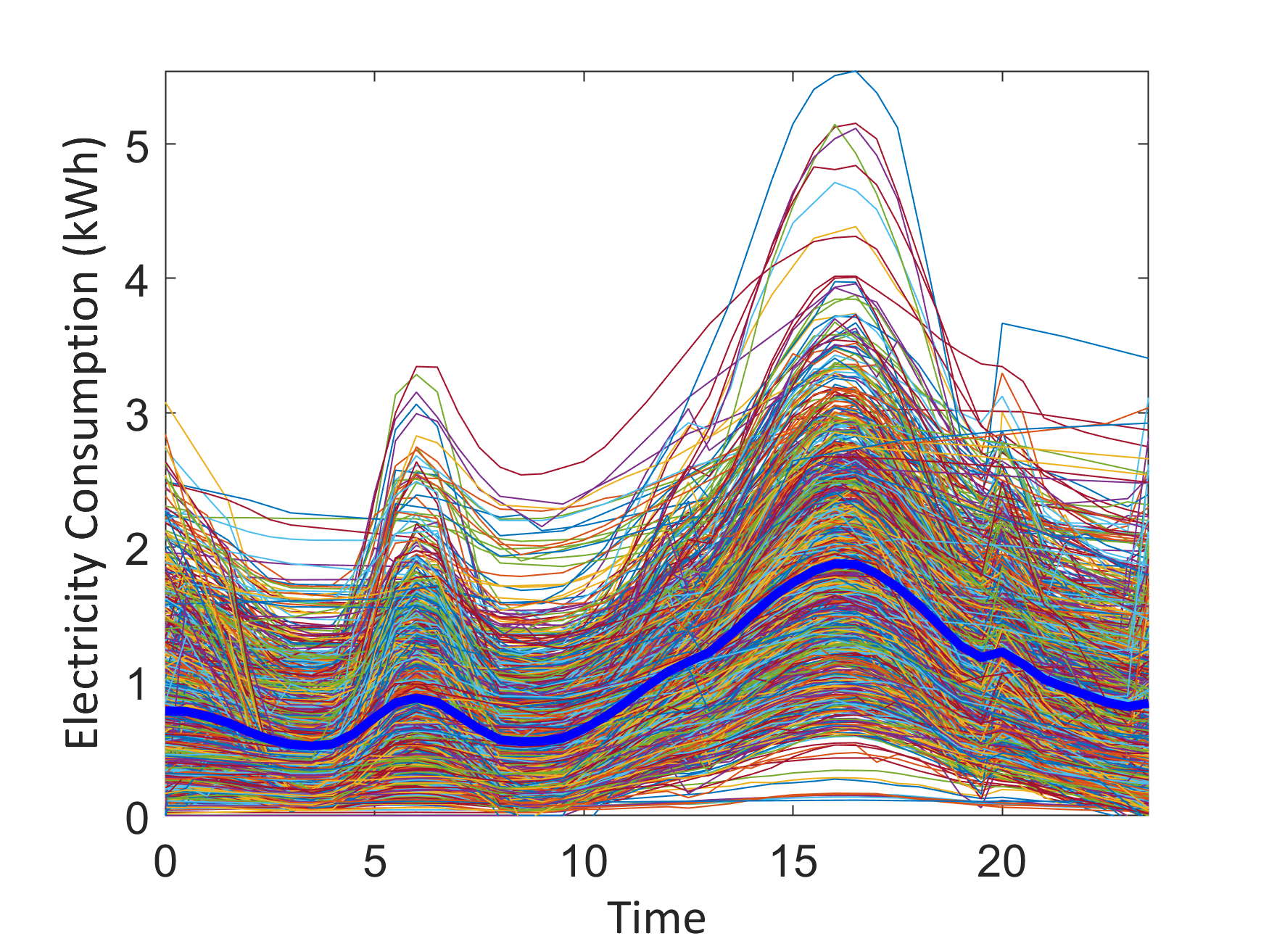}}
    \hspace{-0.2in}
    \subfloat[]{\includegraphics[height = 1.2in]{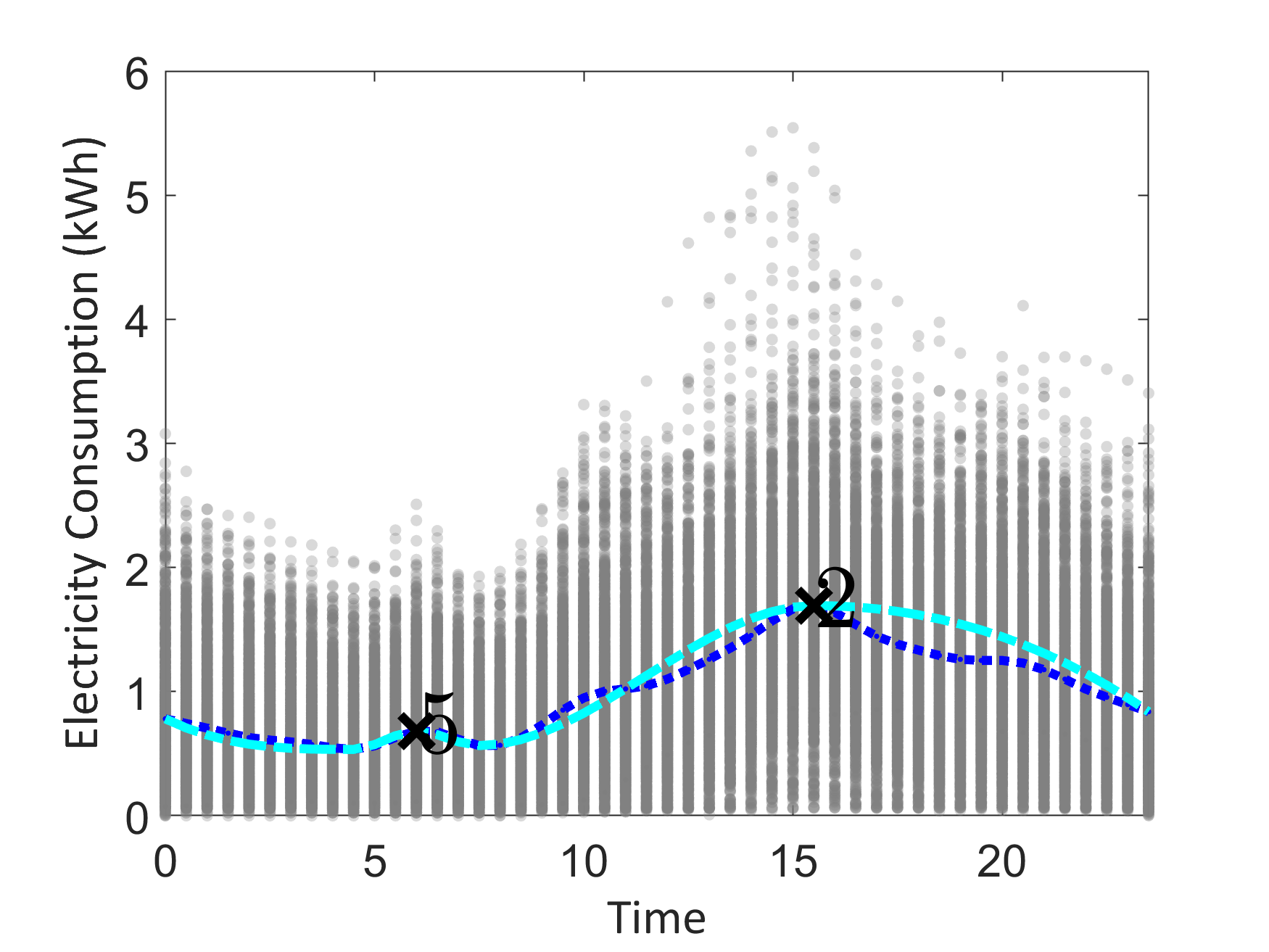}}
    \includegraphics[height = 1in]{fig/legend-rd-1.png}
    \hspace{0in}\\
    \hspace{-0.2in}
    \subfloat[]{\includegraphics[height = 1.2in]{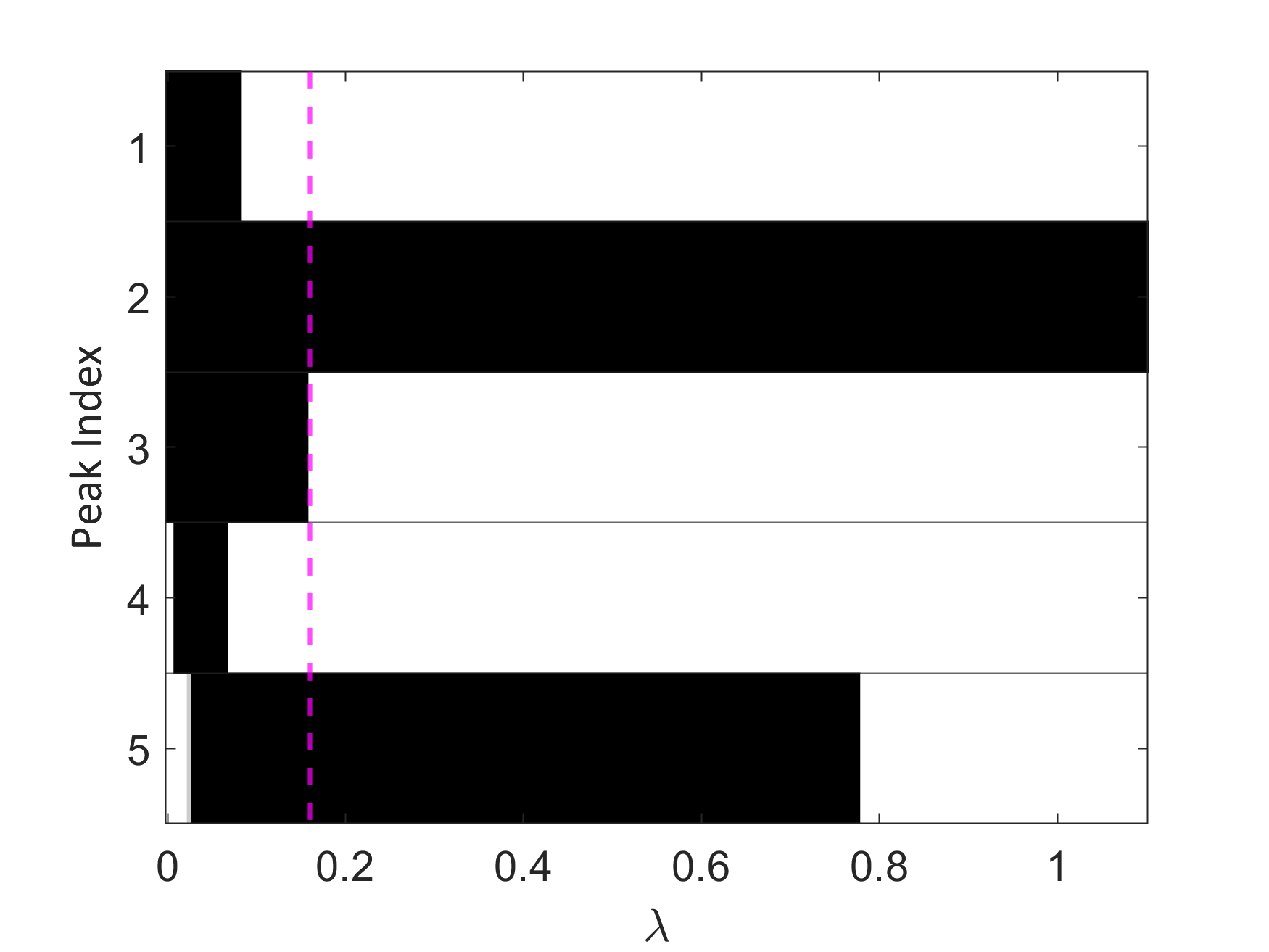}}
    \hspace{-0.2in}
    \subfloat[]{\includegraphics[height = 1.2in]{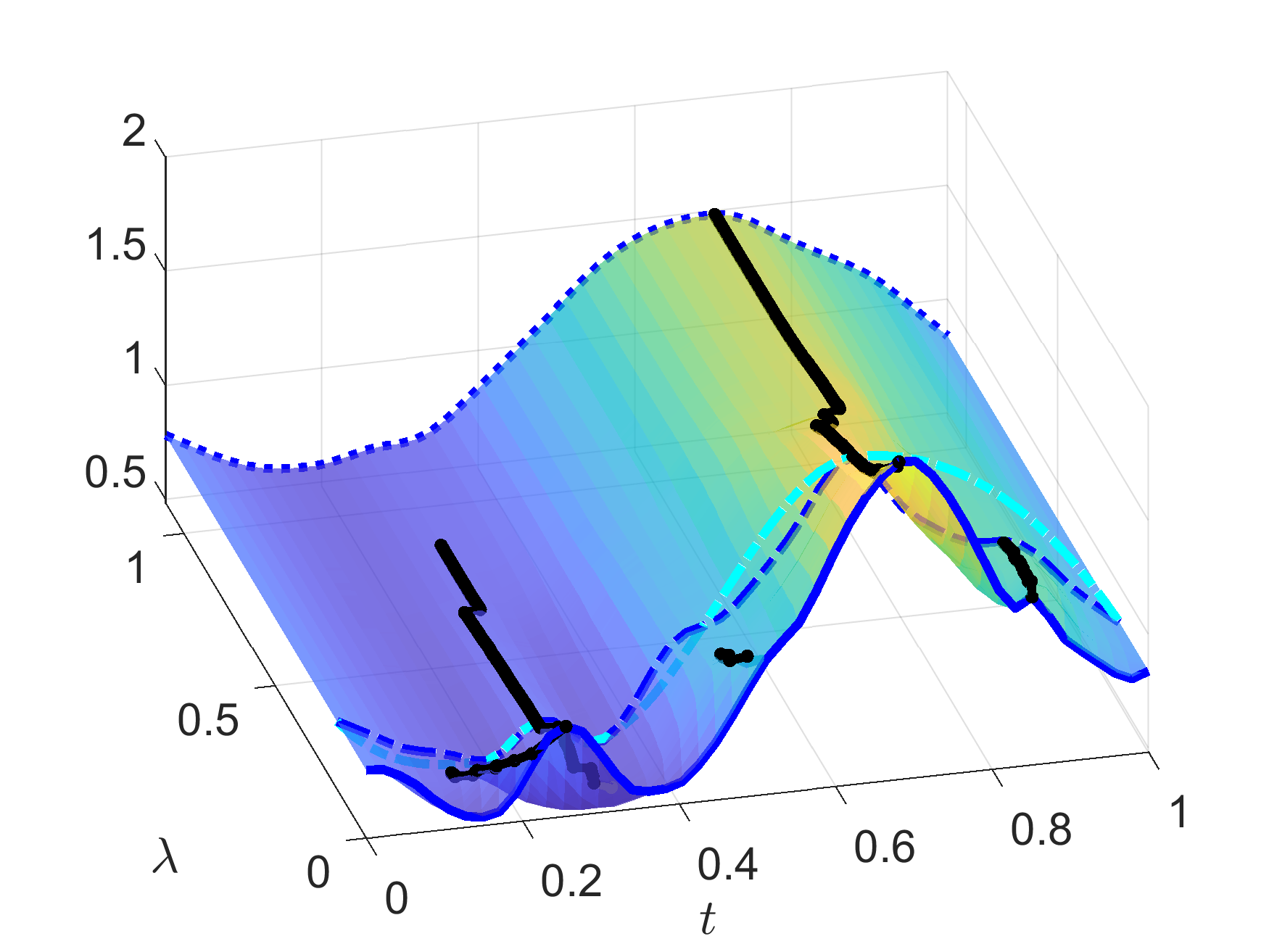}}
    \hspace{-0.2in}
    \subfloat[]{\includegraphics[height = 1.2in]{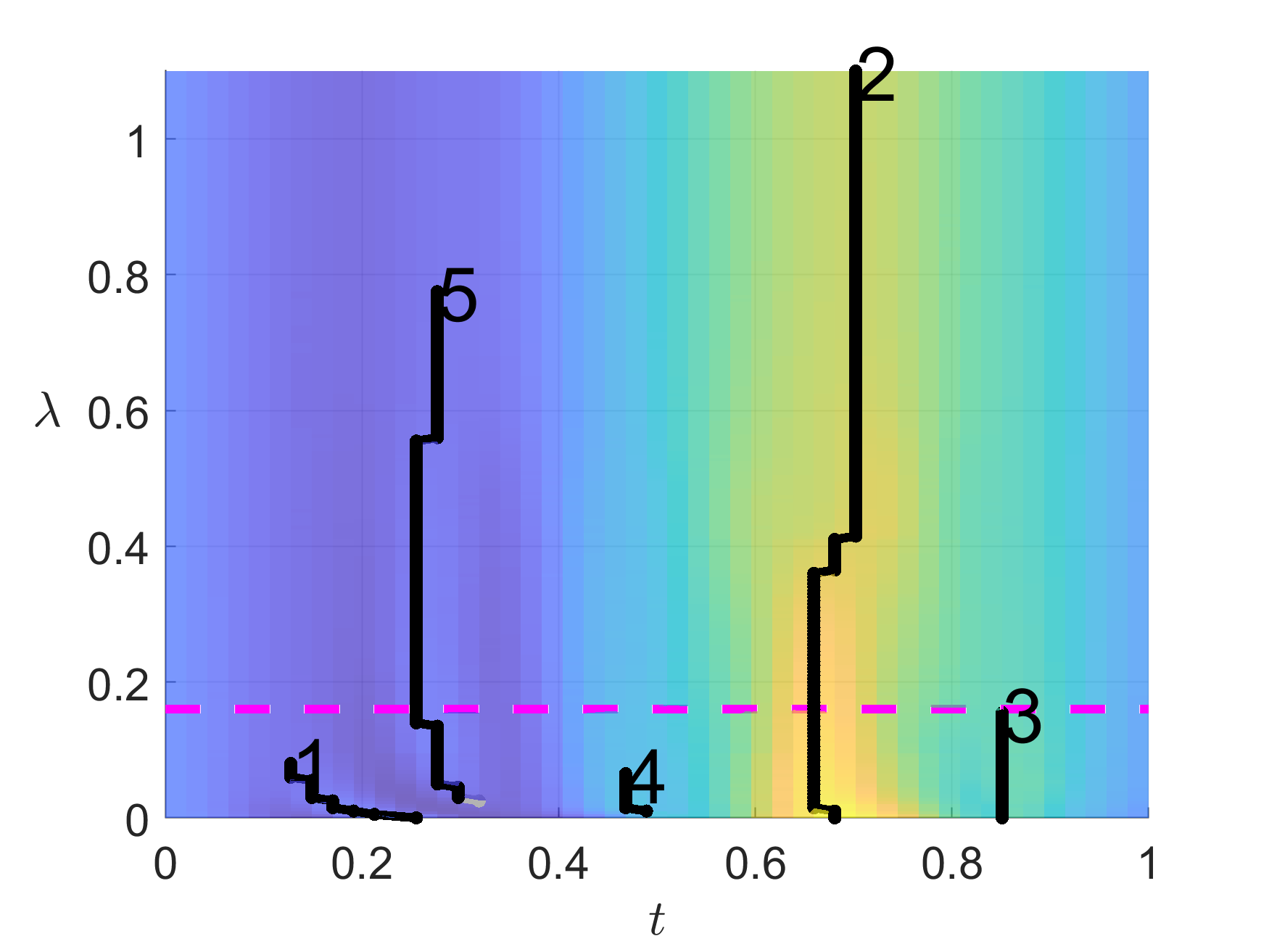}}
    \hspace{-0.2in}
    \subfloat[]{\includegraphics[height = 1.2in]{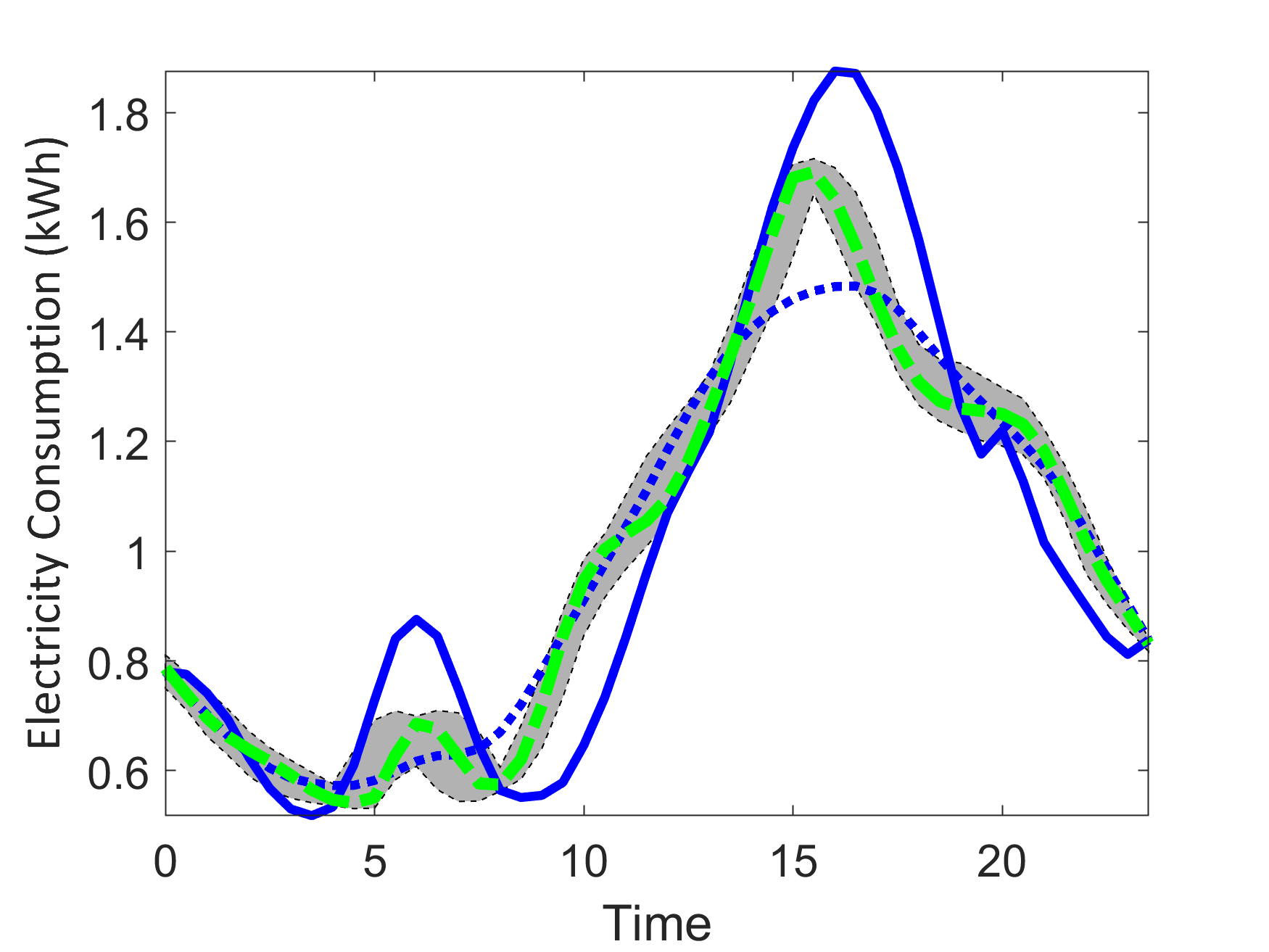}}
    \caption{(Electric Consumption in Tallahassee in July 2015) Plots (a) and (b) illustrate original data and fully-aligned functions, respectively. PPD in (d) provides $\lambda^* = 0.18$, and two significant peaks (2, 5). The final estimation, $\hat g$, (in green)in (g), is distinctive from the others, $\hat g_\infty$, and $\hat g_0$.}
    \label{fig: real5(2)}
\end{figure*}


The objective of next experiment is to analyze the half-hourly electricity consumption data in domestic households in a specific neighborhood in Tallahasee, FL~\citet{dasgupta_2019_ieee}. This study focuses on two subsets of the data, corresponding to the months of January and July, consisting of $968$ and $1,219$ functions, respectively. Fig. \ref{fig: real5(1)} (a) and (b) present the original and aligned functions of the January data. The PPD barchart reveals that five peaks (5, 7, 10, 11, and 12) are significant, with $\lambda^* = 0.245$. Panel (g) shows that there are vast differences among different estimators: $\hat g$ (in green), $\hat g_0$ (in solid blue), and $\hat g_\infty$ (in dotted blue).
Figure~\ref{fig: real5(2)} presents results for the July subset. The PPD barchart reveals two prominent peaks (2 and 5) with $\lambda^* = 0.18$, while $g_0$ and $g_\infty$ suggest three and one peaks, respectively.

\section{Discussion} \label{sec:Discussion}

The experimental results presented in this paper provide evidence that our approach is successful in: (1) estimating the number of peaks in functional data, and (2) estimating the underlying unknown function in a shape constrained manner. An important accomplishment here is the automated selection of the tuning parameter $\lambda$ using PPDs. 
The use of PPD is not only intuitive but also effective in determining the number and locations of peaks in $g$, allowing for the selection of a reasonable value for the smoothing parameter, $\lambda$. Moreover, shape-constrained functional estimation refines the estimate, $\hat g$, by eliminating insignificant peaks as determined by PPD. 

\begin{figure*}[htbp]
    \centering
    \subfloat[]{\includegraphics[height = 1.15in]{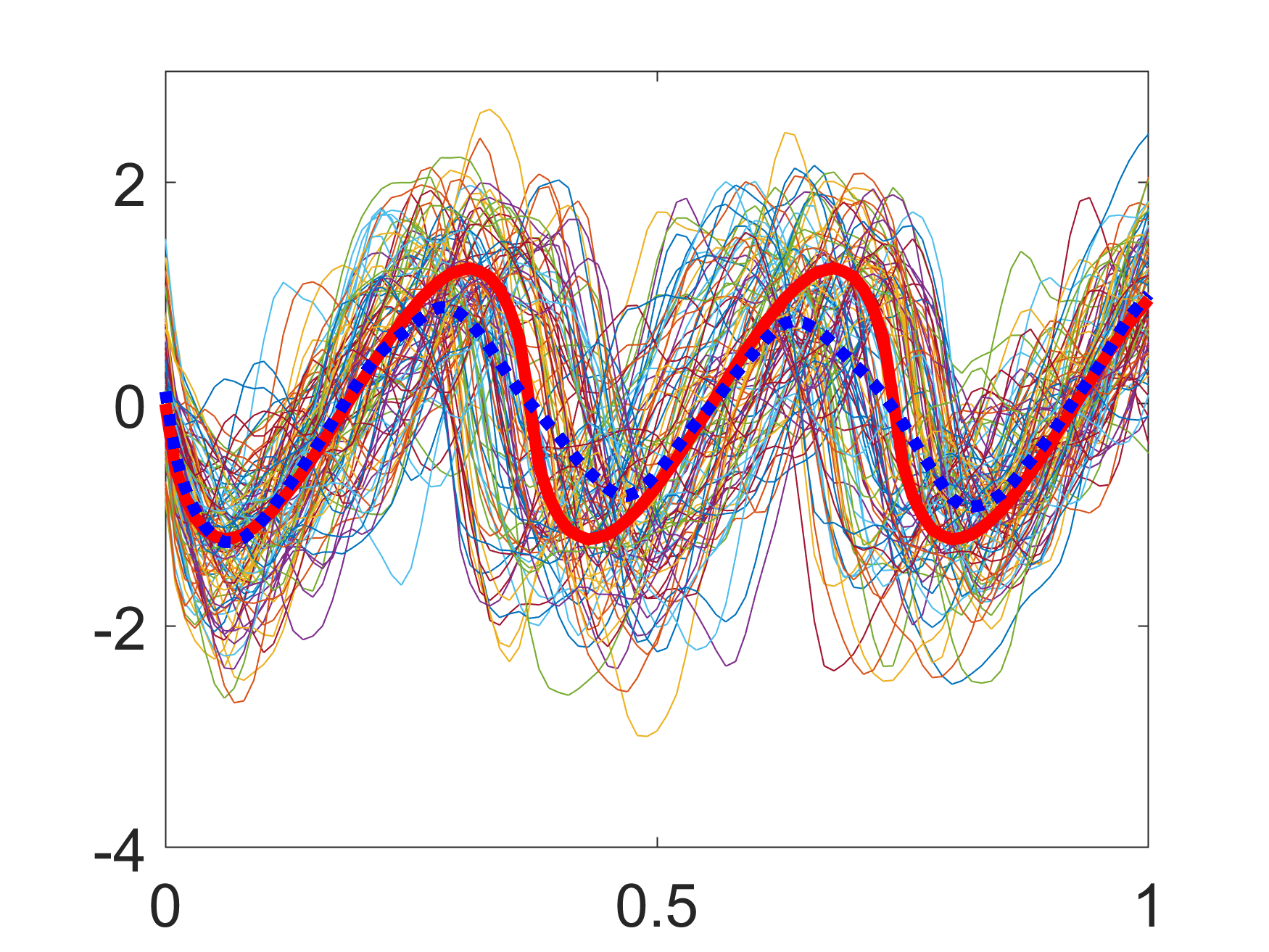}}
    \hspace{-0.2in}
    \subfloat[]{\includegraphics[height = 1.15in]{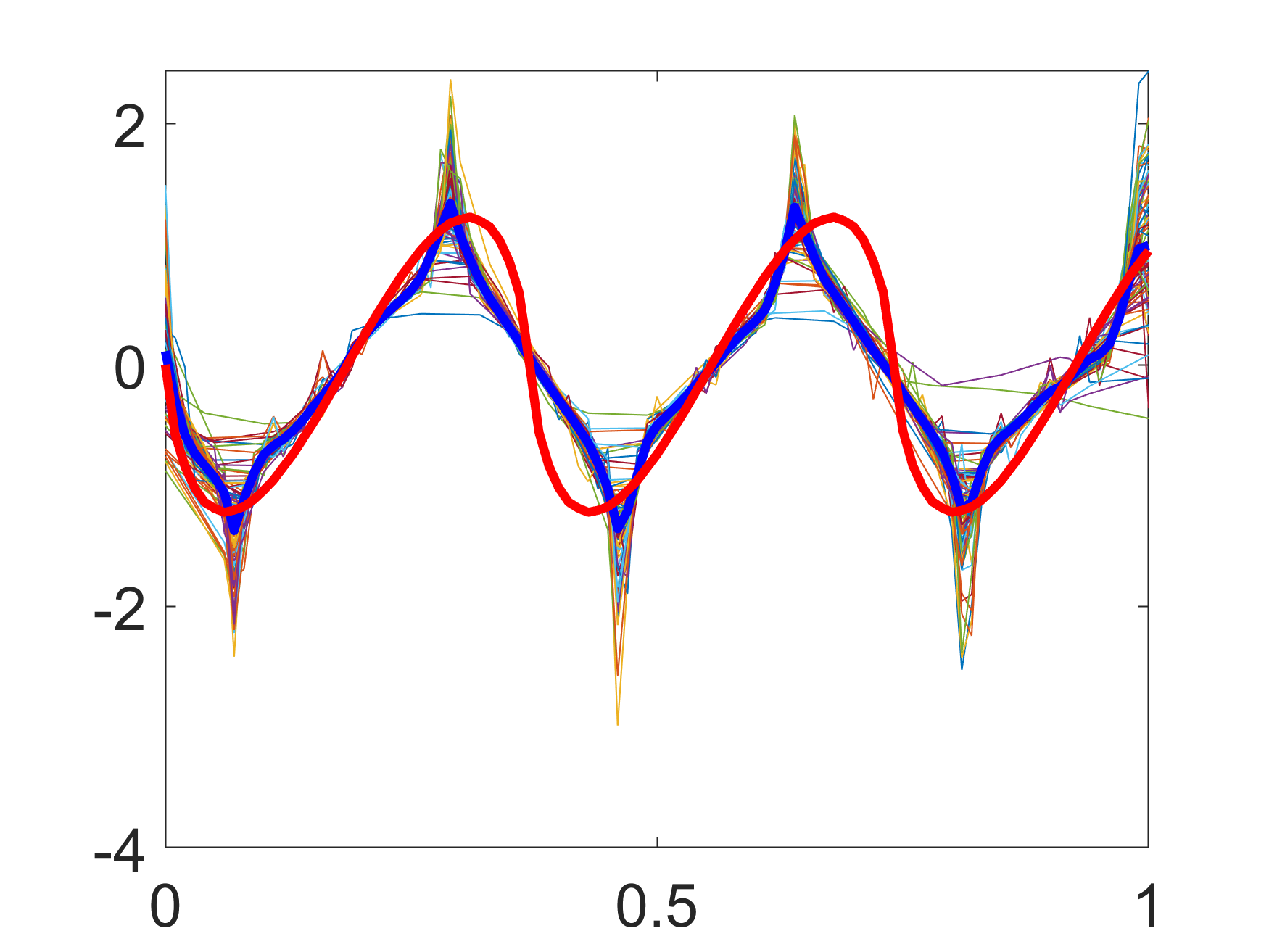}}
    \hspace{-0.2in}
    \subfloat[]{\includegraphics[height = 1.15in]{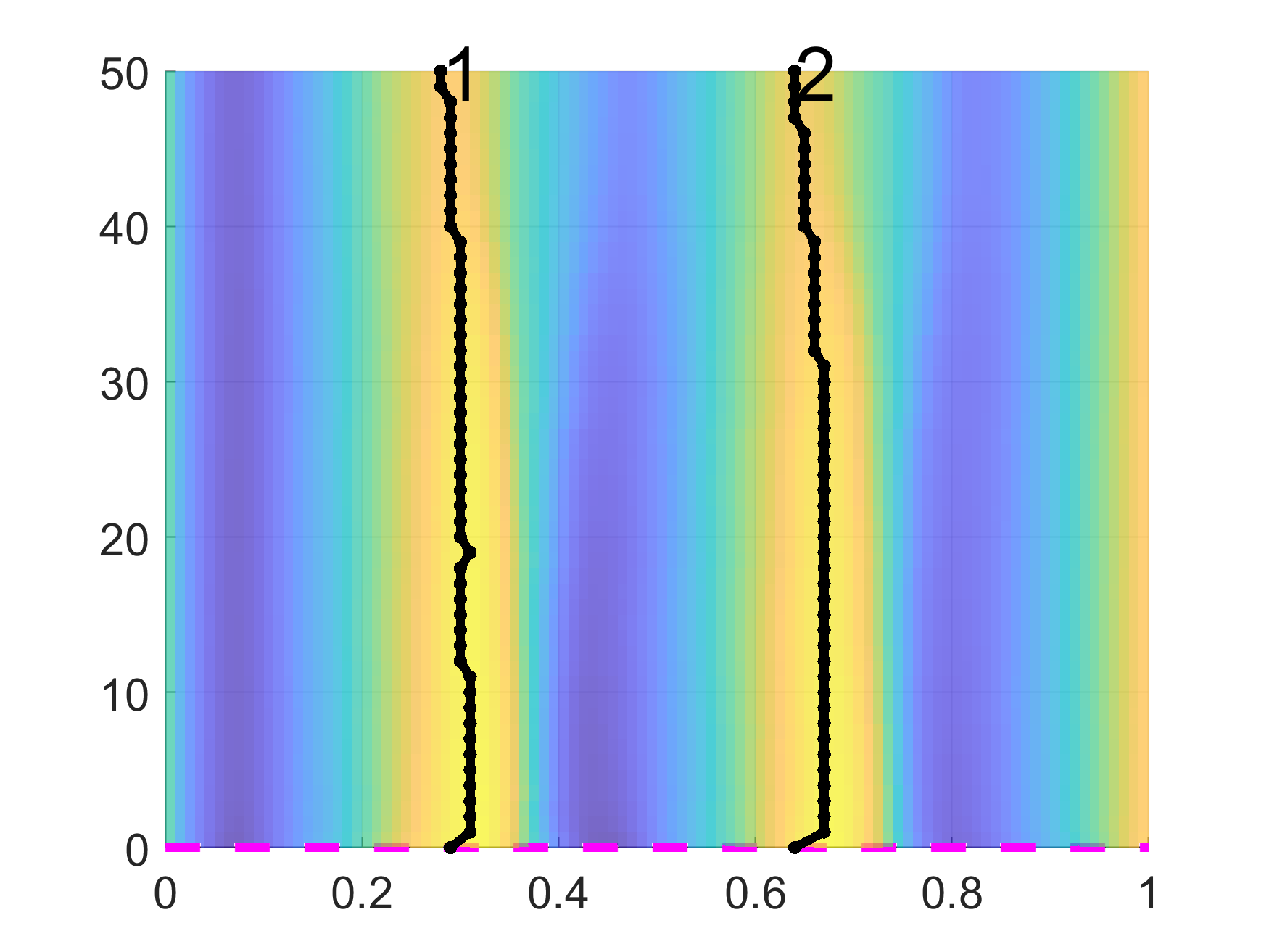}}
    \hspace{-0.2in}
    \subfloat[]{\includegraphics[height = 1.15in]{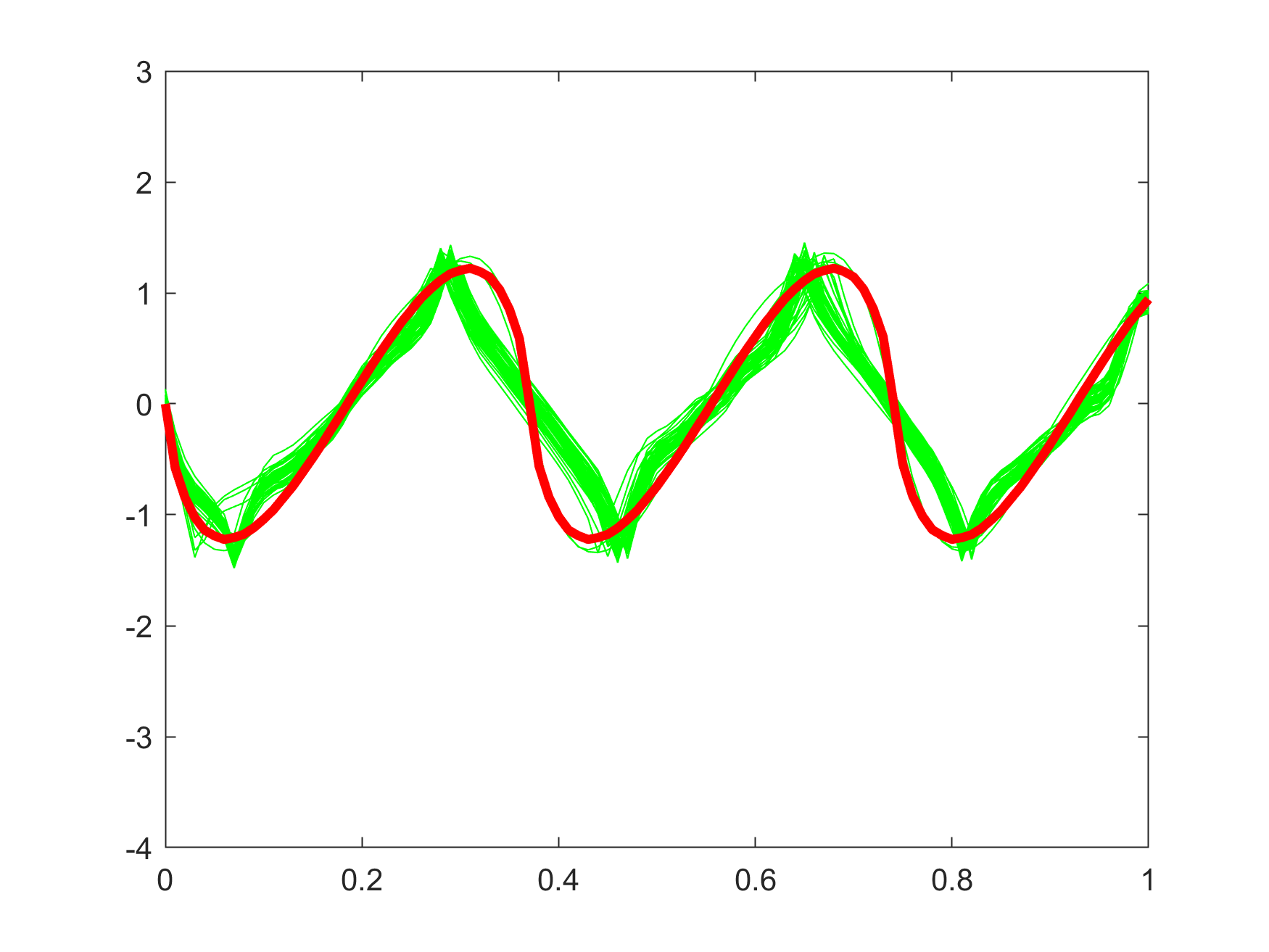}}
    \caption{(PPD on for $\hat g_{L2}$) Panel (a) displays the raw functions with the cross-sectional mean (in dotted blue) and $\hat g_{L2}$ (in red). Panel (b) shows the alignment of the functions using $\mathbb L^2$ metric with its mean (in blue) when $\kappa=0$. Panel (c) displays the posterior probability density (PPD) plot, which suggests that $\kappa = 0$ is the optimal value. Panel (d) presents the results of 50 replicated simulations with the estimates, $\hat g_{L2}$, plotted in green.}
    \label{fig: discussion}
\end{figure*}

An pertinent question here is: Why not define the concept of PPDs in the original function space, why use the SRVF representation instead? 
Fig. \ref{fig: discussion} illustrates as example to answer this question. It tries to form a PPD on one of the previously studied simulated data, and concludes that $\kappa = 0$ is optimal. This, of course, is incorrect as the pinching effect is clearly visible at $\kappa = 0$. Under the SRVF representation,  the pinching effect is completely avoided, even for $\lambda = 0$.

\section{Conclusion \& Future Work} \label{sec:Conclusion}
Understanding the population behavior of the sampled functional data requires estimating the actual underlying signal $g$. In many cases, the number, locations, or heights of extrema can be of direct interest themselves. In the presence of phase and additive noise, the classical unaligned mean $\hat{g}_{\infty}$ loses the geometric characteristics of $g$, while fully elastic mean $\hat{g}_0$ generates spurious peaks. This paper presents a geometric approach that estimates the shape features and the graph of function $g$. 
This approach explores the solution space by studying geometry of $\hat{g}_ {\lambda}$, for a range of smoothing parameter $\lambda \in [0,\infty)$. It introduces a novel tool called peak persistence diagram (PPD) for investigating this geometric space and for focusing on persistent peaks. This persistence of peaks helps us to discard insignificant peaks, estimate the shape of $g$ and reach an optimal $\lambda^*$. The latter two quantities lead to a shape-constrained estimation of $g$. This estimation refines $\hat g_{\lambda^*}$ and produces an optimal estimate, $\hat g$ under penalized MLE. Most importantly, the heights, the locations, and the number of extrema in $\hat g$ are interpretable and supported by data. In contrast, $\hat g_\infty$ underestimates the peak heights, and $\hat g_0$ overestimates heights as well as the number of extrema.

As a follow-up study, we are interested in separating the trend and the seasonality of the true underlying signal, i.e., $f_i(t) = g_1(t)+(g_2\circ\gamma_i)(t)+\epsilon_i(t)$ where $g_1$ and $g_2$ are trend and seasonality functions with time-warping, $\{\gamma_i\}$, and additive noise,$\{\epsilon_i\}$. Separating the data into two main signals may allow one to take a deeper look at it. For example, when investigating climate change datasets such as $\text{CO}_2$ emissions or global temperatures, one can expect a trend, $g_1$, with seasonal volatility, $g_2$. By separating the trend from the seasonality, one can test the trends statistically.
In another direction, we can extend our assumption that there exists a single true source. For instance, in real-world settings, it is difficult to ascertain if the sampled functions, $\{f_i\}$, were generated from a single signal $g$.

\bibliographystyle{rss}
\bibliography{ref}

\end{document}